\documentclass[twocolumn,aps,prd,unsortedaddress,%
superscriptaddress,a4paper,nofootinbib,balancelastpage,preprintnumbers,showpacs,longbibliography,floatfix]{revtex4-1}
\usepackage[utf8]{inputenc}
\usepackage{hyperref}
\usepackage{latexsym,amsbsy,amsmath}
\usepackage[pdftex]{graphicx}
\usepackage{xcolor}
\usepackage{longtable}
\usepackage{isotope}
\usepackage{microtype}
\usepackage[bitstream-charter]{mathdesign}
 \DeclareGraphicsExtensions{.eps, .jpg,.png}
\renewcommand{\vec}[1]{\boldsymbol{#1}}
\def\ss#1#2{\boldsymbol{\sigma}_{#1}.\boldsymbol{\sigma}_{#2}}
\def\ll#1#2{\tilde{\lambda}_{#1}.\tilde{\lambda}_{#2}}
\def\llss#1#2{\tilde{\lambda}_{#1}.\tilde{\lambda}_{#2}\,\boldsymbol{\sigma}_{#1}\boldsymbol{\sigma}_{#2}}
\def\parstar#1{#1{}^{{\scriptscriptstyle (}*{\scriptscriptstyle )}}}
\def\dst{\parstar{D}}
\def\dbst{\parstar{\bar D}}
\def\dbsst{\parstar{{\bar D}_s}}
\makeatletter
\newcommand*{\rom}[1]{\expandafter\@slowromancap\romannumeral #1@}
\makeatother
\usepackage{palatino}
\begin{document}
\title{Exotic hadrons: review and perspectives}
%
%
\author{J.-M.~Richard}
\email{j-m.richard@ipnl.in2p3.fr}
\affiliation{Universit\'e de Lyon, Institut de Physique Nucl\'eaire de Lyon,
IN2P3-CNRS--UCBL,\\
4 rue Enrico Fermi, 69622  Villeurbanne, France}
\date{\emph{Version of }\today}
\begin{abstract}
The physics of exotic hadrons is revisited and reviewed, with emphasis on flavour configurations which have not yet been investigated. The constituent quark model of multiquark states is discussed in some detail, as it can serve as a guide for more elaborate approaches.
\end{abstract}
\maketitle
\section{Introduction}
\label{se:intro}
The physics of hadrons began about one century ago, when it became appealing to build nuclei from protons and  neutrons. The experimental search for the neutron was soon resumed along with the search for the particle predicted by Yukawa as being responsible for the tight binding of protons and neutrons. See, e.g., \cite{Segre:100978,*Pais:107749,*Ezhela:317696} for the history of this early period.

When interacting with nucleons or among themselves, the pions produce  many resonances. The exchange of meson resonances, and the formation of nucleon resonances in the intermediate states are crucial for our understanding of the nuclear forces beyond the single-pion exchange proposed by Yukawa. For instance,  the existence of the vector mesons, nowadays known as $\rho$ and $\omega$, was somewhat anticipated from the need for isospin-dependent spin-orbit components in the nucleon-nucleon interaction. See, e.g.,~\cite{Breit:1962zz} for an introduction into the early literature on this subject. 
The discovery of numerous meson or baryon resonances  stimulated intense efforts to understand the underlying mechanisms that generate the hadrons. For instance, the $\rho$ meson can be described as a $\pi\pi$ resonance in a $p$-wave, with the $s$-wave counterpart being the scalar meson responsible for the spin-isospin averaged attraction observed in nuclei. In 1955, Chew and Low described the $\Delta(1232)$ as the result of a strong $\pi N$ interaction \cite{Chew:1955zz}, and today, the scattering of pions off nuclei is still pictured as series of $\Delta$-hole formation and annihilation.\footnote{The author thanks G.F.~Chew for an enlightening  discussion on the subject many years ago.}\@  

An ever more ambitious picture was proposed, called ``nuclear democracy'' or ``bootstrap'', in which every hadron consists of all possible sets of hadrons compatible with its quantum numbers~\cite{Chew:107781}. Unfortunately, neither the models based on nucleons and pions as elementary bricks, nor the bootstrap, succeeded in providing an overall picture. For instance, the degeneracy of isospin $I=0$ and $I=1$  mesons such as $\rho$ and $\omega$, does not emerge naturally, if one assumes that their content is dominated by the nucleon-antinucleon component~\cite{Ball:1965sa}. 

From the very beginning, the attention was focused on hadrons with properties that look unusual, in contrast to the regularity patterns of the rest of the spectrum. The exotic character of a hadron is of course a time-dependent concept: once a striking property of a hadron is confirmed by further measurements, and understood within a plausible theoretical framework, the hadron  ceases to be exotic. 

Among the most famous examples, there is the $\phi$ meson, a peak of mass 1.02\,GeV  with a remarkably small width of about 4\,MeV \cite{Agashe:2014kda}, decaying preferentially into a pair of kaons, though the decay into pions is energetically much easier. The decay patterns of the $\phi$ are now understood from its quark content $s\bar s$, where $s$ denotes the strange quark, with very little mixing. The suppression of the internal annihilation of $s\bar s$ into light mesons has been codified in terms of the so-called ``Zweig rule'', eventually justified from the asymptotic-freedom limit of Quantum Chromo-Dynamics (QCD). 

The process took, however, several years, and was not mature enough when the $J/\psi$ particle was discovered at the end of 1974; it is a meson of mass about 3.1\,GeV and width much less than 1\,MeV~\cite{Agashe:2014kda}. 
Why the $J/\psi$ was not immediately identified with $c\bar c$? The great surprise was that the Zweig rule works too well! As compared to the $\phi(1020)$, the $J/\psi$ is three times heavier, so its phase-space for decay into light mesons is considerably larger, and yet its width is smaller! For the $\Upsilon$, the effect is even more pronounced, but when the $\Upsilon$ was discovered in 1977~\cite{Agashe:2014kda}, the lesson was learnt, and the $\Upsilon$ was never considered as exotic.

Today, any attempt to provide with a  definition of an ``exotic'' hadron starts from the quark model. A meson with quantum numbers that cannot be matched by any spin-orbital state of a quark and an antiquark is obviously exotic. States with unusual properties but ordinary quantum numbers are named
``crypto-exotic'' states.  In particular, ordinary baryons can reach any set of values of the spin-parity $J^P$ with half-integer spin $J$ and parity $P=\pm 1$, so a baryon cannot be exotic except for its flavour content. 
A consensus has thus been reached that a hadron is exotic if it cannot be understood as mainly made of quark-antiquark nor quark-quark-quark, but ``mainly'' remains vague. While  simple quark models will propose a reference with a frozen number of constituents, more advance pictures include an expansion in the Fock space, with additional quark-antiquark pairs in the wave function. The hope is that the expansion converges, and that a clean separation emerges between exotic and non-exotics.

This review is organised as follows. In Sec.\ \ref{se:mesons}, we shall review states with unusual properties in the light (Sec. \ref{sse:mesons-light}), open heavy-flavour (Sec.\ \ref{sse:mesons-open1}), double heavy flavour (Sec.\  \ref{sse:mesons-open2}), and hidden heavy-flavour (Sec.\ \ref{sse:mesons-hidden}) sectors.  The baryon spectrum is discussed in  Sec.\ \ref{se:baryons}, with, again a separation between the non-exotic light sector (Sec.\ \ref{sse:baryons-light-ns} for non strange and Sec.~\ref{sse:baryons-light-negs} for strange baryons), the exotic light pentaquark with positive strangeness in Sec.~\ref{sse:baryons-light-poss}, the case of open heavy-flavour, single  (Sec. \ref{sse:baryons-open1}) or double single  (Sec. \ref{sse:baryons-open2}), the anticharmed pentaquark in Sec.~\ref{sse:antich-baryons}, and the case of  hidden heavy-flavour  (Sec.\ \ref{sse:baryons-hidden}) with in, particular, the recent pentaquark states found by the LHCb collaboration. The  multibaryon systems will be 
discussed in  Sec.\ \ref{se:multibaryons}. The various theoretical models will be reviewed in Sec.\ \ref{se:theory}, with some emphasis on string dynamics and potential models. 

Some limits have to be set. The case of light hadrons will be just outlined, to focus more on the most recent states containing heavy quarks or antiquarks. In the former sector, the literature is enormous, and can be traced back from the conference proceedings that will be cited in the next sections, from some well-known reviews such as \cite{Agashe:2014kda,Montanet:1980te,Amsler:2004ps,Klempt:2007cp}, or from study groups for future facilities \cite{Dalpiaz:1987qk,Amsler:1997nf,vonHarrach:1998xz,Destefanis:2013xpa}.  For the recent exotics with hidden charm, there exists already excellent and even comprehensive reviews, such as \cite{Swanson:2006st,Nielsen:2009uh,Hambrock:2013tpa,Chen:2016qju}. Our aim here is not to compete with these reviews, nor to duplicate them, but rather to discuss critically the recent findings and the models that they inspired. So, this paper will be more subjective. Hopefully, it will trigger some discussion, and, perhaps, attract newcomers into the field, where innovative ideas 
are welcome.

The bibliography for such a subject cannot be comprehensive. Some past review articles are cited, which give access to the early papers. We apologise in advance in the event where some important contributions are inadvertently omitted. 
\section{Mesons}\label{se:mesons}
\subsection{Light mesons}\label{sse:mesons-light}
There are many hot issues on light mesons: why are the pseudoscalars so light? what is the status of the many scalar mesons? why ideal mixing and Zweig rule work so well in some sectors, and not in other sectors? We refer, e.g., to \cite{Klempt:2007cp} for a critical survey of the claims for exotics, and to~\cite{narison2004qcd} for a thorough discussion within QCD and the bibliography, and to the last hadron conferences for an update \cite{Proceedings:2016efk,Proceedings:2015era,baryon2016}. 

One of the puzzles arises from the  $\gamma\gamma \to \rho\rho$\footnote{I thank Bernard Pire for calling my attention on this channel}, measured in several experiments, in particular at L3 \cite{Achard:2005pb,*Achard:2003qa}: the comparison of  the different charge modes suggest a sizeable isospin $I=2$ contribution. There are interesting theoretical studies on the production mechanism \cite{Anikin:2005ur}, but missing is a simple understanding of why there is  strong attraction in that channel in terms of quark dynamics.  

There is an intense activity in the search of glueballs and light hybrids, either as supernumerary states with ordinary quantum numbers $J^{PC}$, or as genuine exotics with $J^{PC}$ that cannot be matched by ordinary quantum numbers. In particular, evidence has been found for $1^{-+}$ states  between 1 and 2\,GeV$^2$ by VES at Serpukhov\cite{Dorofeev:2001xu}, by E852 at Brookhaven \cite{Thompson:1997bs}, and by COMPASS at CERN \cite{Alekseev:2009aa}, and the search is resumed actively, in particular by GLUEX at Jlab \cite{Ghoul:2015ifw}. For a thorough review of the other contributions, see, e.g., \cite{Meyer:2015eta}.

\subsection{Mesons with single heavy flavour}\label{sse:mesons-open1}
The charmed mesons $D$, $D^*$ have been discovered shortly after the charmonium, and the mesons with charm and strangeness $D_s$, $D_s^*$ have followed. Relevant for this review on exotics are the so-called $D_{s,J}$ states. As summarised in \cite{AmslerHanhart}, the $D^*_{s,0}(2317)$ was discovered at BaBar \cite{Aubert:2003fg} and the $D^*_{s,1}(2460)$ at CLEO \cite{Besson:2003cp}, with a mass remarkably lower than predicted in the ``bible'' \cite{Godfrey:1985xj} and in similar quark-model calculations. Note that the tuning of the spin-dependent forces in  such models has been often debated, and no consensus has ever been reached,  see, e.g., \cite{Pignon:1978ba,*Schnitzer:1981zg,*Cahn:2003cw,*Matsuki:2007zza}. To explain the low mass and the properties of the $D^*_{sJ}$ states, the coupling to the nearby $D^{(*)} K^{(*)}$ thresholds certainly plays a role. These mesons have even been considered as molecules, for instance, in~\cite{Nielsen:2006zz,*Vijande:2007zza,*Segovia:2015dia,*Godfrey:2015dva}.
\subsection{Mesons with double heavy flavour}\label{sse:mesons-open2}
After, the discovery of charm, it has been early recognized that configurations including both light and heavy quarks or antiquarks might be particularly interesting, for instance, in the case of baryon-antibaryon  or multibaryon  systems~\cite{Dover:1977jw,*Dover:1977hs}. Another pioneering contribution was due to Isgur and Lipkin \cite{Isgur:1980en} who considered the flavour exotic $(cs\bar q\bar q)$ configuration. 

Shortly after, even more exotic mesons were studied, $(QQ'\bar q\bar q' )$,  with two heavy quarks and two light antiquarks. 
As explained later in this review, their existence  is a prediction shared by many theorists using different approaches, which will be discussed in the next sections. Doubly-heavy exotics, which are surprisingly omitted in some ``reviews'' on multiquarks, have not yet been searched for in any experiment, but will be one of  the goals of future experiments, and several theorists estimated the rate of production at hadron \cite{Cho:2010db,*Yuqi:2011gm} or electron \cite{Hyodo:2012pm} colliders, or in the decay of hadrons with beauty~\cite{Esposito:2013fma}.\footnote{I thank 
Angelo Esposito and Alessandro Pilloni for a reminder about this reference.} The production of two charm-anticharm pairs is somewhat disconcerting, with the good surprise of a rather abundant production of double charmonium,  leading for instance to the identification of the $\eta_c(2s)$ recoiling against the $J/\psi$ \cite{Abe:2004ww}, and, on the other hand, the double-charm baryons have not been confirmed in recent experiments, see Sec.~\ref{sse:baryons-open2}.
\subsection{Mesons with hidden charm or beauty}\label{sse:mesons-hidden}
The spectrum is quarkonia, $(c\bar c)$ and $(b\bar b)$, is now rather well known and well understood. For a while,  the most delicate experimental sector was the one of spin-singlet states, such as $\eta_{c,b}(ns)$, $h_{c,b}(np)$. For instance, a first candidate for the $\eta_c(1s)$ was announced about 300\,MeV below the $J/\psi$, triggering some premature announcements on the quark dynamics, but eventually replaced by a more reasonable $\eta_c(1s)$ about 112\,MeV below $J/\psi$, see, e.g., the reviews~\cite{Martin:2003rr,*Barnes:2009zza}. A first indication for the $h_c$ was given a by $\bar p p$ formation experiment at CERN \cite{Baglin:1986yd}, with another indication in a similar experiment at Fermilab \cite{Armstrong:1992ae}.  It was eventually firmly identified at CLEO \cite{Rubin:2005px}. The $\eta_c(2s)$ was seen at Belle in $B$ decay \cite{Choi:2002na} and in double-charmonium production \cite{Abe:2004ww}. It also took some time until the spin-singlet states of $(b\bar b)$ were identified~\cite{
Agashe:2014kda}.

Above the threshold, $D\bar D$ for charmonium and $B\bar B$ for bottomonium, the spectrum was known to be more  intricate, with broader peaks, and an interplay of $s$- and $d$-states mixed with the $(Q\bar q)(\bar Q q)$ continuum, and it was a somewhat a daily grind to improve the measurements. 

Then came the shock of the discovery of the $X(3872)$ by Belle \cite{Choi:2003ue}, in the $\pi^+\pi^-J/\psi$ mass spectrum of the reaction $B^\pm\to K^\pm \pi^+\pi^-J/\psi$. Neither the mass nor the decay properties were according to the expectations of the charmonium models, in particular for the $2{}^3\mathrm{P}_2$ level. On the other hand,  such a state was announced in the molecular models, as discussed in Sec.~\ref{sse:molec-XYZ}.  Not only the experts on quark- or hadron-dynamics commented this new state, but also some non-experts became attracted to the subject.

As well summarised in a recent review \cite{Lebed:2016cna}, a second shock was the discovery of the $Y(4260)$ by BaBar~\cite{Aubert:2005rm}, in $e^+e^-$ with ISR.\footnote{Initial State Radiation, this means that the reaction is $e^+e^-\to\gamma+e^+e^-\to \gamma + Y(4260)$.}\@ 
This $1^{--}$ sector of charmonium was already thoroughly explored, with radial excitations, with $s$- and $d$-waves mixed between them and with the continuum. So this new state was immediately associated with an exotic, or, say crypto-exotic structure. The $Y(4260)$ was seen also at CLEO and Belle~\cite{Agashe:2014kda}.

A third shock was the discovery of charged states, with minimal content $c\bar c u\bar d$ or c.c. The first announcement was the $Z^+(4430)$ by the Belle collaboration, in the $\pi^\pm\psi(2s)$ mass distribution of the decay $B\to K\pi^\pm\psi(2s)$~\cite{Choi:2007wga}.

It is hardly possible to list all the states of $X,Y$ or $Z$ type without duplicating the very comprehensive reviews such as \cite{Hambrock:2013tpa,Chen:2016qju}. Let us stress a few points:
\begin{itemize}
 \item The $X(3940)$ and $X(4160)$ have been seen in double-charmonium production \cite{Abe:2007jna,*Abe:2007sya}, which gives a clear signature of their $c\bar c$ content,
 \item Some states have been seen in $\gamma\gamma$, indicating a charge conjugation  $C=+1$, and very likely $J^{PC}=0^{++}$ or $2^{++}$. This is the case for $Z(3930)$, $X(3915)$ and $X(4350)$ \cite{Uehara:2005qd,*Uehara:2009tx,*Shen:2009vs}.
 \item Some states, in particular among the above ones,  are just candidates for radial excitations of charmonium. If confirmed, they indicate the supernumerary character of other states sharing the same quantum numbers. 
\item The $Y(4630)$ has been seen as an enhancement in the $\Lambda_c\Lambda_c$ spectrum of the ISR reaction $e^+e^-\to\gamma \Lambda_c\bar\Lambda{}_c$ \cite{Pakhlova:2008vn}. Many references about this state can be found in a recent paper~\cite{Sonnenschein:2016ibx}. As stressed in this literature, the decay into a baryon and an antibaryon might indicate a ``baryonium'' type of structure, though other scenarios have been proposed~\cite{Guo:2010tk}.
 \item There is sometimes a clustering of states with similar masses. It might the same state with a peak that depends on the production and decay process. 
 \item Most quark models and, to a lesser extent, molecular models predict an isospin $I=1$ partner of the $I=0$ state, analogous to $\rho$ vs.\ $\omega$.  This stimulated searches for charged partner, such as \cite{Aubert:2004zr}. 
 \item While the $X(3872)$ has been seen in many experiments, $Y(4260)$ in a few experiments, some states have been seen in just a single experiment. In between, $Z_c(3900)^\pm$ has been seen by Belle and \mbox{BESIII}~\cite{Ablikim:2013mio,*Liu:2013dau}, but these experiments share many collaborators, and used the same reaction $e^+e^-\to Y(4260)\to \pi^\mp Z_c^\pm$, with a slight variant for the confirmation by CLEO \cite{Xiao:2013iha}. So, the negative result of a search  by COMPASS in  photoproduction is particularly important for the existence or at least the coupling properties of this state~\cite{Nerling:2016uwm}.
 \item
 The lightest $XYZ$ states have minimal quark content $c\bar c q \bar q$, where $q=u,\,d$. In several models, their $c\bar c s \bar s$ with hidden strangeness are expected. Recently, the LHCB collaboration re-analysed the  $J/\psi\,\phi$ mass spectrum from the decay $B\to J/\psi\,\phi K$~\cite{Aaij:2016iza}, and found evidence for $X(4140)$ and $X(4264)$ both with $J^{PC}=1^{++}$, in disagreement with some models (cited in \cite{Aaij:2016iza}) and  in partial or full agreement with others, in particular the extension by Stancu \cite{Stancu:2009ka} of the chromomagnetic model proposed for the $X(3872)$~\cite{Hogaasen:2005jv}.
 \item The models  somewhat differ about the existence and location of the $b\bar b$ analogues of the exotic mesons with hidden charm: the heavier $b$-quark mass, for instance, suppresses the chromomagnetic effect but enhances the chromoelectric ones, while the coupling schemes are sensitive to the gap between bare quarkonia and meson pairs. The list of candidates is rather scarce for the time being, and are due to Belle only. The charged $Z_b^\pm(10610)$ and $Z_b^\pm(10650)$ have been seen in $\Upsilon(5s)\to  (b\bar b) Z_b^\pm \pi^\mp$, where $(b\bar b)$ stands here for some lower $\Upsilon(ns)$ or $h_b(1p)$ or $h_b(2p)$. Later, Belle also got some candidate for the neutral $Z_b(10610)^0$. On the other hand, a search for the heavy partner of $X(3872)$, decaying into $\omega \Upsilon(1s)$ or $\pi^+\pi^-\Upsilon(1s)$, has been so far unsuccessful. 
\end{itemize}

\section{Baryons}\label{se:baryons}
The history of the quark model is intimately linked to baryon spectroscopy. One of the motivation for SU(3) flavour symmetry, nowadays denoted  SU(3)$_\text{F}$, was the observation that the hyperons have properties similar to that of non-strange baryons.
The quark model was developed first in the baryon sector, with the pioneering works by Greenberg, Dalitz, etc., and pushed very far by the subsequent works by Dalitz et al., Hey et al., Isgur and Karl, and many others. For a review, see, e.g., \cite{Crede:2013sze}.
Obviously, the field has been much enriched by the study of baryons with charm or beauty, and baryons carrying two or three heavy flavoured quarks are eagerly awaited.
\subsection{Non strange light baryons}\label{sse:baryons-light-ns}
There are many issues there, that have been presented in conferences  such as ``Baryons'', ``Nstar'', etc.  For years, one of the most striking issues was the signature for the ``symmetric quark model'' with two degrees of freedom, vs.\ the quark--diquark model. Clearly, the recent progress, as e.g., summarised at the recent conference in Tallahassee \cite{baryon2016}, confirms the existence of states such as $N(1900)$ which cannot be accommodated in a naive quark-diquark picture with a frozen diquark.  

%

The lightest state with an excitation in each of the Jacobi variables, say, $\vec x=\vec r_2-\vec r_1$ and $\vec y=(2 \vec r_3 - \vec r_1-\vec r_2)/\sqrt3$,   has isospin $I=1/2$, spin $s=1/2$ and orbital momentum and parity $\ell=1^+$.  In the particular case of the harmonic oscillator, its wave function reads
\begin{equation}\label{eq:wf1plus}
 \Psi(\vec x,\vec y)=\vec x\times \vec y\,\exp(-a(\vec x^2+\vec y^2)/2)~,
\end{equation}
and is antisymmetric under the permutations. Thus, if associated with an antisymmetric colour wave function $(3\times3\times3\to1)$ and an antisymmetric spin-isospin wave function, it fits perfectly the requirement of Fermi statistics. Note that an antisymmetric spin-isospin wave function is possible for $I=1/2$ ($uud$), but not for $I=3/2$ ($uuu$) nor $I=0$ ($ccc$). 

Experimentally, the search for doubly-excited baryons is difficult. As often underlined (see, e.g., \cite{Isgur:1992td}) photon or meson scattering off a nuclear target favours the production of states with a single quark being excited. And in amplitude analyses, doubly excited baryons, if any, come in a region that is already well populated!

Another hot issue is the Roper resonance, an excitation with the same quantum numbers as the ground state. In potential models, this is mainly a hyper-radial excitation, but in explicit model calculations, it is found higher than the negative-parity excitation, while the experimental Roper state lies degenerate or even slightly below the lowest negative-parity excitation. Chiral dynamics, as implemented in specific models or in lattice simulations, accounts for this property, which sets limits to naive potential models.
\subsection{Light baryons with negative strangeness}\label{sse:baryons-light-negs}
Here, again, is the problem of missing states, with slow but continuous progress to fill the holes. A very striking property  is the low mass of  the $\Lambda(1405)$, discussed first by Dalitz \cite{Dalitz:1960du} and now in an abundant literature (see, e.g., \cite{Oset} for a recent summary). In the case of the nucleon spectrum, the spin-orbit splittings, such as between $N(1520)$ with $(1/2)^-$ and $N(1535)$ with $(3/2)^-$, are very small, and Isgur and Karl, for instance, have set the spin-orbit potential to zero in their fit \cite{Isgur:1978xj}. Hence the large $\Lambda(1520)-\Lambda(1405)$ splitting of the strange analogues appears as somewhat a surprise in a pure quark model. 

A minimal solution explains the discrepancy  by a shift due the coupling to the nearby $KN$ threshold. Similarly, the $\psi'$ is shifted down by the nearby $D\bar D$ threshold, and the hyperfine splitting $\psi'-\eta_c(2s)$ is slightly smaller than expected. Similarly, the $2\,{}^3{\rm P}_2$ state of charmonium gets some admixture of $D\bar{D}{}+\mathrm{c.c.}$ in conservative models of the $X(3872)$, etc.

Now, a more appealing solution is also suggested, in which there are two poles with $J^P=(1/2)^-$ near $\Lambda(1405)$~\cite{Kamiya:2016jqc}, a mixing of a $\bar K N$ molecule and a quark-model state. As often underlined in the case of mesons \cite{Close:2001zp}, the mixing scheme might be more involved than in ordinary quantum mechanics, with the absorptive part, i.e., the coupling the decay channels, playing the major role. 
\subsection{Light baryons with positive strangeness}\label{sse:baryons-light-poss}
For years, there has been debates about baryons carrying positive strangeness. 
In the 60s and 70s, much effort was devoted to exploit the analyticity properties of the $S$-matrix, and to reconstruct the amplitudes from the scattering data. This program was rather successful in the case of the $\pi N$ and $\bar K N$ reactions, and led to the identification of a number of baryon resonances with strangeness $S=0$ and $S=-1$, respectively. The case of $K N$ scattering with $S=+1$ was much more delicate and controversial. Some early analyses, based only on differential cross-sections (and just a few data on the recoil polarisation), suggested the possibility of resonances, see, for instance, \cite{PhysRevLett.18.801,*PhysRev.165.1770,*Wilson:1972uh}. The resonances were named $Z_I$ or  $Z_I^*$, where $I=0$ or $1$ is the isospin.

Then came data on  the analysing power, obtained using a polarised target, as in \cite{Albrow:1971qm}. Not surprisingly, the uncertainties were narrowed \cite{Martin:1975gs}, and the $Z$ resonances tended to fade away. For years, the Particle Data Group (PDG), summarised the state of art, as in the 1980 edition \cite{Kelly:1980kr}, where one finds candidates for several exotic baryons with $S=+1$. There are $Z_0(1780)$ and $Z_0(1865)$ in the isospin $I=0$ sector, and $Z_1(1900)$, $Z_1(2150)$ and $Z_1(2500)$ for $I=1$.  Then the $Z$ baryon disappeared from the review, and the letter $Z$ was given to another particle. Note that other risky amplitudes analyses were done later, for instance for $\bar p p\to \pi^+\pi^-$ or $K^+ K^-$ before the advent of data taken on the polarised target, or for the elastic $\bar p p$ scattering with just data on the polarisation, while the simpler $pp$ case notoriously requires several spin observables.

Act two of the play is better known. Meanwhile the name ``pentaquark'' was introduced in the anticharmed sector, as will be discussed shortly, in Sec.~\ref{sse:antich-baryons}.
It reflects that the minimal content of the $Z_I$ baryon is $(\bar s qqqq)$. 
In 2003, Nakano et al.\ found a narrow peak in the $K^+n$ distribution of the reaction $\gamma n\to K^- K^+ n$~\cite{Nakano:2003qx}. This state, named $\theta^+(1540)$, was predicted in a model by Diakonov et al.~\cite{Diakonov:1997mm} based on some developments of the chiral dynamics. 
It was of course rather remarkable to have a relatively small experiment revealing a major discovery. But the sequels quickly  went out of control. On the theory side, a flurry of papers, with hasty calculations and ad-hoc assumptions. On the experimental side, some collaborations discovered that they had data on tape, that were never analysed, and contained in fact the pentaquark! Other positive results came, both for the $\theta^+$, and for other members of the anti-decuplet of SU(3)$_\text{F}$ that was predicted. However,  the trend became inverted, and experiments with better statistics and particle identification did not confirm the existence of the these light pentaquarks. 
This is reviewed for instance in \cite{Wohl:2008st,*Hicks:2012zz}. The theory part also came under better scrutiny. In Ref.~\cite{Tariq:2007ck}, some lattice QCD calculations following the discovery of the $\theta^+$ are criticised. In, e.g., \cite{Jaffe:2004qj,*Weigel:2007yx}, some inconsistencies are pointed out in the estimate of the width in the model of Diakonov et al., a crucial property of the putative light pentaquark.\footnote{I thank Herbert Weigel for an informative  correspondence on this point.}
\subsection{Baryons with single heavy flavour}\label{sse:baryons-open1}
In the last 20 years, much experimental progress has been done in the sector of baryons with single charm or beauty. When available, the comparison of $b$- and $c$-baryons is compatible with the idea of a flavour-independent confinement. For instance, the first evidence for $\Omega_b(bss)$ came from D0 at Fermilab, with a mass about $6165\,$MeV, corresponding to a double-strangeness excitation $\Omega_b-\Lambda_b$ substantially larger than its charm analogue. Further measurements by CDF at Fermilab and LHCb at CERN gave a mass around $6050\,$ in closer agreement with the expectations \cite{Abazov:2008qm,*Aaltonen:2009ny,*Aaij:2013qja}.

A remarkable feature of charmed baryons is their narrowness, with, e.g., $\Gamma\sim 6\,$MeV for $\Lambda_c(2880)$, $\sim 17\,$MeV for $\Lambda_c(2940)$, $\sim 4\,$MeV for $\Xi_c(3123)$ \cite{Agashe:2014kda}. 
Of course, this gives a better view on the spectrum than in the case of overlapping resonances. An optimistic columnist of PDG \cite{Amsler:2008zzb} even stated: ``\dots models of baryon-resonance spectroscopy should now \emph{start} with the narrow charmed baryons, and work back to those broad old resonances.''\@  Among the current discussions, there is the suggestion that $\Lambda_c(2980)$ is mainly a $D^*p$ molecule  \cite{Dong:2011ys}. 

The same narrowness is observed for $b$-baryons, as listed in the latest PDG review \cite{Agashe:2014kda} or revealed in the latest LHCb papers \cite{Aaij:2014yka,*Aaij:2016jnn}. 
\subsection{Baryons with double charm}\label{sse:baryons-open2}
Thirty years ago, double charm baryons were considered as just a straightforward extension of the quark model, to be checked just for the sake of completeness. At most, some experts stressed the co-existence within the same hadron of the slow, adiabatic, charmonium-like motion of the two heavy quarks, and the  fast, $D$-meson-like motion of the light quark around them. In the literature, $(ccq)$ is often described as a quark-diquark \cite{Ebert:2002ig}; however,  the first excitations occur between the two heavy quarks and thus one has to introduce ``excited diquarks''.  An alternative is the Born-Oppenheimer method \cite{Fleck:1989mb},  which, as discussed later, might also be adequate in the $XYZ$ sector and for hybrids. 

Some evidence for double-charm baryons was found at Fermilab, in the SELEX experiment \cite{Mattson:2002vu,*Ocherashvili:2004hi}, but unfortunately not confirmed in other experiments at Fermilab or elsewhere \cite{Ratti:2003ez,*Aubert:2006qw,*Kato:2013ynr,*Aaij:2013voa}. This becomes particularly annoying. Double-charm baryons will certainly receive more attention in the near future \cite{Broadsky:2012rw,*Karliner:2014gca}.

It is an important question to understand why double-charm baryons have not yet been seen. If the golden modes such as $\Lambda_c+K$ or $\Lambda+D$ are suppressed, the detection scheme should be revised. And, perhaps, new production mechanisms should be exploited.

Our ability to detect double-charm baryons and measure their properties is eagerly watched by people waiting for exotic tetraquarks and dibaryons. As pointed out in \cite{Gelman:2002wf,*Cohen:2006jg}, the binding and the correlation of the two charmed quarks in $(ccq)$ gives some idea about $(cc\bar q\bar q)$ and $(ccqqqq)$. And, in most production experiments, they share the same triggers, so that a simultaneous search for double-charm baryons, mesons and dibaryons can be envisaged.
\subsection{Anticharmed baryons}\label{sse:antich-baryons}
Two simultaneous papers by Gignoux et al.\ and Lipkin \cite{Gignoux:1987cn,*Lipkin:1987sk}, in which the word ``pentaquark'' was first used,
suggested the existence of stable or metastable states $(\bar Q qqqq)$ with $qqqq$ denoting $uuds$, $ddsu$ or $ssud$. This pentaquark was searched for in an experiment at Fermilab, which was not conclusive \cite{Aitala:1997ja,*Aitala:1999ij}. At the time of the light pentaquark,  the H1 collaboration at HERA looked at the $\bar D{}^*p$ mass spectrum, i.e., states of minimal content $(\bar c uudd)$, without strangeness contrary to the prediction in \cite{Gignoux:1987cn,*Lipkin:1987sk}, and found a peak  near 3.1\,GeV with a width of about 12\,MeV \cite{Aktas:2004qf}, which was  confirmed neither by Zeus \cite{Chekanov:2004qm}, also operating at HERA, nor by BaBar \cite{Aubert:2006qu} at SLAC, nor in the analysis by H1 of their HERA\,\rom{2} data.\footnote{I thank Katja Krüger and Stefan Schmitt for a clarifying correspondence.}\@  The anticharmed pentaquark was initially one the goals of the COMPASS experiment \cite{Baum:298433}, but it has disappeared from the agenda. 
\subsection{Baryons with hidden charm or beauty}\label{sse:baryons-hidden}
This is one of the latest findings, inducing an intense phenomenological activity supplementing the discussions about $XYZ$ mesons. The rumour is that, due to the bad reputation of the light pentaquark, the LHCb collaboration tried during months to kill the evidence, with more checks than usual for this type of discovery. The result published in \cite{Aaij:2015tga} deals with  the weak decay 
\begin{equation}
 \label{eq:Lambdab1}
 \Lambda_b\to J/\psi+p+K~,
\end{equation}
of the lowest baryon with beauty. 
The Dalitz plot, as well as the $(p,J/\psi)$ mass spectrum, indicates two peaks, one with   a mass of $4380\pm 8\pm 29\,$MeV and a width of $205\pm 18\pm 86\,$MeV, an another with a mass of $4449.8\pm 1.7\pm 2.5\,$MeV and a width of $39\pm 5\pm 19\,$MeV. As the $\Lambda_b$ mass is always reconstructed, some ambiguities are removed. The background is fitted as $J/\psi+ \Lambda^{(*)}$, with many $\Lambda^*$ excitations, up to high mass and high spin. When two resonances are added, the preferred quantum numbers are spin $J=3/2$ and  $5/2$, and opposite parities, namely $(3/2)^\pm$ and $(5/2)^\mp$. See, also, \cite{Aaij:2016phn}.

\begin{figure}[ht!]
 \centering
\includegraphics[width=.45\columnwidth]{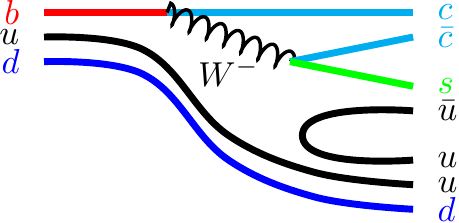}\qquad
 \includegraphics[width=.45\columnwidth]{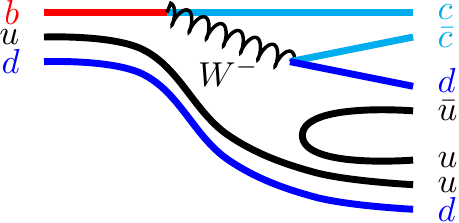}
 \caption{$\Lambda_b\to J/\psi+p+K$ decay (left) and $\Lambda_b\to J/\psi+p+\pi$ decay (right)}
 \label{fig:LHCb-penta1a}
\end{figure}

\begin{figure}[ht!]
 \centering
 \includegraphics[width=.45\columnwidth]{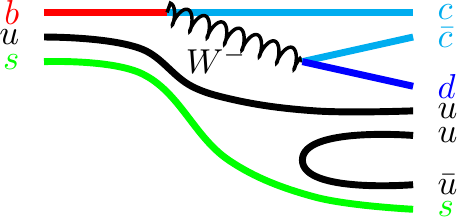}\qquad
 \includegraphics[width=.45\columnwidth]{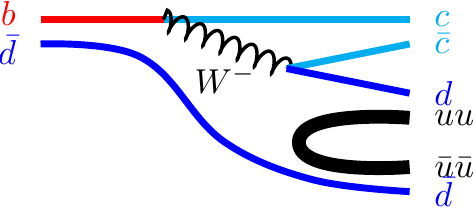}
 \caption{Other weak decays of $b$-hadrons producing a hidden-charm pentaquark.}
 \label{fig:LHCb-penta2}
\end{figure}

The process is schematically described in Fig.~\ref{fig:LHCb-penta1a}, as well as the companion reaction (Cabibbo suppressed) $\Lambda_b\to J/\psi+p+\pi$.
This process $\Lambda_b\to J/\psi+p+\pi$ has been measured by LHCb and published before the discovery of hidden-charm pentaquarks \cite{Aaij:2014zoa}, but the statistics was not sufficient to identify peaks. This reaction has been discussed in a few papers, as well as other weak decays involving the hidden-charm pentaquark \cite{Burns:2015dwa,Wang:2015pcn}, shown in Fig.~\ref{fig:LHCb-penta2}.  A more recent analysis of the reaction $\Lambda_b\to J/\psi+p+\pi$ provides some support for the contribution of hidden-charm pentaquarks~\cite{Aaij:2016ymb}.

It has been proposed \cite{Kubarovsky:2015aaa,*Kubarovsky:2016whd,*Blin:2016dlf,*Karliner:2015voa,*Meziani:2016lhg} to check the existence of the LHCb pentaquarks and provide more information on their properties by studying the photoproduction reaction
$\gamma+p\to J/\psi+p$, and the experiment will be done at Jlab.
\section{Multibaryons}\label{se:multibaryons}
\subsection{Dibaryons}
The situation in the non-strange sector has been recently reviewed at the Conference Baryon 2016 \cite{Mattione}.
There are regularly some claims for resonant behaviour of some partial waves \cite{Arndt:1986jb,*Arndt:1993nd,*Arndt:1994bs,*Tatischeff:1999jh}. Negative results include: \cite{Parker:1989hw} in the isospin $I=2$ sector, \cite{DeBoer:1984nc} for multineutrons bound by a few $\pi^-$. The most serious candidates are the state of mass 2380\,MeV and width $70\,$MeV by the WASA collaboration, in the ${}^3\mathrm{D}_3{}^3\mathrm{G}_3$ coupled partial waves, corresponding to $3^{-}$ \cite{Adlarson:2014pxj,*Adlarson:2014ozl}, which was expected by various models \cite{Goldman:1989zj}, and the more recent state at 2.15\,GeV which is under investigation at Jlab \cite{Mattione}. 

In the sector of single strangeness $S=-1$, there are recurrent discussions about the possibility of dibaryons. See, e.g., \cite{Ohnishi:2011wb,*Hashimoto:2014cri} and refs.~there. Note that a study of the atomic $(K^-,p,p)$ system could provide  some indirect information. In the two-body case, say $(a^+,b^-)$, the atomic spectrum is much modified if the strong-interaction $(a,b)$ potential generates a bound state near threshold. This was realised by Zel'dovich in the 50's and has been much studied \cite{ISI:A1960WT32800001,*Gal:1996pr,*Combescure:2007ki}. The case of three-body exotic atoms with a combination of Coulomb interaction and short-range hadronic interaction remains to be clarified, but probably deserves a detailed study.\footnote{I thank Claude Fayard for several discussions on this topics.}\@

The most advertised case deals with strangeness $S=-2$, following the work by Jaffe\cite{Jaffe:1976yi} on the coherent chromomagnetic attraction in $H=(uuddss)$ with isospin and spin $I=J=0$.  More than 20 experiments have searched without success for the $H$, ranging from the absence of  $H$ in the  decay of double hypernuclei \cite{Dalitz:1989kt} to the mass spectra emerging from heavy-ion collisions \cite{Kuhn:2002xf}. New experiments are currently under construction.  The status of theoretical studies will be reviewed in Sec.~\ref{ssse:chromo}. 

The speculations have been extended to the $S=-3$ sector, see, e.g., \cite{Goldman:1987ma,*Huang:2015yza}.
\subsection{Light hypernuclei, neutron-rich hypernuclei}
If the $H$, considered as $\Lambda\Lambda$ system, is close to binding, as well as some other baryon-baryon systems, it is natural to guess that some multi-baryon configurations will be bound, as an illustration of the quantum phenomenon of Borromean binding \cite{Thomas:1935zz,*Zhukov:1993aw,*Richard:2003nn}. There are for instance debates about the existence of $(\Lambda,\Lambda, N)$ or $(\Lambda,\Lambda, n,n)$ \cite{Hiyama:2012gx,*Filikhin:2001yh,*Garcilazo:2014mra,*Richard:2014pwa}, as well as $(NN\Xi)$ or $(NN\Xi\Xi)$~\cite{Garcilazo:2016ams}.  The field of hypernuclear physics is rather active, with new facilities contributing, and is reviewed regularly at dedicated conferences.

Almost immediately after the discovery of charm, the possibility of new species of hypernuclei was envisaged, with hyperons replaced by charmed baryons \cite{Dover:1977jw}.
\section{Theory}\label{se:theory}
In atomic physics, the study of multi-electron systems looks at first rather straightforward: the same non relativistic dynamics is at work, and one has ``just'' to solve the $N$-body equations with increasing $N$. Even there, the progress is slow, and the calculations, which are very delicate,  have to be cross-checked. For instance, the last bound state of H$_2{}^+$,  the excitations of the positronium molecule or the existence of Borromean molecules were discovered only recently \cite{2003EL.....64..316C,*1998PhRvL..80.1876V,*2009NuPhA.827..541C,*2003PhRvA..67c4702R}. In nuclear physics, the situation is even more delicate, as the basic two-body interaction is presumably modified by the presence of other nucleons, and supplemented by 3-body forces. 

In QCD, the situation is even worse. If one uses simple constituent models to describe mesons and baryons, their extension to multiquark states is hazardous. If one uses more sophisticated approaches to ordinary hadrons, their application to higher configurations becomes technically very difficult. Before discussing in some details a few specific schemes, we review the early history of exotic hadrons. 
\subsection{Duality}
In the 60s, much attention was paid to \emph{crossing symmetry}, which relates, e.g., $a+b\to c+d \quad (s)$ to $a+\bar c\to \bar b+d \quad (t)$. The reaction $(t)$ describes what is exchanged to monitor the direct reaction $(s)$. It is reasonable to assume an equivalence between the content of the two reactions, as well as that of the third reaction,  $a+\bar d\to \bar b+c\quad (u)$. This principle of \emph{duality} was often used as a constraint in building models. For instance, if one describes $\pi N$ elastic scattering with several $N^{(*)}$ and $\Delta^{(*)}$ resonances, duality dictates to refrain from exchanging $\sigma$, $\rho$, \dots, in the $t$-channel.

\begin{figure}[ht!]
 \centering
 \raisebox{.2cm}{\includegraphics[width=.35\columnwidth]{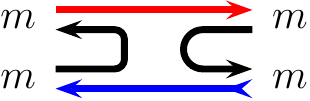}}\qquad
\includegraphics[width=.35\columnwidth]{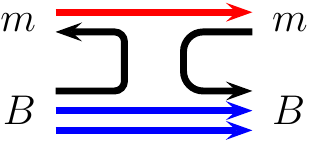}
 \caption{Duality diagram for meson-meson scattering (left) and meson-baryon scattering (right).}
 \label{fig:dual1}
\end{figure}

A consequence of duality is that is the reaction $(s)$ is weak, i.e., without resonances, it does not benefit from coherent $t$-channel exchanges. Though duality was first formulated at the hadron level, it is nowadays explained in terms of quark diagrams. We refer to \cite{Roy:2003hk} for a review and references. For instance, in the case of $\pi\pi$ scattering with isospin $I=0$, the diagram of Fig.~\ref{fig:dual1} (left) shows that the possibility of annihilating a quar-antiquark pair opens the way for $(q\bar q)$ exchange in the $t$-channel and $(q\bar q)$ resonances in the $s$-channel. This is why there is a strong $\pi\pi$ interaction with $I=0$. On the other hand, for isospin $I=2$, there is $s$-channel resonance, and the $t$-channel dynamics suffers from a more intricate topology. 

Similarly, $\bar K N$ scattering, shown in Fig.~\ref{fig:dual1} (right), benefits from both $(qqq)$ resonances in the $s$-channel and $(q\bar q)$ exchanges in the $t$-channel.  This is not the case for $KN$ scattering which is notoriously less strong. 

\begin{figure}[ht!]
 \centering
\includegraphics[width=.40\columnwidth]{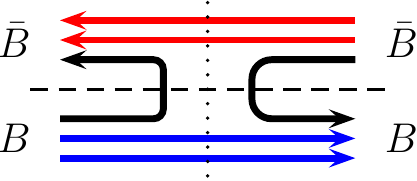}
 \caption{Duality diagram for baryon-antibaryon scattering. The intermediate state in the $s$-channel (intersecting the dotted line) contains two quarks and two antiquarks. The intermediate state in the $t$-channel (crossing the dashed line) is made of a quark and antiquark. }
 \label{fig:dual2}
\end{figure}

This led Rosner to address the problem of baryon-antibaryon scattering \cite{Rosner:1968si}, which is depicted in Fig.~\ref{fig:dual2}. The interaction is very strong, and benefits, indeed, from the exchanges of mesons in the $t$-channel. Hence there should be resonances in the $s$-channel. This is how baryonium was predicted. 

\begin{figure}[ht!]
\centerline{%
\includegraphics[width=.40\columnwidth]{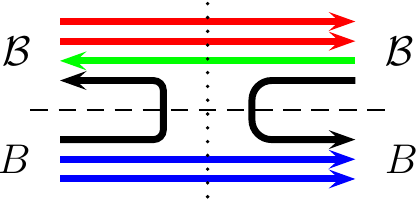}\qquad
\includegraphics[width=.40\columnwidth]{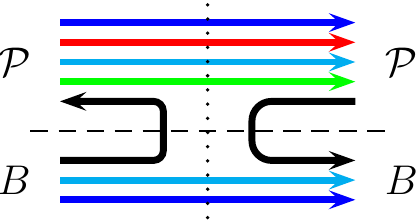}}
 \caption{Pandora box syndrome for duality. Left: pentaquarks (four quarks and an antiquark in the $s$-channel) dual of mesons (quark-antiquark in the $t$-channel) in baryonium-baryon scattering. Right: dibaryon (six quarks in the $s$-channel) dual of mesons (quark-antiquark in the $t$-channel) in pentaquark-baryon scattering.}
 \label{fig:dual3}
\end{figure}

As noted by Roy \cite{Roy:2003hk}, once one accepts baryonium, in a \textsl{Gedankenexperiment} colliding a baryonium against a baryon, meson-exchange is dual of a pentaquark. In turn, the analysis of pentaquark-baryon scattering implies the existence of dibaryons! This is illustrated in Fig.~\ref{fig:dual3}.
\subsection{String dynamics}
In line with duality, a string model was developed for hadrons, explaining why the squared masses increase linearly with the spin $J$. This model was echoed by the elongated version of the bag model, the flux-tube model, or some potential models with multi-body forces which will be mentioned later in this section. 

\begin{figure}[ht!]
\centering
\raisebox{.5cm}{\includegraphics[width=.25\columnwidth]{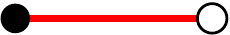}}\qquad\qquad
\includegraphics[width=.25\columnwidth]{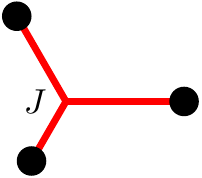}
 \caption{String model of ordinary mesons (left) and baryons (right).}
 \label{fig:string1}
\end{figure}

\begin{figure}[ht!]
\centering
\includegraphics[width=.25\columnwidth]{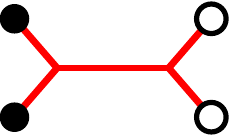}
 \caption{String model of  baryonium.}
 \label{fig:string2}
\end{figure}

The string model was first rather empirical, but later received a  support from some topological and non-perturbative properties of QCD, as reviewed recently \cite{Rossi:2016szw}. The mesons are pictured as a string linking a quark to antiquark, see Fig.~\ref{fig:string1} (left).
The picture of baryons was suggested by Artru, and rediscovered in several approaches \cite{Artru:1974zn,*Dosch:1975gf,*Hasenfratz:1980ka,*Carlson:1982xi,*Bagan:1985zn,*Fabre:1997gf}. It consists of three strings linking each quark to a junction $J$, which plays somewhat the role of a fourth constituent of each baryon, and, in particular, is submitted to an extension of the Zweig rule, which prohibits any internal junction-junction annihilation. In this scheme, the baryonium of Rosner contains two junctions, see Fig.~\ref{fig:string2} and thus widely differs from a set of two disconnected mesons. 
The decay of baryonium occurs by string breaking via pair creation: in the case of an external leg, it corresponds to the pionic cascade $\mathcal{B}^*\to \mathcal{B}+\pi$, while the breaking of the central string corresponds to the decay into baryon and antibaryon. Several experimental states were good candidates for baryonium \cite{Evangelista:1977ni,Montanet:1980te}, but they were not confirmed in experiments using better antiproton beams. Perhaps the nucleon-antinucleon interaction suffers from too strong an annihilation, while the heavy quark sector gives better opportunities. 

\begin{figure}[ht!!]
\centering
\includegraphics[width=.20\columnwidth]{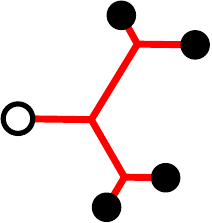}
\qquad\qquad
\includegraphics[width=.29\columnwidth]{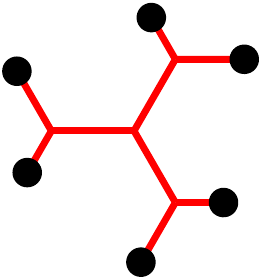}
 \caption{Possible string structures for pentaquarks (left) and dibaryons (right).}
 \label{fig:string3}
\end{figure}

The model can be extended to more complicated structures, as to include pentaquarks and dibaryons. Some examples are given in Fig.~\ref{fig:string3}.
\subsection{Colour chemistry}
The above string pictures of baryonium was challenged by other schemes: the nuclear model which is reviewed in the next section, and some quark models. One of them  describes baryonium states as a diquark and an antidiquark separated by an orbital barrier which suppresses the decay into mesons \cite{Jaffe:1977cv}.  In colour chemistry, there is also the possibility of colour-sextet diquark, and thus of $(qq)-(\bar q\bar q)$ states, named `mock baryonia'' which are predicted to be rather narrow \cite{Chan:1978nk}. This was, indeed, an interesting challenge to identify spectroscopic signatures of the colour degree of freedom. 
\subsection{Molecular binding}
The name has evolved. It used to be \emph{quasi-nuclear}, and became \emph{molecular} model.
It starts with the observation that the meson-exchange interaction of Yukawa, which successfully binds the deuteron and other nuclei, holds for other hadrons, in particular these containing light quarks.
\subsubsection{Baryon-antibaryon}
In the 60s, attempts have been made to describe mesons as baryon-antibaryon states, a simplified version of the bootstrap. Fermi and Yang \cite{Fermi:1959sa} had noticed that the $N\bar N$ potential is often more attractive than the $NN$ one, as the cancellations in the latter become  coherences in the former. But this approach does not account for some important properties of the meson spectrum \cite{Ball:1965sa}, such as exchange degeneracy, e.g., the $\rho$ and $\omega$ having the same mass.

In the late 70s and early 80s, another point of view was adopted, namely that baryon-antibaryon bound states and resonances correspond to new type of mesons, lying higher in mass than ordinary mesons, and preferentially coupled to the  baryon-antibaryon channel. In practice, the $N\bar N$ potential is deduced from the $NN$ one by the $G$-parity rule: if the exchange of a meson $m$ (or set of mesons $m$) contributes as $V_I(m)$ to the $NN$ potential in isospin $I$, then its contributes as $G_m\,V_I(m)$ to the $N\bar N$ in isospin $I$ \cite{Fermi:1959sa}. When annihilation is neglected, a rich spectrum of bound states and resonances is found \cite{Shapiro:1978wi,*Buck:1977rt}, with the hope than some of them will not become too broad when annihilation is switched on.  Before the discovery of the antiproton, annihilation was believed to be be a short-range correction. However, the annihilation part of the $N\bar N$ integrated cross-section was measured to be about twice the elastic one. To fit the cross-
sections, one needs an annihilation acting up to about $1\,$fm, and this makes most quasi-nuclear baryonium states very broad. The molecular model might have better chances to predict exotics in channels with heavy hadrons whose interaction is less absorptive.
\subsubsection{Binding of heavy hadrons, first attempts}
The Yukawa model made a come back in the heavy quark sector. The $\dst\dst$ or $\dst\dbst$ potential, where $\dst=(c\bar q)$, is perhaps weaker than the $NN$ potential, but being experienced by heavier particles, it can produce binding as well. The first papers were by Okun and Voloshin  and Glashow et al.\ \cite{Voloshin:1976ap,*DeRujula:1976qd}. In the latter case, the idea was to explain some intriguing properties of some high-lying charmonium states, but it was eventually realised that these properties were due to the nodes of the wave function when the states are described as radial excitations \cite{LeYaouanc:1977ux,*LeYaouanc:1977gm,*Eichten:1979ms}. 
\subsubsection{Binding of heavy hadrons, $XYZ$}\label{sse:molec-XYZ}
The best known application of the molecular model deals with $XYZ$ states. Almost simultaneously, a few papers came, stressing the attraction due to one-pion-exchange in some $\dst\dst$ or $\dst\dbst$ channels \cite{Tornqvist:1991ks,Manohar:1992nd,Ericson:1993wy,Tornqvist:1993ng}. Of course, the $X(3872)$ discovery at Belle was greeted as a success of this approach. Today, the $X(3872)$ is generally considered as having both a quarkonium and a multiquark component \cite{Ferretti:2013faa}. 
For a clarification of the molecular approach, see \cite{Swanson:2006st,*Thomas:2008ja,*Ohkoda:2011vj,*Cleven:2015era}.
\subsubsection{Molecular pentaquark}
The anticharmed pentaquark has been studied with molecular dynamics, as a light baryon linked to an anticharmed meson. See, for instance, \cite{Hofmann:2005mn}. Note that the $
uuds$ or $ddsu$ and $ssud$ content of the light sector in the chromomagnetic model is not necessarily the same when mesons are exchanged  between a $\dbst$ or a $\dbsst$ and a baryon.

After the publication of the LHCb results on the hidden-charm pentaquark \cite{Aaij:2015tga}, it was reminded or discovered that there were some predictions of bound states of a charmed baryon and an anticharmed meson, in particular by Hofman and Lutz \cite{Hofmann:2005sw}, Oset and collaborators \cite{Xiao:2013yca}, and Z.C.~Yang et al.\ \cite{Yang:2011wz}. The game was even initiated earlier for the case of hidden-strangeness \cite{Landsberg:1993dh}. Of course, the model has been refined recently and compared to the LHCb results. The  literature is comprehensively reviewed in \cite{Chen:2016qju}.

One might wonder why the hidden-charm pentaquark was not searched for before, given the above predictions. 
\begin{enumerate}\item
  There are many predictions vs.\  just a few experiments, which are not primarily devoted to spectroscopy.                                                                                               \item In the molecular model, there are many states of this type, and it is not easy to focus on the configuration giving the best signature for the molecular dynamics. For instance, the binding of a charmed baryon and an anticharmed antibaryon was already envisaged in 1977 \cite{Dover:1977hs}, shortly after the meson-meson case \cite{Voloshin:1976ap,*DeRujula:1976qd}, but he meson-baryon case was unfortunately overlooked in the 70s.  
  \item There is rather generally more effort and more interest  in the meson than in the baryon sector.                                                                                                      \end{enumerate}

The negative parity states, either $(3/2)^-$ or $(5/2)^-$, can be accommodated in the molecular model, with a main $s$-wave between a meson and a baryon, and a small admixture of $d$-wave, as in the more familiar case of the deuteron. The orbital excitation with positive parity is probably much higher in mass.
\subsubsection{Binding other heavy hadrons}
In $\dst\dbst$, the annihilation and creation of light quarks opens a coupling to $c\bar c$, leading to description of some isoscalar $XYZ$ states as a mixture of molecule and quarkonium states. Isovector states do not mix with quarkonium, except for some minor effects of isospin violation. The sector of double-charm molecules has also received much attention, from the very beginning. See for instance, \cite{Manohar:1992nd,Wong:2003xk,*Ohkoda:2012hv}.

It is immediately realised that if the exchange of mesons bind $\dst\dst$ , $\dst\dbst$, and $\Sigma_c \dbst$ their beauty or mixed charm-beauty analogues and other configurations could as well be formed by this mechanism, such as some meson-baryon and baryon-baryon states~\cite{Karliner:2015ina}.

We already mentioned the prediction of hidden-charm pentaquarks as molecules such as $\bar D{}^*\Sigma_c$, before their discovery by LHCb. One could also include $\bar D N$ or $\bar D$-nuclei states, see, e.g., \cite{Yasui:2009bz}, where the possibility of $\bar D NN$ states is also envisaged. The non-strange heavy pentaquark $\bar DN$ is related to the non-confirmed state claimed by H1 \cite{Aktas:2004qf}. 

In particular, there are many studies of bound states of charmed or doubly-charmed baryons with nucleons or with other singly or doubly charmed baryons \cite{Froemel:2004ea,*Liu:2011xc,*Meguro:2011nr,*Huang:2013rla,*Lee:2011rka,*Li:2012bt}.
One can even speculate about a new periodic table, where the proton is replaced by $\Xi_{cc}^{++}$ and the neutron by $\Xi_{cc}^+$ \cite{JuliaDiaz:2004rf}. Of course some rearrangement into triple charm and single charm should be taken into account. 

In several model calculations, many interesting hadron-hadron systems are estimated at the edge between binding and non-binding. This means that 3-body or larger systems are probably bound as long as they are not suppressed by the Fermi statistics.  So,  accounting for the $XYZ$ states implies many other states.
\subsection{Gluonium}
In QCD, the gluon field carries a colour charge and coupled to itself. This lead to speculations about the existence of new neutral states, made essentially of gluons, except for $g\leftrightarrow q\bar q$ loop corrections. This was sometimes hotly debated. I remember for instance that in his talk at the Thessaloniki conference on antiprotons, Van Hove expressed doubts about the necessity of glueballs within QCD, though the written version was somewhat softened \cite{VanHove:1986qx}. 

The estimates for the mass and even the quantum numbers of the lightest glueball or gluonium has somewhat changed during the years. The situation is now stabilised, and is reviewed, e.g., in \cite{Ochs:2013gi}. Note that the lowest glueballs with unusual quantum numbers, such as $0^{+- }$, are sometimes found with rather high masses~\cite{Qiao:2015iea}.
\subsection{Hybrids}
Another sector that used to be rather fashionable is the one of hybrids. Very schematically, hybrids combine quark(s), antiquark(s) and gluons. There are even explicit constituent models in which the constituent gluon is given a mass, and interacts via an effective potential, see, e.g.,~\cite{Buisseret:2008zza}. 

The first approaches describes hybrids as an excitation of the string or, say, of the gluon field linking the quarks.  Some pioneering attempts can be found in~\cite{Giles:1977mp,*Horn:1977rq}. In \cite{Hasenfratz:1980jv}, a Born-Oppenheimer treatment was made, with a variant of the bag model used to estimate the energy of the gluon field. For a given interquark separation, the shape of the bag is adjusted as to give the lowest gluon energy. The ground state of the gluon field provide with the usual quarkonium potential. The first excited state of the gluon field is linked to the effective $Q\bar Q$ potential within a hybrid. The first $(c\bar c g) $ hybrid was predicted near 4.0\,GeV, and this estimate has been much refined with the bag model replaced by  more elaborate lattice calculations \cite{Mandula:1983wc,*Juge:2003qd}, or effective field theory \cite{Berwein:2015vca}.

Note that in the late 70s and early 80s, the motivation was that the hybrids would show up clearly atop a clean and well-understood charmonium spectrum! With the proliferation of $XYZ$ states, we realise today that identifying hybrids is a little more challenging. 
\subsection{QCD sum rules}
The method of QCD sum rules (QCDSR) is a very ambitious and efficient approach, initiated by  Shifman, Vainshtein and Zakharov \cite{Shifman:1978bx,*Shifman:1978by,*Shifman:1978bw}, allowing an interpolation between the rigorous perturbative regime and the long-range regime. For a review, see, e.g., \cite{narison2004qcd}.  It was first applied to ground-state mesons. There is no limitation, in principle, but the numerical stability become more and more delicate for excitations with the same quantum numbers as the ground state (say, radial excitations). For baryons, the choice of the most appropriate operator is not unique, and triggered some debates. This is of course even more delicate for multiquarks. Hence, the choice of the strategy is often dictated by models, such a diquark-antidiquark.

A few groups have applied QCDSR to exotic states, in particular at Montpellier, S\~{a}o Paulo and Peking, see, e.g., \cite{Matheus:2006xi,*Chen:2008ej} for examples of early attempts, and \cite{Nielsen:2009uh,Narison:2010ak,Chen:2016qju} for some reviews of the state of art over the years. Needless to say that QCDSR calculations of multiquark exotics are extremely delicate and require weeks of cross-checks. So any overnight analysis published just after the release of a new experimental state should have been anticipated long in advance. 
\subsection{Lattice QCD}
Lattice QCD (LQCD) is well known and well advertised, and hardly needs to be presented. For an introduction to the recent literature, see, e.g., \cite{DeTar:2015orc}. Dramatic progress have been accomplished on various aspects of strong interaction. There are interesting recent developments, for instance to study how the results behave numerically and have to depend theoretically on the pion mass. Another extension deals with \textsl{ab-initio} nuclear physics, at least for light nuclei, and multiquark physics \cite{Fiebig:1999hs,*Inoue:2010es,*Beane:2010hg,*Suganuma:2011ci,*Aoki:2012tk,*Kirscher:2015yda,*Dudek:2016bxq,*Francis:2016hui}, with in particular, a renewed interest in the study of the $H$ dibaryon.  

In the sector of heavy quark spectroscopy, LQCD has confirmed many important results obtained by simpler, but empirical, tools. For instance, the $c\bar c$ interaction of charmonium, including its spin dependence, has been estimated, in agreement with the current potential models. Latter, the charmonium spectrum was calculated directly. 

As already mentioned,  some  decisive contributions were made for the study of hybrid mesons with hidden heavy flavour \cite{Mandula:1983wc,*Juge:2003qd}.  A Born-Oppenheimer potential can be computed for other configurations, e.g., double-charm baryons, and LQCD is very suited for the treatment of the light degrees of freedom. 

For multiquarks, pure \textsl{ab-initio} studies are preceded by preliminary investigations inspired by models. For instance, in the sector of doubly-heavy mesons, an effective $QQ$ potential or an effective $D^{(*)}D^{(*)}$ or $B^{(*)}B^{(*)}$  interaction has been  studied see, e.g., \cite{Green:1998nt,*Michael:1999nq,*Bali:2010xa,*Bicudo:2015kna}. Nowadays, multiquark states are often studied directly on the lattice, without the intermediate step of an effective potential. Needless to say, LQCD faces difficulties as QCDSR, for the study of multiquark states. In the early days, the quantum numbers were not always clearly identified, and the genuine states were cleanly separated from the artefacts due to lattice boundaries. Nowadays, the field is mature, and an impressive program of multiquark and multihadron physics has been developed by several groups \cite{Fiebig:1999hs,*Inoue:2010es,*Beane:2010hg,*Suganuma:2011ci,*Aoki:2012tk,*Kirscher:2015yda,*Dudek:2016bxq,*Francis:2016hui}.
\subsection{Constituent models}
In this section, we briefly survey some phenomenological models which have been used to predict some exotic hadrons or describe them after their discovery. It should be stressed, indeed, that if these models are more empirical than QCDSR and LQCD, they can be adapted more flexibly to describe the world of ordinary hadrons and used to extrapolate towards novel structures.
\subsubsection{Chromomagnetic binding}\label{ssse:chromo}
In the 70s, the hyperfine splitting among hadron multiplet was interpreted as the QCD analogue of the Breit-Fermi interaction in QED, namely the spin dependent part of the one-gluon exchange \cite{DeRujula:1975ge}. The corresponding hyperfine interaction reads 
\begin{equation}\label{eq:DGG}
 V_\text{hyp}=-\sum_{i<j}\frac{C}{m_i\,m_j}\, \llss{i}{j} \,\delta^{(3)}(\vec r_{ij})~,
\end{equation}
to be treated at first order, otherwise it requires some regularisation in the channels where it is attractive. A tensor term also contributes beyond $s$-waves. 

In 1977, Jaffe applied this Hamiltonian \eqref{eq:DGG} to the configuration $H=(uuddss)$ with $I=J=0$ assuming \textsl{i)} SU(3)$_\text{F}$ flavour symmetry, and \textsl{ii)} that the short-range correlation $\langle \delta^{(3)}(\vec r_{ij})\rangle$ has the same value as for ordinary baryons~\cite{Jaffe:1976yi}. He found a remarkable coherence in the spin-colour operator, with $\langle\sum \llss{i}{j}\rangle = 24$ for $H$, to be compared to $8$ for $N$ or $\Lambda$, i.e., $16$ for the $\Lambda\Lambda$ threshold. This means that the $H$ is bound below the $\Lambda\Lambda$ threshold by half the $\Delta$-$N$ splitting, about $150\,$MeV.  The $H$ was searched for in many experiments, without success so far. For instance, the double hypernucleus $\isotope[6][\Lambda\Lambda]{He}$ \cite{Takahashi:2001nm} was not seen decaying into $H+\alpha$. 

The model has been reexamined thoroughly \cite{Oka:1983ku,*Rosner:1985yh,*Karl:1987cg,*Faessler:1989si}, with the result that any correction tends to reduce the binding, and eventually, the $H$ is pushed in the continuum. These corrections are: SU(3)$_\text{F}$ breaking, more realistic estimate of the short-short correlation (a dibaryon is more dilute than a baryon), and proper inclusion of the full Hamiltonian (kinetic energy and spin-independent confinement besides $V_\text{hyp}$). Of course, some additional attraction such a scalar-isoscalar exchange between the two $\Lambda$ can restore the binding.

Meanwhile, some other configurations were found, for which the chromomagnetic interaction, if alone, induces binding. In  1987, the Grenoble group and H.J.~Lipkin \cite{Gignoux:1987cn,*Lipkin:1987sk} proposed independently and simultaneously a heavy pentaquark $P=(\bar Q qqqq)$, where $qqqq=uuds, ddsu$ or $ssud$. With the same hypotheses as Jaffe's for the $H$, and assuming an infinite mass for the heavy antiquark $\bar Q$, they found an optimal $\langle\sum \llss{i}{j}\rangle = 16$, to be compared to $8$ for the threshold $(\bar Q q)+(qqq)$, i.e., the same binding energy $150\,$MeV. 
In a series of papers \cite{Leandri:1989su,*SilvestreBrac:1992yg,*Leandri:1995zm}, Leandri and Silvestre-Brac applied the chromomagnetic operator 
\begin{equation}\label{eq:CM-op1}
-\sum_{i<j}\frac{C}{m_i\,m_j}\, \llss{i}{j}~,
\end{equation}
with suitable changes for antiquarks, to a variety of configurations, and found several cases, for which the chromomagnetic interaction, if alone, induces binding. As noted by Lichtenberg \cite{Lichtenberg:1992ji}, the $(m_i\,m_j)^{-1}$ dependence is somewhat extreme, as heavy quarks give more compact wave-functions with larger short-range correlation factor.  This led some authors to adopt a more empirical prescription, 
\begin{equation}\label{eq:CM-op2}
-\sum_{i<j} c_{ij}\, \llss{i}{j}~,
\end{equation}
where the coefficients are taken from a fit to ordinary mesons and baryons \cite{Hogaasen:2005jv,*Buccella:2006fn,*Stancu:2009ka}.
For instance, in the case of $(c\bar c q\bar q)$ with $J^{PC}=1^{++}$, the diagonalisation of \eqref{eq:CM-op2}, supplemented by effective constituent masses, gives a state which corresponds nicely to the mass and properties of the $X(3872)$. Its narrowness comes from being almost a pure octet-octet state of colour in the channels $(c\bar c)-(q \bar q)$.  As most approaches based on quark dynamics, the chromomagnetic model predicts an isospin $I=1$ partner of $X(3872)$, with $J=1$ and $C=+1$ for the neutral, which is not yet seen experimentally. 
\subsubsection{Chromoelectric binding}\label{sse:chromoelectric}
\paragraph{The role of flavour independence}
In constituent models, one can switch off the spin-dependent part of the interaction, and study the limit of a pure central potential. Its main property is \emph{flavour independence}, with some interesting consequences, which are independent of the detailed shape of the interaction. In particular, if a constituent mass increases, the kinetic energy decreases, and one gets a lower energy, or, say, more binding. This explains why the $\Upsilon$ family has more bound states than the $\Psi$ family below the open-flavour threshold. 

Another well-studied consequence deals with doubly-heavy tetraquarks $(QQ\bar q\bar q)$. The 4-body system benefits from the heavy-heavy interaction, which is absent in the $(Q\bar q)+(Q\bar q)$ threshold. Hence, for a large enough $Q$-to-$q$ mass ratio, the system becomes stable. Explicit calculations with a colour-additive model
\begin{equation}\label{eq:col-add}
 V=-\frac{16}{3}\, \sum_{i<j}\ll{i}{j}\,v(r_{ij)}~,
\end{equation}
normalised to the quarkonium potential $v(r)$, indicates that in such  spin-independent interaction, $(cc\bar q\bar q)$ is probably unbound, but $(bb\bar q\bar q)$ stable \cite{Ader:1981db,*Heller:1986bt,*Zouzou:1986qh,*Carlson:1988hh}.

One hardly finds similar configurations. If, for instance, one considers $(QQqqqq)$ within \eqref{eq:col-add}, there is no obvious gain with respect to the best threshold made of two baryons, but flavour independence dictates that before any spin corrections, the threshold $(QQq)+(qqq)$ is lower than $2\,(Qqq)$. 

The stability of $(QQ\bar q\bar q)$ for large mass ratios is intimately related to the observation that the hydrogen molecule is more deeply bound (in units of the threshold binding energy) than the equal-mass molecules such as Ps$_2$. The Hamiltonian for $(M^+,M^+,m^-,m^-)$ reads
\begin{multline}
\label{eq:Ps2-H2}
H=H_0+H_1=\left[\left(\frac{1}{4\,M}+\frac{1}{4\,m}\right)\sum \vec{p}_i^2+V\right]\\
{}+\left(\frac{1}{4\,M}-\frac{1}{4\,m}\right)
\left(\vec{p}_1^2+\vec{p}_2^2-\vec{p}_3^2-\vec{p}_4^2\right)~. 
\end{multline} 
Its $C$-breaking term $H_1$ lowers the ground-state energy with respect to the $C$-even part $H_0$. Since $H_0$, a rescaled version of the Ps$_2$ molecule, and $H$, have the same threshold, and since Ps$_2$ is stable, H$_2$ is even more stable. Clearly, the Coulomb character of $V$ hardly matters in this reasoning. The key property is that the potential does not change when the masses are modified. In the case of a simple spin-independent potential with the colour-additive rule~\eqref{eq:col-add}, one does not get binding for $M=m$, but binding shows up when $M/m$ increases.

Note that, if one breaks the symmetry of Ps$_2$ differently and considers instead $(M^+,m^+,M^-,m^-)$ configurations, stability is lost for $M/m\gtrsim 2.2$, or, of course, $M/m\lesssim1/2.2$ 
\cite{PhysRevA.57.4956,*1999fbpp.conf...11V}. A protonium atom cannot polarise strongly enough a positronium atom and attach it. Similarly a  chromoelectric potential cannot produce a bound $(Q q\bar Q \bar q)$ tetraquark: one needs chromomagnetism or long-range Yukawa interaction. 
\paragraph{More detailed comparison of Ps$_2$ and tetraquark}
Another problem is to understand why the positronium molecule Ps$_2$ (neglecting annihilation) lies slightly below its dissociation threshold, while a chromo-electric model, associated with the colour-additive rule~\eqref{eq:col-add}, does not bind (at least according to most computations, an exception being \cite{Lloyd:2003yc}). This is due to a larger disorder in the colour coefficients than in the electrostatic strength factors $q_i\,q_j$ entering the Coulomb potential in Ps$_2$. 
Consider, indeed, a 4-body Hamiltonian scaled to 
\begin{equation}\label{eq:H-toy}
H[G]=\vec p_1^2+\vec p_2^2+\vec p_3^2+\vec p_4^2 +\sum_{i<j} g_{ij}\,v(r_{ij})~,
\qquad G=\{g_{ij}\}~,
\end{equation}
with $\sum_{i<j} g_{ij}=2$, having at least one bound state in the symmetric case $G=G_0$ for which $g_{ij}=1/3$ $\forall i,j$. 
The variational principle, with the solution of this symmetric case as a trial function, implies that
\begin{equation}\label{eq:H-toy-ineg1}
 \min(H[G])\le \min(H[G_0])~,
\end{equation}
so that, schematically, the more asymmetric the distribution $G$, the lower the ground-state energy.  For the cases of interest, one could be more precise, as the distribution is of the type $G(\lambda)=(1-\lambda)\,G_0 + \lambda \, G_1$. The ground-state energy $E(\lambda)$ is convex as a function of $\lambda$, and, since maximal at $\lambda=0$, it is convexly decreasing as a function of $|\lambda|$ for both $\lambda\ge0$ and $\lambda \le0$.  If one compares two values of $\lambda$ with opposite sign, then it is a good guess, but not a rigorous statement, to say that the larger $|\lambda|$, the lower the energy, as $E(\lambda)$  is nearly parabolic near $\lambda=0$. 

Now consider the actual values in Table~\ref{tab:coeff-tetra}.

\begin{table}[!htb]
\caption{Strength coefficients for the positronium molecule and the tetraquark states with colour wave-function $\bar 3 3$ or $6\bar 6$, as well as for their threshold. The list of coefficients is of the type $\{g',g',g'',g'',g'',g''\}$ after suitable renumbering, and summarised as $G=(1-\lambda)\,G_0+\lambda\, G_1$, where $G_0$ is the symmetric case $g'=g''=1/3$. \label{tab:coeff-tetra} }
\begin{ruledtabular}
\begin{tabular}{ccccc}
$(abcd)$ & $v(r)$ & $g'$ & $g''$ & $\lambda$\\
\hline
$\mathrm{Ps}+\mathrm{Ps}$  & $-1/r$ & 1 & 0 &1\\
Ps$_2$ &  $-1/r$ &  $-1$  & 1 & $-2$ \\
$(q\bar q)+(q\bar q)$ & $-1/r, r$ & 1 & 0 &1\\
$[(qq)_{\bar 3}(\bar q\bar q)_3]$ & $-1/r, r$ &$1/2$ & $1/4$& $1/4$\\
$[(qq)_{6}(\bar q\bar q)_{\bar 6}]$ & $-1/r, r$ & $-1/4$  & $5/8$ &  $-7/8$\\
\end{tabular}
\end{ruledtabular}
\end{table}

The asymmetry parameter $|\lambda|$ is much larger for Ps$_2$ than for $\mathrm{Ps}+\mathrm{Ps}$. This explains why the positronium molecule is bound.\footnote{Of course, this is not exactly what is written in the textbooks on quantum chemistry, but a large $|\lambda|$ means that Ps$_2$ benefits from correlation and anticorrelation effects in the various pairs.}

A tetraquark with a pure $\bar3 3$ colour structure has its asymmetry parameter smaller than that of the threshold and 
has the same sign, so this toy model is rigorously unbound. For $6\bar 6$, the signs are different, and the magnitude are comparable, so the analysis is more delicate. We note that $|\lambda|=2$ for Ps$_2$ results in a fragile binding, so it is not surprising that $|\lambda|=7/8$ leaves $6\bar 6$ unbound in numerical studies.  

To summarise, the tetraquark is penalised by the non-abelian character of the colour algebra in the simple model \eqref{eq:col-add}, and its stability  cannot rely on the asymmetries of the potential energy. It should use other asymmetries, in particular through the masses entering the kinetic energy,  
or spin effects, or mixing of $\bar3 3$ and $6\bar6$, or the coupling to decay channels, etc.
\paragraph{String dynamics}
The prescription \eqref{eq:col-add} is of course rather crude, as it treats colour as a global variable similar to isospin, and assumes arbitrarily the choice of a pairwise interaction. It can be justified in the limiting cases were $(N-1)$ quarks are clustered and the last quark isolated. So it might provide a reasonable interpolation for other spatial configurations.

As mentioned earlier, an alternative is the  $Y$-shape interaction of Fig.~\ref{fig:string1}, now used to estimate a potential which reads
\begin{equation}\label{eq:Y}
 V_Y^\text{conf}=\lambda \,\min_J(r_{1J}+r_{2J}+r_{3J})~,
\end{equation}
if $\lambda \,r_{12}$ is the linear confinement of mesons. Note that \eqref{eq:Y} features the minimal path linking three points, an old problem first raised by Fermat and Torricelli.\footnote{This minimal distance is also intimately linked to a theorem by Napoleon: by adding external equilateral triangle on each side of the quark triangle, one easily construct the location of the junction $J$ and the value of the minimal sum of distances~\cite{Ay:2009zp}.}\@ For baryons alone, solving the 3-quark problem with \eqref{eq:Y} instead of the perimeter-based confinement
\begin{equation}\label{eq:peri}
 V_P^\text{conf}=\frac{\lambda}{2}\,(r_{12}+r_{23}+r_{31})~,
\end{equation}
is technically more involved, but does not change the patterns of the excitation spectrum. 

\begin{figure}[ht!]
 \centering
 \includegraphics[scale=.8]{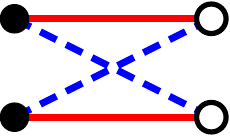}
\qquad
\includegraphics[scale=.8]{./strings-fig4.pdf}
 \caption{Potential for the tetraquark in a simple string model. The minimum is taken of the connected diagram (right) and the two possible meson-meson diagrams (left).}
 \label{fig:tetra-string}
\end{figure}

 For tetraquarks, a precursor of non-pairwise potential was the flip-flop model \cite{Lenz:1985jk}, with quadratic confinement. In the more recent linear version, one takes the minimum of the two possible meson-meson pairing configurations and the connected diagram shown in Fig.~\ref{fig:tetra-string}.

\begin{figure}[ht!]
 \centering
 \includegraphics[scale=.8]{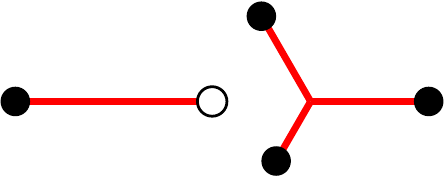}
\qquad
\includegraphics[scale=.8]{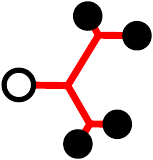}
 \caption{Some contributions to the pentaquark potential: for each set of quark positions, the minimum is taken of the connected diagrams such as the one on the right and  disconnected diagrams such as the one on the left.}
 \label{fig:penta-string}
\end{figure}

\begin{figure}[ht!]
 \centering
 \includegraphics[scale=.8]{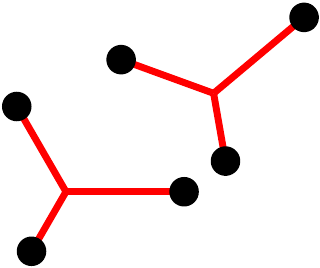}
\qquad
\includegraphics[scale=.8]{./strings-fig9.pdf}
 \caption{Some contributions to the hexaquark potential: for each set of quark positions, the minimum is taken of the connected diagrams such as the one on the right and  disconnected diagrams such as the one on the left.}
 \label{fig:hexa-string}
\end{figure}

\begin{figure}[ht!]
 \centering
 \raisebox{.5cm}{\includegraphics[scale=.8]{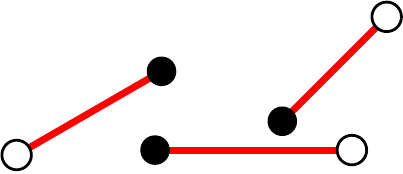}} 
\qquad
\includegraphics[scale=.8]{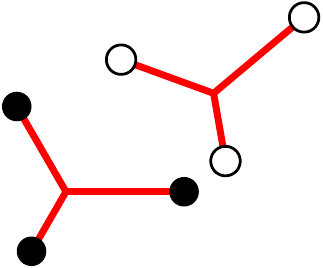}
\\
\raisebox{1cm}{\includegraphics[scale=.8]{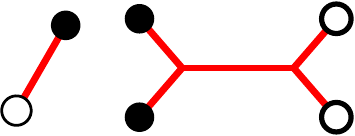}}
\qquad
\includegraphics[scale=.8,angle=90]{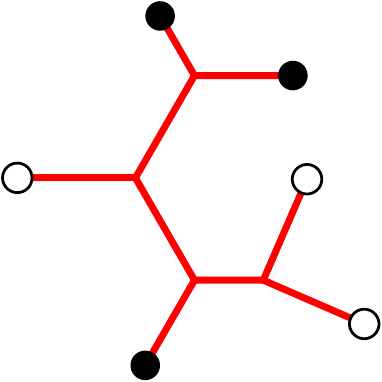}
 \caption{Some contributions to the  potential for 3 quarks and 3 antiquarks: for each set of quark positions, the minimum is taken of  connected diagrams (bottom, right) and  disconnected diagrams (top and bottom left).}
 \label{fig:3q-3qbar-string}
\end{figure}

The good surprise is that this potential is more attractive than the one deduced from the colour-additive rule \eqref{eq:col-add} and thus is more favourable for binding \cite{Vijande:2007ix,Ay:2009zp}. The same property is observed for the pentaquark \cite{Richard:2009rp}, Fig.~\ref{fig:penta-string}, the dibaryon, Fig.~\ref{fig:hexa-string} and the system made of 3 quarks and 3 antiquarks \cite{Vijande:2011im}, Fig.~\ref{fig:3q-3qbar-string}. However, this string potential corresponds to a kind of Born-Oppenheimer approximation for the colour fluxes at given quark positions. It mixes colour $\bar 3$ and $6$ for any pair of quark, to optimise the overall energy. Thus it holds for quarks bearing different flavours. In \cite{Vijande:2013qr}, an attempt is proposed to combine quark statistics and string dynamics in the case of tetraquarks with identical quarks or antiquarks: not surprisingly, the possibility of binding is reduced.

Note that in the above potential models, there is no explicit junction \textsl{\`a la} Rossi and Veneziano: the minimal length is the only guidance, and nothing refrain a transition from a disconnected flip-flop to a connected string. It is observed, however, that the ground-state uses mainly the flip-flop term, so there are probably two classes of states: the lowest have the same number of junctions as the thrshold, and thus are rather broad if the fall-apart decay is energetically allowed. The second class might be higher in mass, but benefits from a kind of metastability due to its junction content.

These string-inspired potentials, however simplistic, have motivated some studies. For instance, the literature on Steiner trees helps in optimising the location of the junctions and the estimate of the minimal strength~\cite{SilvestreBrac:2003sa,*Bicudo:2008yr,*Dmitrasinovic:2009dy,Ay:2009zp}.

\subsubsection{Combining chromoelectric and chromomagnetic effects}
As discussed at length in Sec.~\ref{sse:chromoelectric}, the tetraquark $(QQ\bar q\bar q)$ might become stable by the sole effect of chromomagnetic forces, but the stability requires a very large $M/m$ mass ratio.  It was also noticed \cite{Semay:1994ht,Janc:2004qn}, that some configurations such as  $J^{PC}=1^{++}$, there is a favourable spin-spin interaction in the light sector that benefits the tetraquark and not the threshold. A calculation by Rosina et al.\ \cite{Janc:2004qn}, using a potential tuned to fit the ordinary hadrons, and the colour-additive rule, demonstrated stability for $(cc\bar u\bar d)$, and this was confirmed by Vijande et al.~\cite{Vijande:2007rf}. 

There are not too many similar configurations, in which both chromoelectric and chromomagnetic effects are favourable. For instance, for the doubly-flavoured dibaryons $(QQqqqq)$, the two thresholds $(QQq)+(qqq)$ and $(Qqq)+(Qqq)$ are nearly degenerate, at least in case the charm, $Q=c$. The former threshold benefits from the chromoelectric attraction between the two heavy quarks, while the second can receive a favourable chromomagnetic interaction. It is, however, difficult for the 6-quark configuration to combine astutely the two effects, and double-charm baryons seem hardly stable in constituent models~\cite{Vijande:2016nzk}. 
\subsubsection{Solving the Schr\"odinger equation}
Once a model is adopted, typically
\begin{equation}\label{eq:gen-pot}
 V=-\frac{16}{3}\, \sum_{i<j}\ll{i}{j}\left[v(r_{ij)}+ v_{ss}(r_{ij})\,\ss{i}{j}\right]~,
\end{equation}
or a more complicated model involving multi-body forces, one is left with solving the wave equation to obtain the energies and the wavefunctions. 

In the case of quarkonium, there are dozens of reliable numerical methods, starting with the one proposed by Hartree \cite{Hartree:112712}, consisting of matching a regular outgoing solution and  a regular incoming solution of the radial equation. Most methods can be extended to the case of coupled equations in the unnatural-parity channels, similar to the $s-d$ coupling of the deuteron in nuclear physics. Many rigorous results have been revisited and developed for studies related to the quarkonium: scaling law, virial-like theorems, level order, value of wave-function at the origin for the successive radial excitations, etc.\@ This is reviewed in \cite{Quigg:1979vr,*Grosse:1979xm,*Grosse:847188}.

The 3-body problem one has to solve for baryons is often a psychological barrier that refrain to envisage a detailed and rigorous treatment beyond the harmonic oscillator, and there is sometimes a confusion between the ``constituent model of baryons'' and the ''harmonic oscillator model of baryons''. For instance, the negative-parity excitation being half-way of the positive-parity excitation is a property of the latter, but not necessarily of the former. 

For baryons, there are just a few rigorous results, for instance dealing the level order of states when the degeneracy obtained in the harmonic oscillator or hyperscalar limit are lifted at first order \cite{Richard:1989ra,*Stancu:1991cz}. As for the numerical solution, various methods have been used, such as hyperspherical expansion, Faddeev equations, or variational methods, with a comparison in \cite{Richard:1992uk}. Note that among the latter, the expansion on a harmonic-oscillator basis is less and less adopted, as it does not give good convergence for the short-range correlation parameters which enter some decays and some production mechanisms. 

For multiquark systems, the hyperspherical expansion works rather well \cite{Vijande:2007rf}, especially for deeply bound states, if any, for which the various r.m.s.\ radii are of the same order of magnitude. The tendency is towards the method of Gaussian expansion, already used in quantum chemistry. If $\tilde{\vec x}^\dagger=\{\vec x_1,\ldots, \vec{x}_{N-1}\}$ are Jacobi variable describing the relative motion of the $N$-body system, then the wave-function is sought as 
\begin{equation}
 \label{eq:Gauss-exp}
 \Psi=\sum_{a,i} \gamma_{a,i} \exp[-\tilde{\vec x}^\dagger.A_{a,i}.\tilde{\vec x}/2]\,|a\rangle~,
\end{equation}
where $|a\rangle$ denotes the basis of internal colour-spin-flavour states needed to construct the most general colour singlet of given overall spin and flavour, and $A_{a,i}$ is a $(N-1)\times(N-1)$ definite positive symmetric matrix. For a given set of matrices $A_{a,i}$, the weight factors $\gamma_{a,i}$ are given by a generalised eigenvalue equation. Then the range parameters entering the  $A_{a,i}$ are minimised numerically. This latter task is not obvious, and sophisticated techniques have been developed, such as the stochastic search\ \cite{RevModPhys.85.693}, and the method of Hiyama, Kino and Kamimura \cite{Hiyama:2003cu}, in which a specific choice of Jacobi variables is chosen in each channel. The variational energies below the threshold are interpreted as bound states. The scaling properties of the states above the threshold are scrutinised to decide whether they are genuine resonances or artefacts of the finite variational expansion. This is somewhat similar to the analyses of 
states 
above the threshold in QCDSR or LQCD.
\subsubsection{Limits and extensions}
\paragraph{Diquarks}
Hadron spectroscopy with diquarks has ups and downs, or, say, pros and cons. The art of physics is to identity effective degrees of freedom at a certain scale, and any simplification is welcome. The concept of diquark is very useful to describe many phenomenons such a multiple production \cite{Anselmino:1992vg}, and might be legitimated by boson-fermion symmetry \cite{Brodsky:2016rvj}. In spectroscopy, the diquark model provides a simple physical picture that might be obscured in complicated three-body calculations, and it provides an efficient tool for investigating exotics \cite{Lichtenberg:1997bq}. Some reservations are however expressed about the abundant literature on diquarks, which is sometimes confusing, as some recent papers do refer to the old ones:
\begin{itemize}
 \item Explicit quark model calculations find diquark correlations, but the effect depends on a somewhat arbitrary regularisation of the chromomagnetic interaction \cite{Grach:1993mj}. In other studies, no diquark clustering is found for the ground state \cite{Fleck:1988vm}.  This means that the diquark picture of low-lying baryons implies a change of dynamics. On the other hand, a dynamical diquark formation is obtained for highly excited baryons: if the total angular momentum $L$ increases, it is first shared equally among the 3 quarks, but at larger $L$ the third quark takes $\ell_y\sim L$ while the two first are mainly in a relative $s$-wave (of course modulo permutations) \cite{Martin:1985hw}.

 \item The diquark model becomes a good approximation for the ground state of double-charm baryons $(QQq)$ provided it is handled and interpreted properly, since a part of the $QQ$ interaction comes from the third quark. 
 In the case of the harmonic oscillator, with the usual definition of the Jacobi variables $\vec x=\vec r_2-\vec r_1$ and $\vec y=(2\,\vec r_3-\vec r_1-\vec r_2)/\sqrt3$, one notices that $\sum r_{ij}^2=(3/2)(x^2+y^2)$, so the direct $QQ$ interaction $x^2$ gets  a 50\% increase from the light quark. Moreover, the first excitations of $(QQq)$ (besides spin) occur with $(QQ)$, so a new diquark has to be estimated for each level. In fact a Born-Oppenheimer treatment \`a la H$_2{}^+$ is more appropriate. 
\end{itemize}
\paragraph{Born--Oppenheimer}
Along this review, we stressed the importance of the Born-Oppenheimer point of view: the quarkonium potential is just an effective interaction integrating out the gluon and light-quark degrees of freedom surrounding a heavy quark and a heavy antiquark; doubly-heavy baryons $(QQq)$ have typically two scales of energy for its spectrum, the excitation energy of the $QQ$ pair, and the excitation energy of the light quark, and this comes very naturally in the Born-Oppenheimer treatment, as it does in atomic physics for H$_2{}^+$.

We also mentioned that hybrid mesons, schematically noted $(Q\bar Q g)$ have been studied by estimating an ``excited quarkonium potential'': for a given $Q\bar Q$ separation, the energy of the excited gluon field is computed either from a simple bag model, or from LQCD. A Born-Oppenheimer treatment of $(QQ\bar q\bar q)$ would also be feasible, to cross-check the 4-body calculations discussed in Sec.~\ref{sse:chromoelectric}.

Recently, a very comprehensive and unifying Born-Oppenheimer treatment of the $XYZ$ states was presented \cite{Braaten:2014qka}, with a series of effective $Q\bar Q$ potentials adapted to the quantum states of the light constituents (gluon and light quarks). This work could of course be generalised as to include $qqq$ instead of $q\bar q$  in the light sector, and applied to the LHCb pentaquarks. 

There are interesting alternatives for the coupling of $Q\bar Q$ to light quarks and antiquarks. For instance, a  relatively weak interaction between $J/\psi$ or $\eta_c$ with nuclei has been proposed, but, once more, this potential is probed by a particle which is heavy enough to get bound. In \emph{hadrocharmonium}, the compact $Q\bar Q$ is trapped inside an extended light hadron. See \cite{Dubynskiy:2008mq,*Voloshin:2013dpa,Cleven:2015era}.

\section{Super-exotics}
In the future, the field of exotic hadrons might be extended very far, if some new constituents are discovered, for instance at LHC. As stressed by many authors (see, for instance, \cite{Wu:2015bfr}), some anomalies in the Higgs branching ratios, and the other recent observations at CERN might indicate the existence of heavy scalars. Among the many possible scenarios, there is the possibility of scalar quarks with colour charge $3$, having the same colour-mediated interaction as the ordinary quarks, except  of course for spin and statistics. Perhaps the lowest of theses scalars will live long enough to hadronise, and give rise to new hadrons. 

The lifetime of the hypothetical scalar quarks would  of course be crucial. Remember that for many years the  potential model of charmonium and bottomonium was extrapolated towards topomonium. Then it was realised that, when the mass of $t$ quark is too large,  its weak decay becomes too fast for hadronisation to take place~\cite{Bigi:1986jk}. If this is the case for any new super-heavy quark or squark, then the  realm of exotic hadrons will remain the privilege of low-energy hadron physics.

Let us consider a scenario with heavy scalar quarks of colour $3$, as predicted by supersymmetry or any other BSM (Beyond Standard Model) theory. (There are speculations about colour $6$ or other gauge groups.)\@ We note them ``squarks'' without necessarily endorsing supersymmetry. Then, one gets squarkonia ($\varsigma\bar{\varsigma}$), and perhaps the scalar ground state could be responsible for the excess tentatively seen at 750\,GeV. See, e.g., \cite{Luo:2015yio}. I apologise for the lack of space needed to cite the other $\gtrsim 1000$ relevant references, and the lack of expertise to separate the really innovative papers from the ``ambulance chasing'' opportunities \cite{Backovic:2016xno}.

The lightest charged hadron is likely $\underline{d}u$ for $\varsigma=\underline{d}$ with charge $-1/3$ or $\underline{u}ud$ of $\Lambda_Q$ type, but spin 0, if $\varsigma=\underline{u}$ with charge $2/3$.

The baryon sector becomes entertaining if several squarks are put together. For instance $(\varsigma\varsigma q)$ has quantum numbers $(1/2)^-$ as the $\varsigma\varsigma$ disquark contains one unit of orbital excitation, to fulfill the Bose statistics. The triply-superexotic $(\varsigma\varsigma\varsigma)$ requires a fully antisymmetric wave function $\propto\vec\rho\times \vec\lambda$, as per Eq.~\eqref{eq:wf1plus} that was proposed 50 years ago for the last member of the $N=2$ multiplet for light baryons, the so-called $[20,1^+]$ multiplet in the notation of, e.g.,~\cite{Hey:1982aj}. Fifty years after the first baryon conference, the physics has evolved, but one uses the same wave function (besides some scaling). 

\section{Outlook}
Exotic hadrons are coming up and down, or back and forth if one prefers. For outsiders, there is somewhat a discredit on a field in which many states regularly show up rapidly and then disappear within a few years, with many papers hastily written to explain the new hadrons, and never any deep understanding on why the models fail. 

This is of course, a superficial view. The truth is that the experiments are very difficult, due to a low statistics for a signal above an intricate background. The situation is even worse for broad resonances. As often stressed  \cite{Dytman:2004kk,Briceno:2015rlt}, the location of the peak might depend on the production process and on the final state, so that there is often the question on whether two or more candidates correspond to the same ``state''.  So far, several mass ranges and flavour sectors have been investigated with already a few good candidates for exotics such as $X(3872)$. The main lesson of the last decades is clearly the richness of the configurations mixing heavy and light constituents. The hidden-charm sector is already much explored, and even if the recent discoveries have to be checked, refined and completed, there is clearly an urgent need for investigating the double charm and other flavor exotic sectors. Hopefully, there will be an emulation between electron-positron machines, 
proton colliders, heavy-ion collisions and intense hadron facilities to develop this program.

On the theory side, the fundamental approaches such as LQCD or QCDSR involve lengthy calculations and usually come second, to clarify some predictions. First appear the speculations based on  more empirical models, inspired by QCD. 

In phenomenological QCD, \textsl{Arx tarpeia Capitoli proxima}:\footnote{``The Tarpeian Rock is close to the Capitol''. In ancient Rom, some prominent wining generals had a triumph procession, the Fifth-Avenue parade of that time, ending to the Capitol Temple, and the worse criminals were plunged to death from the nearby Tarpeian Rock. This Latin sentence is probably  posterior to the Roman period. I thank the historian Fran\c{c}ois Richard for a correspondence on this point.}\ Physicists inventing  molecules or quark clusters that account for the latest exotics are praised for their inspiration, invited to many workshops and conferences, hired in committees, etc.,  but, when the states disappear, the same model-builders are pointed out as lacking rigour and corrupting the field. To be honest, it often happens that the pioneers have written  their predictions rather carefully, with many warnings about the width of states in 
the continuum, but that the followers have developed the models beyond any reasonable control, with tables and tables of states which simply fall part.

For one of the models of multiquarks, a warning was set \cite{Jaffe:1978bu} that only a few states will show up as peaks, the others, being very broad, just paving the continuum, but the warning was often forgotten. This \emph{Pandora box syndrome} affects practically all approaches. For instance, the molecular picture was first focused on one-pion exchange, with strict selection rules, that makes the interaction repulsive in any second channel. With the exchange of vector mesons, and the old-fashioned potentials replaced by effective theories, the molecular picture has gained some respectability, but has dramatically widened the spectrum of predictions.   In the diquark model, Frederikson and Jandel  \cite{Fredriksson:1981mh} discussed the existence of possible dibaryons, and the problem was rediscovered recently \cite{Maiani:2015iaa}, that the picture of $XYZ$ as diquark-antidiquark states probably implies higher configurations. This is why genuinely exotic states, lying below their dissociation threshold, 
will always play a privileged role, and the  searches in the sector of double heavy flavour should be given the highest priority.


In any case, the studies on  the mechanisms of confinement and the resulting nuclear forces should be pursued.  The strong-coupling limit of QCD   and lattice simulations explain how a linear confinement is developed between a quark and an antiquark. It is already more delicate to understand its extension to baryons, with possible interplay between $\Delta$-shape and $Y$-shape configurations for the gluon density, and even more speculative to write down a confining Hamiltonian for multiquarks. What remain to be further investigated, based on the already rich literature is how clustering sometimes occur in baryons and higher multiquarks, and how the linear confinement breaks to lead to the actual decay of resonances and the virtual hadron-hadron components, i.e., to operate the matching of the quark dynamics and the long-range hadron-hadron interaction.


\begin{acknowledgments}
This review benefited from stimulating discussions with my past and present collaborators and with many colleagues. It was completed during a visit at RIKEN, in the framework of a JSPS-CNRS agreement, and finalised during a visit at ECT*. The hospitality provided by Emiko Hiyama,  by Jochen Wambach,  and by their colleagues is gratefully acknowledged. Thanks are also due to M.~Asghar for very interesting comments on the manuscript.
\end{acknowledgments}


\begin{thebibliography}{300}%
\makeatletter
\providecommand \@ifxundefined [1]{%
 \@ifx{#1\undefined}
}%
\providecommand \@ifnum [1]{%
 \ifnum #1\expandafter \@firstoftwo
 \else \expandafter \@secondoftwo
 \fi
}%
\providecommand \@ifx [1]{%
 \ifx #1\expandafter \@firstoftwo
 \else \expandafter \@secondoftwo
 \fi
}%
\providecommand \natexlab [1]{#1}%
\providecommand \enquote  [1]{``#1''}%
\providecommand \bibnamefont  [1]{#1}%
\providecommand \bibfnamefont [1]{#1}%
\providecommand \citenamefont [1]{#1}%
\providecommand \href@noop [0]{\@secondoftwo}%
\providecommand \href [0]{\begingroup \@sanitize@url \@href}%
\providecommand \@href[1]{\@@startlink{#1}\@@href}%
\providecommand \@@href[1]{\endgroup#1\@@endlink}%
\providecommand \@sanitize@url [0]{\catcode `\\12\catcode `\$12\catcode
  `\&12\catcode `\#12\catcode `\^12\catcode `\_12\catcode `\%12\relax}%
\providecommand \@@startlink[1]{}%
\providecommand \@@endlink[0]{}%
\providecommand \url  [0]{\begingroup\@sanitize@url \@url }%
\providecommand \@url [1]{\endgroup\@href {#1}{\urlprefix }}%
\providecommand \urlprefix  [0]{URL }%
\providecommand \Eprint [0]{\href }%
\providecommand \doibase [0]{http://dx.doi.org/}%
\providecommand \selectlanguage [0]{\@gobble}%
\providecommand \bibinfo  [0]{\@secondoftwo}%
\providecommand \bibfield  [0]{\@secondoftwo}%
\providecommand \translation [1]{[#1]}%
\providecommand \BibitemOpen [0]{}%
\providecommand \bibitemStop [0]{}%
\providecommand \bibitemNoStop [0]{.\EOS\space}%
\providecommand \EOS [0]{\spacefactor3000\relax}%
\providecommand \BibitemShut  [1]{\csname bibitem#1\endcsname}%
\let\auto@bib@innerbib\@empty
\bibitem [{\citenamefont {Segr{\`e}}(1980)}]{Segre:100978}%
  \BibitemOpen
  \bibfield  {author} {\bibinfo {author} {\bibfnamefont {Emilio}\ \bibnamefont
  {Segr{\`e}}},\ }\href {http://cds.cern.ch/record/100978} {\emph {\bibinfo
  {title} {{From X-rays to quarks: modern physicists and their discoveries}}}}\
  (\bibinfo  {publisher} {Freeman},\ \bibinfo {address} {San Francisco, CA},\
  \bibinfo {year} {1980})\ \bibinfo {note} {trans. of : Personaggi e scoperte
  nella fisica contemporanea. Milano : Mondadori, 1976}\BibitemShut {NoStop}%
\bibitem [{\citenamefont {Pais}(1986)}]{Pais:107749}%
  \BibitemOpen
  \bibfield  {author} {\bibinfo {author} {\bibfnamefont {Abraham}\ \bibnamefont
  {Pais}},\ }\href {http://cds.cern.ch/record/107749} {\emph {\bibinfo {title}
  {{Inward bound: of matter and forces in the physical world}}}}\ (\bibinfo
  {publisher} {Clarendon Press},\ \bibinfo {address} {Oxford},\ \bibinfo {year}
  {1986})\BibitemShut {NoStop}%
\bibitem [{\citenamefont {Ezhela}\ \emph {et~al.}(1996)\citenamefont {Ezhela},
  \citenamefont {Filimonov}, \citenamefont {Lugovsky}, \citenamefont
  {Polishchuk}, \citenamefont {Striganov}, \citenamefont {Stroganov},
  \citenamefont {Armstrong}, \citenamefont {Barnett}, \citenamefont {Groom},
  \citenamefont {Gee}, \citenamefont {Trippe}, \citenamefont {Wohl},\ and\
  \citenamefont {Jackson}}]{Ezhela:317696}%
  \BibitemOpen
  \bibfield  {author} {\bibinfo {author} {\bibfnamefont {Vladimir~V}\
  \bibnamefont {Ezhela}}, \bibinfo {author} {\bibfnamefont {B~B}\ \bibnamefont
  {Filimonov}}, \bibinfo {author} {\bibfnamefont {S~B}\ \bibnamefont
  {Lugovsky}}, \bibinfo {author} {\bibfnamefont {B~V}\ \bibnamefont
  {Polishchuk}}, \bibinfo {author} {\bibfnamefont {S~I}\ \bibnamefont
  {Striganov}}, \bibinfo {author} {\bibfnamefont {Y~G}\ \bibnamefont
  {Stroganov}}, \bibinfo {author} {\bibfnamefont {Betty}\ \bibnamefont
  {Armstrong}}, \bibinfo {author} {\bibfnamefont {Richard~Michael}\
  \bibnamefont {Barnett}}, \bibinfo {author} {\bibfnamefont {D~E}\ \bibnamefont
  {Groom}}, \bibinfo {author} {\bibfnamefont {P~S}\ \bibnamefont {Gee}},
  \bibinfo {author} {\bibfnamefont {Thomas~G}\ \bibnamefont {Trippe}}, \bibinfo
  {author} {\bibfnamefont {Charles~G}\ \bibnamefont {Wohl}}, \ and\ \bibinfo
  {author} {\bibfnamefont {John~David}\ \bibnamefont {Jackson}},\ }\href
  {http://cds.cern.ch/record/317696} {\emph {\bibinfo {title} {{Particle
  physics, one hundred years of discoveries: an annotated chronological
  bibliography}}}}\ (\bibinfo  {publisher} {AIP},\ \bibinfo {address} {New
  York, NY},\ \bibinfo {year} {1996})\BibitemShut {NoStop}%
\bibitem [{\citenamefont {Breit}(1962)}]{Breit:1962zz}%
  \BibitemOpen
  \bibfield  {author} {\bibinfo {author} {\bibfnamefont {G.}~\bibnamefont
  {Breit}},\ }\bibfield  {title} {\enquote {\bibinfo {title} {{Aspects of
  Nucleon-Nucleon Scattering Theory}},}\ }\href {\doibase
  10.1103/RevModPhys.34.766} {\bibfield  {journal} {\bibinfo  {journal} {Rev.
  Mod. Phys.}\ }\textbf {\bibinfo {volume} {34}},\ \bibinfo {pages} {766--812}
  (\bibinfo {year} {1962})}\BibitemShut {NoStop}%
\bibitem [{\citenamefont {Chew}\ and\ \citenamefont {Low}(1956)}]{Chew:1955zz}%
  \BibitemOpen
  \bibfield  {author} {\bibinfo {author} {\bibfnamefont {G.~F.}\ \bibnamefont
  {Chew}}\ and\ \bibinfo {author} {\bibfnamefont {F.~E.}\ \bibnamefont {Low}},\
  }\bibfield  {title} {\enquote {\bibinfo {title} {{Effective range approach to
  the low-energy p wave pion - nucleon interaction}},}\ }\href {\doibase
  10.1103/PhysRev.101.1570} {\bibfield  {journal} {\bibinfo  {journal} {Phys.
  Rev.}\ }\textbf {\bibinfo {volume} {101}},\ \bibinfo {pages} {1570--1579}
  (\bibinfo {year} {1956})}\BibitemShut {NoStop}%
\bibitem [{\citenamefont {Chew}(1966)}]{Chew:107781}%
  \BibitemOpen
  \bibfield  {author} {\bibinfo {author} {\bibfnamefont {Geoffrey~Foucar}\
  \bibnamefont {Chew}},\ }\href {http://cds.cern.ch/record/107781} {\emph
  {\bibinfo {title} {{The analytic S-matrix: a basis for nuclear democracy}}}}\
  (\bibinfo  {publisher} {Benjamin},\ \bibinfo {address} {New York, NY},\
  \bibinfo {year} {1966})\BibitemShut {NoStop}%
\bibitem [{\citenamefont {Ball}\ \emph {et~al.}(1966)\citenamefont {Ball},
  \citenamefont {Scotti},\ and\ \citenamefont {Wong}}]{Ball:1965sa}%
  \BibitemOpen
  \bibfield  {author} {\bibinfo {author} {\bibfnamefont {James~S.}\
  \bibnamefont {Ball}}, \bibinfo {author} {\bibfnamefont {Antonio}\
  \bibnamefont {Scotti}}, \ and\ \bibinfo {author} {\bibfnamefont {David~Y.}\
  \bibnamefont {Wong}},\ }\bibfield  {title} {\enquote {\bibinfo {title}
  {{One-boson-exchange model of $NN$ and $N\bar N$ interaction}},}\ }\href
  {\doibase 10.1103/PhysRev.142.1000} {\bibfield  {journal} {\bibinfo
  {journal} {Phys. Rev.}\ }\textbf {\bibinfo {volume} {142}},\ \bibinfo {pages}
  {1000--1012} (\bibinfo {year} {1966})}\BibitemShut {NoStop}%
\bibitem [{\citenamefont {Olive}\ \emph {et~al.}(2014)\citenamefont {Olive}
  \emph {et~al.}}]{Agashe:2014kda}%
  \BibitemOpen
  \bibfield  {author} {\bibinfo {author} {\bibfnamefont {K.~A.}\ \bibnamefont
  {Olive}} \emph {et~al.} (\bibinfo {collaboration} {Particle Data Group}),\
  }\bibfield  {title} {\enquote {\bibinfo {title} {{Review of Particle
  Physics}},}\ }\href {\doibase 10.1088/1674-1137/38/9/090001} {\bibfield
  {journal} {\bibinfo  {journal} {Chin. Phys.}\ }\textbf {\bibinfo {volume}
  {C38}},\ \bibinfo {pages} {090001} (\bibinfo {year} {2014})}\BibitemShut
  {NoStop}%
\bibitem [{\citenamefont {Montanet}\ \emph {et~al.}(1980)\citenamefont
  {Montanet}, \citenamefont {Rossi},\ and\ \citenamefont
  {Veneziano}}]{Montanet:1980te}%
  \BibitemOpen
  \bibfield  {author} {\bibinfo {author} {\bibfnamefont {L.}~\bibnamefont
  {Montanet}}, \bibinfo {author} {\bibfnamefont {G.~C.}\ \bibnamefont {Rossi}},
  \ and\ \bibinfo {author} {\bibfnamefont {G.}~\bibnamefont {Veneziano}},\
  }\bibfield  {title} {\enquote {\bibinfo {title} {Baryonium physics},}\
  }\href@noop {} {\bibfield  {journal} {\bibinfo  {journal} {Phys. Rept.}\
  }\textbf {\bibinfo {volume} {63}},\ \bibinfo {pages} {149--222} (\bibinfo
  {year} {1980})}\BibitemShut {NoStop}%
\bibitem [{\citenamefont {Amsler}\ and\ \citenamefont
  {Tornqvist}(2004)}]{Amsler:2004ps}%
  \BibitemOpen
  \bibfield  {author} {\bibinfo {author} {\bibfnamefont {C.}~\bibnamefont
  {Amsler}}\ and\ \bibinfo {author} {\bibfnamefont {N.~A.}\ \bibnamefont
  {Tornqvist}},\ }\bibfield  {title} {\enquote {\bibinfo {title} {Mesons beyond
  the naive quark model},}\ }\href@noop {} {\bibfield  {journal} {\bibinfo
  {journal} {Phys. Rept.}\ }\textbf {\bibinfo {volume} {389}},\ \bibinfo
  {pages} {61--117} (\bibinfo {year} {2004})}\BibitemShut {NoStop}%
\bibitem [{\citenamefont {Klempt}\ and\ \citenamefont
  {Zaitsev}(2007)}]{Klempt:2007cp}%
  \BibitemOpen
  \bibfield  {author} {\bibinfo {author} {\bibfnamefont {Eberhard}\
  \bibnamefont {Klempt}}\ and\ \bibinfo {author} {\bibfnamefont {Alexander}\
  \bibnamefont {Zaitsev}},\ }\bibfield  {title} {\enquote {\bibinfo {title}
  {{Glueballs, Hybrids, Multiquarks. Experimental facts versus QCD inspired
  concepts}},}\ }\href {\doibase 10.1016/j.physrep.2007.07.006} {\bibfield
  {journal} {\bibinfo  {journal} {Phys. Rept.}\ }\textbf {\bibinfo {volume}
  {454}},\ \bibinfo {pages} {1--202} (\bibinfo {year} {2007})},\ \Eprint
  {http://arxiv.org/abs/0708.4016} {arXiv:0708.4016 [hep-ph]} \BibitemShut
  {NoStop}%
\bibitem [{\citenamefont {Dalpiaz}\ \emph {et~al.}(1987)\citenamefont
  {Dalpiaz}, \citenamefont {Klapisch}, \citenamefont {Lefevre}, \citenamefont
  {Macri}, \citenamefont {Montanet}, \citenamefont {Mohl}, \citenamefont
  {Martin}, \citenamefont {Richard}, \citenamefont {Pirner},\ and\
  \citenamefont {Tecchio}}]{Dalpiaz:1987qk}%
  \BibitemOpen
  \bibfield  {author} {\bibinfo {author} {\bibfnamefont {P.}~\bibnamefont
  {Dalpiaz}}, \bibinfo {author} {\bibfnamefont {R.}~\bibnamefont {Klapisch}},
  \bibinfo {author} {\bibfnamefont {P.}~\bibnamefont {Lefevre}}, \bibinfo
  {author} {\bibfnamefont {M.}~\bibnamefont {Macri}}, \bibinfo {author}
  {\bibfnamefont {L.}~\bibnamefont {Montanet}}, \bibinfo {author}
  {\bibfnamefont {Dieter}\ \bibnamefont {Mohl}}, \bibinfo {author}
  {\bibfnamefont {Andr{\'e}}\ \bibnamefont {Martin}}, \bibinfo {author}
  {\bibfnamefont {J.-M.}\ \bibnamefont {Richard}}, \bibinfo {author}
  {\bibfnamefont {H.J.}\ \bibnamefont {Pirner}}, \ and\ \bibinfo {author}
  {\bibfnamefont {L.}~\bibnamefont {Tecchio}},\ }\bibfield  {title} {\enquote
  {\bibinfo {title} {{Physics at SuperLEAR}},}\ }in\ \href
  {http://alice.cern.ch/format/showfull?sysnb=0105412} {\emph {\bibinfo
  {booktitle} {{Les Houches Workshop 1987:0414}}}}\ (\bibinfo {year} {1987})\
  p.\ \bibinfo {pages} {0414}\BibitemShut {NoStop}%
\bibitem [{\citenamefont {Amsler}\ \emph {et~al.}(1997)\citenamefont {Amsler}
  \emph {et~al.}}]{Amsler:1997nf}%
  \BibitemOpen
  \bibfield  {author} {\bibinfo {author} {\bibfnamefont {Claude}\ \bibnamefont
  {Amsler}} \emph {et~al.},\ }\bibfield  {title} {\enquote {\bibinfo {title}
  {{Perspectives in hadron and quark dynamics}},}\ }\bibfield  {booktitle}
  {\emph {\bibinfo {booktitle} {{Prospects of hadron and quark physics with
  electromagnetic probes. Proceedings, 2nd ELFE Workshop on Hadronic Physics,
  Saint Malo, France, September 23-27, 1996}}},\ }\href {\doibase
  10.1016/S0375-9474(97)00348-5} {\bibfield  {journal} {\bibinfo  {journal}
  {Nucl. Phys.}\ }\textbf {\bibinfo {volume} {A622}},\ \bibinfo {pages}
  {315C--354C} (\bibinfo {year} {1997})}\BibitemShut {NoStop}%
\bibitem [{\citenamefont {von Harrach}(1998)}]{vonHarrach:1998xz}%
  \BibitemOpen
  \bibfield  {author} {\bibinfo {author} {\bibfnamefont {D.}~\bibnamefont {von
  Harrach}} (\bibinfo {collaboration} {COMPASS}),\ }\bibfield  {title}
  {\enquote {\bibinfo {title} {{The COMPASS experiment at CERN}},}\ }\bibfield
  {booktitle} {\emph {\bibinfo {booktitle} {{Quark lepton nuclear physics.
  Proceedings, International Conference, QULEN'97, Osaka, Japan, May 20-23,
  1997}}},\ }\href {\doibase 10.1016/S0375-9474(97)00694-5} {\bibfield
  {journal} {\bibinfo  {journal} {Nucl. Phys.}\ }\textbf {\bibinfo {volume}
  {A629}},\ \bibinfo {pages} {245C--254C} (\bibinfo {year} {1998})}\BibitemShut
  {NoStop}%
\bibitem [{\citenamefont {Destefanis}(2013)}]{Destefanis:2013xpa}%
  \BibitemOpen
  \bibfield  {author} {\bibinfo {author} {\bibfnamefont {M.}~\bibnamefont
  {Destefanis}} (\bibinfo {collaboration} {PANDA}),\ }\bibfield  {title}
  {\enquote {\bibinfo {title} {{The PANDA experiment at FAIR}},}\ }\bibfield
  {booktitle} {\emph {\bibinfo {booktitle} {{Proceedings, 7th Joint
  International Hadron Structure'13 Conference (HS 13)}}},\ }\href {\doibase
  10.1016/j.nuclphysbps.2013.10.040} {\bibfield  {journal} {\bibinfo  {journal}
  {Nucl. Phys. Proc. Suppl.}\ }\textbf {\bibinfo {volume} {245}},\ \bibinfo
  {pages} {199--206} (\bibinfo {year} {2013})}\BibitemShut {NoStop}%
\bibitem [{\citenamefont {Swanson}(2006)}]{Swanson:2006st}%
  \BibitemOpen
  \bibfield  {author} {\bibinfo {author} {\bibfnamefont {Eric~S.}\ \bibnamefont
  {Swanson}},\ }\bibfield  {title} {\enquote {\bibinfo {title} {The new heavy
  mesons: A status report},}\ }\href@noop {} {\bibfield  {journal} {\bibinfo
  {journal} {Phys. Rep.}\ }\textbf {\bibinfo {volume} {429}},\ \bibinfo {pages}
  {243--405} (\bibinfo {year} {2006})},\ \Eprint
  {http://arxiv.org/abs/hep-ph/0601110} {hep-ph/0601110} \BibitemShut {NoStop}%
\bibitem [{\citenamefont {Nielsen}\ \emph {et~al.}(2010)\citenamefont
  {Nielsen}, \citenamefont {Navarra},\ and\ \citenamefont
  {Lee}}]{Nielsen:2009uh}%
  \BibitemOpen
  \bibfield  {author} {\bibinfo {author} {\bibfnamefont {Marina}\ \bibnamefont
  {Nielsen}}, \bibinfo {author} {\bibfnamefont {Fernando~S.}\ \bibnamefont
  {Navarra}}, \ and\ \bibinfo {author} {\bibfnamefont {Su~Houng}\ \bibnamefont
  {Lee}},\ }\bibfield  {title} {\enquote {\bibinfo {title} {{New Charmonium
  States in QCD Sum Rules: A Concise Review}},}\ }\href {\doibase
  10.1016/j.physrep.2010.07.005} {\bibfield  {journal} {\bibinfo  {journal}
  {Phys.Rept.}\ }\textbf {\bibinfo {volume} {497}},\ \bibinfo {pages} {41--83}
  (\bibinfo {year} {2010})},\ \Eprint {http://arxiv.org/abs/0911.1958}
  {arXiv:0911.1958 [hep-ph]} \BibitemShut {NoStop}%
\bibitem [{\citenamefont {Hambrock}(2013)}]{Hambrock:2013tpa}%
  \BibitemOpen
  \bibfield  {author} {\bibinfo {author} {\bibfnamefont {Christian}\
  \bibnamefont {Hambrock}},\ }\bibfield  {title} {\enquote {\bibinfo {title}
  {{Exotic Heavy Quark Spectroscopy: Theory Interpretation vs.\ Data}},}\
  }\bibfield  {booktitle} {\emph {\bibinfo {booktitle} {{Proceedings, 14th
  International Conference on B-Physics at Hadron Machines (Beauty 2013)}}},\
  }\href@noop {} {\bibfield  {journal} {\bibinfo  {journal} {PoS}\ }\textbf
  {\bibinfo {volume} {Beauty2013}},\ \bibinfo {pages} {044} (\bibinfo {year}
  {2013})},\ \Eprint {http://arxiv.org/abs/1306.0695} {arXiv:1306.0695
  [hep-ph]} \BibitemShut {NoStop}%
\bibitem [{\citenamefont {Chen}\ \emph {et~al.}(2016)\citenamefont {Chen},
  \citenamefont {Chen}, \citenamefont {Liu},\ and\ \citenamefont
  {Zhu}}]{Chen:2016qju}%
  \BibitemOpen
  \bibfield  {author} {\bibinfo {author} {\bibfnamefont {Hua-Xing}\
  \bibnamefont {Chen}}, \bibinfo {author} {\bibfnamefont {Wei}\ \bibnamefont
  {Chen}}, \bibinfo {author} {\bibfnamefont {Xiang}\ \bibnamefont {Liu}}, \
  and\ \bibinfo {author} {\bibfnamefont {Shi-Lin}\ \bibnamefont {Zhu}},\
  }\bibfield  {title} {\enquote {\bibinfo {title} {{The hidden-charm pentaquark
  and tetraquark states}},}\ }\href {\doibase 10.1016/j.physrep.2016.05.004} {\
   (\bibinfo {year} {2016}),\ 10.1016/j.physrep.2016.05.004},\ \Eprint
  {http://arxiv.org/abs/1601.02092} {arXiv:1601.02092 [hep-ph]} \BibitemShut
  {NoStop}%
\bibitem [{\citenamefont {Narison}(2004)}]{narison2004qcd}%
  \BibitemOpen
  \bibfield  {author} {\bibinfo {author} {\bibfnamefont {S.}~\bibnamefont
  {Narison}},\ }\href {https://books.google.fr/books?id=r\_27t1yxXqAC} {\emph
  {\bibinfo {title} {QCD as a Theory of Hadrons: From Partons to
  Confinement}}},\ Cambridge Monographs on Particle Physics, Nuclear Physics
  and Cosmology\ (\bibinfo  {publisher} {Cambridge University Press},\ \bibinfo
  {year} {2004})\BibitemShut {NoStop}%
\bibitem [{\citenamefont {Pennington}(2016)}]{Proceedings:2016efk}%
  \BibitemOpen
  \bibinfo {editor} {\bibfnamefont {Michael~R.}\ \bibnamefont {Pennington}},\
  ed.,\ \href {http://scitation.aip.org/content/aip/proceeding/aipcp/1735}
  {\emph {\bibinfo {title} {{Proceedings, 16th International Conference on
  Hadron Spectroscopy (Hadron 2015)}}}},\ Vol.\ \bibinfo {volume} {1735}\
  (\bibinfo {year} {2016})\BibitemShut {NoStop}%
\bibitem [{\citenamefont {Bicudo}\ \emph {et~al.}(2015)\citenamefont {Bicudo},
  \citenamefont {Dubnicka}, \citenamefont {Giacosa}, \citenamefont {Kaminski},\
  and\ \citenamefont {Marinkovic}}]{Proceedings:2015era}%
  \BibitemOpen
  \bibinfo {editor} {\bibfnamefont {Pedro}\ \bibnamefont {Bicudo}}, \bibinfo
  {editor} {\bibfnamefont {Stanislav}\ \bibnamefont {Dubnicka}}, \bibinfo
  {editor} {\bibfnamefont {Francesco}\ \bibnamefont {Giacosa}}, \bibinfo
  {editor} {\bibfnamefont {Robert}\ \bibnamefont {Kaminski}}, \ and\ \bibinfo
  {editor} {\bibfnamefont {Marina~Krstic}\ \bibnamefont {Marinkovic}},\ eds.,\
  \href {http://www.actaphys.uj.edu.pl/findarticle?series=Sup&vol=8&page=503}
  {\emph {\bibinfo {title} {{Proceedings, International Meeting of Excited QCD
  2015}}}},\ Vol.~\bibinfo {volume} {8}\ (\bibinfo {year} {2015})\BibitemShut
  {NoStop}%
\bibitem [{bar()}]{baryon2016}%
  \BibitemOpen
  \href@noop {} {}\bibinfo {note}
  {\url{https://baryons2016.physics.fsu.edu/indico/event/0/contributions}}\BibitemShut
  {NoStop}%
\bibitem [{\citenamefont {Achard}\ \emph {et~al.}(2005)\citenamefont {Achard}
  \emph {et~al.}}]{Achard:2005pb}%
  \BibitemOpen
  \bibfield  {author} {\bibinfo {author} {\bibfnamefont {P.}~\bibnamefont
  {Achard}} \emph {et~al.} (\bibinfo {collaboration} {L3}),\ }\bibfield
  {title} {\enquote {\bibinfo {title} {{Measurement of exclusive $\rho^+
  \rho^-$ production in mid-virtuality two-photon interactions and study of the
  $\gamma\gamma^*\to\rho\rho$ process at LEP}},}\ }\href {\doibase
  10.1016/j.physletb.2005.04.011} {\bibfield  {journal} {\bibinfo  {journal}
  {Phys. Lett.}\ }\textbf {\bibinfo {volume} {B615}},\ \bibinfo {pages}
  {19--30} (\bibinfo {year} {2005})},\ \Eprint
  {http://arxiv.org/abs/hep-ex/0504016} {arXiv:hep-ex/0504016 [hep-ex]}
  \BibitemShut {NoStop}%
\bibitem [{\citenamefont {Achard}\ \emph {et~al.}(2003)\citenamefont {Achard}
  \emph {et~al.}}]{Achard:2003qa}%
  \BibitemOpen
  \bibfield  {author} {\bibinfo {author} {\bibfnamefont {P.}~\bibnamefont
  {Achard}} \emph {et~al.} (\bibinfo {collaboration} {L3}),\ }\bibfield
  {title} {\enquote {\bibinfo {title} {{Measurement of exclusive $\rho^0
  \rho^0$ production in two photon collisions at high $Q^{2}$ at LEP}},}\
  }\href {\doibase 10.1016/j.physletb.2003.05.003} {\bibfield  {journal}
  {\bibinfo  {journal} {Phys. Lett.}\ }\textbf {\bibinfo {volume} {B568}},\
  \bibinfo {pages} {11--22} (\bibinfo {year} {2003})},\ \Eprint
  {http://arxiv.org/abs/hep-ex/0305082} {arXiv:hep-ex/0305082 [hep-ex]}
  \BibitemShut {NoStop}%
\bibitem [{\citenamefont {Anikin}\ \emph {et~al.}(2005)\citenamefont {Anikin},
  \citenamefont {Pire},\ and\ \citenamefont {Teryaev}}]{Anikin:2005ur}%
  \BibitemOpen
  \bibfield  {author} {\bibinfo {author} {\bibfnamefont {I.~V.}\ \bibnamefont
  {Anikin}}, \bibinfo {author} {\bibfnamefont {B.}~\bibnamefont {Pire}}, \ and\
  \bibinfo {author} {\bibfnamefont {O.~V.}\ \bibnamefont {Teryaev}},\
  }\bibfield  {title} {\enquote {\bibinfo {title} {{Search for isotensor exotic
  meson and twist-4 contribution to $\gamma^* \gamma\to\rho\rho$}},}\ }\href
  {\doibase 10.1016/j.physletb.2005.08.113} {\bibfield  {journal} {\bibinfo
  {journal} {Phys. Lett.}\ }\textbf {\bibinfo {volume} {B626}},\ \bibinfo
  {pages} {86--94} (\bibinfo {year} {2005})},\ \Eprint
  {http://arxiv.org/abs/hep-ph/0506277} {arXiv:hep-ph/0506277 [hep-ph]}
  \BibitemShut {NoStop}%
\bibitem [{\citenamefont {Dorofeev}\ \emph {et~al.}(2002)\citenamefont
  {Dorofeev} \emph {et~al.}}]{Dorofeev:2001xu}%
  \BibitemOpen
  \bibfield  {author} {\bibinfo {author} {\bibfnamefont {Valery}\ \bibnamefont
  {Dorofeev}} \emph {et~al.} (\bibinfo {collaboration} {VES}),\ }\bibfield
  {title} {\enquote {\bibinfo {title} {{The $J^{PC} = 1^{-+}$ hunting season at
  VES}},}\ }\bibfield  {booktitle} {\emph {\bibinfo {booktitle} {{Proceedings,
  9th International Conference on Hadron Spectroscopy (Hadron 2001)}}},\ }\href
  {\doibase 10.1063/1.1482444} {\bibfield  {journal} {\bibinfo  {journal} {AIP
  Conf. Proc.}\ }\textbf {\bibinfo {volume} {619}},\ \bibinfo {pages}
  {143--154} (\bibinfo {year} {2002})},\ \bibinfo {note} {[,143(2001)]},\
  \Eprint {http://arxiv.org/abs/hep-ex/0110075} {arXiv:hep-ex/0110075 [hep-ex]}
  \BibitemShut {NoStop}%
\bibitem [{\citenamefont {Thompson}\ \emph {et~al.}(1997)\citenamefont
  {Thompson} \emph {et~al.}}]{Thompson:1997bs}%
  \BibitemOpen
  \bibfield  {author} {\bibinfo {author} {\bibfnamefont {D.~R.}\ \bibnamefont
  {Thompson}} \emph {et~al.} (\bibinfo {collaboration} {E852}),\ }\bibfield
  {title} {\enquote {\bibinfo {title} {{Evidence for exotic meson production in
  the reaction pi- p ---> eta pi- p at 18-GeV/c}},}\ }\href {\doibase
  10.1103/PhysRevLett.79.1630} {\bibfield  {journal} {\bibinfo  {journal}
  {Phys. Rev. Lett.}\ }\textbf {\bibinfo {volume} {79}},\ \bibinfo {pages}
  {1630--1633} (\bibinfo {year} {1997})},\ \Eprint
  {http://arxiv.org/abs/hep-ex/9705011} {arXiv:hep-ex/9705011 [hep-ex]}
  \BibitemShut {NoStop}%
\bibitem [{\citenamefont {Alekseev}\ \emph {et~al.}(2010)\citenamefont
  {Alekseev} \emph {et~al.}}]{Alekseev:2009aa}%
  \BibitemOpen
  \bibfield  {author} {\bibinfo {author} {\bibfnamefont {M.}~\bibnamefont
  {Alekseev}} \emph {et~al.} (\bibinfo {collaboration} {COMPASS}),\ }\bibfield
  {title} {\enquote {\bibinfo {title} {{Observation of a $J^{PC} = 1^{-+}$
  exotic resonance in diffractive dissociation of 190\,GeV/$c$ $\pi^-$ into
  $\pi^-\pi^- \pi^+$}},}\ }\href {\doibase 10.1103/PhysRevLett.104.241803}
  {\bibfield  {journal} {\bibinfo  {journal} {Phys. Rev. Lett.}\ }\textbf
  {\bibinfo {volume} {104}},\ \bibinfo {pages} {241803} (\bibinfo {year}
  {2010})},\ \Eprint {http://arxiv.org/abs/0910.5842} {arXiv:0910.5842
  [hep-ex]} \BibitemShut {NoStop}%
\bibitem [{\citenamefont {Ghoul}\ \emph {et~al.}(2015)\citenamefont {Ghoul}
  \emph {et~al.}}]{Ghoul:2015ifw}%
  \BibitemOpen
  \bibfield  {author} {\bibinfo {author} {\bibfnamefont {H.~Al}\ \bibnamefont
  {Ghoul}} \emph {et~al.} (\bibinfo {collaboration} {GlueX}),\ }\bibfield
  {title} {\enquote {\bibinfo {title} {{First Results from The GlueX
  Experiment}},}\ }in\ \href
  {https://inspirehep.net/record/1409299/files/arXiv:1512.03699.pdf} {\emph
  {\bibinfo {booktitle} {{16th International Conference on Hadron Spectroscopy
  (Hadron 2015) Newport News, Virginia, USA, September 13-18, 2015}}}}\
  (\bibinfo {year} {2015})\ \Eprint {http://arxiv.org/abs/1512.03699}
  {arXiv:1512.03699 [nucl-ex]} \BibitemShut {NoStop}%
\bibitem [{\citenamefont {Meyer}\ and\ \citenamefont
  {Swanson}(2015)}]{Meyer:2015eta}%
  \BibitemOpen
  \bibfield  {author} {\bibinfo {author} {\bibfnamefont {C.~A.}\ \bibnamefont
  {Meyer}}\ and\ \bibinfo {author} {\bibfnamefont {E.~S.}\ \bibnamefont
  {Swanson}},\ }\bibfield  {title} {\enquote {\bibinfo {title} {{Hybrid
  Mesons}},}\ }\href {\doibase 10.1016/j.ppnp.2015.03.001} {\bibfield
  {journal} {\bibinfo  {journal} {Prog. Part. Nucl. Phys.}\ }\textbf {\bibinfo
  {volume} {82}},\ \bibinfo {pages} {21--58} (\bibinfo {year} {2015})},\
  \Eprint {http://arxiv.org/abs/1502.07276} {arXiv:1502.07276 [hep-ph]}
  \BibitemShut {NoStop}%
\bibitem [{Ams()}]{AmslerHanhart}%
  \BibitemOpen
  \href@noop {} {}\bibinfo {note} {{C.~Amsler and C.~Hanhart, {Non-$q\bar q$
  mesons}, in \cite{Agashe:2014kda}}}\BibitemShut {NoStop}%
\bibitem [{\citenamefont {Aubert}\ \emph {et~al.}(2003)\citenamefont {Aubert}
  \emph {et~al.}}]{Aubert:2003fg}%
  \BibitemOpen
  \bibfield  {author} {\bibinfo {author} {\bibfnamefont {B.}~\bibnamefont
  {Aubert}} \emph {et~al.} (\bibinfo {collaboration} {BaBar}),\ }\bibfield
  {title} {\enquote {\bibinfo {title} {{Observation of a narrow meson decaying
  to $D_s^+ \pi^0$ at a mass of $2.32\,$GeV$/c^2$}},}\ }\href {\doibase
  10.1103/PhysRevLett.90.242001} {\bibfield  {journal} {\bibinfo  {journal}
  {Phys. Rev. Lett.}\ }\textbf {\bibinfo {volume} {90}},\ \bibinfo {pages}
  {242001} (\bibinfo {year} {2003})},\ \Eprint
  {http://arxiv.org/abs/hep-ex/0304021} {arXiv:hep-ex/0304021 [hep-ex]}
  \BibitemShut {NoStop}%
\bibitem [{\citenamefont {Besson}\ \emph {et~al.}(2003)\citenamefont {Besson}
  \emph {et~al.}}]{Besson:2003cp}%
  \BibitemOpen
  \bibfield  {author} {\bibinfo {author} {\bibfnamefont {D.}~\bibnamefont
  {Besson}} \emph {et~al.} (\bibinfo {collaboration} {CLEO}),\ }\bibfield
  {title} {\enquote {\bibinfo {title} {{Observation of a narrow resonance of
  mass $2.46\,$GeV$/c^2$ decaying to $D*+_s \pi^0$ and confirmation of the
  $D^*_{sJ}(2317)$ state}},}\ }\href {\doibase 10.1103/PhysRevD.68.032002,
  10.1103/PhysRevD.75.119908} {\bibfield  {journal} {\bibinfo  {journal} {Phys.
  Rev.}\ }\textbf {\bibinfo {volume} {D68}},\ \bibinfo {pages} {032002}
  (\bibinfo {year} {2003})},\ \bibinfo {note} {[Erratum: Phys.
  Rev.D75,119908(2007)]},\ \Eprint {http://arxiv.org/abs/hep-ex/0305100}
  {arXiv:hep-ex/0305100 [hep-ex]} \BibitemShut {NoStop}%
\bibitem [{\citenamefont {Godfrey}\ and\ \citenamefont
  {Isgur}(1985)}]{Godfrey:1985xj}%
  \BibitemOpen
  \bibfield  {author} {\bibinfo {author} {\bibfnamefont {S.}~\bibnamefont
  {Godfrey}}\ and\ \bibinfo {author} {\bibfnamefont {Nathan}\ \bibnamefont
  {Isgur}},\ }\bibfield  {title} {\enquote {\bibinfo {title} {{Mesons in a
  Relativized Quark Model with Chromodynamics}},}\ }\href {\doibase
  10.1103/PhysRevD.32.189} {\bibfield  {journal} {\bibinfo  {journal} {Phys.
  Rev.}\ }\textbf {\bibinfo {volume} {D32}},\ \bibinfo {pages} {189--231}
  (\bibinfo {year} {1985})}\BibitemShut {NoStop}%
\bibitem [{\citenamefont {Pignon}\ and\ \citenamefont
  {Piketty}(1979)}]{Pignon:1978ba}%
  \BibitemOpen
  \bibfield  {author} {\bibinfo {author} {\bibfnamefont {D.}~\bibnamefont
  {Pignon}}\ and\ \bibinfo {author} {\bibfnamefont {C.~A.}\ \bibnamefont
  {Piketty}},\ }\bibfield  {title} {\enquote {\bibinfo {title} {{Charmed meson
  spectra}},}\ }\href {\doibase 10.1016/0370-2693(79)90347-2} {\bibfield
  {journal} {\bibinfo  {journal} {Phys. Lett.}\ }\textbf {\bibinfo {volume}
  {B81}},\ \bibinfo {pages} {334--338} (\bibinfo {year} {1979})}\BibitemShut
  {NoStop}%
\bibitem [{\citenamefont {Schnitzer}(1982)}]{Schnitzer:1981zg}%
  \BibitemOpen
  \bibfield  {author} {\bibinfo {author} {\bibfnamefont {Howard~J.}\
  \bibnamefont {Schnitzer}},\ }\bibfield  {title} {\enquote {\bibinfo {title}
  {{Spin Dependence and Glueball Mixing With $\theta(1640)$ in Ordinary Meson
  Spectroscopy}},}\ }\href {\doibase 10.1016/0550-3213(82)90140-7} {\bibfield
  {journal} {\bibinfo  {journal} {Nucl. Phys.}\ }\textbf {\bibinfo {volume}
  {B207}},\ \bibinfo {pages} {131--156} (\bibinfo {year} {1982})}\BibitemShut
  {NoStop}%
\bibitem [{\citenamefont {Cahn}\ and\ \citenamefont
  {Jackson}(2003)}]{Cahn:2003cw}%
  \BibitemOpen
  \bibfield  {author} {\bibinfo {author} {\bibfnamefont {Robert~N.}\
  \bibnamefont {Cahn}}\ and\ \bibinfo {author} {\bibfnamefont {J.~David}\
  \bibnamefont {Jackson}},\ }\bibfield  {title} {\enquote {\bibinfo {title}
  {{Spin-orbit and tensor forces in heavy-quark light-quark mesons:
  Implications of the new $D_s$ state at $2.32\,$GeV}},}\ }\href@noop {}
  {\bibfield  {journal} {\bibinfo  {journal} {Phys. Rev.}\ }\textbf {\bibinfo
  {volume} {D68}},\ \bibinfo {pages} {037502} (\bibinfo {year} {2003})},\
  \Eprint {http://arxiv.org/abs/hep-ph/0305012} {hep-ph/0305012} \BibitemShut
  {NoStop}%
\bibitem [{\citenamefont {Matsuki}\ \emph {et~al.}(2007)\citenamefont
  {Matsuki}, \citenamefont {Morii},\ and\ \citenamefont
  {Sudoh}}]{Matsuki:2007zza}%
  \BibitemOpen
  \bibfield  {author} {\bibinfo {author} {\bibfnamefont {Takayuki}\
  \bibnamefont {Matsuki}}, \bibinfo {author} {\bibfnamefont {Toshiyuki}\
  \bibnamefont {Morii}}, \ and\ \bibinfo {author} {\bibfnamefont {Kazutaka}\
  \bibnamefont {Sudoh}},\ }\bibfield  {title} {\enquote {\bibinfo {title} {{New
  heavy-light mesons $Q\bar q$}},}\ }\href {\doibase 10.1143/PTP.117.1077}
  {\bibfield  {journal} {\bibinfo  {journal} {Prog. Theor. Phys.}\ }\textbf
  {\bibinfo {volume} {117}},\ \bibinfo {pages} {1077--1098} (\bibinfo {year}
  {2007})},\ \Eprint {http://arxiv.org/abs/hep-ph/0605019}
  {arXiv:hep-ph/0605019 [hep-ph]} \BibitemShut {NoStop}%
\bibitem [{\citenamefont {Nielsen}\ \emph {et~al.}(2007)\citenamefont
  {Nielsen}, \citenamefont {Matheus}, \citenamefont {Navarra},\ and\
  \citenamefont {Bracco}}]{Nielsen:2006zz}%
  \BibitemOpen
  \bibfield  {author} {\bibinfo {author} {\bibfnamefont {M.}~\bibnamefont
  {Nielsen}}, \bibinfo {author} {\bibfnamefont {Ricardo~D'Elia}\ \bibnamefont
  {Matheus}}, \bibinfo {author} {\bibfnamefont {F.~S.}\ \bibnamefont
  {Navarra}}, \ and\ \bibinfo {author} {\bibfnamefont {M.~E.}\ \bibnamefont
  {Bracco}},\ }\bibfield  {title} {\enquote {\bibinfo {title} {{Testing the
  nature of the $D_{sJ}^+(2317)$ and $X(3782)$ states using QCD sum rules}},}\
  }\bibfield  {booktitle} {\emph {\bibinfo {booktitle} {{Proceedings, 18th
  International IUPAP Conference on Few-body problems in physics (FB18)}}},\
  }\href {\doibase 10.1016/j.nuclphysa.2007.03.173} {\bibfield  {journal}
  {\bibinfo  {journal} {Nucl. Phys.}\ }\textbf {\bibinfo {volume} {A790}},\
  \bibinfo {pages} {526--529} (\bibinfo {year} {2007})}\BibitemShut {NoStop}%
\bibitem [{\citenamefont {Vijande}\ \emph
  {et~al.}(2007{\natexlab{a}})\citenamefont {Vijande}, \citenamefont
  {Fernandez},\ and\ \citenamefont {Valcarce}}]{Vijande:2007zza}%
  \BibitemOpen
  \bibfield  {author} {\bibinfo {author} {\bibfnamefont {J.}~\bibnamefont
  {Vijande}}, \bibinfo {author} {\bibfnamefont {F.}~\bibnamefont {Fernandez}},
  \ and\ \bibinfo {author} {\bibfnamefont {A.}~\bibnamefont {Valcarce}},\
  }\bibfield  {title} {\enquote {\bibinfo {title} {{The puzzle of the $D$ and
  $D_s$ mesons}},}\ }\bibfield  {booktitle} {\emph {\bibinfo {booktitle}
  {{Quarks and nuclear physics. Proceedings, 4th International Conference, QNP
  2006, Madrid, Spain, June 5-10, 2006}}},\ }\href {\doibase
  10.1140/epja/i2006-10199-0} {\bibfield  {journal} {\bibinfo  {journal} {Eur.
  Phys. J.}\ }\textbf {\bibinfo {volume} {A31}},\ \bibinfo {pages} {722--724}
  (\bibinfo {year} {2007}{\natexlab{a}})}\BibitemShut {NoStop}%
\bibitem [{\citenamefont {Segovia}\ \emph {et~al.}(2015)\citenamefont
  {Segovia}, \citenamefont {Entem},\ and\ \citenamefont
  {Fernandez}}]{Segovia:2015dia}%
  \BibitemOpen
  \bibfield  {author} {\bibinfo {author} {\bibfnamefont {Jorge}\ \bibnamefont
  {Segovia}}, \bibinfo {author} {\bibfnamefont {David~R.}\ \bibnamefont
  {Entem}}, \ and\ \bibinfo {author} {\bibfnamefont {Francisco}\ \bibnamefont
  {Fernandez}},\ }\bibfield  {title} {\enquote {\bibinfo {title}
  {{Charmed-strange Meson Spectrum: Old and New Problems}},}\ }\href {\doibase
  10.1103/PhysRevD.91.094020} {\bibfield  {journal} {\bibinfo  {journal} {Phys.
  Rev.}\ }\textbf {\bibinfo {volume} {D91}},\ \bibinfo {pages} {094020}
  (\bibinfo {year} {2015})},\ \Eprint {http://arxiv.org/abs/1502.03827}
  {arXiv:1502.03827 [hep-ph]} \BibitemShut {NoStop}%
\bibitem [{\citenamefont {Godfrey}\ and\ \citenamefont
  {Moats}(2016)}]{Godfrey:2015dva}%
  \BibitemOpen
  \bibfield  {author} {\bibinfo {author} {\bibfnamefont {Stephen}\ \bibnamefont
  {Godfrey}}\ and\ \bibinfo {author} {\bibfnamefont {Kenneth}\ \bibnamefont
  {Moats}},\ }\bibfield  {title} {\enquote {\bibinfo {title} {{Properties of
  Excited Charm and Charm-Strange Mesons}},}\ }\href {\doibase
  10.1103/PhysRevD.93.034035} {\bibfield  {journal} {\bibinfo  {journal} {Phys.
  Rev.}\ }\textbf {\bibinfo {volume} {D93}},\ \bibinfo {pages} {034035}
  (\bibinfo {year} {2016})},\ \Eprint {http://arxiv.org/abs/1510.08305}
  {arXiv:1510.08305 [hep-ph]} \BibitemShut {NoStop}%
\bibitem [{\citenamefont {Dover}\ and\ \citenamefont
  {Kahana}(1977)}]{Dover:1977jw}%
  \BibitemOpen
  \bibfield  {author} {\bibinfo {author} {\bibfnamefont {C.~B.}\ \bibnamefont
  {Dover}}\ and\ \bibinfo {author} {\bibfnamefont {S.~H.}\ \bibnamefont
  {Kahana}},\ }\bibfield  {title} {\enquote {\bibinfo {title} {{Possibility of
  Charmed Hypernuclei}},}\ }\href {\doibase 10.1103/PhysRevLett.39.1506}
  {\bibfield  {journal} {\bibinfo  {journal} {Phys. Rev. Lett.}\ }\textbf
  {\bibinfo {volume} {39}},\ \bibinfo {pages} {1506--1509} (\bibinfo {year}
  {1977})}\BibitemShut {NoStop}%
\bibitem [{\citenamefont {Dover}\ \emph {et~al.}(1977)\citenamefont {Dover},
  \citenamefont {Kahana},\ and\ \citenamefont {Trueman}}]{Dover:1977hs}%
  \BibitemOpen
  \bibfield  {author} {\bibinfo {author} {\bibfnamefont {C.~B.}\ \bibnamefont
  {Dover}}, \bibinfo {author} {\bibfnamefont {S.~H.}\ \bibnamefont {Kahana}}, \
  and\ \bibinfo {author} {\bibfnamefont {T.~L.}\ \bibnamefont {Trueman}},\
  }\bibfield  {title} {\enquote {\bibinfo {title} {{Bound States of Charmed
  Baryons and anti-Baryons}},}\ }\href {\doibase 10.1103/PhysRevD.16.799}
  {\bibfield  {journal} {\bibinfo  {journal} {Phys. Rev.}\ }\textbf {\bibinfo
  {volume} {D16}},\ \bibinfo {pages} {799--815} (\bibinfo {year}
  {1977})}\BibitemShut {NoStop}%
\bibitem [{\citenamefont {Isgur}\ and\ \citenamefont
  {Lipkin}(1981)}]{Isgur:1980en}%
  \BibitemOpen
  \bibfield  {author} {\bibinfo {author} {\bibfnamefont {Nathan}\ \bibnamefont
  {Isgur}}\ and\ \bibinfo {author} {\bibfnamefont {Harry~J.}\ \bibnamefont
  {Lipkin}},\ }\bibfield  {title} {\enquote {\bibinfo {title} {O\lowercase{N
  THE POSSIBLE EXISTENCE OF STABLE FOUR QUARK SCALAR MESONS WITH CHARM AND
  STRANGENESS}},}\ }\href@noop {} {\bibfield  {journal} {\bibinfo  {journal}
  {Phys. Lett.}\ }\textbf {\bibinfo {volume} {B99}},\ \bibinfo {pages} {151}
  (\bibinfo {year} {1981})}\BibitemShut {NoStop}%
\bibitem [{\citenamefont {Cho}\ \emph {et~al.}(2011)\citenamefont {Cho} \emph
  {et~al.}}]{Cho:2010db}%
  \BibitemOpen
  \bibfield  {author} {\bibinfo {author} {\bibfnamefont {Sungtae}\ \bibnamefont
  {Cho}} \emph {et~al.} (\bibinfo {collaboration} {ExHIC}),\ }\bibfield
  {title} {\enquote {\bibinfo {title} {{Multi-quark hadrons from Heavy Ion
  Collisions}},}\ }\href {\doibase 10.1103/PhysRevLett.106.212001} {\bibfield
  {journal} {\bibinfo  {journal} {Phys. Rev. Lett.}\ }\textbf {\bibinfo
  {volume} {106}},\ \bibinfo {pages} {212001} (\bibinfo {year} {2011})},\
  \Eprint {http://arxiv.org/abs/1011.0852} {arXiv:1011.0852 [nucl-th]}
  \BibitemShut {NoStop}%
\bibitem [{\citenamefont {Yu-qi}\ and\ \citenamefont
  {Su-zhi}(2011)}]{Yuqi:2011gm}%
  \BibitemOpen
  \bibfield  {author} {\bibinfo {author} {\bibfnamefont {Chen}\ \bibnamefont
  {Yu-qi}}\ and\ \bibinfo {author} {\bibfnamefont {Wu}~\bibnamefont {Su-zhi}},\
  }\bibfield  {title} {\enquote {\bibinfo {title} {{Production of four-quark
  states with double heavy quarks at LHC}},}\ }\href {\doibase
  10.1016/j.physletb.2011.09.096} {\bibfield  {journal} {\bibinfo  {journal}
  {Phys. Lett.}\ }\textbf {\bibinfo {volume} {B705}},\ \bibinfo {pages}
  {93--97} (\bibinfo {year} {2011})},\ \Eprint {http://arxiv.org/abs/1101.4568}
  {arXiv:1101.4568 [hep-ph]} \BibitemShut {NoStop}%
\bibitem [{\citenamefont {Hyodo}\ \emph {et~al.}(2012)\citenamefont {Hyodo},
  \citenamefont {Liu}, \citenamefont {Oka}, \citenamefont {Sudoh},\ and\
  \citenamefont {Yasui}}]{Hyodo:2012pm}%
  \BibitemOpen
  \bibfield  {author} {\bibinfo {author} {\bibfnamefont {Tetsuo}\ \bibnamefont
  {Hyodo}}, \bibinfo {author} {\bibfnamefont {Yan-Rui}\ \bibnamefont {Liu}},
  \bibinfo {author} {\bibfnamefont {Makoto}\ \bibnamefont {Oka}}, \bibinfo
  {author} {\bibfnamefont {Kazutaka}\ \bibnamefont {Sudoh}}, \ and\ \bibinfo
  {author} {\bibfnamefont {Shigehiro}\ \bibnamefont {Yasui}},\ }\bibfield
  {title} {\enquote {\bibinfo {title} {{Production of doubly charmed
  tetraquarks with exotic color configurations in electron-positron
  collisions}},}\ }\href@noop {} {\  (\bibinfo {year} {2012})},\ \Eprint
  {http://arxiv.org/abs/1209.6207} {arXiv:1209.6207 [hep-ph]} \BibitemShut
  {NoStop}%
\bibitem [{\citenamefont {Esposito}\ \emph {et~al.}(2013)\citenamefont
  {Esposito}, \citenamefont {Papinutto}, \citenamefont {Pilloni}, \citenamefont
  {Polosa},\ and\ \citenamefont {Tantalo}}]{Esposito:2013fma}%
  \BibitemOpen
  \bibfield  {author} {\bibinfo {author} {\bibfnamefont {A.}~\bibnamefont
  {Esposito}}, \bibinfo {author} {\bibfnamefont {M.}~\bibnamefont {Papinutto}},
  \bibinfo {author} {\bibfnamefont {A.}~\bibnamefont {Pilloni}}, \bibinfo
  {author} {\bibfnamefont {A.~D.}\ \bibnamefont {Polosa}}, \ and\ \bibinfo
  {author} {\bibfnamefont {N.}~\bibnamefont {Tantalo}},\ }\bibfield  {title}
  {\enquote {\bibinfo {title} {{Doubly charmed tetraquarks in $B_c$ and
  $\Xi_{bc}$ decays}},}\ }\href {\doibase 10.1103/PhysRevD.88.054029}
  {\bibfield  {journal} {\bibinfo  {journal} {Phys. Rev.}\ }\textbf {\bibinfo
  {volume} {D88}},\ \bibinfo {pages} {054029} (\bibinfo {year} {2013})},\
  \Eprint {http://arxiv.org/abs/1307.2873} {arXiv:1307.2873 [hep-ph]}
  \BibitemShut {NoStop}%
\bibitem [{\citenamefont {Abe}\ \emph {et~al.}(2004)\citenamefont {Abe} \emph
  {et~al.}}]{Abe:2004ww}%
  \BibitemOpen
  \bibfield  {author} {\bibinfo {author} {\bibfnamefont {Kazuo}\ \bibnamefont
  {Abe}} \emph {et~al.} (\bibinfo {collaboration} {Belle}),\ }\bibfield
  {title} {\enquote {\bibinfo {title} {{Study of double charmonium production
  in e+ e- annihilation at $\sqrt{s}\sim10.6\,$GeV}},}\ }\href {\doibase
  10.1103/PhysRevD.70.071102} {\bibfield  {journal} {\bibinfo  {journal} {Phys.
  Rev.}\ }\textbf {\bibinfo {volume} {D70}},\ \bibinfo {pages} {071102}
  (\bibinfo {year} {2004})},\ \Eprint {http://arxiv.org/abs/hep-ex/0407009}
  {arXiv:hep-ex/0407009 [hep-ex]} \BibitemShut {NoStop}%
\bibitem [{\citenamefont {Martin}\ and\ \citenamefont
  {Richard}(2003)}]{Martin:2003rr}%
  \BibitemOpen
  \bibfield  {author} {\bibinfo {author} {\bibfnamefont {A.}~\bibnamefont
  {Martin}}\ and\ \bibinfo {author} {\bibfnamefont {J.~M.}\ \bibnamefont
  {Richard}},\ }\bibfield  {title} {\enquote {\bibinfo {title} {{The eventful
  story of charmonium singlet states}},}\ }\href@noop {} {\bibfield  {journal}
  {\bibinfo  {journal} {CERN Cour.}\ }\textbf {\bibinfo {volume} {43N3}},\
  \bibinfo {pages} {17--18} (\bibinfo {year} {2003})}\BibitemShut {NoStop}%
\bibitem [{\citenamefont {Barnes}\ and\ \citenamefont
  {Olsen}(2009)}]{Barnes:2009zza}%
  \BibitemOpen
  \bibfield  {author} {\bibinfo {author} {\bibfnamefont {T.}~\bibnamefont
  {Barnes}}\ and\ \bibinfo {author} {\bibfnamefont {S.~L.}\ \bibnamefont
  {Olsen}},\ }\bibfield  {title} {\enquote {\bibinfo {title} {{Charmonium
  spectroscopy}},}\ }\href {\doibase 10.1142/S0217751X09046576} {\bibfield
  {journal} {\bibinfo  {journal} {Int. J. Mod. Phys.}\ }\textbf {\bibinfo
  {volume} {A24S1}},\ \bibinfo {pages} {305--325} (\bibinfo {year}
  {2009})}\BibitemShut {NoStop}%
\bibitem [{\citenamefont {Baglin}\ \emph {et~al.}(1986)\citenamefont {Baglin}
  \emph {et~al.}}]{Baglin:1986yd}%
  \BibitemOpen
  \bibfield  {author} {\bibinfo {author} {\bibfnamefont {C.}~\bibnamefont
  {Baglin}} \emph {et~al.} (\bibinfo {collaboration} {R704,
  Annecy(LAPP)-CERN-Genoa-Lyon-Oslo-Rome-Strasbourg-Turin}),\ }\bibfield
  {title} {\enquote {\bibinfo {title} {{Search for the $p$ Wave Singlet
  Charmonium State in $\bar{p} p$ Annihilations at the {CERN} Intersecting
  Storage Rings}},}\ }\href {\doibase 10.1016/0370-2693(86)91013-0} {\bibfield
  {journal} {\bibinfo  {journal} {Phys. Lett.}\ }\textbf {\bibinfo {volume}
  {B171}},\ \bibinfo {pages} {135--141} (\bibinfo {year} {1986})}\BibitemShut
  {NoStop}%
\bibitem [{\citenamefont {Armstrong}\ \emph {et~al.}(1992)\citenamefont
  {Armstrong} \emph {et~al.}}]{Armstrong:1992ae}%
  \BibitemOpen
  \bibfield  {author} {\bibinfo {author} {\bibfnamefont {T.~A.}\ \bibnamefont
  {Armstrong}} \emph {et~al.},\ }\bibfield  {title} {\enquote {\bibinfo {title}
  {{Observation of the p wave singlet state of charmonium}},}\ }\href {\doibase
  10.1103/PhysRevLett.69.2337} {\bibfield  {journal} {\bibinfo  {journal}
  {Phys. Rev. Lett.}\ }\textbf {\bibinfo {volume} {69}},\ \bibinfo {pages}
  {2337--2340} (\bibinfo {year} {1992})}\BibitemShut {NoStop}%
\bibitem [{\citenamefont {Rubin}\ \emph {et~al.}(2005)\citenamefont {Rubin}
  \emph {et~al.}}]{Rubin:2005px}%
  \BibitemOpen
  \bibfield  {author} {\bibinfo {author} {\bibfnamefont {P.}~\bibnamefont
  {Rubin}} \emph {et~al.} (\bibinfo {collaboration} {CLEO}),\ }\bibfield
  {title} {\enquote {\bibinfo {title} {{Observation of the $^1\mathrm{P}_1$
  state of charmonium}},}\ }\href {\doibase 10.1103/PhysRevD.72.092004}
  {\bibfield  {journal} {\bibinfo  {journal} {Phys. Rev.}\ }\textbf {\bibinfo
  {volume} {D72}},\ \bibinfo {pages} {092004} (\bibinfo {year} {2005})},\
  \Eprint {http://arxiv.org/abs/hep-ex/0508037} {arXiv:hep-ex/0508037 [hep-ex]}
  \BibitemShut {NoStop}%
\bibitem [{\citenamefont {Choi}\ \emph {et~al.}(2002)\citenamefont {Choi} \emph
  {et~al.}}]{Choi:2002na}%
  \BibitemOpen
  \bibfield  {author} {\bibinfo {author} {\bibfnamefont {S.~K.}\ \bibnamefont
  {Choi}} \emph {et~al.} (\bibinfo {collaboration} {Belle}),\ }\bibfield
  {title} {\enquote {\bibinfo {title} {{Observation of the $\eta_c(2s)$ in
  exclusive $B\to K K_s K^- \pi^+$ decays}},}\ }\href {\doibase
  10.1103/PhysRevLett.89.102001} {\bibfield  {journal} {\bibinfo  {journal}
  {Phys. Rev. Lett.}\ }\textbf {\bibinfo {volume} {89}},\ \bibinfo {pages}
  {102001} (\bibinfo {year} {2002})},\ \bibinfo {note} {[Erratum: Phys. Rev.
  Lett.89,129901(2002)]},\ \Eprint {http://arxiv.org/abs/hep-ex/0206002}
  {arXiv:hep-ex/0206002 [hep-ex]} \BibitemShut {NoStop}%
\bibitem [{\citenamefont {Choi}\ \emph {et~al.}(2003)\citenamefont {Choi} \emph
  {et~al.}}]{Choi:2003ue}%
  \BibitemOpen
  \bibfield  {author} {\bibinfo {author} {\bibfnamefont {S.~K.}\ \bibnamefont
  {Choi}} \emph {et~al.} (\bibinfo {collaboration} {Belle}),\ }\bibfield
  {title} {\enquote {\bibinfo {title} {{Observation of a new narrow charmonium
  state in exclusive $B^\pm\to K^\pm \pi^+ \pi^- J/\psi$ decays}},}\
  }\href@noop {} {\bibfield  {journal} {\bibinfo  {journal} {Phys. Rev. Lett.}\
  }\textbf {\bibinfo {volume} {91}},\ \bibinfo {pages} {262001} (\bibinfo
  {year} {2003})},\ \Eprint {http://arxiv.org/abs/hep-ex/0309032}
  {hep-ex/0309032} \BibitemShut {NoStop}%
\bibitem [{\citenamefont {Lebed}(2016)}]{Lebed:2016cna}%
  \BibitemOpen
  \bibfield  {author} {\bibinfo {author} {\bibfnamefont {Richard~F.}\
  \bibnamefont {Lebed}},\ }\bibfield  {title} {\enquote {\bibinfo {title}
  {{Exotic Discoveries in Familiar Places: Theory of the Onia and Exotics}},}\
  \ }(\bibinfo {year} {2016})\ \bibinfo {note} {16th International Conference
  on B-Physics at Frontier Machines (Beauty 2016), 2-6 May 2016. Marseille,
  France, to appear in PoS},\ \Eprint {http://arxiv.org/abs/1605.07975}
  {arXiv:1605.07975 [hep-ph]} \BibitemShut {NoStop}%
\bibitem [{\citenamefont {Aubert}\ \emph
  {et~al.}(2005{\natexlab{a}})\citenamefont {Aubert} \emph
  {et~al.}}]{Aubert:2005rm}%
  \BibitemOpen
  \bibfield  {author} {\bibinfo {author} {\bibfnamefont {B.}~\bibnamefont
  {Aubert}} \emph {et~al.} (\bibinfo {collaboration} {BABAR}),\ }\bibfield
  {title} {\enquote {\bibinfo {title} {{Observation of a broad structure in the
  $\pi^+\pi^- J/\psi$ mass spectrum around $4.26\,$GeV$/c^2$}},}\ }\href@noop
  {} {\bibfield  {journal} {\bibinfo  {journal} {Phys. Rev. Lett.}\ }\textbf
  {\bibinfo {volume} {95}},\ \bibinfo {pages} {142001} (\bibinfo {year}
  {2005}{\natexlab{a}})},\ \Eprint {http://arxiv.org/abs/hep-ex/0506081}
  {hep-ex/0506081} \BibitemShut {NoStop}%
\bibitem [{\citenamefont {Choi}\ \emph {et~al.}(2008)\citenamefont {Choi} \emph
  {et~al.}}]{Choi:2007wga}%
  \BibitemOpen
  \bibfield  {author} {\bibinfo {author} {\bibfnamefont {S.K.}\ \bibnamefont
  {Choi}} \emph {et~al.} (\bibinfo {collaboration} {BELLE Collaboration}),\
  }\bibfield  {title} {\enquote {\bibinfo {title} {{Observation of a
  resonance-like structure in the $\pi^\pm \psi'$ mass distribution in
  exclusive $B\to K \pi^\pm\psi'$ decays}},}\ }\href {\doibase
  10.1103/PhysRevLett.100.142001} {\bibfield  {journal} {\bibinfo  {journal}
  {Phys.Rev.Lett.}\ }\textbf {\bibinfo {volume} {100}},\ \bibinfo {pages}
  {142001} (\bibinfo {year} {2008})},\ \Eprint {http://arxiv.org/abs/0708.1790}
  {arXiv:0708.1790 [hep-ex]} \BibitemShut {NoStop}%
\bibitem [{\citenamefont {Abe}\ \emph {et~al.}(2007)\citenamefont {Abe} \emph
  {et~al.}}]{Abe:2007jna}%
  \BibitemOpen
  \bibfield  {author} {\bibinfo {author} {\bibfnamefont {Kazuo}\ \bibnamefont
  {Abe}} \emph {et~al.} (\bibinfo {collaboration} {Belle}),\ }\bibfield
  {title} {\enquote {\bibinfo {title} {{Observation of a new charmonium state
  in double charmonium production in $e^+ e^-$ annihilation at $\sqrt{s}\sim
  10.6\,$GeV}},}\ }\href {\doibase 10.1103/PhysRevLett.98.082001} {\bibfield
  {journal} {\bibinfo  {journal} {Phys. Rev. Lett.}\ }\textbf {\bibinfo
  {volume} {98}},\ \bibinfo {pages} {082001} (\bibinfo {year} {2007})},\
  \Eprint {http://arxiv.org/abs/hep-ex/0507019} {arXiv:hep-ex/0507019 [hep-ex]}
  \BibitemShut {NoStop}%
\bibitem [{\citenamefont {Pakhlov}\ \emph {et~al.}(2008)\citenamefont {Pakhlov}
  \emph {et~al.}}]{Abe:2007sya}%
  \BibitemOpen
  \bibfield  {author} {\bibinfo {author} {\bibfnamefont {P.}~\bibnamefont
  {Pakhlov}} \emph {et~al.} (\bibinfo {collaboration} {Belle}),\ }\bibfield
  {title} {\enquote {\bibinfo {title} {{Production of New Charmoniumlike States
  in $e^+ e^-\to J/\psi D^{(*)} \bar{D}{}^{(*)}$ at $\sqrt{s} \sim
  10\,$GeV}},}\ }\href {\doibase 10.1103/PhysRevLett.100.202001} {\bibfield
  {journal} {\bibinfo  {journal} {Phys. Rev. Lett.}\ }\textbf {\bibinfo
  {volume} {100}},\ \bibinfo {pages} {202001} (\bibinfo {year} {2008})},\
  \Eprint {http://arxiv.org/abs/0708.3812} {arXiv:0708.3812 [hep-ex]}
  \BibitemShut {NoStop}%
\bibitem [{\citenamefont {Uehara}\ \emph {et~al.}(2006)\citenamefont {Uehara}
  \emph {et~al.}}]{Uehara:2005qd}%
  \BibitemOpen
  \bibfield  {author} {\bibinfo {author} {\bibfnamefont {S.}~\bibnamefont
  {Uehara}} \emph {et~al.} (\bibinfo {collaboration} {Belle}),\ }\bibfield
  {title} {\enquote {\bibinfo {title} {{Observation of a $\chi_2'$ candidate in
  $\gamma\gamma\to D \bar D$ production at BELLE}},}\ }\href {\doibase
  10.1103/PhysRevLett.96.082003} {\bibfield  {journal} {\bibinfo  {journal}
  {Phys. Rev. Lett.}\ }\textbf {\bibinfo {volume} {96}},\ \bibinfo {pages}
  {082003} (\bibinfo {year} {2006})},\ \Eprint
  {http://arxiv.org/abs/hep-ex/0512035} {arXiv:hep-ex/0512035 [hep-ex]}
  \BibitemShut {NoStop}%
\bibitem [{\citenamefont {Uehara}\ \emph {et~al.}(2010)\citenamefont {Uehara}
  \emph {et~al.}}]{Uehara:2009tx}%
  \BibitemOpen
  \bibfield  {author} {\bibinfo {author} {\bibfnamefont {S.}~\bibnamefont
  {Uehara}} \emph {et~al.} (\bibinfo {collaboration} {Belle}),\ }\bibfield
  {title} {\enquote {\bibinfo {title} {{Observation of a charmonium-like
  enhancement in the $\gamma \gamma \to \omega J/\psi$ process}},}\ }\href
  {\doibase 10.1103/PhysRevLett.104.092001} {\bibfield  {journal} {\bibinfo
  {journal} {Phys. Rev. Lett.}\ }\textbf {\bibinfo {volume} {104}},\ \bibinfo
  {pages} {092001} (\bibinfo {year} {2010})},\ \Eprint
  {http://arxiv.org/abs/0912.4451} {arXiv:0912.4451 [hep-ex]} \BibitemShut
  {NoStop}%
\bibitem [{\citenamefont {Shen}\ \emph {et~al.}(2010)\citenamefont {Shen} \emph
  {et~al.}}]{Shen:2009vs}%
  \BibitemOpen
  \bibfield  {author} {\bibinfo {author} {\bibfnamefont {C.~P.}\ \bibnamefont
  {Shen}} \emph {et~al.} (\bibinfo {collaboration} {Belle}),\ }\bibfield
  {title} {\enquote {\bibinfo {title} {{Evidence for a new resonance and search
  for the $Y(4140)$ in the $\gamma \gamma \to\phi J/\psi$ process}},}\ }\href
  {\doibase 10.1103/PhysRevLett.104.112004} {\bibfield  {journal} {\bibinfo
  {journal} {Phys. Rev. Lett.}\ }\textbf {\bibinfo {volume} {104}},\ \bibinfo
  {pages} {112004} (\bibinfo {year} {2010})},\ \Eprint
  {http://arxiv.org/abs/0912.2383} {arXiv:0912.2383 [hep-ex]} \BibitemShut
  {NoStop}%
\bibitem [{\citenamefont {Pakhlova}\ \emph {et~al.}(2008)\citenamefont
  {Pakhlova} \emph {et~al.}}]{Pakhlova:2008vn}%
  \BibitemOpen
  \bibfield  {author} {\bibinfo {author} {\bibfnamefont {G.}~\bibnamefont
  {Pakhlova}} \emph {et~al.} (\bibinfo {collaboration} {Belle}),\ }\bibfield
  {title} {\enquote {\bibinfo {title} {{Observation of a near-threshold
  enhancement in the $e^+e^- \to\Lambda^+_c\bar{\Lambda}{}^-_c$ cross section
  using initial-state radiation}},}\ }\href {\doibase
  10.1103/PhysRevLett.101.172001} {\bibfield  {journal} {\bibinfo  {journal}
  {Phys. Rev. Lett.}\ }\textbf {\bibinfo {volume} {101}},\ \bibinfo {pages}
  {172001} (\bibinfo {year} {2008})},\ \Eprint {http://arxiv.org/abs/0807.4458}
  {arXiv:0807.4458 [hep-ex]} \BibitemShut {NoStop}%
\bibitem [{\citenamefont {Sonnenschein}\ and\ \citenamefont
  {Weissman}(2016)}]{Sonnenschein:2016ibx}%
  \BibitemOpen
  \bibfield  {author} {\bibinfo {author} {\bibfnamefont {Jacob}\ \bibnamefont
  {Sonnenschein}}\ and\ \bibinfo {author} {\bibfnamefont {Dorin}\ \bibnamefont
  {Weissman}},\ }\bibfield  {title} {\enquote {\bibinfo {title} {{A tetraquark
  or not a tetraquark: A holography inspired stringy hadron (HISH)
  perspective}},}\ }\href@noop {} {\  (\bibinfo {year} {2016})},\ \Eprint
  {http://arxiv.org/abs/1606.02732} {arXiv:1606.02732 [hep-ph]} \BibitemShut
  {NoStop}%
\bibitem [{\citenamefont {Guo}\ \emph {et~al.}(2010)\citenamefont {Guo},
  \citenamefont {Haidenbauer}, \citenamefont {Hanhart},\ and\ \citenamefont
  {Meissner}}]{Guo:2010tk}%
  \BibitemOpen
  \bibfield  {author} {\bibinfo {author} {\bibfnamefont {Feng-Kun}\
  \bibnamefont {Guo}}, \bibinfo {author} {\bibfnamefont {Johann}\ \bibnamefont
  {Haidenbauer}}, \bibinfo {author} {\bibfnamefont {Christoph}\ \bibnamefont
  {Hanhart}}, \ and\ \bibinfo {author} {\bibfnamefont {Ulf-G.}\ \bibnamefont
  {Meissner}},\ }\bibfield  {title} {\enquote {\bibinfo {title} {{Reconciling
  the X(4630) with the Y(4660)}},}\ }\href {\doibase
  10.1103/PhysRevD.82.094008} {\bibfield  {journal} {\bibinfo  {journal} {Phys.
  Rev.}\ }\textbf {\bibinfo {volume} {D82}},\ \bibinfo {pages} {094008}
  (\bibinfo {year} {2010})},\ \Eprint {http://arxiv.org/abs/1005.2055}
  {arXiv:1005.2055 [hep-ph]} \BibitemShut {NoStop}%
\bibitem [{\citenamefont {Aubert}\ \emph
  {et~al.}(2005{\natexlab{b}})\citenamefont {Aubert} \emph
  {et~al.}}]{Aubert:2004zr}%
  \BibitemOpen
  \bibfield  {author} {\bibinfo {author} {\bibfnamefont {Bernard}\ \bibnamefont
  {Aubert}} \emph {et~al.} (\bibinfo {collaboration} {BaBar}),\ }\bibfield
  {title} {\enquote {\bibinfo {title} {{Search for a charged partner of the
  X(3872) in the $B$ meson decay $B \to X^- K$, $X^- \to J/\psi \pi^-
  \pi^0$}},}\ }\href {\doibase 10.1103/PhysRevD.71.031501} {\bibfield
  {journal} {\bibinfo  {journal} {Phys. Rev.}\ }\textbf {\bibinfo {volume}
  {D71}},\ \bibinfo {pages} {031501} (\bibinfo {year} {2005}{\natexlab{b}})},\
  \Eprint {http://arxiv.org/abs/hep-ex/0412051} {arXiv:hep-ex/0412051 [hep-ex]}
  \BibitemShut {NoStop}%
\bibitem [{\citenamefont {Ablikim}\ \emph {et~al.}(2013)\citenamefont {Ablikim}
  \emph {et~al.}}]{Ablikim:2013mio}%
  \BibitemOpen
  \bibfield  {author} {\bibinfo {author} {\bibfnamefont {M.}~\bibnamefont
  {Ablikim}} \emph {et~al.} (\bibinfo {collaboration} {BESIII Collaboration}),\
  }\bibfield  {title} {\enquote {\bibinfo {title} {{Observation of a charged
  charmoniumlike structure in $e^+e^- \to \pi^+\pi^-J/\psi$ at $\sqrt{s}=4.26$
  GeV}},}\ }\href@noop {} {\bibfield  {journal} {\bibinfo  {journal}
  {Phys.Rev.Lett.}\ }\textbf {\bibinfo {volume} {110}},\ \bibinfo {pages}
  {252001} (\bibinfo {year} {2013})},\ \Eprint {http://arxiv.org/abs/1303.5949}
  {arXiv:1303.5949 [hep-ex]} \BibitemShut {NoStop}%
\bibitem [{\citenamefont {Liu}\ \emph {et~al.}(2013)\citenamefont {Liu} \emph
  {et~al.}}]{Liu:2013dau}%
  \BibitemOpen
  \bibfield  {author} {\bibinfo {author} {\bibfnamefont {Z.Q.}\ \bibnamefont
  {Liu}} \emph {et~al.} (\bibinfo {collaboration} {Belle Collaboration}),\
  }\bibfield  {title} {\enquote {\bibinfo {title} {{Study of $e^+ e^- \to \pi^+
  \pi^- J/\psi$ and Observation of a Charged Charmonium-like State at
  Belle}},}\ }\href {\doibase 10.1103/PhysRevLett.110.252002} {\bibfield
  {journal} {\bibinfo  {journal} {Phys.Rev.Lett.}\ }\textbf {\bibinfo {volume}
  {110}},\ \bibinfo {pages} {252002} (\bibinfo {year} {2013})},\ \Eprint
  {http://arxiv.org/abs/1304.0121} {arXiv:1304.0121 [hep-ex]} \BibitemShut
  {NoStop}%
\bibitem [{\citenamefont {Xiao}\ \emph
  {et~al.}(2013{\natexlab{a}})\citenamefont {Xiao}, \citenamefont {Dobbs},
  \citenamefont {Tomaradze},\ and\ \citenamefont {Seth}}]{Xiao:2013iha}%
  \BibitemOpen
  \bibfield  {author} {\bibinfo {author} {\bibfnamefont {T.}~\bibnamefont
  {Xiao}}, \bibinfo {author} {\bibfnamefont {S.}~\bibnamefont {Dobbs}},
  \bibinfo {author} {\bibfnamefont {A.}~\bibnamefont {Tomaradze}}, \ and\
  \bibinfo {author} {\bibfnamefont {Kamal~K.}\ \bibnamefont {Seth}},\
  }\bibfield  {title} {\enquote {\bibinfo {title} {{Observation of the Charged
  Hadron $Z_c^{\pm}(3900)$ and Evidence for the Neutral $Z_c^0(3900)$ in
  $e^+e^-\to \pi\pi J/\psi$ at $\sqrt{s}=4170$ MeV}},}\ }\href {\doibase
  10.1016/j.physletb.2013.10.041} {\bibfield  {journal} {\bibinfo  {journal}
  {Phys. Lett.}\ }\textbf {\bibinfo {volume} {B727}},\ \bibinfo {pages}
  {366--370} (\bibinfo {year} {2013}{\natexlab{a}})},\ \Eprint
  {http://arxiv.org/abs/1304.3036} {arXiv:1304.3036 [hep-ex]} \BibitemShut
  {NoStop}%
\bibitem [{\citenamefont {Nerling}(2016)}]{Nerling:2016uwm}%
  \BibitemOpen
  \bibfield  {author} {\bibinfo {author} {\bibfnamefont {Frank}\ \bibnamefont
  {Nerling}} (\bibinfo {collaboration} {COMPASS}),\ }\bibfield  {title}
  {\enquote {\bibinfo {title} {{Highlights from the COMPASS experiment at CERN
  -- Hadron spectroscopy and excitations}},}\ }in\ \href
  {http://inspirehep.net/record/1415952/files/arXiv:1601.05025.pdf} {\emph
  {\bibinfo {booktitle} {{4th International Conference on New Frontiers in
  Physics (ICNFP 2015) Kolymbari, Greece, August 23-30, 2015}}}}\ (\bibinfo
  {year} {2016})\ \Eprint {http://arxiv.org/abs/1601.05025} {arXiv:1601.05025
  [hep-ex]} \BibitemShut {NoStop}%
\bibitem [{\citenamefont {Aaij}\ \emph
  {et~al.}(2016{\natexlab{a}})\citenamefont {Aaij} \emph
  {et~al.}}]{Aaij:2016iza}%
  \BibitemOpen
  \bibfield  {author} {\bibinfo {author} {\bibfnamefont {Roel}\ \bibnamefont
  {Aaij}} \emph {et~al.} (\bibinfo {collaboration} {LHCb}),\ }\bibfield
  {title} {\enquote {\bibinfo {title} {{Observation of $J/\psi\phi$ structures
  consistent with exotic states from amplitude analysis of $B^+\to J/\psi \phi
  K^+$ decays}},}\ }\href@noop {} {\  (\bibinfo {year} {2016}{\natexlab{a}})},\
  \Eprint {http://arxiv.org/abs/1606.07895} {arXiv:1606.07895 [hep-ex]}
  \BibitemShut {NoStop}%
\bibitem [{\citenamefont {Stancu}(2010)}]{Stancu:2009ka}%
  \BibitemOpen
  \bibfield  {author} {\bibinfo {author} {\bibfnamefont {Fl.}\ \bibnamefont
  {Stancu}},\ }\bibfield  {title} {\enquote {\bibinfo {title} {{Can $Y(4140)$
  be a $c \bar c s \bar s$ tetraquark?}}}\ }\href {\doibase
  10.1088/0954-3899/37/7/075017} {\bibfield  {journal} {\bibinfo  {journal} {J.
  Phys.}\ }\textbf {\bibinfo {volume} {G37}},\ \bibinfo {pages} {075017}
  (\bibinfo {year} {2010})},\ \Eprint {http://arxiv.org/abs/0906.2485}
  {arXiv:0906.2485 [hep-ph]} \BibitemShut {NoStop}%
\bibitem [{\citenamefont {H\o{}g\aa{}sen}\ \emph {et~al.}(2006)\citenamefont
  {H\o{}g\aa{}sen}, \citenamefont {Richard},\ and\ \citenamefont
  {Sorba}}]{Hogaasen:2005jv}%
  \BibitemOpen
  \bibfield  {author} {\bibinfo {author} {\bibfnamefont {H.}~\bibnamefont
  {H\o{}g\aa{}sen}}, \bibinfo {author} {\bibfnamefont {J.-M.}\ \bibnamefont
  {Richard}}, \ and\ \bibinfo {author} {\bibfnamefont {P.}~\bibnamefont
  {Sorba}},\ }\bibfield  {title} {\enquote {\bibinfo {title} {{A chromomagnetic
  mechanism for the X(3872) resonance}},}\ }\href@noop {} {\bibfield  {journal}
  {\bibinfo  {journal} {Phys. Rev.}\ }\textbf {\bibinfo {volume} {D73}},\
  \bibinfo {pages} {054013} (\bibinfo {year} {2006})},\ \Eprint
  {http://arxiv.org/abs/hep-ph/0511039} {hep-ph/0511039} \BibitemShut {NoStop}%
\bibitem [{\citenamefont {Crede}\ and\ \citenamefont
  {Roberts}(2013)}]{Crede:2013sze}%
  \BibitemOpen
  \bibfield  {author} {\bibinfo {author} {\bibfnamefont {V.}~\bibnamefont
  {Crede}}\ and\ \bibinfo {author} {\bibfnamefont {W.}~\bibnamefont
  {Roberts}},\ }\bibfield  {title} {\enquote {\bibinfo {title} {{Progress
  towards understanding baryon resonances}},}\ }\href {\doibase
  10.1088/0034-4885/76/7/076301} {\bibfield  {journal} {\bibinfo  {journal}
  {Rept. Prog. Phys.}\ }\textbf {\bibinfo {volume} {76}},\ \bibinfo {pages}
  {076301} (\bibinfo {year} {2013})},\ \Eprint {http://arxiv.org/abs/1302.7299}
  {arXiv:1302.7299 [nucl-ex]} \BibitemShut {NoStop}%
\bibitem [{\citenamefont {Isgur}(1992)}]{Isgur:1992td}%
  \BibitemOpen
  \bibfield  {author} {\bibinfo {author} {\bibfnamefont {Nathan}\ \bibnamefont
  {Isgur}},\ }\bibfield  {title} {\enquote {\bibinfo {title} {{An introduction
  to the quark model for baryons}},}\ }\href {\doibase
  10.1142/S0218301392000242} {\bibfield  {journal} {\bibinfo  {journal} {Int.
  J. Mod. Phys.}\ }\textbf {\bibinfo {volume} {E1}},\ \bibinfo {pages}
  {465--490} (\bibinfo {year} {1992})}\BibitemShut {NoStop}%
\bibitem [{\citenamefont {Dalitz}\ and\ \citenamefont
  {Tuan}(1960)}]{Dalitz:1960du}%
  \BibitemOpen
  \bibfield  {author} {\bibinfo {author} {\bibfnamefont {R.~H.}\ \bibnamefont
  {Dalitz}}\ and\ \bibinfo {author} {\bibfnamefont {S.~F.}\ \bibnamefont
  {Tuan}},\ }\bibfield  {title} {\enquote {\bibinfo {title} {{The
  phenomenological description of $K$-nucleon reaction processes}},}\ }\href
  {\doibase 10.1016/0003-4916(60)90001-4} {\bibfield  {journal} {\bibinfo
  {journal} {Annals Phys.}\ }\textbf {\bibinfo {volume} {10}},\ \bibinfo
  {pages} {307--351} (\bibinfo {year} {1960})}\BibitemShut {NoStop}%
\bibitem [{Ose()}]{Oset}%
  \BibitemOpen
  \href@noop {} {}\bibinfo {note} {{Eulogio Oset, in
  \protect\cite{baryon2016}}}\BibitemShut {NoStop}%
\bibitem [{\citenamefont {Isgur}\ and\ \citenamefont
  {Karl}(1978)}]{Isgur:1978xj}%
  \BibitemOpen
  \bibfield  {author} {\bibinfo {author} {\bibfnamefont {Nathan}\ \bibnamefont
  {Isgur}}\ and\ \bibinfo {author} {\bibfnamefont {Gabriel}\ \bibnamefont
  {Karl}},\ }\bibfield  {title} {\enquote {\bibinfo {title} {{P-Wave Baryons in
  the Quark Model}},}\ }\href {\doibase 10.1103/PhysRevD.18.4187} {\bibfield
  {journal} {\bibinfo  {journal} {Phys. Rev.}\ }\textbf {\bibinfo {volume}
  {D18}},\ \bibinfo {pages} {4187} (\bibinfo {year} {1978})}\BibitemShut
  {NoStop}%
\bibitem [{\citenamefont {Kamiya}\ \emph {et~al.}(2016)\citenamefont {Kamiya},
  \citenamefont {Miyahara}, \citenamefont {Ohnishi}, \citenamefont {Ikeda},
  \citenamefont {Hyodo}, \citenamefont {Oset},\ and\ \citenamefont
  {Weise}}]{Kamiya:2016jqc}%
  \BibitemOpen
  \bibfield  {author} {\bibinfo {author} {\bibfnamefont {Yuki}\ \bibnamefont
  {Kamiya}}, \bibinfo {author} {\bibfnamefont {Kenta}\ \bibnamefont
  {Miyahara}}, \bibinfo {author} {\bibfnamefont {Shota}\ \bibnamefont
  {Ohnishi}}, \bibinfo {author} {\bibfnamefont {Yoichi}\ \bibnamefont {Ikeda}},
  \bibinfo {author} {\bibfnamefont {Tetsuo}\ \bibnamefont {Hyodo}}, \bibinfo
  {author} {\bibfnamefont {Eulogio}\ \bibnamefont {Oset}}, \ and\ \bibinfo
  {author} {\bibfnamefont {Wolfram}\ \bibnamefont {Weise}},\ }\bibfield
  {title} {\enquote {\bibinfo {title} {{Antikaon-nucleon interaction and
  $\Lambda$(1405) in chiral SU(3) dynamics}},}\ }\href {\doibase
  10.1016/j.nuclphysa.2016.04.013} {\bibfield  {journal} {\bibinfo  {journal}
  {Nucl. Phys.}\ }\textbf {\bibinfo {volume} {A954}},\ \bibinfo {pages}
  {41--57} (\bibinfo {year} {2016})},\ \Eprint
  {http://arxiv.org/abs/1602.08852} {arXiv:1602.08852 [hep-ph]} \BibitemShut
  {NoStop}%
\bibitem [{\citenamefont {Close}(2002)}]{Close:2001zp}%
  \BibitemOpen
  \bibfield  {author} {\bibinfo {author} {\bibfnamefont {Frank~E.}\
  \bibnamefont {Close}},\ }\bibfield  {title} {\enquote {\bibinfo {title}
  {{Light hadron spectroscopy: Theory and experiment}},}\ }\bibfield
  {booktitle} {\emph {\bibinfo {booktitle} {{Lepton and photon interactions at
  high energies. Proceedings, 20th International Symposium, LP 2001, Rome,
  Italy, July 23-28, 2001}}},\ }\href {\doibase 10.1142/S0217751X02012715}
  {\bibfield  {journal} {\bibinfo  {journal} {Int. J. Mod. Phys.}\ }\textbf
  {\bibinfo {volume} {A17}},\ \bibinfo {pages} {3239--3258} (\bibinfo {year}
  {2002})},\ \bibinfo {note} {[,327(2001)]},\ \Eprint
  {http://arxiv.org/abs/hep-ph/0110081} {arXiv:hep-ph/0110081 [hep-ph]}
  \BibitemShut {NoStop}%
\bibitem [{\citenamefont {Carter}(1967)}]{PhysRevLett.18.801}%
  \BibitemOpen
  \bibfield  {author} {\bibinfo {author} {\bibfnamefont {A.~A.}\ \bibnamefont
  {Carter}},\ }\bibfield  {title} {\enquote {\bibinfo {title} {{Evidence for an
  $S=+1$, $I=0$ Resonance in the ${K}^{+}$-Nucleon System}},}\ }\href {\doibase
  10.1103/PhysRevLett.18.801} {\bibfield  {journal} {\bibinfo  {journal} {Phys.
  Rev. Lett.}\ }\textbf {\bibinfo {volume} {18}},\ \bibinfo {pages} {801--803}
  (\bibinfo {year} {1967})}\BibitemShut {NoStop}%
\bibitem [{\citenamefont {Lea}\ \emph {et~al.}(1968)\citenamefont {Lea},
  \citenamefont {Martin},\ and\ \citenamefont {Oades}}]{PhysRev.165.1770}%
  \BibitemOpen
  \bibfield  {author} {\bibinfo {author} {\bibfnamefont {A.~T.}\ \bibnamefont
  {Lea}}, \bibinfo {author} {\bibfnamefont {B.~R.}\ \bibnamefont {Martin}}, \
  and\ \bibinfo {author} {\bibfnamefont {G.~C.}\ \bibnamefont {Oades}},\
  }\bibfield  {title} {\enquote {\bibinfo {title} {${K}^{+}p$ phase-shift
  analysis below $1500\,\mathrm{MeV}/c$},}\ }\href {\doibase
  10.1103/PhysRev.165.1770} {\bibfield  {journal} {\bibinfo  {journal} {Phys.
  Rev.}\ }\textbf {\bibinfo {volume} {165}},\ \bibinfo {pages} {1770--1786}
  (\bibinfo {year} {1968})}\BibitemShut {NoStop}%
\bibitem [{\citenamefont {Wilson}\ \emph {et~al.}(1972)\citenamefont {Wilson}
  \emph {et~al.}}]{Wilson:1972uh}%
  \BibitemOpen
  \bibfield  {author} {\bibinfo {author} {\bibfnamefont {B.~C.}\ \bibnamefont
  {Wilson}} \emph {et~al.},\ }\bibfield  {title} {\enquote {\bibinfo {title}
  {{Phase-shift analysis of the $K^+$-nucleon interaction in the $I=0$ state up
  to 1.5\,GeV$/c$}},}\ }\href {\doibase 10.1016/0550-3213(72)90491-9}
  {\bibfield  {journal} {\bibinfo  {journal} {Nucl. Phys.}\ }\textbf {\bibinfo
  {volume} {B42}},\ \bibinfo {pages} {445--453} (\bibinfo {year}
  {1972})}\BibitemShut {NoStop}%
\bibitem [{\citenamefont {Albrow}\ \emph {et~al.}(1971)\citenamefont {Albrow},
  \citenamefont {Andersson-Almehed}, \citenamefont {Bosnjakovic}, \citenamefont
  {Daum}, \citenamefont {Erne}, \citenamefont {Kimura}, \citenamefont
  {Lagnaux}, \citenamefont {Sens}, \citenamefont {Udo},\ and\ \citenamefont
  {Wagner}}]{Albrow:1971qm}%
  \BibitemOpen
  \bibfield  {author} {\bibinfo {author} {\bibfnamefont {M.~G.}\ \bibnamefont
  {Albrow}}, \bibinfo {author} {\bibfnamefont {S.}~\bibnamefont
  {Andersson-Almehed}}, \bibinfo {author} {\bibfnamefont {B.}~\bibnamefont
  {Bosnjakovic}}, \bibinfo {author} {\bibfnamefont {C.}~\bibnamefont {Daum}},
  \bibinfo {author} {\bibfnamefont {F.~C.}\ \bibnamefont {Erne}}, \bibinfo
  {author} {\bibfnamefont {Y.}~\bibnamefont {Kimura}}, \bibinfo {author}
  {\bibfnamefont {J.~P.}\ \bibnamefont {Lagnaux}}, \bibinfo {author}
  {\bibfnamefont {J.~C.}\ \bibnamefont {Sens}}, \bibinfo {author}
  {\bibfnamefont {F.}~\bibnamefont {Udo}}, \ and\ \bibinfo {author}
  {\bibfnamefont {F.}~\bibnamefont {Wagner}},\ }\bibfield  {title} {\enquote
  {\bibinfo {title} {{Elastic scattering of positive kaons on polarized protons
  between 0.87 and 2.74\,Gev$/c$. Results and phase-shift analysis}},}\ }\href
  {\doibase 10.1016/0550-3213(71)90291-4} {\bibfield  {journal} {\bibinfo
  {journal} {Nucl. Phys.}\ }\textbf {\bibinfo {volume} {B30}},\ \bibinfo
  {pages} {273--305} (\bibinfo {year} {1971})}\BibitemShut {NoStop}%
\bibitem [{\citenamefont {Martin}(1975)}]{Martin:1975gs}%
  \BibitemOpen
  \bibfield  {author} {\bibinfo {author} {\bibfnamefont {B.~R.}\ \bibnamefont
  {Martin}},\ }\bibfield  {title} {\enquote {\bibinfo {title} {{Kaon-Nucleon
  Partial Wave Amplitudes Below $1.5\,$GeV$/c$ for $I=0$ and 1}},}\ }\href
  {\doibase 10.1016/0550-3213(75)90104-2} {\bibfield  {journal} {\bibinfo
  {journal} {Nucl. Phys.}\ }\textbf {\bibinfo {volume} {B94}},\ \bibinfo
  {pages} {413--430} (\bibinfo {year} {1975})}\BibitemShut {NoStop}%
\bibitem [{\citenamefont {Kelly}\ \emph {et~al.}(1980)\citenamefont {Kelly}
  \emph {et~al.}}]{Kelly:1980kr}%
  \BibitemOpen
  \bibfield  {author} {\bibinfo {author} {\bibfnamefont {R.~L.}\ \bibnamefont
  {Kelly}} \emph {et~al.} (\bibinfo {collaboration} {Particle Data Group}),\
  }\bibfield  {title} {\enquote {\bibinfo {title} {{Review of Particle
  Properties. Particle Data Group}},}\ }\href {\doibase
  10.1103/RevModPhys.52.S1} {\bibfield  {journal} {\bibinfo  {journal} {Rev.
  Mod. Phys.}\ }\textbf {\bibinfo {volume} {52}},\ \bibinfo {pages} {S1--S286}
  (\bibinfo {year} {1980})}\BibitemShut {NoStop}%
\bibitem [{\citenamefont {Nakano}\ \emph {et~al.}(2003)\citenamefont {Nakano}
  \emph {et~al.}}]{Nakano:2003qx}%
  \BibitemOpen
  \bibfield  {author} {\bibinfo {author} {\bibfnamefont {T.}~\bibnamefont
  {Nakano}} \emph {et~al.} (\bibinfo {collaboration} {LEPS}),\ }\bibfield
  {title} {\enquote {\bibinfo {title} {{Evidence for Narrow $S=+1$ Baryon
  Resonance in Photo- production from Neutron}},}\ }\href {\doibase
  10.1103/PhysRevLett.91.012002} {\bibfield  {journal} {\bibinfo  {journal}
  {Phys. Rev. Lett.}\ }\textbf {\bibinfo {volume} {91}},\ \bibinfo {pages}
  {012002} (\bibinfo {year} {2003})},\ \Eprint
  {http://arxiv.org/abs/hep-ex/0301020} {arXiv:hep-ex/0301020} \BibitemShut
  {NoStop}%
\bibitem [{\citenamefont {Diakonov}\ \emph {et~al.}(1997)\citenamefont
  {Diakonov}, \citenamefont {Petrov},\ and\ \citenamefont
  {Polyakov}}]{Diakonov:1997mm}%
  \BibitemOpen
  \bibfield  {author} {\bibinfo {author} {\bibfnamefont {Dmitri}\ \bibnamefont
  {Diakonov}}, \bibinfo {author} {\bibfnamefont {Victor}\ \bibnamefont
  {Petrov}}, \ and\ \bibinfo {author} {\bibfnamefont {Maxim~V.}\ \bibnamefont
  {Polyakov}},\ }\bibfield  {title} {\enquote {\bibinfo {title} {{Exotic
  anti-decuplet of baryons: Prediction from chiral solitons}},}\ }\href
  {\doibase 10.1007/s002180050406} {\bibfield  {journal} {\bibinfo  {journal}
  {Z. Phys.}\ }\textbf {\bibinfo {volume} {A359}},\ \bibinfo {pages} {305--314}
  (\bibinfo {year} {1997})},\ \Eprint {http://arxiv.org/abs/hep-ph/9703373}
  {arXiv:hep-ph/9703373} \BibitemShut {NoStop}%
\bibitem [{\citenamefont {Wohl}(2008)}]{Wohl:2008st}%
  \BibitemOpen
  \bibfield  {author} {\bibinfo {author} {\bibfnamefont {C.~G.}\ \bibnamefont
  {Wohl}},\ }\href@noop {} {\bibfield  {journal} {\bibinfo  {journal}
  {{Pentaquarks}, in \protect\cite{Amsler:2008zz}}\ } (\bibinfo {year}
  {2008})}\BibitemShut {NoStop}%
\bibitem [{\citenamefont {Hicks}(2012)}]{Hicks:2012zz}%
  \BibitemOpen
  \bibfield  {author} {\bibinfo {author} {\bibfnamefont {Kenneth~H.}\
  \bibnamefont {Hicks}},\ }\bibfield  {title} {\enquote {\bibinfo {title} {{On
  the conundrum of the pentaquark}},}\ }\href {\doibase
  10.1140/epjh/e2012-20032-0} {\bibfield  {journal} {\bibinfo  {journal} {Eur.
  Phys. J.}\ }\textbf {\bibinfo {volume} {H37}},\ \bibinfo {pages} {1--31}
  (\bibinfo {year} {2012})}\BibitemShut {NoStop}%
\bibitem [{\citenamefont {Tariq}(2007)}]{Tariq:2007ck}%
  \BibitemOpen
  \bibfield  {author} {\bibinfo {author} {\bibfnamefont {Abdullah Shams~Bin}\
  \bibnamefont {Tariq}},\ }\bibfield  {title} {\enquote {\bibinfo {title}
  {{Revisiting the pentaquark episode for lattice QCD}},}\ }\bibfield
  {booktitle} {\emph {\bibinfo {booktitle} {{Proceedings, 25th International
  Symposium on Lattice field theory (Lattice 2007)}}},\ }\href@noop {}
  {\bibfield  {journal} {\bibinfo  {journal} {PoS}\ }\textbf {\bibinfo {volume}
  {LAT2007}},\ \bibinfo {pages} {136} (\bibinfo {year} {2007})},\ \Eprint
  {http://arxiv.org/abs/0711.0566} {arXiv:0711.0566 [hep-lat]} \BibitemShut
  {NoStop}%
\bibitem [{\citenamefont {Jaffe}(2004)}]{Jaffe:2004qj}%
  \BibitemOpen
  \bibfield  {author} {\bibinfo {author} {\bibfnamefont {R.~L.}\ \bibnamefont
  {Jaffe}},\ }\bibfield  {title} {\enquote {\bibinfo {title} {{The width of the
  Theta+ exotic baryon in the chiral soliton model}},}\ }\href {\doibase
  10.1140/epjc/s2004-01815-4} {\bibfield  {journal} {\bibinfo  {journal} {Eur.
  Phys. J.}\ }\textbf {\bibinfo {volume} {C35}},\ \bibinfo {pages} {221--222}
  (\bibinfo {year} {2004})},\ \Eprint {http://arxiv.org/abs/hep-ph/0401187}
  {arXiv:hep-ph/0401187 [hep-ph]} \BibitemShut {NoStop}%
\bibitem [{\citenamefont {Weigel}(2007)}]{Weigel:2007yx}%
  \BibitemOpen
  \bibfield  {author} {\bibinfo {author} {\bibfnamefont {H.}~\bibnamefont
  {Weigel}},\ }\bibfield  {title} {\enquote {\bibinfo {title} {{Axial current
  matrix elements and pentaquark decay widths in chiral soliton models}},}\
  }\href {\doibase 10.1103/PhysRevD.75.114018} {\bibfield  {journal} {\bibinfo
  {journal} {Phys. Rev.}\ }\textbf {\bibinfo {volume} {D75}},\ \bibinfo {pages}
  {114018} (\bibinfo {year} {2007})},\ \Eprint
  {http://arxiv.org/abs/hep-ph/0703072} {arXiv:hep-ph/0703072 [hep-ph]}
  \BibitemShut {NoStop}%
\bibitem [{\citenamefont {Abazov}\ \emph {et~al.}(2008)\citenamefont {Abazov}
  \emph {et~al.}}]{Abazov:2008qm}%
  \BibitemOpen
  \bibfield  {author} {\bibinfo {author} {\bibfnamefont {V.~M.}\ \bibnamefont
  {Abazov}} \emph {et~al.} (\bibinfo {collaboration} {D0}),\ }\bibfield
  {title} {\enquote {\bibinfo {title} {{Observation of the doubly strange $b$
  baryon $\Omega_b^-$}},}\ }\href {\doibase 10.1103/PhysRevLett.101.232002}
  {\bibfield  {journal} {\bibinfo  {journal} {Phys. Rev. Lett.}\ }\textbf
  {\bibinfo {volume} {101}},\ \bibinfo {pages} {232002} (\bibinfo {year}
  {2008})},\ \Eprint {http://arxiv.org/abs/0808.4142} {arXiv:0808.4142
  [hep-ex]} \BibitemShut {NoStop}%
\bibitem [{\citenamefont {Aaltonen}\ \emph {et~al.}(2009)\citenamefont
  {Aaltonen} \emph {et~al.}}]{Aaltonen:2009ny}%
  \BibitemOpen
  \bibfield  {author} {\bibinfo {author} {\bibfnamefont {T.}~\bibnamefont
  {Aaltonen}} \emph {et~al.} (\bibinfo {collaboration} {CDF}),\ }\bibfield
  {title} {\enquote {\bibinfo {title} {{Observation of the $\Omega_b^-$ and
  Measurement of the Properties of the $\Xi_b^-$ and $\Omega_b^-$}},}\ }\href
  {\doibase 10.1103/PhysRevD.80.072003} {\bibfield  {journal} {\bibinfo
  {journal} {Phys. Rev.}\ }\textbf {\bibinfo {volume} {D80}},\ \bibinfo {pages}
  {072003} (\bibinfo {year} {2009})},\ \Eprint {http://arxiv.org/abs/0905.3123}
  {arXiv:0905.3123 [hep-ex]} \BibitemShut {NoStop}%
\bibitem [{\citenamefont {Aaij}\ \emph
  {et~al.}(2013{\natexlab{a}})\citenamefont {Aaij} \emph
  {et~al.}}]{Aaij:2013qja}%
  \BibitemOpen
  \bibfield  {author} {\bibinfo {author} {\bibfnamefont {R}~\bibnamefont
  {Aaij}} \emph {et~al.} (\bibinfo {collaboration} {LHCb}),\ }\bibfield
  {title} {\enquote {\bibinfo {title} {{Measurement of the $\Lambda_b^0$,
  $\Xi_b^-$ and $\Omega_b^-$ baryon masses}},}\ }\href {\doibase
  10.1103/PhysRevLett.110.182001} {\bibfield  {journal} {\bibinfo  {journal}
  {Phys. Rev. Lett.}\ }\textbf {\bibinfo {volume} {110}},\ \bibinfo {pages}
  {182001} (\bibinfo {year} {2013}{\natexlab{a}})},\ \Eprint
  {http://arxiv.org/abs/1302.1072} {arXiv:1302.1072 [hep-ex]} \BibitemShut
  {NoStop}%
\bibitem [{\citenamefont {Amsler}\ \emph
  {et~al.}(2008{\natexlab{a}})\citenamefont {Amsler} \emph
  {et~al.}}]{Amsler:2008zzb}%
  \BibitemOpen
  \bibfield  {author} {\bibinfo {author} {\bibfnamefont {Claude}\ \bibnamefont
  {Amsler}} \emph {et~al.} (\bibinfo {collaboration} {Particle Data Group}),\
  }\bibfield  {title} {\enquote {\bibinfo {title} {{Review of Particle
  Physics}},}\ }\href {\doibase 10.1016/j.physletb.2008.07.018} {\bibfield
  {journal} {\bibinfo  {journal} {Phys. Lett.}\ }\textbf {\bibinfo {volume}
  {B667}},\ \bibinfo {pages} {1--1340} (\bibinfo {year}
  {2008}{\natexlab{a}})}\BibitemShut {NoStop}%
\bibitem [{\citenamefont {Dong}\ \emph {et~al.}(2011)\citenamefont {Dong},
  \citenamefont {Faessler}, \citenamefont {Gutsche}, \citenamefont {Kumano},\
  and\ \citenamefont {Lyubovitskij}}]{Dong:2011ys}%
  \BibitemOpen
  \bibfield  {author} {\bibinfo {author} {\bibfnamefont {Yubing}\ \bibnamefont
  {Dong}}, \bibinfo {author} {\bibfnamefont {Amand}\ \bibnamefont {Faessler}},
  \bibinfo {author} {\bibfnamefont {Thomas}\ \bibnamefont {Gutsche}}, \bibinfo
  {author} {\bibfnamefont {S.}~\bibnamefont {Kumano}}, \ and\ \bibinfo {author}
  {\bibfnamefont {Valery~E.}\ \bibnamefont {Lyubovitskij}},\ }\bibfield
  {title} {\enquote {\bibinfo {title} {{Strong three-body decays of
  $\Lambda_c(2940)^+$}},}\ }\href {\doibase 10.1103/PhysRevD.83.094005}
  {\bibfield  {journal} {\bibinfo  {journal} {Phys. Rev.}\ }\textbf {\bibinfo
  {volume} {D83}},\ \bibinfo {pages} {094005} (\bibinfo {year} {2011})},\
  \Eprint {http://arxiv.org/abs/1103.4762} {arXiv:1103.4762 [hep-ph]}
  \BibitemShut {NoStop}%
\bibitem [{\citenamefont {Aaij}\ \emph
  {et~al.}(2015{\natexlab{a}})\citenamefont {Aaij} \emph
  {et~al.}}]{Aaij:2014yka}%
  \BibitemOpen
  \bibfield  {author} {\bibinfo {author} {\bibfnamefont {Roel}\ \bibnamefont
  {Aaij}} \emph {et~al.} (\bibinfo {collaboration} {LHCb}),\ }\bibfield
  {title} {\enquote {\bibinfo {title} {{Observation of two new $\Xi_b^-$ baryon
  resonances}},}\ }\href {\doibase 10.1103/PhysRevLett.114.062004} {\bibfield
  {journal} {\bibinfo  {journal} {Phys. Rev. Lett.}\ }\textbf {\bibinfo
  {volume} {114}},\ \bibinfo {pages} {062004} (\bibinfo {year}
  {2015}{\natexlab{a}})},\ \Eprint {http://arxiv.org/abs/1411.4849}
  {arXiv:1411.4849 [hep-ex]} \BibitemShut {NoStop}%
\bibitem [{\citenamefont {Aaij}\ \emph
  {et~al.}(2016{\natexlab{b}})\citenamefont {Aaij} \emph
  {et~al.}}]{Aaij:2016jnn}%
  \BibitemOpen
  \bibfield  {author} {\bibinfo {author} {\bibfnamefont {Roel}\ \bibnamefont
  {Aaij}} \emph {et~al.} (\bibinfo {collaboration} {LHCb}),\ }\bibfield
  {title} {\enquote {\bibinfo {title} {{Measurement of the properties of the
  $\Xi_b^{*0}$ baryon}},}\ }\href@noop {} {\  (\bibinfo {year}
  {2016}{\natexlab{b}})},\ \Eprint {http://arxiv.org/abs/1604.03896}
  {arXiv:1604.03896 [hep-ex]} \BibitemShut {NoStop}%
\bibitem [{\citenamefont {Ebert}\ \emph {et~al.}(2002)\citenamefont {Ebert},
  \citenamefont {Faustov}, \citenamefont {Galkin},\ and\ \citenamefont
  {Martynenko}}]{Ebert:2002ig}%
  \BibitemOpen
  \bibfield  {author} {\bibinfo {author} {\bibfnamefont {D.}~\bibnamefont
  {Ebert}}, \bibinfo {author} {\bibfnamefont {R.~N.}\ \bibnamefont {Faustov}},
  \bibinfo {author} {\bibfnamefont {V.~O.}\ \bibnamefont {Galkin}}, \ and\
  \bibinfo {author} {\bibfnamefont {A.~P.}\ \bibnamefont {Martynenko}},\
  }\bibfield  {title} {\enquote {\bibinfo {title} {{Mass spectra of doubly
  heavy baryons in the relativistic quark model}},}\ }\href {\doibase
  10.1103/PhysRevD.66.014008} {\bibfield  {journal} {\bibinfo  {journal} {Phys.
  Rev.}\ }\textbf {\bibinfo {volume} {D66}},\ \bibinfo {pages} {014008}
  (\bibinfo {year} {2002})},\ \Eprint {http://arxiv.org/abs/hep-ph/0201217}
  {arXiv:hep-ph/0201217 [hep-ph]} \BibitemShut {NoStop}%
\bibitem [{\citenamefont {Fleck}\ and\ \citenamefont
  {Richard}(1989)}]{Fleck:1989mb}%
  \BibitemOpen
  \bibfield  {author} {\bibinfo {author} {\bibfnamefont {S.}~\bibnamefont
  {Fleck}}\ and\ \bibinfo {author} {\bibfnamefont {J.~M.}\ \bibnamefont
  {Richard}},\ }\bibfield  {title} {\enquote {\bibinfo {title} {Baryons with
  double charm},}\ }\href@noop {} {\bibfield  {journal} {\bibinfo  {journal}
  {Prog. Theor. Phys.}\ }\textbf {\bibinfo {volume} {82}},\ \bibinfo {pages}
  {760--774} (\bibinfo {year} {1989})}\BibitemShut {NoStop}%
\bibitem [{\citenamefont {Mattson}\ \emph {et~al.}(2002)\citenamefont {Mattson}
  \emph {et~al.}}]{Mattson:2002vu}%
  \BibitemOpen
  \bibfield  {author} {\bibinfo {author} {\bibfnamefont {M.}~\bibnamefont
  {Mattson}} \emph {et~al.} (\bibinfo {collaboration} {SELEX}),\ }\bibfield
  {title} {\enquote {\bibinfo {title} {{First observation of the doubly charmed
  baryon $\Xi_{cc}^+$}},}\ }\href@noop {} {\bibfield  {journal} {\bibinfo
  {journal} {Phys. Rev. Lett.}\ }\textbf {\bibinfo {volume} {89}},\ \bibinfo
  {pages} {112001} (\bibinfo {year} {2002})},\ \Eprint
  {http://arxiv.org/abs/hep-ex/0208014} {hep-ex/0208014} \BibitemShut {NoStop}%
\bibitem [{\citenamefont {Ocherashvili}\ \emph {et~al.}(2005)\citenamefont
  {Ocherashvili} \emph {et~al.}}]{Ocherashvili:2004hi}%
  \BibitemOpen
  \bibfield  {author} {\bibinfo {author} {\bibfnamefont {A.}~\bibnamefont
  {Ocherashvili}} \emph {et~al.} (\bibinfo {collaboration} {SELEX}),\
  }\bibfield  {title} {\enquote {\bibinfo {title} {{Confirmation of the double
  charm baryon $\Xi_{cc}^+(3520)$ via its decay to $pD^+ K^-$}},}\ }\href@noop
  {} {\bibfield  {journal} {\bibinfo  {journal} {Phys. Lett.}\ }\textbf
  {\bibinfo {volume} {B628}},\ \bibinfo {pages} {18--24} (\bibinfo {year}
  {2005})},\ \Eprint {http://arxiv.org/abs/hep-ex/0406033} {hep-ex/0406033}
  \BibitemShut {NoStop}%
\bibitem [{\citenamefont {Ratti}(2003)}]{Ratti:2003ez}%
  \BibitemOpen
  \bibfield  {author} {\bibinfo {author} {\bibfnamefont {S.~P.}\ \bibnamefont
  {Ratti}},\ }\bibfield  {title} {\enquote {\bibinfo {title} {{New results on
  c-baryons and a search for cc-baryons in FOCUS}},}\ }\bibfield  {booktitle}
  {\emph {\bibinfo {booktitle} {{Proceedings, 5th International Conference on
  Hyperons, charm and beauty hadrons (BEACH 2002)}}},\ }\href {\doibase
  10.1016/S0920-5632(02)01948-5} {\bibfield  {journal} {\bibinfo  {journal}
  {Nucl. Phys. Proc. Suppl.}\ }\textbf {\bibinfo {volume} {115}},\ \bibinfo
  {pages} {33--36} (\bibinfo {year} {2003})},\ \bibinfo {note}
  {[,33(2003)]}\BibitemShut {NoStop}%
\bibitem [{\citenamefont {Aubert}\ \emph
  {et~al.}(2006{\natexlab{a}})\citenamefont {Aubert} \emph
  {et~al.}}]{Aubert:2006qw}%
  \BibitemOpen
  \bibfield  {author} {\bibinfo {author} {\bibfnamefont {Bernard}\ \bibnamefont
  {Aubert}} \emph {et~al.} (\bibinfo {collaboration} {BaBar}),\ }\bibfield
  {title} {\enquote {\bibinfo {title} {{Search for doubly charmed baryons
  $\Xi_{cc}^+$ and $\Xi_{cc}^{++}$ in BaBar}},}\ }\href {\doibase
  10.1103/PhysRevD.74.011103} {\bibfield  {journal} {\bibinfo  {journal} {Phys.
  Rev.}\ }\textbf {\bibinfo {volume} {D74}},\ \bibinfo {pages} {011103}
  (\bibinfo {year} {2006}{\natexlab{a}})},\ \Eprint
  {http://arxiv.org/abs/hep-ex/0605075} {arXiv:hep-ex/0605075 [hep-ex]}
  \BibitemShut {NoStop}%
\bibitem [{\citenamefont {Kato}\ \emph {et~al.}(2014)\citenamefont {Kato} \emph
  {et~al.}}]{Kato:2013ynr}%
  \BibitemOpen
  \bibfield  {author} {\bibinfo {author} {\bibfnamefont {Y.}~\bibnamefont
  {Kato}} \emph {et~al.} (\bibinfo {collaboration} {Belle}),\ }\bibfield
  {title} {\enquote {\bibinfo {title} {{Search for doubly charmed baryons and
  study of charmed strange baryons at Belle}},}\ }\href {\doibase
  10.1103/PhysRevD.89.052003} {\bibfield  {journal} {\bibinfo  {journal} {Phys.
  Rev.}\ }\textbf {\bibinfo {volume} {D89}},\ \bibinfo {pages} {052003}
  (\bibinfo {year} {2014})},\ \Eprint {http://arxiv.org/abs/1312.1026}
  {arXiv:1312.1026 [hep-ex]} \BibitemShut {NoStop}%
\bibitem [{\citenamefont {Aaij}\ \emph
  {et~al.}(2013{\natexlab{b}})\citenamefont {Aaij} \emph
  {et~al.}}]{Aaij:2013voa}%
  \BibitemOpen
  \bibfield  {author} {\bibinfo {author} {\bibfnamefont {R}~\bibnamefont
  {Aaij}} \emph {et~al.} (\bibinfo {collaboration} {LHCb}),\ }\bibfield
  {title} {\enquote {\bibinfo {title} {{Search for the doubly charmed baryon
  $\Xi_{cc}^+$}},}\ }\href {\doibase 10.1007/JHEP12(2013)090} {\bibfield
  {journal} {\bibinfo  {journal} {JHEP}\ }\textbf {\bibinfo {volume} {12}},\
  \bibinfo {pages} {090} (\bibinfo {year} {2013}{\natexlab{b}})},\ \Eprint
  {http://arxiv.org/abs/1310.2538} {arXiv:1310.2538 [hep-ex]} \BibitemShut
  {NoStop}%
\bibitem [{\citenamefont {Brodsky}\ \emph {et~al.}(2012)\citenamefont
  {Brodsky}, \citenamefont {de~Teramond},\ and\ \citenamefont
  {Karliner}}]{Broadsky:2012rw}%
  \BibitemOpen
  \bibfield  {author} {\bibinfo {author} {\bibfnamefont {Stanley}\ \bibnamefont
  {Brodsky}}, \bibinfo {author} {\bibfnamefont {Guy}\ \bibnamefont
  {de~Teramond}}, \ and\ \bibinfo {author} {\bibfnamefont {Marek}\ \bibnamefont
  {Karliner}},\ }\bibfield  {title} {\enquote {\bibinfo {title} {{Puzzles in
  Hadronic Physics and Novel Quantum Chromodynamics Phenomenology}},}\ }\href
  {\doibase 10.1146/annurev-nucl-102711-094949} {\bibfield  {journal} {\bibinfo
   {journal} {Ann. Rev. Nucl. Part. Sci.}\ }\textbf {\bibinfo {volume} {62}},\
  \bibinfo {pages} {1--35} (\bibinfo {year} {2012})},\ \bibinfo {note} {[Ann.
  Rev. Nucl. Part. Sci.62,2082(2011)]},\ \Eprint
  {http://arxiv.org/abs/1302.5684} {arXiv:1302.5684 [hep-ph]} \BibitemShut
  {NoStop}%
\bibitem [{\citenamefont {Karliner}\ and\ \citenamefont
  {Rosner}(2014)}]{Karliner:2014gca}%
  \BibitemOpen
  \bibfield  {author} {\bibinfo {author} {\bibfnamefont {Marek}\ \bibnamefont
  {Karliner}}\ and\ \bibinfo {author} {\bibfnamefont {Jonathan~L.}\
  \bibnamefont {Rosner}},\ }\bibfield  {title} {\enquote {\bibinfo {title}
  {{Baryons with two heavy quarks: Masses, production, decays, and
  detection}},}\ }\href {\doibase 10.1103/PhysRevD.90.094007} {\bibfield
  {journal} {\bibinfo  {journal} {Phys. Rev.}\ }\textbf {\bibinfo {volume}
  {D90}},\ \bibinfo {pages} {094007} (\bibinfo {year} {2014})},\ \Eprint
  {http://arxiv.org/abs/1408.5877} {arXiv:1408.5877 [hep-ph]} \BibitemShut
  {NoStop}%
\bibitem [{\citenamefont {Gelman}\ and\ \citenamefont
  {Nussinov}(2003)}]{Gelman:2002wf}%
  \BibitemOpen
  \bibfield  {author} {\bibinfo {author} {\bibfnamefont {Boris~A.}\
  \bibnamefont {Gelman}}\ and\ \bibinfo {author} {\bibfnamefont {Shmuel}\
  \bibnamefont {Nussinov}},\ }\bibfield  {title} {\enquote {\bibinfo {title}
  {{Does a narrow tetraquark $(cc\bar u \bar d)$ state exist?}}}\ }\href
  {\doibase 10.1016/S0370-2693(02)03069-1} {\bibfield  {journal} {\bibinfo
  {journal} {Phys. Lett.}\ }\textbf {\bibinfo {volume} {B551}},\ \bibinfo
  {pages} {296--304} (\bibinfo {year} {2003})},\ \Eprint
  {http://arxiv.org/abs/hep-ph/0209095} {arXiv:hep-ph/0209095 [hep-ph]}
  \BibitemShut {NoStop}%
\bibitem [{\citenamefont {Cohen}\ and\ \citenamefont
  {Hohler}(2006)}]{Cohen:2006jg}%
  \BibitemOpen
  \bibfield  {author} {\bibinfo {author} {\bibfnamefont {Thomas~D.}\
  \bibnamefont {Cohen}}\ and\ \bibinfo {author} {\bibfnamefont {Paul~M.}\
  \bibnamefont {Hohler}},\ }\bibfield  {title} {\enquote {\bibinfo {title}
  {{Doubly heavy hadrons and the domain of validity of doubly heavy
  diquark-anti-quark symmetry}},}\ }\href {\doibase 10.1103/PhysRevD.74.094003}
  {\bibfield  {journal} {\bibinfo  {journal} {Phys. Rev.}\ }\textbf {\bibinfo
  {volume} {D74}},\ \bibinfo {pages} {094003} (\bibinfo {year} {2006})},\
  \Eprint {http://arxiv.org/abs/hep-ph/0606084} {arXiv:hep-ph/0606084 [hep-ph]}
  \BibitemShut {NoStop}%
\bibitem [{\citenamefont {Gignoux}\ \emph {et~al.}(1987)\citenamefont
  {Gignoux}, \citenamefont {Silvestre-Brac},\ and\ \citenamefont
  {Richard}}]{Gignoux:1987cn}%
  \BibitemOpen
  \bibfield  {author} {\bibinfo {author} {\bibfnamefont {C.}~\bibnamefont
  {Gignoux}}, \bibinfo {author} {\bibfnamefont {B.}~\bibnamefont
  {Silvestre-Brac}}, \ and\ \bibinfo {author} {\bibfnamefont {J.~M.}\
  \bibnamefont {Richard}},\ }\bibfield  {title} {\enquote {\bibinfo {title}
  {{Possibility of Stable Multi - Quark Baryons}},}\ }\href {\doibase
  10.1016/0370-2693(87)91244-5} {\bibfield  {journal} {\bibinfo  {journal}
  {Phys. Lett.}\ }\textbf {\bibinfo {volume} {B193}},\ \bibinfo {pages} {323}
  (\bibinfo {year} {1987})}\BibitemShut {NoStop}%
\bibitem [{\citenamefont {Lipkin}(1987)}]{Lipkin:1987sk}%
  \BibitemOpen
  \bibfield  {author} {\bibinfo {author} {\bibfnamefont {Harry~J.}\
  \bibnamefont {Lipkin}},\ }\bibfield  {title} {\enquote {\bibinfo {title}
  {{New Possibilities for Exotic Hadrons: Anticharmed Strange Baryons}},}\
  }\href {\doibase 10.1016/0370-2693(87)90055-4} {\bibfield  {journal}
  {\bibinfo  {journal} {Phys. Lett.}\ }\textbf {\bibinfo {volume} {B195}},\
  \bibinfo {pages} {484} (\bibinfo {year} {1987})}\BibitemShut {NoStop}%
\bibitem [{\citenamefont {Aitala}\ \emph {et~al.}(1998)\citenamefont {Aitala}
  \emph {et~al.}}]{Aitala:1997ja}%
  \BibitemOpen
  \bibfield  {author} {\bibinfo {author} {\bibfnamefont {E.~M.}\ \bibnamefont
  {Aitala}} \emph {et~al.} (\bibinfo {collaboration} {E791}),\ }\bibfield
  {title} {\enquote {\bibinfo {title} {{Search for the pentaquark via the
  $P^0(\bar c s)$ decay}},}\ }\href {\doibase 10.1103/PhysRevLett.81.44}
  {\bibfield  {journal} {\bibinfo  {journal} {Phys. Rev. Lett.}\ }\textbf
  {\bibinfo {volume} {81}},\ \bibinfo {pages} {44--48} (\bibinfo {year}
  {1998})},\ \Eprint {http://arxiv.org/abs/hep-ex/9709013}
  {arXiv:hep-ex/9709013 [hep-ex]} \BibitemShut {NoStop}%
\bibitem [{\citenamefont {Aitala}\ \emph {et~al.}(1999)\citenamefont {Aitala}
  \emph {et~al.}}]{Aitala:1999ij}%
  \BibitemOpen
  \bibfield  {author} {\bibinfo {author} {\bibfnamefont {E.~M.}\ \bibnamefont
  {Aitala}} \emph {et~al.} (\bibinfo {collaboration} {E791}),\ }\bibfield
  {title} {\enquote {\bibinfo {title} {{Search for the pentaquark via the
  $P^0(\bar c s) \to K^{0,*} K^- p$ decay}},}\ }\href {\doibase
  10.1016/S0370-2693(99)00058-1} {\bibfield  {journal} {\bibinfo  {journal}
  {Phys. Lett.}\ }\textbf {\bibinfo {volume} {B448}},\ \bibinfo {pages}
  {303--310} (\bibinfo {year} {1999})}\BibitemShut {NoStop}%
\bibitem [{\citenamefont {Aktas}\ \emph {et~al.}(2004)\citenamefont {Aktas}
  \emph {et~al.}}]{Aktas:2004qf}%
  \BibitemOpen
  \bibfield  {author} {\bibinfo {author} {\bibfnamefont {A.}~\bibnamefont
  {Aktas}} \emph {et~al.} (\bibinfo {collaboration} {H1}),\ }\bibfield  {title}
  {\enquote {\bibinfo {title} {{Evidence for a narrow anti-charmed baryon
  state}},}\ }\href {\doibase 10.1016/j.physletb.2004.03.012} {\bibfield
  {journal} {\bibinfo  {journal} {Phys. Lett.}\ }\textbf {\bibinfo {volume}
  {B588}},\ \bibinfo {pages} {17} (\bibinfo {year} {2004})},\ \Eprint
  {http://arxiv.org/abs/hep-ex/0403017} {arXiv:hep-ex/0403017 [hep-ex]}
  \BibitemShut {NoStop}%
\bibitem [{\citenamefont {Chekanov}\ \emph {et~al.}(2004)\citenamefont
  {Chekanov} \emph {et~al.}}]{Chekanov:2004qm}%
  \BibitemOpen
  \bibfield  {author} {\bibinfo {author} {\bibfnamefont {S.}~\bibnamefont
  {Chekanov}} \emph {et~al.} (\bibinfo {collaboration} {ZEUS}),\ }\bibfield
  {title} {\enquote {\bibinfo {title} {{Search for a narrow charmed baryonic
  state decaying to $D^{*\pm} p^\mp$ in $ep$ collisions at HERA}},}\ }\href
  {\doibase 10.1140/epjc/s2004-02042-9} {\bibfield  {journal} {\bibinfo
  {journal} {Eur. Phys. J.}\ }\textbf {\bibinfo {volume} {C38}},\ \bibinfo
  {pages} {29--41} (\bibinfo {year} {2004})},\ \Eprint
  {http://arxiv.org/abs/hep-ex/0409033} {arXiv:hep-ex/0409033 [hep-ex]}
  \BibitemShut {NoStop}%
\bibitem [{\citenamefont {Aubert}\ \emph
  {et~al.}(2006{\natexlab{b}})\citenamefont {Aubert} \emph
  {et~al.}}]{Aubert:2006qu}%
  \BibitemOpen
  \bibfield  {author} {\bibinfo {author} {\bibfnamefont {Bernard}\ \bibnamefont
  {Aubert}} \emph {et~al.} (\bibinfo {collaboration} {BaBar}),\ }\bibfield
  {title} {\enquote {\bibinfo {title} {{Search for the charmed pentaquark
  candidate $\Theta_c(3100)$ in $e^+e^-$ annihilations at $\sqrt{s} =
  10.58\,$GeV}},}\ }\href {\doibase 10.1103/PhysRevD.73.091101} {\bibfield
  {journal} {\bibinfo  {journal} {Phys. Rev.}\ }\textbf {\bibinfo {volume}
  {D73}},\ \bibinfo {pages} {091101} (\bibinfo {year} {2006}{\natexlab{b}})},\
  \Eprint {http://arxiv.org/abs/hep-ex/0604006} {arXiv:hep-ex/0604006 [hep-ex]}
  \BibitemShut {NoStop}%
\bibitem [{\citenamefont {Baum}\ \emph {et~al.}(1996)\citenamefont {Baum},
  \citenamefont {Kyynäräinen},\ and\ \citenamefont {Tripet}}]{Baum:298433}%
  \BibitemOpen
  \bibfield  {author} {\bibinfo {author} {\bibfnamefont {G}~\bibnamefont
  {Baum}}, \bibinfo {author} {\bibfnamefont {J}~\bibnamefont {Kyynäräinen}},
  \ and\ \bibinfo {author} {\bibfnamefont {A}~\bibnamefont {Tripet}} (\bibinfo
  {collaboration} {NA58 Collaboration}),\ }\href
  {https://cds.cern.ch/record/298433} {\emph {\bibinfo {title} {{COMPASS: a
  proposal for a common muon and proton apparatus for structure and
  spectroscopy}}}},\ \bibinfo {type} {Tech. Rep.}\ \bibinfo {number}
  {CERN-SPSLC-96-14. SPSLC-P-297}\ (\bibinfo  {institution} {CERN},\ \bibinfo
  {address} {Geneva},\ \bibinfo {year} {1996})\BibitemShut {NoStop}%
\bibitem [{\citenamefont {Aaij}\ \emph
  {et~al.}(2015{\natexlab{b}})\citenamefont {Aaij} \emph
  {et~al.}}]{Aaij:2015tga}%
  \BibitemOpen
  \bibfield  {author} {\bibinfo {author} {\bibfnamefont {Roel}\ \bibnamefont
  {Aaij}} \emph {et~al.} (\bibinfo {collaboration} {LHCb}),\ }\bibfield
  {title} {\enquote {\bibinfo {title} {{Observation of $J/\psi p$ Resonances
  Consistent with Pentaquark States in $\Lambda_b^0\to J/\psi K^-p$ Decays}},}\
  }\href {\doibase 10.1103/PhysRevLett.115.072001} {\bibfield  {journal}
  {\bibinfo  {journal} {Phys. Rev. Lett.}\ }\textbf {\bibinfo {volume} {115}},\
  \bibinfo {pages} {072001} (\bibinfo {year} {2015}{\natexlab{b}})},\ \Eprint
  {http://arxiv.org/abs/1507.03414} {arXiv:1507.03414 [hep-ex]} \BibitemShut
  {NoStop}%
\bibitem [{\citenamefont {Aaij}\ \emph
  {et~al.}(2016{\natexlab{c}})\citenamefont {Aaij} \emph
  {et~al.}}]{Aaij:2016phn}%
  \BibitemOpen
  \bibfield  {author} {\bibinfo {author} {\bibfnamefont {Roel}\ \bibnamefont
  {Aaij}} \emph {et~al.} (\bibinfo {collaboration} {LHCb}),\ }\bibfield
  {title} {\enquote {\bibinfo {title} {{Model-independent evidence for $J/\psi
  p$ contributions to $\Lambda_b^0\to J/\psi p K^-$ decays}},}\ }\href@noop {}
  {\  (\bibinfo {year} {2016}{\natexlab{c}})},\ \Eprint
  {http://arxiv.org/abs/1604.05708} {arXiv:1604.05708 [hep-ex]} \BibitemShut
  {NoStop}%
\bibitem [{\citenamefont {Aaij}\ \emph {et~al.}(2014)\citenamefont {Aaij} \emph
  {et~al.}}]{Aaij:2014zoa}%
  \BibitemOpen
  \bibfield  {author} {\bibinfo {author} {\bibfnamefont {Roel}\ \bibnamefont
  {Aaij}} \emph {et~al.} (\bibinfo {collaboration} {LHCb}),\ }\bibfield
  {title} {\enquote {\bibinfo {title} {{Observation of the $\Lambda_b^0
  \rightarrow J/\psi p \pi^-$ decay}},}\ }\href {\doibase
  10.1007/JHEP07(2014)103} {\bibfield  {journal} {\bibinfo  {journal} {JHEP}\
  }\textbf {\bibinfo {volume} {07}},\ \bibinfo {pages} {103} (\bibinfo {year}
  {2014})},\ \Eprint {http://arxiv.org/abs/1406.0755} {arXiv:1406.0755
  [hep-ex]} \BibitemShut {NoStop}%
\bibitem [{\citenamefont {Burns}(2015)}]{Burns:2015dwa}%
  \BibitemOpen
  \bibfield  {author} {\bibinfo {author} {\bibfnamefont {T.~J.}\ \bibnamefont
  {Burns}},\ }\bibfield  {title} {\enquote {\bibinfo {title} {{Phenomenology of
  P$_{c}$(4380)$^{+}$, P$_{c}$(4450)$^{+}$ and related states}},}\ }\href
  {\doibase 10.1140/epja/i2015-15152-6} {\bibfield  {journal} {\bibinfo
  {journal} {Eur. Phys. J.}\ }\textbf {\bibinfo {volume} {A51}},\ \bibinfo
  {pages} {152} (\bibinfo {year} {2015})},\ \Eprint
  {http://arxiv.org/abs/1509.02460} {arXiv:1509.02460 [hep-ph]} \BibitemShut
  {NoStop}%
\bibitem [{\citenamefont {Wang}\ \emph {et~al.}(2016)\citenamefont {Wang},
  \citenamefont {Chen}, \citenamefont {Geng}, \citenamefont {Li},\ and\
  \citenamefont {Oset}}]{Wang:2015pcn}%
  \BibitemOpen
  \bibfield  {author} {\bibinfo {author} {\bibfnamefont {En}~\bibnamefont
  {Wang}}, \bibinfo {author} {\bibfnamefont {Hua-Xing}\ \bibnamefont {Chen}},
  \bibinfo {author} {\bibfnamefont {Li-Sheng}\ \bibnamefont {Geng}}, \bibinfo
  {author} {\bibfnamefont {De-Min}\ \bibnamefont {Li}}, \ and\ \bibinfo
  {author} {\bibfnamefont {Eulogio}\ \bibnamefont {Oset}},\ }\bibfield  {title}
  {\enquote {\bibinfo {title} {{A hidden-charm pentaquark state in $\Lambda^0_b
  \to J/\psi p \pi^-$ decay}},}\ }\href {\doibase 10.1103/PhysRevD.93.094001}
  {\bibfield  {journal} {\bibinfo  {journal} {Phys. Rev.}\ }\textbf {\bibinfo
  {volume} {D93}},\ \bibinfo {pages} {094001} (\bibinfo {year} {2016})},\
  \Eprint {http://arxiv.org/abs/1512.01959} {arXiv:1512.01959 [hep-ph]}
  \BibitemShut {NoStop}%
\bibitem [{\citenamefont {Aaij}\ \emph
  {et~al.}(2016{\natexlab{d}})\citenamefont {Aaij} \emph
  {et~al.}}]{Aaij:2016ymb}%
  \BibitemOpen
  \bibfield  {author} {\bibinfo {author} {\bibfnamefont {Roel}\ \bibnamefont
  {Aaij}} \emph {et~al.} (\bibinfo {collaboration} {LHCb}),\ }\bibfield
  {title} {\enquote {\bibinfo {title} {{Evidence for exotic hadron
  contributions to $\Lambda_b^0 \to J/\psi p \pi^-$ decays}},}\ }\href
  {\doibase 10.1103/PhysRevLett.117.082003, 10.1103/PhysRevLett.117.109902}
  {\bibfield  {journal} {\bibinfo  {journal} {Phys. Rev. Lett.}\ }\textbf
  {\bibinfo {volume} {117}},\ \bibinfo {pages} {082003} (\bibinfo {year}
  {2016}{\natexlab{d}})},\ \bibinfo {note} {[Addendum: Phys. Rev.
  Lett.117,no.10,109902(2016)]},\ \Eprint {http://arxiv.org/abs/1606.06999}
  {arXiv:1606.06999 [hep-ex]} \BibitemShut {NoStop}%
\bibitem [{\citenamefont {Kubarovsky}\ and\ \citenamefont
  {Voloshin}(2015)}]{Kubarovsky:2015aaa}%
  \BibitemOpen
  \bibfield  {author} {\bibinfo {author} {\bibfnamefont {V.}~\bibnamefont
  {Kubarovsky}}\ and\ \bibinfo {author} {\bibfnamefont {M.~B.}\ \bibnamefont
  {Voloshin}},\ }\bibfield  {title} {\enquote {\bibinfo {title} {{Formation of
  hidden-charm pentaquarks in photon-nucleon collisions}},}\ }\href {\doibase
  10.1103/PhysRevD.92.031502} {\bibfield  {journal} {\bibinfo  {journal} {Phys.
  Rev.}\ }\textbf {\bibinfo {volume} {D92}},\ \bibinfo {pages} {031502}
  (\bibinfo {year} {2015})},\ \Eprint {http://arxiv.org/abs/1508.00888}
  {arXiv:1508.00888 [hep-ph]} \BibitemShut {NoStop}%
\bibitem [{\citenamefont {Kubarovsky}\ and\ \citenamefont
  {Voloshin}(2016)}]{Kubarovsky:2016whd}%
  \BibitemOpen
  \bibfield  {author} {\bibinfo {author} {\bibfnamefont {V.}~\bibnamefont
  {Kubarovsky}}\ and\ \bibinfo {author} {\bibfnamefont {M.~B.}\ \bibnamefont
  {Voloshin}},\ }\bibfield  {title} {\enquote {\bibinfo {title} {{Search for
  Hidden-Charm Pentaquark with CLAS12}},}\ }\href@noop {} {\  (\bibinfo {year}
  {2016})},\ \Eprint {http://arxiv.org/abs/1609.00050} {arXiv:1609.00050
  [hep-ph]} \BibitemShut {NoStop}%
\bibitem [{\citenamefont {Blin}\ \emph {et~al.}(2016)\citenamefont {Blin},
  \citenamefont {Fernández-Ramírez}, \citenamefont {Jackura}, \citenamefont
  {Mathieu}, \citenamefont {Mokeev}, \citenamefont {Pilloni},\ and\
  \citenamefont {Szczepaniak}}]{Blin:2016dlf}%
  \BibitemOpen
  \bibfield  {author} {\bibinfo {author} {\bibfnamefont {A.~N.~Hiller}\
  \bibnamefont {Blin}}, \bibinfo {author} {\bibfnamefont {C.}~\bibnamefont
  {Fernández-Ramírez}}, \bibinfo {author} {\bibfnamefont {A.}~\bibnamefont
  {Jackura}}, \bibinfo {author} {\bibfnamefont {V.}~\bibnamefont {Mathieu}},
  \bibinfo {author} {\bibfnamefont {V.~I.}\ \bibnamefont {Mokeev}}, \bibinfo
  {author} {\bibfnamefont {A.}~\bibnamefont {Pilloni}}, \ and\ \bibinfo
  {author} {\bibfnamefont {A.~P.}\ \bibnamefont {Szczepaniak}},\ }\bibfield
  {title} {\enquote {\bibinfo {title} {{Studying the P$_c$(4450) resonance in
  J/$\psi$ photoproduction off protons}},}\ }\href {\doibase
  10.1103/PhysRevD.94.034002} {\bibfield  {journal} {\bibinfo  {journal} {Phys.
  Rev.}\ }\textbf {\bibinfo {volume} {D94}},\ \bibinfo {pages} {034002}
  (\bibinfo {year} {2016})},\ \Eprint {http://arxiv.org/abs/1606.08912}
  {arXiv:1606.08912 [hep-ph]} \BibitemShut {NoStop}%
\bibitem [{\citenamefont {Karliner}\ and\ \citenamefont
  {Rosner}(2016)}]{Karliner:2015voa}%
  \BibitemOpen
  \bibfield  {author} {\bibinfo {author} {\bibfnamefont {Marek}\ \bibnamefont
  {Karliner}}\ and\ \bibinfo {author} {\bibfnamefont {Jonathan~L.}\
  \bibnamefont {Rosner}},\ }\bibfield  {title} {\enquote {\bibinfo {title}
  {{Photoproduction of Exotic Baryon Resonances}},}\ }\href {\doibase
  10.1016/j.physletb.2015.11.068} {\bibfield  {journal} {\bibinfo  {journal}
  {Phys. Lett.}\ }\textbf {\bibinfo {volume} {B752}},\ \bibinfo {pages}
  {329--332} (\bibinfo {year} {2016})},\ \Eprint
  {http://arxiv.org/abs/1508.01496} {arXiv:1508.01496 [hep-ph]} \BibitemShut
  {NoStop}%
\bibitem [{\citenamefont {Meziani}\ \emph {et~al.}(2016)\citenamefont {Meziani}
  \emph {et~al.}}]{Meziani:2016lhg}%
  \BibitemOpen
  \bibfield  {author} {\bibinfo {author} {\bibfnamefont {Z.~E.}\ \bibnamefont
  {Meziani}} \emph {et~al.},\ }\bibfield  {title} {\enquote {\bibinfo {title}
  {{A Search for the LHCb Charmed 'Pentaquark' using Photo-Production of
  $J/{\psi}$ at Threshold in Hall C at Jefferson Lab}},}\ }\href@noop {} {\
  (\bibinfo {year} {2016})},\ \Eprint {http://arxiv.org/abs/1609.00676}
  {arXiv:1609.00676 [hep-ex]} \BibitemShut {NoStop}%
\bibitem [{Mat()}]{Mattione}%
  \BibitemOpen
  \href@noop {} {}\bibinfo {note} {{Paul Mattione, in
  \protect\cite{baryon2016}}}\BibitemShut {NoStop}%
\bibitem [{\citenamefont {Arndt}\ \emph {et~al.}(1987)\citenamefont {Arndt},
  \citenamefont {Hyslop},\ and\ \citenamefont {Roper}}]{Arndt:1986jb}%
  \BibitemOpen
  \bibfield  {author} {\bibinfo {author} {\bibfnamefont {Richard~A.}\
  \bibnamefont {Arndt}}, \bibinfo {author} {\bibfnamefont {John~S.}\
  \bibnamefont {Hyslop}, \bibfnamefont {III}}, \ and\ \bibinfo {author}
  {\bibfnamefont {L.~David}\ \bibnamefont {Roper}},\ }\bibfield  {title}
  {\enquote {\bibinfo {title} {{Nucleon-Nucleon Partial Wave Analysis to
  1100\,MeV}},}\ }\href {\doibase 10.1103/PhysRevD.35.128} {\bibfield
  {journal} {\bibinfo  {journal} {Phys. Rev.}\ }\textbf {\bibinfo {volume}
  {D35}},\ \bibinfo {pages} {128} (\bibinfo {year} {1987})}\BibitemShut
  {NoStop}%
\bibitem [{\citenamefont {Arndt}\ \emph {et~al.}(1993)\citenamefont {Arndt},
  \citenamefont {Strakovsky}, \citenamefont {Workman},\ and\ \citenamefont
  {Bugg}}]{Arndt:1993nd}%
  \BibitemOpen
  \bibfield  {author} {\bibinfo {author} {\bibfnamefont {Richard~A.}\
  \bibnamefont {Arndt}}, \bibinfo {author} {\bibfnamefont {Igor~I.}\
  \bibnamefont {Strakovsky}}, \bibinfo {author} {\bibfnamefont {Ron~L.}\
  \bibnamefont {Workman}}, \ and\ \bibinfo {author} {\bibfnamefont {David~V.}\
  \bibnamefont {Bugg}},\ }\bibfield  {title} {\enquote {\bibinfo {title}
  {{Analysis of the reaction $\pi^+ d\to p p$ to 500\,MeV}},}\ }\href {\doibase
  10.1103/PhysRevC.48.1926} {\bibfield  {journal} {\bibinfo  {journal} {Phys.
  Rev.}\ }\textbf {\bibinfo {volume} {C48}},\ \bibinfo {pages} {1926--1938}
  (\bibinfo {year} {1993})}\BibitemShut {NoStop}%
\bibitem [{\citenamefont {Arndt}\ \emph {et~al.}(1994)\citenamefont {Arndt},
  \citenamefont {Strakovsky},\ and\ \citenamefont {Workman}}]{Arndt:1994bs}%
  \BibitemOpen
  \bibfield  {author} {\bibinfo {author} {\bibfnamefont {Richard~A.}\
  \bibnamefont {Arndt}}, \bibinfo {author} {\bibfnamefont {Igor~I.}\
  \bibnamefont {Strakovsky}}, \ and\ \bibinfo {author} {\bibfnamefont {Ron~L.}\
  \bibnamefont {Workman}},\ }\bibfield  {title} {\enquote {\bibinfo {title}
  {{Analysis of $\pi d$ elastic scattering data to 500\,MeV}},}\ }\href
  {\doibase 10.1103/PhysRevC.50.1796} {\bibfield  {journal} {\bibinfo
  {journal} {Phys. Rev.}\ }\textbf {\bibinfo {volume} {C50}},\ \bibinfo {pages}
  {1796--1806} (\bibinfo {year} {1994})},\ \Eprint
  {http://arxiv.org/abs/nucl-th/9407032} {arXiv:nucl-th/9407032 [nucl-th]}
  \BibitemShut {NoStop}%
\bibitem [{\citenamefont {Tatischeff}\ \emph {et~al.}(1999)\citenamefont
  {Tatischeff} \emph {et~al.}}]{Tatischeff:1999jh}%
  \BibitemOpen
  \bibfield  {author} {\bibinfo {author} {\bibfnamefont {B.}~\bibnamefont
  {Tatischeff}} \emph {et~al.},\ }\bibfield  {title} {\enquote {\bibinfo
  {title} {{Evidence for narrow dibaryons at 2050\,MeV, 2122\,MeV, and
  2150\,MeV observed in inelastic $p p$ scattering}},}\ }\href {\doibase
  10.1103/PhysRevC.59.1878} {\bibfield  {journal} {\bibinfo  {journal} {Phys.
  Rev.}\ }\textbf {\bibinfo {volume} {C59}},\ \bibinfo {pages} {1878--1889}
  (\bibinfo {year} {1999})}\BibitemShut {NoStop}%
\bibitem [{\citenamefont {Parker}\ \emph {et~al.}(1989)\citenamefont {Parker},
  \citenamefont {Seth}, \citenamefont {Ginsburg}, \citenamefont {O'Reilly},\
  and\ \citenamefont {Sarmiento}}]{Parker:1989hw}%
  \BibitemOpen
  \bibfield  {author} {\bibinfo {author} {\bibfnamefont {B.}~\bibnamefont
  {Parker}}, \bibinfo {author} {\bibfnamefont {Kamal~K.}\ \bibnamefont {Seth}},
  \bibinfo {author} {\bibfnamefont {C.~M.}\ \bibnamefont {Ginsburg}}, \bibinfo
  {author} {\bibfnamefont {B.}~\bibnamefont {O'Reilly}}, \ and\ \bibinfo
  {author} {\bibfnamefont {M.}~\bibnamefont {Sarmiento}},\ }\bibfield  {title}
  {\enquote {\bibinfo {title} {{Search for a $T=2$ Dibaryon}},}\ }\href
  {\doibase 10.1103/PhysRevLett.63.1570} {\bibfield  {journal} {\bibinfo
  {journal} {Phys. Rev. Lett.}\ }\textbf {\bibinfo {volume} {63}},\ \bibinfo
  {pages} {1570} (\bibinfo {year} {1989})}\BibitemShut {NoStop}%
\bibitem [{\citenamefont {De~Boer}\ \emph {et~al.}(1984)\citenamefont {De~Boer}
  \emph {et~al.}}]{DeBoer:1984nc}%
  \BibitemOpen
  \bibfield  {author} {\bibinfo {author} {\bibfnamefont {F.~W.~N.}\
  \bibnamefont {De~Boer}} \emph {et~al.},\ }\bibfield  {title} {\enquote
  {\bibinfo {title} {{S\lowercase{EARCH FOR BOUND STATES OF NEUTRONS AND
  NEGATIVE PIONS}}},}\ }\href {\doibase 10.1103/PhysRevLett.53.423} {\bibfield
  {journal} {\bibinfo  {journal} {Phys. Rev. Lett.}\ }\textbf {\bibinfo
  {volume} {53}},\ \bibinfo {pages} {423--426} (\bibinfo {year}
  {1984})}\BibitemShut {NoStop}%
\bibitem [{\citenamefont {Adlarson}\ \emph
  {et~al.}(2014{\natexlab{a}})\citenamefont {Adlarson} \emph
  {et~al.}}]{Adlarson:2014pxj}%
  \BibitemOpen
  \bibfield  {author} {\bibinfo {author} {\bibfnamefont {P.}~\bibnamefont
  {Adlarson}} \emph {et~al.} (\bibinfo {collaboration} {WASA-at-COSY}),\
  }\bibfield  {title} {\enquote {\bibinfo {title} {{Evidence for a New
  Resonance from Polarized Neutron-Proton Scattering}},}\ }\href {\doibase
  10.1103/PhysRevLett.112.202301} {\bibfield  {journal} {\bibinfo  {journal}
  {Phys. Rev. Lett.}\ }\textbf {\bibinfo {volume} {112}},\ \bibinfo {pages}
  {202301} (\bibinfo {year} {2014}{\natexlab{a}})},\ \Eprint
  {http://arxiv.org/abs/1402.6844} {arXiv:1402.6844 [nucl-ex]} \BibitemShut
  {NoStop}%
\bibitem [{\citenamefont {Adlarson}\ \emph
  {et~al.}(2014{\natexlab{b}})\citenamefont {Adlarson} \emph
  {et~al.}}]{Adlarson:2014ozl}%
  \BibitemOpen
  \bibfield  {author} {\bibinfo {author} {\bibfnamefont {P.}~\bibnamefont
  {Adlarson}} \emph {et~al.} (\bibinfo {collaboration} {WASA-at-COSY}),\
  }\bibfield  {title} {\enquote {\bibinfo {title} {{Neutron-proton scattering
  in the context of the d* (2380) resonance}},}\ }\href {\doibase
  10.1103/PhysRevC.90.035204} {\bibfield  {journal} {\bibinfo  {journal} {Phys.
  Rev.}\ }\textbf {\bibinfo {volume} {C90}},\ \bibinfo {pages} {035204}
  (\bibinfo {year} {2014}{\natexlab{b}})},\ \Eprint
  {http://arxiv.org/abs/1408.4928} {arXiv:1408.4928 [nucl-ex]} \BibitemShut
  {NoStop}%
\bibitem [{\citenamefont {Goldman}\ \emph {et~al.}(1989)\citenamefont
  {Goldman}, \citenamefont {Maltman}, \citenamefont {Stephenson}, \citenamefont
  {Schmidt},\ and\ \citenamefont {Wang}}]{Goldman:1989zj}%
  \BibitemOpen
  \bibfield  {author} {\bibinfo {author} {\bibfnamefont {J.~Terrance}\
  \bibnamefont {Goldman}}, \bibinfo {author} {\bibfnamefont {Kim}\ \bibnamefont
  {Maltman}}, \bibinfo {author} {\bibfnamefont {G.~J.}\ \bibnamefont
  {Stephenson}, \bibfnamefont {Jr.}}, \bibinfo {author} {\bibfnamefont {K.~E.}\
  \bibnamefont {Schmidt}}, \ and\ \bibinfo {author} {\bibfnamefont {Fan}\
  \bibnamefont {Wang}},\ }\bibfield  {title} {\enquote {\bibinfo {title} {{An
  'inevitable' nonstrange dibaryon}},}\ }\href {\doibase
  10.1103/PhysRevC.39.1889} {\bibfield  {journal} {\bibinfo  {journal} {Phys.
  Rev.}\ }\textbf {\bibinfo {volume} {C39}},\ \bibinfo {pages} {1889--1895}
  (\bibinfo {year} {1989})}\BibitemShut {NoStop}%
\bibitem [{\citenamefont {Ohnishi}\ \emph {et~al.}(2013)\citenamefont
  {Ohnishi}, \citenamefont {Ikeda}, \citenamefont {Kamano},\ and\ \citenamefont
  {Sato}}]{Ohnishi:2011wb}%
  \BibitemOpen
  \bibfield  {author} {\bibinfo {author} {\bibfnamefont {Shota}\ \bibnamefont
  {Ohnishi}}, \bibinfo {author} {\bibfnamefont {Yoichi}\ \bibnamefont {Ikeda}},
  \bibinfo {author} {\bibfnamefont {Hiroyuki}\ \bibnamefont {Kamano}}, \ and\
  \bibinfo {author} {\bibfnamefont {Toru}\ \bibnamefont {Sato}},\ }\bibfield
  {title} {\enquote {\bibinfo {title} {{Signature of strange dibaryon in
  kaon-induced reaction}},}\ }\bibfield  {booktitle} {\emph {\bibinfo
  {booktitle} {{Proceedings, 5th Asia-Pacific Conference on Few-Body Problems
  in Physics 2011 (APFB2011)}}},\ }\href {\doibase 10.1007/s00601-012-0390-6}
  {\bibfield  {journal} {\bibinfo  {journal} {Few Body Syst.}\ }\textbf
  {\bibinfo {volume} {54}},\ \bibinfo {pages} {347--351} (\bibinfo {year}
  {2013})},\ \Eprint {http://arxiv.org/abs/1109.4724} {arXiv:1109.4724
  [nucl-th]} \BibitemShut {NoStop}%
\bibitem [{\citenamefont {Hashimoto}\ \emph {et~al.}(2014)\citenamefont
  {Hashimoto} \emph {et~al.}}]{Hashimoto:2014cri}%
  \BibitemOpen
  \bibfield  {author} {\bibinfo {author} {\bibfnamefont {T.}~\bibnamefont
  {Hashimoto}} \emph {et~al.} (\bibinfo {collaboration} {J-PARC E15}),\
  }\bibfield  {title} {\enquote {\bibinfo {title} {{Search for the deeply bound
  $K^-pp$ state from the semi-inclusive forward-neutron spectrum in the
  in-flight $K^-$ reaction on ${}^3$He}},}\ }\href {\doibase
  10.1093/ptep/ptv076} {\bibfield  {journal} {\bibinfo  {journal} {PTEP}\
  }\textbf {\bibinfo {volume} {2015}},\ \bibinfo {pages} {061D01} (\bibinfo
  {year} {2014})},\ \Eprint {http://arxiv.org/abs/1408.5637} {arXiv:1408.5637
  [nucl-ex]} \BibitemShut {NoStop}%
\bibitem [{\citenamefont {Zel'dovich}({1960})}]{ISI:A1960WT32800001}%
  \BibitemOpen
  \bibfield  {author} {\bibinfo {author} {\bibfnamefont {Y.B.}\ \bibnamefont
  {Zel'dovich}},\ }\bibfield  {title} {\enquote {\bibinfo {title} {{Energy
  levels in a distorted Coulomb Field}},}\ }\href@noop {} {\bibfield  {journal}
  {\bibinfo  {journal} {{Soviet Physics-Solid State}}\ }\textbf {\bibinfo
  {volume} {{1}}},\ \bibinfo {pages} {{1497--1501}} (\bibinfo {year}
  {{1960}})}\BibitemShut {NoStop}%
\bibitem [{\citenamefont {Gal}\ \emph {et~al.}(1996)\citenamefont {Gal},
  \citenamefont {Friedman},\ and\ \citenamefont {Batty}}]{Gal:1996pr}%
  \BibitemOpen
  \bibfield  {author} {\bibinfo {author} {\bibfnamefont {A.}~\bibnamefont
  {Gal}}, \bibinfo {author} {\bibfnamefont {E.}~\bibnamefont {Friedman}}, \
  and\ \bibinfo {author} {\bibfnamefont {C.~J.}\ \bibnamefont {Batty}},\
  }\bibfield  {title} {\enquote {\bibinfo {title} {{On the interplay between
  Coulomb and nuclear states in exotic atoms}},}\ }\href {\doibase
  10.1016/0375-9474(96)00211-4} {\bibfield  {journal} {\bibinfo  {journal}
  {Nucl. Phys.}\ }\textbf {\bibinfo {volume} {A606}},\ \bibinfo {pages}
  {283--291} (\bibinfo {year} {1996})}\BibitemShut {NoStop}%
\bibitem [{\citenamefont {Combescure}\ \emph {et~al.}(2007)\citenamefont
  {Combescure}, \citenamefont {Khare}, \citenamefont {Raina}, \citenamefont
  {Richard},\ and\ \citenamefont {Weydert}}]{Combescure:2007ki}%
  \BibitemOpen
  \bibfield  {author} {\bibinfo {author} {\bibfnamefont {Monique}\ \bibnamefont
  {Combescure}}, \bibinfo {author} {\bibfnamefont {Avinash}\ \bibnamefont
  {Khare}}, \bibinfo {author} {\bibfnamefont {Ashok}\ \bibnamefont {Raina}},
  \bibinfo {author} {\bibfnamefont {Jean-Marc}\ \bibnamefont {Richard}}, \ and\
  \bibinfo {author} {\bibfnamefont {Carole}\ \bibnamefont {Weydert}},\
  }\bibfield  {title} {\enquote {\bibinfo {title} {{Level rearrangement in
  exotic atoms and quantum dots}},}\ }\href {\doibase
  10.1142/S0217979207037855} {\bibfield  {journal} {\bibinfo  {journal} {Int.
  J. Mod. Phys.}\ }\textbf {\bibinfo {volume} {B21}},\ \bibinfo {pages}
  {3765--3781} (\bibinfo {year} {2007})},\ \Eprint
  {http://arxiv.org/abs/cond-mat/0701006} {arXiv:cond-mat/0701006 [cond-mat]}
  \BibitemShut {NoStop}%
\bibitem [{\citenamefont {Jaffe}(1977)}]{Jaffe:1976yi}%
  \BibitemOpen
  \bibfield  {author} {\bibinfo {author} {\bibfnamefont {Robert~L.}\
  \bibnamefont {Jaffe}},\ }\bibfield  {title} {\enquote {\bibinfo {title}
  {{Perhaps a Stable Dihyperon}},}\ }\href {\doibase
  10.1103/PhysRevLett.38.195} {\bibfield  {journal} {\bibinfo  {journal} {Phys.
  Rev. Lett.}\ }\textbf {\bibinfo {volume} {38}},\ \bibinfo {pages} {195--198}
  (\bibinfo {year} {1977})},\ \bibinfo {note} {[Erratum: Phys. Rev.
  Lett.38,617(1977)]}\BibitemShut {NoStop}%
\bibitem [{\citenamefont {Dalitz}\ \emph {et~al.}(1989)\citenamefont {Dalitz},
  \citenamefont {Davis}, \citenamefont {Fowler}, \citenamefont {Montwill},
  \citenamefont {Pniewski},\ and\ \citenamefont {Zakrzewski}}]{Dalitz:1989kt}%
  \BibitemOpen
  \bibfield  {author} {\bibinfo {author} {\bibfnamefont {R.~H.}\ \bibnamefont
  {Dalitz}}, \bibinfo {author} {\bibfnamefont {D.~H.}\ \bibnamefont {Davis}},
  \bibinfo {author} {\bibfnamefont {P.~H.}\ \bibnamefont {Fowler}}, \bibinfo
  {author} {\bibfnamefont {A.}~\bibnamefont {Montwill}}, \bibinfo {author}
  {\bibfnamefont {J.}~\bibnamefont {Pniewski}}, \ and\ \bibinfo {author}
  {\bibfnamefont {Janusz~Andrzej}\ \bibnamefont {Zakrzewski}},\ }\bibfield
  {title} {\enquote {\bibinfo {title} {{The Identified $\Lambda \Lambda$
  Hypernuclei and the Predicted $H$ Particle}},}\ }\href {\doibase
  10.1098/rspa.1989.0115} {\bibfield  {journal} {\bibinfo  {journal} {Proc.
  Roy. Soc. Lond.}\ }\textbf {\bibinfo {volume} {A426}},\ \bibinfo {pages}
  {1--17} (\bibinfo {year} {1989})}\BibitemShut {NoStop}%
\bibitem [{\citenamefont {Kuhn}\ \emph {et~al.}(2002)\citenamefont {Kuhn},
  \citenamefont {Hippolyte}, \citenamefont {Coffin}, \citenamefont {Baudot},
  \citenamefont {Belikov}, \citenamefont {Dietrich}, \citenamefont {Germain},\
  and\ \citenamefont {Suire}}]{Kuhn:2002xf}%
  \BibitemOpen
  \bibfield  {author} {\bibinfo {author} {\bibfnamefont {C.}~\bibnamefont
  {Kuhn}}, \bibinfo {author} {\bibfnamefont {B.}~\bibnamefont {Hippolyte}},
  \bibinfo {author} {\bibfnamefont {J.~P.}\ \bibnamefont {Coffin}}, \bibinfo
  {author} {\bibfnamefont {J.}~\bibnamefont {Baudot}}, \bibinfo {author}
  {\bibfnamefont {I.}~\bibnamefont {Belikov}}, \bibinfo {author} {\bibfnamefont
  {D.}~\bibnamefont {Dietrich}}, \bibinfo {author} {\bibfnamefont
  {M.}~\bibnamefont {Germain}}, \ and\ \bibinfo {author} {\bibfnamefont
  {C.}~\bibnamefont {Suire}},\ }\bibfield  {title} {\enquote {\bibinfo {title}
  {{Search for strange dibaryons in STAR and ALICE}},}\ }\bibfield  {booktitle}
  {\emph {\bibinfo {booktitle} {{Strange quarks in matter. Proceedings, 6th
  International Conference, SQM 2001, Frankfurt, Germany, September 24-29,
  2001}}},\ }\href {\doibase 10.1088/0954-3899/28/7/323} {\bibfield  {journal}
  {\bibinfo  {journal} {J. Phys.}\ }\textbf {\bibinfo {volume} {G28}},\
  \bibinfo {pages} {1707--1714} (\bibinfo {year} {2002})}\BibitemShut {NoStop}%
\bibitem [{\citenamefont {Goldman}\ \emph {et~al.}(1987)\citenamefont
  {Goldman}, \citenamefont {Maltman}, \citenamefont {Stephenson}, \citenamefont
  {Schmidt},\ and\ \citenamefont {Wang}}]{Goldman:1987ma}%
  \BibitemOpen
  \bibfield  {author} {\bibinfo {author} {\bibfnamefont {J.~Terrance}\
  \bibnamefont {Goldman}}, \bibinfo {author} {\bibfnamefont {Kim}\ \bibnamefont
  {Maltman}}, \bibinfo {author} {\bibfnamefont {G.~J.}\ \bibnamefont
  {Stephenson}, \bibfnamefont {Jr.}}, \bibinfo {author} {\bibfnamefont {K.~E.}\
  \bibnamefont {Schmidt}}, \ and\ \bibinfo {author} {\bibfnamefont {Fan}\
  \bibnamefont {Wang}},\ }\bibfield  {title} {\enquote {\bibinfo {title}
  {{Strangeness $-3$ dibaryons}},}\ }\href {\doibase
  10.1103/PhysRevLett.59.627} {\bibfield  {journal} {\bibinfo  {journal} {Phys.
  Rev. Lett.}\ }\textbf {\bibinfo {volume} {59}},\ \bibinfo {pages} {627}
  (\bibinfo {year} {1987})}\BibitemShut {NoStop}%
\bibitem [{\citenamefont {Huang}\ \emph {et~al.}(2015)\citenamefont {Huang},
  \citenamefont {Ping},\ and\ \citenamefont {Wang}}]{Huang:2015yza}%
  \BibitemOpen
  \bibfield  {author} {\bibinfo {author} {\bibfnamefont {Hongxia}\ \bibnamefont
  {Huang}}, \bibinfo {author} {\bibfnamefont {Jialun}\ \bibnamefont {Ping}}, \
  and\ \bibinfo {author} {\bibfnamefont {Fan}\ \bibnamefont {Wang}},\
  }\bibfield  {title} {\enquote {\bibinfo {title} {{Further study of the
  $N\Omega$ dibaryon within constituent quark models}},}\ }\href {\doibase
  10.1103/PhysRevC.92.065202} {\bibfield  {journal} {\bibinfo  {journal} {Phys.
  Rev.}\ }\textbf {\bibinfo {volume} {C92}},\ \bibinfo {pages} {065202}
  (\bibinfo {year} {2015})},\ \Eprint {http://arxiv.org/abs/1507.07124}
  {arXiv:1507.07124 [hep-ph]} \BibitemShut {NoStop}%
\bibitem [{\citenamefont {Thomas}(1935)}]{Thomas:1935zz}%
  \BibitemOpen
  \bibfield  {author} {\bibinfo {author} {\bibfnamefont {L.~H.}\ \bibnamefont
  {Thomas}},\ }\bibfield  {title} {\enquote {\bibinfo {title} {{The Interaction
  Between a Neutron and a Proton and the Structure of ${}^3$H}},}\ }\href
  {\doibase 10.1103/PhysRev.47.903} {\bibfield  {journal} {\bibinfo  {journal}
  {Phys. Rev.}\ }\textbf {\bibinfo {volume} {47}},\ \bibinfo {pages} {903--909}
  (\bibinfo {year} {1935})}\BibitemShut {NoStop}%
\bibitem [{\citenamefont {Zhukov}\ \emph {et~al.}(1993)\citenamefont {Zhukov},
  \citenamefont {Danilin}, \citenamefont {Fedorov}, \citenamefont {Bang},
  \citenamefont {Thompson},\ and\ \citenamefont {Vaagen}}]{Zhukov:1993aw}%
  \BibitemOpen
  \bibfield  {author} {\bibinfo {author} {\bibfnamefont {M.~V.}\ \bibnamefont
  {Zhukov}}, \bibinfo {author} {\bibfnamefont {B.~V.}\ \bibnamefont {Danilin}},
  \bibinfo {author} {\bibfnamefont {D.~V.}\ \bibnamefont {Fedorov}}, \bibinfo
  {author} {\bibfnamefont {J.~M.}\ \bibnamefont {Bang}}, \bibinfo {author}
  {\bibfnamefont {I.~J.}\ \bibnamefont {Thompson}}, \ and\ \bibinfo {author}
  {\bibfnamefont {J.~S.}\ \bibnamefont {Vaagen}},\ }\bibfield  {title}
  {\enquote {\bibinfo {title} {{Bound state properties of Borromean Halo
  nuclei: ${}^6$He and ${}^{11}$Li}},}\ }\href {\doibase
  10.1016/0370-1573(93)90141-Y} {\bibfield  {journal} {\bibinfo  {journal}
  {Phys. Rept.}\ }\textbf {\bibinfo {volume} {231}},\ \bibinfo {pages}
  {151--199} (\bibinfo {year} {1993})}\BibitemShut {NoStop}%
\bibitem [{\citenamefont {Richard}(2003)}]{Richard:2003nn}%
  \BibitemOpen
  \bibfield  {author} {\bibinfo {author} {\bibfnamefont {Jean-Marc}\
  \bibnamefont {Richard}},\ }\bibfield  {title} {\enquote {\bibinfo {title}
  {{Borromean binding}},}\ }\href@noop {} {\  (\bibinfo {year} {2003})},\
  \bibinfo {note} {\url{http://theor.jinr.ru/Few-body/Belyaev-70/}},\ \Eprint
  {http://arxiv.org/abs/nucl-th/0305076} {arXiv:nucl-th/0305076 [nucl-th]}
  \BibitemShut {NoStop}%
\bibitem [{\citenamefont {Hiyama}(2012)}]{Hiyama:2012gx}%
  \BibitemOpen
  \bibfield  {author} {\bibinfo {author} {\bibfnamefont {Emiko}\ \bibnamefont
  {Hiyama}},\ }\bibfield  {title} {\enquote {\bibinfo {title} {{Few-body
  aspects of hypernuclear physics}},}\ }\href {\doibase
  10.1007/s00601-011-0296-8} {\bibfield  {journal} {\bibinfo  {journal} {Few
  Body Syst.}\ }\textbf {\bibinfo {volume} {53}},\ \bibinfo {pages} {189--236}
  (\bibinfo {year} {2012})}\BibitemShut {NoStop}%
\bibitem [{\citenamefont {Filikhin}\ and\ \citenamefont
  {Gal}(2002)}]{Filikhin:2001yh}%
  \BibitemOpen
  \bibfield  {author} {\bibinfo {author} {\bibfnamefont {I.~N.}\ \bibnamefont
  {Filikhin}}\ and\ \bibinfo {author} {\bibfnamefont {A.}~\bibnamefont {Gal}},\
  }\bibfield  {title} {\enquote {\bibinfo {title} {{Light $\Lambda\Lambda$
  hypernuclei and the onset of stability for $\Lambda\Xi$ hypernuclei}},}\
  }\href {\doibase 10.1103/PhysRevC.65.041001} {\bibfield  {journal} {\bibinfo
  {journal} {Phys. Rev.}\ }\textbf {\bibinfo {volume} {C65}},\ \bibinfo {pages}
  {041001} (\bibinfo {year} {2002})},\ \Eprint
  {http://arxiv.org/abs/nucl-th/0110008} {arXiv:nucl-th/0110008 [nucl-th]}
  \BibitemShut {NoStop}%
\bibitem [{\citenamefont {Garcilazo}\ \emph {et~al.}(2015)\citenamefont
  {Garcilazo}, \citenamefont {Valcarce},\ and\ \citenamefont
  {Carames}}]{Garcilazo:2014mra}%
  \BibitemOpen
  \bibfield  {author} {\bibinfo {author} {\bibfnamefont {H.}~\bibnamefont
  {Garcilazo}}, \bibinfo {author} {\bibfnamefont {A.}~\bibnamefont {Valcarce}},
  \ and\ \bibinfo {author} {\bibfnamefont {T.~F.}\ \bibnamefont {Carames}},\
  }\bibfield  {title} {\enquote {\bibinfo {title} {{The $N\Lambda\Lambda-\Xi
  NN$ bound state problem with $N\Lambda \Sigma$ and $N\Sigma \Sigma$
  channels}},}\ }\href {\doibase 10.1088/0954-3899/42/2/025103} {\bibfield
  {journal} {\bibinfo  {journal} {J. Phys.}\ }\textbf {\bibinfo {volume}
  {G42}},\ \bibinfo {pages} {025103} (\bibinfo {year} {2015})}\BibitemShut
  {NoStop}%
\bibitem [{\citenamefont {Richard}\ \emph {et~al.}(2015)\citenamefont
  {Richard}, \citenamefont {Wang},\ and\ \citenamefont
  {Zhao}}]{Richard:2014pwa}%
  \BibitemOpen
  \bibfield  {author} {\bibinfo {author} {\bibfnamefont {Jean-Marc}\
  \bibnamefont {Richard}}, \bibinfo {author} {\bibfnamefont {Qian}\
  \bibnamefont {Wang}}, \ and\ \bibinfo {author} {\bibfnamefont {Qiang}\
  \bibnamefont {Zhao}},\ }\bibfield  {title} {\enquote {\bibinfo {title}
  {{Lightest neutral hypernuclei with strangeness $-1$ and $-2$}},}\ }\href
  {\doibase 10.1103/PhysRevC.91.014003} {\bibfield  {journal} {\bibinfo
  {journal} {Phys. Rev.}\ }\textbf {\bibinfo {volume} {C91}},\ \bibinfo {pages}
  {014003} (\bibinfo {year} {2015})},\ \Eprint {http://arxiv.org/abs/1404.3473}
  {arXiv:1404.3473 [nucl-th]} \BibitemShut {NoStop}%
\bibitem [{\citenamefont {Garcilazo}\ \emph {et~al.}(2016)\citenamefont
  {Garcilazo}, \citenamefont {Valcarce},\ and\ \citenamefont
  {Vijande}}]{Garcilazo:2016ams}%
  \BibitemOpen
  \bibfield  {author} {\bibinfo {author} {\bibfnamefont {H.}~\bibnamefont
  {Garcilazo}}, \bibinfo {author} {\bibfnamefont {A.}~\bibnamefont {Valcarce}},
  \ and\ \bibinfo {author} {\bibfnamefont {J.}~\bibnamefont {Vijande}},\
  }\bibfield  {title} {\enquote {\bibinfo {title} {{Maximal isospin few-body
  systems of nucleons and $\Xi$ hyperons}},}\ }\href {\doibase
  10.1103/PhysRevC.94.024002} {\bibfield  {journal} {\bibinfo  {journal} {Phys.
  Rev.}\ }\textbf {\bibinfo {volume} {C94}},\ \bibinfo {pages} {024002}
  (\bibinfo {year} {2016})},\ \Eprint {http://arxiv.org/abs/1608.05192}
  {arXiv:1608.05192 [nucl-th]} \BibitemShut {NoStop}%
\bibitem [{\citenamefont {{Carbonell}}\ \emph {et~al.}(2003)\citenamefont
  {{Carbonell}}, \citenamefont {{Lazauskas}}, \citenamefont {{Delande}},
  \citenamefont {{Hilico}},\ and\ \citenamefont {{Kili{\c
  c}}}}]{2003EL.....64..316C}%
  \BibitemOpen
  \bibfield  {author} {\bibinfo {author} {\bibfnamefont {J.}~\bibnamefont
  {{Carbonell}}}, \bibinfo {author} {\bibfnamefont {R.}~\bibnamefont
  {{Lazauskas}}}, \bibinfo {author} {\bibfnamefont {D.}~\bibnamefont
  {{Delande}}}, \bibinfo {author} {\bibfnamefont {L.}~\bibnamefont {{Hilico}}},
  \ and\ \bibinfo {author} {\bibfnamefont {S.}~\bibnamefont {{Kili{\c c}}}},\
  }\bibfield  {title} {\enquote {\bibinfo {title} {{A new vibrational level of
  the H$_{2}$$^{+}$ molecular ion}},}\ }\href {\doibase
  10.1209/epl/i2003-00176-1} {\bibfield  {journal} {\bibinfo  {journal} {EPL
  (Europhysics Letters)}\ }\textbf {\bibinfo {volume} {64}},\ \bibinfo {pages}
  {316--322} (\bibinfo {year} {2003})},\ \Eprint
  {http://arxiv.org/abs/physics/0207007} {physics/0207007} \BibitemShut
  {NoStop}%
\bibitem [{\citenamefont {{Varga}}\ \emph {et~al.}(1998)\citenamefont
  {{Varga}}, \citenamefont {{Usukura}},\ and\ \citenamefont
  {{Suzuki}}}]{1998PhRvL..80.1876V}%
  \BibitemOpen
  \bibfield  {author} {\bibinfo {author} {\bibfnamefont {K.}~\bibnamefont
  {{Varga}}}, \bibinfo {author} {\bibfnamefont {J.}~\bibnamefont {{Usukura}}},
  \ and\ \bibinfo {author} {\bibfnamefont {Y.}~\bibnamefont {{Suzuki}}},\
  }\bibfield  {title} {\enquote {\bibinfo {title} {{Second Bound State of the
  Positronium Molecule and Biexcitons}},}\ }\href {\doibase
  10.1103/PhysRevLett.80.1876} {\bibfield  {journal} {\bibinfo  {journal}
  {Physical Review Letters}\ }\textbf {\bibinfo {volume} {80}},\ \bibinfo
  {pages} {1876--1879} (\bibinfo {year} {1998})},\ \Eprint
  {http://arxiv.org/abs/cond-mat/9802261} {cond-mat/9802261} \BibitemShut
  {NoStop}%
\bibitem [{\citenamefont {{Czarnecki}}(2009)}]{2009NuPhA.827..541C}%
  \BibitemOpen
  \bibfield  {author} {\bibinfo {author} {\bibfnamefont {A.}~\bibnamefont
  {{Czarnecki}}},\ }\bibfield  {title} {\enquote {\bibinfo {title}
  {{Positronium and Polyelectrons}},}\ }\href {\doibase
  10.1016/j.nuclphysa.2009.05.118} {\bibfield  {journal} {\bibinfo  {journal}
  {Nuclear Physics A}\ }\textbf {\bibinfo {volume} {827}},\ \bibinfo {pages}
  {541--543c} (\bibinfo {year} {2009})}\BibitemShut {NoStop}%
\bibitem [{\citenamefont {{Richard}}(2003)}]{2003PhRvA..67c4702R}%
  \BibitemOpen
  \bibfield  {author} {\bibinfo {author} {\bibfnamefont {J.-M.}\ \bibnamefont
  {{Richard}}},\ }\bibfield  {title} {\enquote {\bibinfo {title} {{Critically
  bound four-body molecules}},}\ }\href {\doibase 10.1103/PhysRevA.67.034702}
  {\bibfield  {journal} {\bibinfo  {journal} {\pra}\ }\textbf {\bibinfo
  {volume} {67}},\ \bibinfo {eid} {034702} (\bibinfo {year} {2003})},\ \Eprint
  {http://arxiv.org/abs/physics/0302004} {physics/0302004} \BibitemShut
  {NoStop}%
\bibitem [{\citenamefont {Roy}(2004)}]{Roy:2003hk}%
  \BibitemOpen
  \bibfield  {author} {\bibinfo {author} {\bibfnamefont {D.~P.}\ \bibnamefont
  {Roy}},\ }\bibfield  {title} {\enquote {\bibinfo {title} {{History of exotic
  meson (4-quark) and baryon (5-quark) states}},}\ }\href@noop {} {\bibfield
  {journal} {\bibinfo  {journal} {J. Phys.}\ }\textbf {\bibinfo {volume}
  {G30}},\ \bibinfo {pages} {R113} (\bibinfo {year} {2004})},\ \Eprint
  {http://arxiv.org/abs/hep-ph/0311207} {arXiv:hep-ph/0311207} \BibitemShut
  {NoStop}%
\bibitem [{\citenamefont {Rosner}(1968)}]{Rosner:1968si}%
  \BibitemOpen
  \bibfield  {author} {\bibinfo {author} {\bibfnamefont {Jonathan~L.}\
  \bibnamefont {Rosner}},\ }\bibfield  {title} {\enquote {\bibinfo {title}
  {{Possibility of baryon - anti-baryon enhancements with unusual quantum
  numbers}},}\ }\href {\doibase 10.1103/PhysRevLett.21.950} {\bibfield
  {journal} {\bibinfo  {journal} {Phys. Rev. Lett.}\ }\textbf {\bibinfo
  {volume} {21}},\ \bibinfo {pages} {950--952} (\bibinfo {year}
  {1968})}\BibitemShut {NoStop}%
\bibitem [{\citenamefont {Rossi}\ and\ \citenamefont
  {Veneziano}(2016)}]{Rossi:2016szw}%
  \BibitemOpen
  \bibfield  {author} {\bibinfo {author} {\bibfnamefont {Giancarlo}\
  \bibnamefont {Rossi}}\ and\ \bibinfo {author} {\bibfnamefont {Gabriele}\
  \bibnamefont {Veneziano}},\ }\bibfield  {title} {\enquote {\bibinfo {title}
  {{The string-junction picture of multiquark states: an update}},}\ }\href
  {\doibase 10.1007/JHEP06(2016)041} {\bibfield  {journal} {\bibinfo  {journal}
  {JHEP}\ }\textbf {\bibinfo {volume} {06}},\ \bibinfo {pages} {041} (\bibinfo
  {year} {2016})},\ \Eprint {http://arxiv.org/abs/1603.05830} {arXiv:1603.05830
  [hep-th]} \BibitemShut {NoStop}%
\bibitem [{\citenamefont {Artru}(1975)}]{Artru:1974zn}%
  \BibitemOpen
  \bibfield  {author} {\bibinfo {author} {\bibfnamefont {X.}~\bibnamefont
  {Artru}},\ }\bibfield  {title} {\enquote {\bibinfo {title} {String model with
  baryons: Topology, classical motion},}\ }\href@noop {} {\bibfield  {journal}
  {\bibinfo  {journal} {Nucl. Phys.}\ }\textbf {\bibinfo {volume} {B85}},\
  \bibinfo {pages} {442} (\bibinfo {year} {1975})}\BibitemShut {NoStop}%
\bibitem [{\citenamefont {Dosch}\ and\ \citenamefont
  {Muller}(1976)}]{Dosch:1975gf}%
  \BibitemOpen
  \bibfield  {author} {\bibinfo {author} {\bibfnamefont {Hans~Gunter}\
  \bibnamefont {Dosch}}\ and\ \bibinfo {author} {\bibfnamefont {V.~F.}\
  \bibnamefont {Muller}},\ }\bibfield  {title} {\enquote {\bibinfo {title} {On
  composite hadrons in nonabelian lattice gauge theories},}\ }\href@noop {}
  {\bibfield  {journal} {\bibinfo  {journal} {Nucl. Phys.}\ }\textbf {\bibinfo
  {volume} {B116}},\ \bibinfo {pages} {470} (\bibinfo {year}
  {1976})}\BibitemShut {NoStop}%
\bibitem [{\citenamefont {Hasenfratz}\ \emph
  {et~al.}(1980{\natexlab{a}})\citenamefont {Hasenfratz}, \citenamefont
  {Horgan}, \citenamefont {Kuti},\ and\ \citenamefont
  {Richard}}]{Hasenfratz:1980ka}%
  \BibitemOpen
  \bibfield  {author} {\bibinfo {author} {\bibfnamefont {P.}~\bibnamefont
  {Hasenfratz}}, \bibinfo {author} {\bibfnamefont {R.~R.}\ \bibnamefont
  {Horgan}}, \bibinfo {author} {\bibfnamefont {J.}~\bibnamefont {Kuti}}, \ and\
  \bibinfo {author} {\bibfnamefont {J.~M.}\ \bibnamefont {Richard}},\
  }\bibfield  {title} {\enquote {\bibinfo {title} {{Heavy Baryon Spectroscopy
  in the {QCD} Bag Model}},}\ }\href {\doibase 10.1016/0370-2693(80)90906-5}
  {\bibfield  {journal} {\bibinfo  {journal} {Phys. Lett.}\ }\textbf {\bibinfo
  {volume} {B94}},\ \bibinfo {pages} {401--404} (\bibinfo {year}
  {1980}{\natexlab{a}})}\BibitemShut {NoStop}%
\bibitem [{\citenamefont {Carlson}\ \emph {et~al.}(1983)\citenamefont
  {Carlson}, \citenamefont {Kogut},\ and\ \citenamefont
  {Pandharipande}}]{Carlson:1982xi}%
  \BibitemOpen
  \bibfield  {author} {\bibinfo {author} {\bibfnamefont {J.}~\bibnamefont
  {Carlson}}, \bibinfo {author} {\bibfnamefont {John~B.}\ \bibnamefont
  {Kogut}}, \ and\ \bibinfo {author} {\bibfnamefont {V.~R.}\ \bibnamefont
  {Pandharipande}},\ }\bibfield  {title} {\enquote {\bibinfo {title} {A quark
  model for baryons based on quantum chromodynamics},}\ }\href@noop {}
  {\bibfield  {journal} {\bibinfo  {journal} {Phys. Rev.}\ }\textbf {\bibinfo
  {volume} {D27}},\ \bibinfo {pages} {233} (\bibinfo {year}
  {1983})}\BibitemShut {NoStop}%
\bibitem [{\citenamefont {Bagan}\ \emph {et~al.}(1985)\citenamefont {Bagan},
  \citenamefont {Latorre}, \citenamefont {Merkurev},\ and\ \citenamefont
  {Tarrach}}]{Bagan:1985zn}%
  \BibitemOpen
  \bibfield  {author} {\bibinfo {author} {\bibfnamefont {E.}~\bibnamefont
  {Bagan}}, \bibinfo {author} {\bibfnamefont {J.~I.}\ \bibnamefont {Latorre}},
  \bibinfo {author} {\bibfnamefont {S.~P.}\ \bibnamefont {Merkurev}}, \ and\
  \bibinfo {author} {\bibfnamefont {R.}~\bibnamefont {Tarrach}},\ }\bibfield
  {title} {\enquote {\bibinfo {title} {{The baryon loop tension from continuum
  QCD}},}\ }\href@noop {} {\bibfield  {journal} {\bibinfo  {journal} {Phys.
  Lett.}\ }\textbf {\bibinfo {volume} {B158}},\ \bibinfo {pages} {145}
  (\bibinfo {year} {1985})}\BibitemShut {NoStop}%
\bibitem [{\citenamefont {Fabre de~la Ripelle}\ and\ \citenamefont
  {Lassaut}(1997)}]{Fabre:1997gf}%
  \BibitemOpen
  \bibfield  {author} {\bibinfo {author} {\bibfnamefont {M.}~\bibnamefont
  {Fabre de~la Ripelle}}\ and\ \bibinfo {author} {\bibfnamefont
  {M.}~\bibnamefont {Lassaut}},\ }\bibfield  {title} {\enquote {\bibinfo
  {title} {{Transformation of a three-body interaction into a sum of pairwise
  potentials: Application to the quark-string- junction and the Urbana
  potentials}},}\ }\href@noop {} {\bibfield  {journal} {\bibinfo  {journal}
  {Few Body Syst.}\ }\textbf {\bibinfo {volume} {23}},\ \bibinfo {pages}
  {75--86} (\bibinfo {year} {1997})}\BibitemShut {NoStop}%
\bibitem [{\citenamefont {Evangelista}\ \emph {et~al.}(1977)\citenamefont
  {Evangelista} \emph {et~al.}}]{Evangelista:1977ni}%
  \BibitemOpen
  \bibfield  {author} {\bibinfo {author} {\bibfnamefont {C.}~\bibnamefont
  {Evangelista}} \emph {et~al.},\ }\bibfield  {title} {\enquote {\bibinfo
  {title} {{Evidence for a Narrow Width Boson of Mass 2.95\,GeV}},}\ }\href
  {\doibase 10.1016/0370-2693(77)90081-8} {\bibfield  {journal} {\bibinfo
  {journal} {Phys. Lett.}\ }\textbf {\bibinfo {volume} {B72}},\ \bibinfo
  {pages} {139} (\bibinfo {year} {1977})}\BibitemShut {NoStop}%
\bibitem [{\citenamefont {Jaffe}(1978)}]{Jaffe:1977cv}%
  \BibitemOpen
  \bibfield  {author} {\bibinfo {author} {\bibfnamefont {R.~L.}\ \bibnamefont
  {Jaffe}},\ }\bibfield  {title} {\enquote {\bibinfo {title} {{$Q^2\bar Q{}^2$
  Resonances in the Baryon - Antibaryon System}},}\ }\href {\doibase
  10.1103/PhysRevD.17.1444} {\bibfield  {journal} {\bibinfo  {journal} {Phys.
  Rev.}\ }\textbf {\bibinfo {volume} {D17}},\ \bibinfo {pages} {1444} (\bibinfo
  {year} {1978})}\BibitemShut {NoStop}%
\bibitem [{\citenamefont {Chan}\ \emph {et~al.}(1978)\citenamefont {Chan},
  \citenamefont {Fukugita}, \citenamefont {Hansson}, \citenamefont {Hoffman},
  \citenamefont {Konishi}, \citenamefont {H\o{}g\aa{}sen},\ and\ \citenamefont
  {Tsou}}]{Chan:1978nk}%
  \BibitemOpen
  \bibfield  {author} {\bibinfo {author} {\bibfnamefont {Hong-Mo}\ \bibnamefont
  {Chan}}, \bibinfo {author} {\bibfnamefont {M.}~\bibnamefont {Fukugita}},
  \bibinfo {author} {\bibfnamefont {T.~H.}\ \bibnamefont {Hansson}}, \bibinfo
  {author} {\bibfnamefont {H.~J.}\ \bibnamefont {Hoffman}}, \bibinfo {author}
  {\bibfnamefont {K.}~\bibnamefont {Konishi}}, \bibinfo {author} {\bibfnamefont
  {H.}~\bibnamefont {H\o{}g\aa{}sen}}, \ and\ \bibinfo {author} {\bibfnamefont
  {Sheung~Tsun}\ \bibnamefont {Tsou}},\ }\bibfield  {title} {\enquote {\bibinfo
  {title} {{Color Chemistry: A Study of Metastable Multi - Quark Molecules}},}\
  }\href {\doibase 10.1016/0370-2693(78)90872-9} {\bibfield  {journal}
  {\bibinfo  {journal} {Phys. Lett.}\ }\textbf {\bibinfo {volume} {B76}},\
  \bibinfo {pages} {634--640} (\bibinfo {year} {1978})}\BibitemShut {NoStop}%
\bibitem [{\citenamefont {Fermi}\ and\ \citenamefont
  {Yang}(1949)}]{Fermi:1959sa}%
  \BibitemOpen
  \bibfield  {author} {\bibinfo {author} {\bibfnamefont {E.}~\bibnamefont
  {Fermi}}\ and\ \bibinfo {author} {\bibfnamefont {Chen-Ning}\ \bibnamefont
  {Yang}},\ }\bibfield  {title} {\enquote {\bibinfo {title} {{A\lowercase{RE
  MESONS ELEMENTARY PARTICLES?}}}}\ }\href {\doibase 10.1103/PhysRev.76.1739}
  {\bibfield  {journal} {\bibinfo  {journal} {Phys. Rev.}\ }\textbf {\bibinfo
  {volume} {76}},\ \bibinfo {pages} {1739--1743} (\bibinfo {year}
  {1949})}\BibitemShut {NoStop}%
\bibitem [{\citenamefont {Shapiro}(1978)}]{Shapiro:1978wi}%
  \BibitemOpen
  \bibfield  {author} {\bibinfo {author} {\bibfnamefont {I.~S.}\ \bibnamefont
  {Shapiro}},\ }\bibfield  {title} {\enquote {\bibinfo {title} {{The Physics of
  Nucleon - anti-Nucleon Systems}},}\ }\href {\doibase
  10.1016/0370-1573(78)90190-4} {\bibfield  {journal} {\bibinfo  {journal}
  {Phys. Rept.}\ }\textbf {\bibinfo {volume} {35}},\ \bibinfo {pages}
  {129--185} (\bibinfo {year} {1978})}\BibitemShut {NoStop}%
\bibitem [{\citenamefont {Buck}\ \emph {et~al.}(1979)\citenamefont {Buck},
  \citenamefont {Dover},\ and\ \citenamefont {Richard}}]{Buck:1977rt}%
  \BibitemOpen
  \bibfield  {author} {\bibinfo {author} {\bibfnamefont {W.W.}\ \bibnamefont
  {Buck}}, \bibinfo {author} {\bibfnamefont {C.B.}\ \bibnamefont {Dover}}, \
  and\ \bibinfo {author} {\bibfnamefont {J.-M.}\ \bibnamefont {Richard}},\
  }\bibfield  {title} {\enquote {\bibinfo {title} {{The Interaction of Nucleons
  with anti-Nucleons. 1. General Features of the $\bar N N$ Spectrum in
  Potential Models}},}\ }\href {\doibase 10.1016/0003-4916(79)90091-5}
  {\bibfield  {journal} {\bibinfo  {journal} {Annals Phys.}\ }\textbf {\bibinfo
  {volume} {121}},\ \bibinfo {pages} {47} (\bibinfo {year} {1979})}\BibitemShut
  {NoStop}%
\bibitem [{\citenamefont {Voloshin}\ and\ \citenamefont
  {Okun}(1976)}]{Voloshin:1976ap}%
  \BibitemOpen
  \bibfield  {author} {\bibinfo {author} {\bibfnamefont {M.~B.}\ \bibnamefont
  {Voloshin}}\ and\ \bibinfo {author} {\bibfnamefont {L.~B.}\ \bibnamefont
  {Okun}},\ }\bibfield  {title} {\enquote {\bibinfo {title} {Hadron molecules
  and charmonium atom},}\ }\href@noop {} {\bibfield  {journal} {\bibinfo
  {journal} {JETP Lett.}\ }\textbf {\bibinfo {volume} {23}},\ \bibinfo {pages}
  {333--336} (\bibinfo {year} {1976})}\BibitemShut {NoStop}%
\bibitem [{\citenamefont {De~Rujula}\ \emph {et~al.}(1977)\citenamefont
  {De~Rujula}, \citenamefont {Georgi},\ and\ \citenamefont
  {Glashow}}]{DeRujula:1976qd}%
  \BibitemOpen
  \bibfield  {author} {\bibinfo {author} {\bibfnamefont {A.}~\bibnamefont
  {De~Rujula}}, \bibinfo {author} {\bibfnamefont {Howard}\ \bibnamefont
  {Georgi}}, \ and\ \bibinfo {author} {\bibfnamefont {S.~L.}\ \bibnamefont
  {Glashow}},\ }\bibfield  {title} {\enquote {\bibinfo {title} {Molecular
  charmonium: a new spectroscopy?}}\ }\href@noop {} {\bibfield  {journal}
  {\bibinfo  {journal} {Phys. Rev. Lett.}\ }\textbf {\bibinfo {volume} {38}},\
  \bibinfo {pages} {317} (\bibinfo {year} {1977})}\BibitemShut {NoStop}%
\bibitem [{\citenamefont {Le~Yaouanc}\ \emph
  {et~al.}(1977{\natexlab{a}})\citenamefont {Le~Yaouanc}, \citenamefont
  {Oliver}, \citenamefont {Pene},\ and\ \citenamefont
  {Raynal}}]{LeYaouanc:1977ux}%
  \BibitemOpen
  \bibfield  {author} {\bibinfo {author} {\bibfnamefont {A.}~\bibnamefont
  {Le~Yaouanc}}, \bibinfo {author} {\bibfnamefont {L.}~\bibnamefont {Oliver}},
  \bibinfo {author} {\bibfnamefont {O.}~\bibnamefont {Pene}}, \ and\ \bibinfo
  {author} {\bibfnamefont {J.~C.}\ \bibnamefont {Raynal}},\ }\bibfield  {title}
  {\enquote {\bibinfo {title} {Strong decays of $\psi(4.028)$ as a radial
  excitation of charmonium},}\ }\href@noop {} {\bibfield  {journal} {\bibinfo
  {journal} {Phys. Lett.}\ }\textbf {\bibinfo {volume} {B71}},\ \bibinfo
  {pages} {397} (\bibinfo {year} {1977}{\natexlab{a}})}\BibitemShut {NoStop}%
\bibitem [{\citenamefont {Le~Yaouanc}\ \emph
  {et~al.}(1977{\natexlab{b}})\citenamefont {Le~Yaouanc}, \citenamefont
  {Oliver}, \citenamefont {Pene},\ and\ \citenamefont
  {Raynal}}]{LeYaouanc:1977gm}%
  \BibitemOpen
  \bibfield  {author} {\bibinfo {author} {\bibfnamefont {A.}~\bibnamefont
  {Le~Yaouanc}}, \bibinfo {author} {\bibfnamefont {L.}~\bibnamefont {Oliver}},
  \bibinfo {author} {\bibfnamefont {O.}~\bibnamefont {Pene}}, \ and\ \bibinfo
  {author} {\bibfnamefont {J.~C.}\ \bibnamefont {Raynal}},\ }\bibfield  {title}
  {\enquote {\bibinfo {title} {Why is $\psi'''(4.414)$ so narrow?}}\
  }\href@noop {} {\bibfield  {journal} {\bibinfo  {journal} {Phys. Lett.}\
  }\textbf {\bibinfo {volume} {B72}},\ \bibinfo {pages} {57} (\bibinfo {year}
  {1977}{\natexlab{b}})}\BibitemShut {NoStop}%
\bibitem [{\citenamefont {Eichten}\ \emph {et~al.}(1980)\citenamefont
  {Eichten}, \citenamefont {Gottfried}, \citenamefont {Kinoshita},
  \citenamefont {Lane},\ and\ \citenamefont {Yan}}]{Eichten:1979ms}%
  \BibitemOpen
  \bibfield  {author} {\bibinfo {author} {\bibfnamefont {E.}~\bibnamefont
  {Eichten}}, \bibinfo {author} {\bibfnamefont {K.}~\bibnamefont {Gottfried}},
  \bibinfo {author} {\bibfnamefont {T.}~\bibnamefont {Kinoshita}}, \bibinfo
  {author} {\bibfnamefont {K.~D.}\ \bibnamefont {Lane}}, \ and\ \bibinfo
  {author} {\bibfnamefont {Tung-Mow}\ \bibnamefont {Yan}},\ }\bibfield  {title}
  {\enquote {\bibinfo {title} {{Charmonium: Comparison with Experiment}},}\
  }\href {\doibase 10.1103/PhysRevD.21.203} {\bibfield  {journal} {\bibinfo
  {journal} {Phys. Rev.}\ }\textbf {\bibinfo {volume} {D21}},\ \bibinfo {pages}
  {203} (\bibinfo {year} {1980})}\BibitemShut {NoStop}%
\bibitem [{\citenamefont {T{\"o}rnqvist}(1991)}]{Tornqvist:1991ks}%
  \BibitemOpen
  \bibfield  {author} {\bibinfo {author} {\bibfnamefont {Nils~A.}\ \bibnamefont
  {T{\"o}rnqvist}},\ }\bibfield  {title} {\enquote {\bibinfo {title} {Possible
  large deuteron - like meson meson states bound by pions},}\ }\href@noop {}
  {\bibfield  {journal} {\bibinfo  {journal} {Phys. Rev. Lett.}\ }\textbf
  {\bibinfo {volume} {67}},\ \bibinfo {pages} {556--559} (\bibinfo {year}
  {1991})}\BibitemShut {NoStop}%
\bibitem [{\citenamefont {Manohar}\ and\ \citenamefont
  {Wise}(1993)}]{Manohar:1992nd}%
  \BibitemOpen
  \bibfield  {author} {\bibinfo {author} {\bibfnamefont {Aneesh~V.}\
  \bibnamefont {Manohar}}\ and\ \bibinfo {author} {\bibfnamefont {Mark~B.}\
  \bibnamefont {Wise}},\ }\bibfield  {title} {\enquote {\bibinfo {title}
  {{Exotic $Q Q \bar q \bar q$ states in QCD}},}\ }\href@noop {} {\bibfield
  {journal} {\bibinfo  {journal} {Nucl. Phys.}\ }\textbf {\bibinfo {volume}
  {B399}},\ \bibinfo {pages} {17--33} (\bibinfo {year} {1993})},\ \Eprint
  {http://arxiv.org/abs/hep-ph/9212236} {hep-ph/9212236} \BibitemShut {NoStop}%
\bibitem [{\citenamefont {Ericson}\ and\ \citenamefont
  {Karl}(1993)}]{Ericson:1993wy}%
  \BibitemOpen
  \bibfield  {author} {\bibinfo {author} {\bibfnamefont {Torleif E.~O.}\
  \bibnamefont {Ericson}}\ and\ \bibinfo {author} {\bibfnamefont
  {G.}~\bibnamefont {Karl}},\ }\bibfield  {title} {\enquote {\bibinfo {title}
  {Strength of pion exchange in hadronic molecules},}\ }\href@noop {}
  {\bibfield  {journal} {\bibinfo  {journal} {Phys. Lett.}\ }\textbf {\bibinfo
  {volume} {B309}},\ \bibinfo {pages} {426--430} (\bibinfo {year}
  {1993})}\BibitemShut {NoStop}%
\bibitem [{\citenamefont {T{\"o}rnqvist}(1994)}]{Tornqvist:1993ng}%
  \BibitemOpen
  \bibfield  {author} {\bibinfo {author} {\bibfnamefont {Nils~A.}\ \bibnamefont
  {T{\"o}rnqvist}},\ }\bibfield  {title} {\enquote {\bibinfo {title} {{From the
  deuteron to deusons, an analysis of deuteron - like meson meson bound
  states}},}\ }\href {\doibase 10.1007/BF01413192} {\bibfield  {journal}
  {\bibinfo  {journal} {Z. Phys.}\ }\textbf {\bibinfo {volume} {C61}},\
  \bibinfo {pages} {525--537} (\bibinfo {year} {1994})},\ \Eprint
  {http://arxiv.org/abs/hep-ph/9310247} {arXiv:hep-ph/9310247 [hep-ph]}
  \BibitemShut {NoStop}%
\bibitem [{\citenamefont {Ferretti}\ \emph {et~al.}(2013)\citenamefont
  {Ferretti}, \citenamefont {Galat{\`a}},\ and\ \citenamefont
  {Santopinto}}]{Ferretti:2013faa}%
  \BibitemOpen
  \bibfield  {author} {\bibinfo {author} {\bibfnamefont {J.}~\bibnamefont
  {Ferretti}}, \bibinfo {author} {\bibfnamefont {G.}~\bibnamefont
  {Galat{\`a}}}, \ and\ \bibinfo {author} {\bibfnamefont {E.}~\bibnamefont
  {Santopinto}},\ }\bibfield  {title} {\enquote {\bibinfo {title}
  {{Interpretation of the X(3872) as a charmonium state plus an extra component
  due to the coupling to the meson-meson continuum}},}\ }\href {\doibase
  10.1103/PhysRevC.88.015207} {\bibfield  {journal} {\bibinfo  {journal} {Phys.
  Rev.}\ }\textbf {\bibinfo {volume} {C88}},\ \bibinfo {pages} {015207}
  (\bibinfo {year} {2013})},\ \Eprint {http://arxiv.org/abs/1302.6857}
  {arXiv:1302.6857 [hep-ph]} \BibitemShut {NoStop}%
\bibitem [{\citenamefont {Thomas}\ and\ \citenamefont
  {Close}(2008)}]{Thomas:2008ja}%
  \BibitemOpen
  \bibfield  {author} {\bibinfo {author} {\bibfnamefont {C.~E.}\ \bibnamefont
  {Thomas}}\ and\ \bibinfo {author} {\bibfnamefont {F.~E.}\ \bibnamefont
  {Close}},\ }\bibfield  {title} {\enquote {\bibinfo {title} {{Is X(3872) a
  molecule?}}}\ }\href {\doibase 10.1103/PhysRevD.78.034007} {\bibfield
  {journal} {\bibinfo  {journal} {Phys. Rev.}\ }\textbf {\bibinfo {volume}
  {D78}},\ \bibinfo {pages} {034007} (\bibinfo {year} {2008})},\ \Eprint
  {http://arxiv.org/abs/0805.3653} {arXiv:0805.3653 [hep-ph]} \BibitemShut
  {NoStop}%
\bibitem [{\citenamefont {Ohkoda}\ \emph
  {et~al.}(2012{\natexlab{a}})\citenamefont {Ohkoda}, \citenamefont
  {Yamaguchi}, \citenamefont {Yasui}, \citenamefont {Sudoh},\ and\
  \citenamefont {Hosaka}}]{Ohkoda:2011vj}%
  \BibitemOpen
  \bibfield  {author} {\bibinfo {author} {\bibfnamefont {Shunsuke}\
  \bibnamefont {Ohkoda}}, \bibinfo {author} {\bibfnamefont {Yasuhiro}\
  \bibnamefont {Yamaguchi}}, \bibinfo {author} {\bibfnamefont {Shigehiro}\
  \bibnamefont {Yasui}}, \bibinfo {author} {\bibfnamefont {Kazutaka}\
  \bibnamefont {Sudoh}}, \ and\ \bibinfo {author} {\bibfnamefont {Atsushi}\
  \bibnamefont {Hosaka}},\ }\bibfield  {title} {\enquote {\bibinfo {title}
  {{Exotic Mesons with Hidden Bottom near Thresholds}},}\ }\href {\doibase
  10.1103/PhysRevD.86.014004} {\bibfield  {journal} {\bibinfo  {journal} {Phys.
  Rev.}\ }\textbf {\bibinfo {volume} {D86}},\ \bibinfo {pages} {014004}
  (\bibinfo {year} {2012}{\natexlab{a}})},\ \Eprint
  {http://arxiv.org/abs/1111.2921} {arXiv:1111.2921 [hep-ph]} \BibitemShut
  {NoStop}%
\bibitem [{\citenamefont {Cleven}\ \emph {et~al.}(2015)\citenamefont {Cleven},
  \citenamefont {Guo}, \citenamefont {Hanhart}, \citenamefont {Wang},\ and\
  \citenamefont {Zhao}}]{Cleven:2015era}%
  \BibitemOpen
  \bibfield  {author} {\bibinfo {author} {\bibfnamefont {Martin}\ \bibnamefont
  {Cleven}}, \bibinfo {author} {\bibfnamefont {Feng-Kun}\ \bibnamefont {Guo}},
  \bibinfo {author} {\bibfnamefont {Christoph}\ \bibnamefont {Hanhart}},
  \bibinfo {author} {\bibfnamefont {Qian}\ \bibnamefont {Wang}}, \ and\
  \bibinfo {author} {\bibfnamefont {Qiang}\ \bibnamefont {Zhao}},\ }\bibfield
  {title} {\enquote {\bibinfo {title} {{Employing spin symmetry to disentangle
  different models for the $XYZ$ states}},}\ }\href {\doibase
  10.1103/PhysRevD.92.014005} {\bibfield  {journal} {\bibinfo  {journal} {Phys.
  Rev.}\ }\textbf {\bibinfo {volume} {D92}},\ \bibinfo {pages} {014005}
  (\bibinfo {year} {2015})},\ \Eprint {http://arxiv.org/abs/1505.01771}
  {arXiv:1505.01771 [hep-ph]} \BibitemShut {NoStop}%
\bibitem [{\citenamefont {Hofmann}\ and\ \citenamefont
  {Lutz}(2005{\natexlab{a}})}]{Hofmann:2005mn}%
  \BibitemOpen
  \bibfield  {author} {\bibinfo {author} {\bibfnamefont {Julian}\ \bibnamefont
  {Hofmann}}\ and\ \bibinfo {author} {\bibfnamefont {Matthias F.~M.}\
  \bibnamefont {Lutz}},\ }\bibfield  {title} {\enquote {\bibinfo {title}
  {{Coupled-channel study of baryon resonances with charm}},}\ }in\ \href@noop
  {} {\emph {\bibinfo {booktitle} {{Proceedings, 11th International Conference
  on Hadron spectroscopy (Hadron 2005)}}}}\ (\bibinfo {year} {2005})\ \Eprint
  {http://arxiv.org/abs/nucl-th/0510091} {arXiv:nucl-th/0510091 [nucl-th]}
  \BibitemShut {NoStop}%
\bibitem [{\citenamefont {Hofmann}\ and\ \citenamefont
  {Lutz}(2005{\natexlab{b}})}]{Hofmann:2005sw}%
  \BibitemOpen
  \bibfield  {author} {\bibinfo {author} {\bibfnamefont {J.}~\bibnamefont
  {Hofmann}}\ and\ \bibinfo {author} {\bibfnamefont {M.~F.~M.}\ \bibnamefont
  {Lutz}},\ }\bibfield  {title} {\enquote {\bibinfo {title} {{Coupled-channel
  study of crypto-exotic baryons with charm}},}\ }\href {\doibase
  10.1016/j.nuclphysa.2005.08.022} {\bibfield  {journal} {\bibinfo  {journal}
  {Nucl. Phys.}\ }\textbf {\bibinfo {volume} {A763}},\ \bibinfo {pages}
  {90--139} (\bibinfo {year} {2005}{\natexlab{b}})},\ \Eprint
  {http://arxiv.org/abs/hep-ph/0507071} {arXiv:hep-ph/0507071 [hep-ph]}
  \BibitemShut {NoStop}%
\bibitem [{\citenamefont {Xiao}\ \emph
  {et~al.}(2013{\natexlab{b}})\citenamefont {Xiao}, \citenamefont {Nieves},\
  and\ \citenamefont {Oset}}]{Xiao:2013yca}%
  \BibitemOpen
  \bibfield  {author} {\bibinfo {author} {\bibfnamefont {C.~W.}\ \bibnamefont
  {Xiao}}, \bibinfo {author} {\bibfnamefont {J.}~\bibnamefont {Nieves}}, \ and\
  \bibinfo {author} {\bibfnamefont {E.}~\bibnamefont {Oset}},\ }\bibfield
  {title} {\enquote {\bibinfo {title} {{Combining heavy quark spin and local
  hidden gauge symmetries in the dynamical generation of hidden charm
  baryons}},}\ }\href {\doibase 10.1103/PhysRevD.88.056012} {\bibfield
  {journal} {\bibinfo  {journal} {Phys. Rev.}\ }\textbf {\bibinfo {volume}
  {D88}},\ \bibinfo {pages} {056012} (\bibinfo {year} {2013}{\natexlab{b}})},\
  \Eprint {http://arxiv.org/abs/1304.5368} {arXiv:1304.5368 [hep-ph]}
  \BibitemShut {NoStop}%
\bibitem [{\citenamefont {Yang}\ \emph {et~al.}(2012)\citenamefont {Yang},
  \citenamefont {Sun}, \citenamefont {He}, \citenamefont {Liu},\ and\
  \citenamefont {Zhu}}]{Yang:2011wz}%
  \BibitemOpen
  \bibfield  {author} {\bibinfo {author} {\bibfnamefont {Zhong-Cheng}\
  \bibnamefont {Yang}}, \bibinfo {author} {\bibfnamefont {Zhi-Feng}\
  \bibnamefont {Sun}}, \bibinfo {author} {\bibfnamefont {Jun}\ \bibnamefont
  {He}}, \bibinfo {author} {\bibfnamefont {Xiang}\ \bibnamefont {Liu}}, \ and\
  \bibinfo {author} {\bibfnamefont {Shi-Lin}\ \bibnamefont {Zhu}},\ }\bibfield
  {title} {\enquote {\bibinfo {title} {{The possible hidden-charm molecular
  baryons composed of anti-charmed meson and charmed baryon}},}\ }\href
  {\doibase 10.1088/1674-1137/36/1/002, 10.1088/1674-1137/36/3/006} {\bibfield
  {journal} {\bibinfo  {journal} {Chin. Phys.}\ }\textbf {\bibinfo {volume}
  {C36}},\ \bibinfo {pages} {6--13} (\bibinfo {year} {2012})},\ \Eprint
  {http://arxiv.org/abs/1105.2901} {arXiv:1105.2901 [hep-ph]} \BibitemShut
  {NoStop}%
\bibitem [{\citenamefont {Landsberg}(1994)}]{Landsberg:1993dh}%
  \BibitemOpen
  \bibfield  {author} {\bibinfo {author} {\bibfnamefont {L.~G.}\ \bibnamefont
  {Landsberg}},\ }\bibfield  {title} {\enquote {\bibinfo {title} {{Do the
  narrow cryptoexotic baryon resonances exist?}}}\ }\href@noop {} {\bibfield
  {journal} {\bibinfo  {journal} {Phys. Atom. Nucl.}\ }\textbf {\bibinfo
  {volume} {57}},\ \bibinfo {pages} {2127--2131} (\bibinfo {year} {1994})},\
  \bibinfo {note} {[Yad. Fiz.57N12,2210(1994)]}\BibitemShut {NoStop}%
\bibitem [{\citenamefont {Wong}(2004)}]{Wong:2003xk}%
  \BibitemOpen
  \bibfield  {author} {\bibinfo {author} {\bibfnamefont {Cheuk-Yin}\
  \bibnamefont {Wong}},\ }\bibfield  {title} {\enquote {\bibinfo {title}
  {{Molecular states of heavy quark mesons}},}\ }\href {\doibase
  10.1103/PhysRevC.69.055202} {\bibfield  {journal} {\bibinfo  {journal} {Phys.
  Rev.}\ }\textbf {\bibinfo {volume} {C69}},\ \bibinfo {pages} {055202}
  (\bibinfo {year} {2004})},\ \Eprint {http://arxiv.org/abs/hep-ph/0311088}
  {arXiv:hep-ph/0311088 [hep-ph]} \BibitemShut {NoStop}%
\bibitem [{\citenamefont {Ohkoda}\ \emph
  {et~al.}(2012{\natexlab{b}})\citenamefont {Ohkoda}, \citenamefont
  {Yamaguchi}, \citenamefont {Yasui}, \citenamefont {Sudoh},\ and\
  \citenamefont {Hosaka}}]{Ohkoda:2012hv}%
  \BibitemOpen
  \bibfield  {author} {\bibinfo {author} {\bibfnamefont {S.}~\bibnamefont
  {Ohkoda}}, \bibinfo {author} {\bibfnamefont {Y.}~\bibnamefont {Yamaguchi}},
  \bibinfo {author} {\bibfnamefont {S.}~\bibnamefont {Yasui}}, \bibinfo
  {author} {\bibfnamefont {K.}~\bibnamefont {Sudoh}}, \ and\ \bibinfo {author}
  {\bibfnamefont {A.}~\bibnamefont {Hosaka}},\ }\bibfield  {title} {\enquote
  {\bibinfo {title} {{Exotic mesons with double charm and bottom flavor}},}\
  }\href {\doibase 10.1103/PhysRevD.86.034019} {\bibfield  {journal} {\bibinfo
  {journal} {Phys. Rev.}\ }\textbf {\bibinfo {volume} {D86}},\ \bibinfo {pages}
  {034019} (\bibinfo {year} {2012}{\natexlab{b}})},\ \Eprint
  {http://arxiv.org/abs/1202.0760} {arXiv:1202.0760 [hep-ph]} \BibitemShut
  {NoStop}%
\bibitem [{\citenamefont {Karliner}\ and\ \citenamefont
  {Rosner}(2015)}]{Karliner:2015ina}%
  \BibitemOpen
  \bibfield  {author} {\bibinfo {author} {\bibfnamefont {Marek}\ \bibnamefont
  {Karliner}}\ and\ \bibinfo {author} {\bibfnamefont {Jonathan~L.}\
  \bibnamefont {Rosner}},\ }\bibfield  {title} {\enquote {\bibinfo {title}
  {{New Exotic Meson and Baryon Resonances from Doubly-Heavy Hadronic
  Molecules}},}\ }\href {\doibase 10.1103/PhysRevLett.115.122001} {\bibfield
  {journal} {\bibinfo  {journal} {Phys. Rev. Lett.}\ }\textbf {\bibinfo
  {volume} {115}},\ \bibinfo {pages} {122001} (\bibinfo {year} {2015})},\
  \Eprint {http://arxiv.org/abs/1506.06386} {arXiv:1506.06386 [hep-ph]}
  \BibitemShut {NoStop}%
\bibitem [{\citenamefont {Yasui}\ and\ \citenamefont
  {Sudoh}(2009)}]{Yasui:2009bz}%
  \BibitemOpen
  \bibfield  {author} {\bibinfo {author} {\bibfnamefont {Shigehiro}\
  \bibnamefont {Yasui}}\ and\ \bibinfo {author} {\bibfnamefont {Kazutaka}\
  \bibnamefont {Sudoh}},\ }\bibfield  {title} {\enquote {\bibinfo {title}
  {{Exotic nuclei with open heavy flavor mesons}},}\ }\href {\doibase
  10.1103/PhysRevD.80.034008} {\bibfield  {journal} {\bibinfo  {journal} {Phys.
  Rev.}\ }\textbf {\bibinfo {volume} {D80}},\ \bibinfo {pages} {034008}
  (\bibinfo {year} {2009})},\ \Eprint {http://arxiv.org/abs/0906.1452}
  {arXiv:0906.1452 [hep-ph]} \BibitemShut {NoStop}%
\bibitem [{\citenamefont {Froemel}\ \emph {et~al.}(2005)\citenamefont
  {Froemel}, \citenamefont {Julia-Diaz},\ and\ \citenamefont
  {Riska}}]{Froemel:2004ea}%
  \BibitemOpen
  \bibfield  {author} {\bibinfo {author} {\bibfnamefont {F.}~\bibnamefont
  {Froemel}}, \bibinfo {author} {\bibfnamefont {B.}~\bibnamefont {Julia-Diaz}},
  \ and\ \bibinfo {author} {\bibfnamefont {D.~O.}\ \bibnamefont {Riska}},\
  }\bibfield  {title} {\enquote {\bibinfo {title} {{Bound states of double
  flavor hyperons}},}\ }\href {\doibase 10.1016/j.nuclphysa.2005.01.022}
  {\bibfield  {journal} {\bibinfo  {journal} {Nucl. Phys.}\ }\textbf {\bibinfo
  {volume} {A750}},\ \bibinfo {pages} {337--356} (\bibinfo {year} {2005})},\
  \Eprint {http://arxiv.org/abs/nucl-th/0410034} {arXiv:nucl-th/0410034
  [nucl-th]} \BibitemShut {NoStop}%
\bibitem [{\citenamefont {Liu}\ and\ \citenamefont {Oka}(2012)}]{Liu:2011xc}%
  \BibitemOpen
  \bibfield  {author} {\bibinfo {author} {\bibfnamefont {Yan-Rui}\ \bibnamefont
  {Liu}}\ and\ \bibinfo {author} {\bibfnamefont {Makoto}\ \bibnamefont {Oka}},\
  }\bibfield  {title} {\enquote {\bibinfo {title} {{$\Lambda_c N$ bound states
  revisited}},}\ }\href {\doibase 10.1103/PhysRevD.85.014015} {\bibfield
  {journal} {\bibinfo  {journal} {Phys. Rev.}\ }\textbf {\bibinfo {volume}
  {D85}},\ \bibinfo {pages} {014015} (\bibinfo {year} {2012})},\ \Eprint
  {http://arxiv.org/abs/1103.4624} {arXiv:1103.4624 [hep-ph]} \BibitemShut
  {NoStop}%
\bibitem [{\citenamefont {Meguro}\ \emph {et~al.}(2011)\citenamefont {Meguro},
  \citenamefont {Liu},\ and\ \citenamefont {Oka}}]{Meguro:2011nr}%
  \BibitemOpen
  \bibfield  {author} {\bibinfo {author} {\bibfnamefont {Wakafumi}\
  \bibnamefont {Meguro}}, \bibinfo {author} {\bibfnamefont {Yan-Rui}\
  \bibnamefont {Liu}}, \ and\ \bibinfo {author} {\bibfnamefont {Makoto}\
  \bibnamefont {Oka}},\ }\bibfield  {title} {\enquote {\bibinfo {title}
  {{Possible $\Lambda_c\Lambda_c$ molecular bound state}},}\ }\href {\doibase
  10.1016/j.physletb.2011.09.088} {\bibfield  {journal} {\bibinfo  {journal}
  {Phys. Lett.}\ }\textbf {\bibinfo {volume} {B704}},\ \bibinfo {pages}
  {547--550} (\bibinfo {year} {2011})},\ \Eprint
  {http://arxiv.org/abs/1105.3693} {arXiv:1105.3693 [hep-ph]} \BibitemShut
  {NoStop}%
\bibitem [{\citenamefont {Huang}\ \emph {et~al.}(2014)\citenamefont {Huang},
  \citenamefont {Ping},\ and\ \citenamefont {Wang}}]{Huang:2013rla}%
  \BibitemOpen
  \bibfield  {author} {\bibinfo {author} {\bibfnamefont {Hongxia}\ \bibnamefont
  {Huang}}, \bibinfo {author} {\bibfnamefont {Jialun}\ \bibnamefont {Ping}}, \
  and\ \bibinfo {author} {\bibfnamefont {Fan}\ \bibnamefont {Wang}},\
  }\bibfield  {title} {\enquote {\bibinfo {title} {{Possible $H$-like dibaryon
  states with heavy quarks}},}\ }\href {\doibase 10.1103/PhysRevC.89.035201}
  {\bibfield  {journal} {\bibinfo  {journal} {Phys. Rev.}\ }\textbf {\bibinfo
  {volume} {C89}},\ \bibinfo {pages} {035201} (\bibinfo {year} {2014})},\
  \Eprint {http://arxiv.org/abs/1311.4732} {arXiv:1311.4732 [hep-ph]}
  \BibitemShut {NoStop}%
\bibitem [{\citenamefont {Lee}\ \emph {et~al.}(2011)\citenamefont {Lee},
  \citenamefont {Luo}, \citenamefont {Chen},\ and\ \citenamefont
  {Zhu}}]{Lee:2011rka}%
  \BibitemOpen
  \bibfield  {author} {\bibinfo {author} {\bibfnamefont {Ning}\ \bibnamefont
  {Lee}}, \bibinfo {author} {\bibfnamefont {Zhi-Gang}\ \bibnamefont {Luo}},
  \bibinfo {author} {\bibfnamefont {Xiao-Lin}\ \bibnamefont {Chen}}, \ and\
  \bibinfo {author} {\bibfnamefont {Shi-Lin}\ \bibnamefont {Zhu}},\ }\bibfield
  {title} {\enquote {\bibinfo {title} {{Possible Deuteron-like Molecular States
  Composed of Heavy Baryons}},}\ }\href {\doibase 10.1103/PhysRevD.84.014031}
  {\bibfield  {journal} {\bibinfo  {journal} {Phys. Rev.}\ }\textbf {\bibinfo
  {volume} {D84}},\ \bibinfo {pages} {014031} (\bibinfo {year} {2011})},\
  \Eprint {http://arxiv.org/abs/1104.4257} {arXiv:1104.4257 [hep-ph]}
  \BibitemShut {NoStop}%
\bibitem [{\citenamefont {Li}\ and\ \citenamefont {Zhu}(2012)}]{Li:2012bt}%
  \BibitemOpen
  \bibfield  {author} {\bibinfo {author} {\bibfnamefont {Ning}\ \bibnamefont
  {Li}}\ and\ \bibinfo {author} {\bibfnamefont {Shi-Lin}\ \bibnamefont {Zhu}},\
  }\bibfield  {title} {\enquote {\bibinfo {title} {{Hadronic Molecular States
  Composed of Heavy Flavor Baryons}},}\ }\href {\doibase
  10.1103/PhysRevD.86.014020} {\bibfield  {journal} {\bibinfo  {journal} {Phys.
  Rev.}\ }\textbf {\bibinfo {volume} {D86}},\ \bibinfo {pages} {014020}
  (\bibinfo {year} {2012})},\ \Eprint {http://arxiv.org/abs/1204.3364}
  {arXiv:1204.3364 [hep-ph]} \BibitemShut {NoStop}%
\bibitem [{\citenamefont {Julia-Diaz}\ and\ \citenamefont
  {Riska}(2005)}]{JuliaDiaz:2004rf}%
  \BibitemOpen
  \bibfield  {author} {\bibinfo {author} {\bibfnamefont {B.}~\bibnamefont
  {Julia-Diaz}}\ and\ \bibinfo {author} {\bibfnamefont {D.~O.}\ \bibnamefont
  {Riska}},\ }\bibfield  {title} {\enquote {\bibinfo {title} {{Nuclei of double
  charm hyperons}},}\ }\bibfield  {booktitle} {\emph {\bibinfo {booktitle}
  {{The structure of baryons. Proceedings, 10th International Conference,
  Baryons'04, Palaiseau, France, October 25-29, 2004}}},\ }\href {\doibase
  10.1016/j.nuclphysa.2005.03.050} {\bibfield  {journal} {\bibinfo  {journal}
  {Nucl. Phys.}\ }\textbf {\bibinfo {volume} {A755}},\ \bibinfo {pages}
  {431--434} (\bibinfo {year} {2005})},\ \Eprint
  {http://arxiv.org/abs/nucl-th/0405061} {arXiv:nucl-th/0405061 [nucl-th]}
  \BibitemShut {NoStop}%
\bibitem [{\citenamefont {Van~Hove}(1986)}]{VanHove:1986qx}%
  \BibitemOpen
  \bibfield  {author} {\bibinfo {author} {\bibfnamefont {L.}~\bibnamefont
  {Van~Hove}},\ }\bibfield  {title} {\enquote {\bibinfo {title}
  {{F\lowercase{UTURE PROSPECTS OF PARTICLE PHYSICS}}},}\ }in\ \href
  {http://alice.cern.ch/format/showfull?sysnb=0082623} {\emph {\bibinfo
  {booktitle} {{7th European Symposium on Nucleon Antinucleon Interactions:
  Antiproton 86, Thessaloniki, Greece, September 1-5, 1986}}}}\ (\bibinfo
  {year} {1986})\BibitemShut {NoStop}%
\bibitem [{\citenamefont {Ochs}(2013)}]{Ochs:2013gi}%
  \BibitemOpen
  \bibfield  {author} {\bibinfo {author} {\bibfnamefont {Wolfgang}\
  \bibnamefont {Ochs}},\ }\bibfield  {title} {\enquote {\bibinfo {title} {{The
  Status of Glueballs}},}\ }\href {\doibase 10.1088/0954-3899/40/4/043001}
  {\bibfield  {journal} {\bibinfo  {journal} {J. Phys.}\ }\textbf {\bibinfo
  {volume} {G40}},\ \bibinfo {pages} {043001} (\bibinfo {year} {2013})},\
  \Eprint {http://arxiv.org/abs/1301.5183} {arXiv:1301.5183 [hep-ph]}
  \BibitemShut {NoStop}%
\bibitem [{\citenamefont {Tang}\ and\ \citenamefont
  {Qiao}(2016)}]{Qiao:2015iea}%
  \BibitemOpen
  \bibfield  {author} {\bibinfo {author} {\bibfnamefont {Liang}\ \bibnamefont
  {Tang}}\ and\ \bibinfo {author} {\bibfnamefont {Cong-Feng}\ \bibnamefont
  {Qiao}},\ }\bibfield  {title} {\enquote {\bibinfo {title} {{Mass Spectra of
  $0^{+-}$, $1^{-+}$, and $2^{+-}$ Exotic Glueballs}},}\ }\href {\doibase
  10.1016/j.nuclphysb.2016.01.017} {\bibfield  {journal} {\bibinfo  {journal}
  {Nucl. Phys.}\ }\textbf {\bibinfo {volume} {B904}},\ \bibinfo {pages}
  {282--296} (\bibinfo {year} {2016})},\ \Eprint
  {http://arxiv.org/abs/1509.00305} {arXiv:1509.00305 [hep-ph]} \BibitemShut
  {NoStop}%
\bibitem [{\citenamefont {Buisseret}\ \emph {et~al.}(2008)\citenamefont
  {Buisseret}, \citenamefont {Semay}, \citenamefont {Mathieu},\ and\
  \citenamefont {Silvestre-Brac}}]{Buisseret:2008zza}%
  \BibitemOpen
  \bibfield  {author} {\bibinfo {author} {\bibfnamefont {F.}~\bibnamefont
  {Buisseret}}, \bibinfo {author} {\bibfnamefont {C.}~\bibnamefont {Semay}},
  \bibinfo {author} {\bibfnamefont {V.}~\bibnamefont {Mathieu}}, \ and\
  \bibinfo {author} {\bibfnamefont {B.}~\bibnamefont {Silvestre-Brac}},\
  }\bibfield  {title} {\enquote {\bibinfo {title} {{Excited string and
  constituent gluon descriptions of hybrid mesons}},}\ }\bibfield  {booktitle}
  {\emph {\bibinfo {booktitle} {{Proceedings, 20th European Conference on
  Few-Body Problems in Physics (EFB20)}}},\ }\href {\doibase
  10.1007/s00601-008-0263-1} {\bibfield  {journal} {\bibinfo  {journal} {Few
  Body Syst.}\ }\textbf {\bibinfo {volume} {44}},\ \bibinfo {pages} {87--89}
  (\bibinfo {year} {2008})}\BibitemShut {NoStop}%
\bibitem [{\citenamefont {Giles}\ and\ \citenamefont
  {Tye}(1977)}]{Giles:1977mp}%
  \BibitemOpen
  \bibfield  {author} {\bibinfo {author} {\bibfnamefont {Roscoe}\ \bibnamefont
  {Giles}}\ and\ \bibinfo {author} {\bibfnamefont {S.~H.~H.}\ \bibnamefont
  {Tye}},\ }\bibfield  {title} {\enquote {\bibinfo {title} {{T\lowercase{HE
  APPLICATION OF THE QUARK - CONFINING STRING TO THE PSI SPECTROSCOPY}}},}\
  }\href@noop {} {\bibfield  {journal} {\bibinfo  {journal} {Phys. Rev.}\
  }\textbf {\bibinfo {volume} {D16}},\ \bibinfo {pages} {1079} (\bibinfo {year}
  {1977})}\BibitemShut {NoStop}%
\bibitem [{\citenamefont {Horn}\ and\ \citenamefont
  {Mandula}(1978)}]{Horn:1977rq}%
  \BibitemOpen
  \bibfield  {author} {\bibinfo {author} {\bibfnamefont {D.}~\bibnamefont
  {Horn}}\ and\ \bibinfo {author} {\bibfnamefont {J.}~\bibnamefont {Mandula}},\
  }\bibfield  {title} {\enquote {\bibinfo {title} {{A \lowercase{MODEL OF
  MESONS WITH CONSTITUENT GLUONS}}},}\ }\href@noop {} {\bibfield  {journal}
  {\bibinfo  {journal} {Phys. Rev.}\ }\textbf {\bibinfo {volume} {D17}},\
  \bibinfo {pages} {898} (\bibinfo {year} {1978})}\BibitemShut {NoStop}%
\bibitem [{\citenamefont {Hasenfratz}\ \emph
  {et~al.}(1980{\natexlab{b}})\citenamefont {Hasenfratz}, \citenamefont
  {Horgan}, \citenamefont {Kuti},\ and\ \citenamefont
  {Richard}}]{Hasenfratz:1980jv}%
  \BibitemOpen
  \bibfield  {author} {\bibinfo {author} {\bibfnamefont {P.}~\bibnamefont
  {Hasenfratz}}, \bibinfo {author} {\bibfnamefont {R.~R.}\ \bibnamefont
  {Horgan}}, \bibinfo {author} {\bibfnamefont {J.}~\bibnamefont {Kuti}}, \ and\
  \bibinfo {author} {\bibfnamefont {J.~M.}\ \bibnamefont {Richard}},\
  }\bibfield  {title} {\enquote {\bibinfo {title} {{The Effects of Colored Glue
  in the QCD Motivated Bag of Heavy Quark - anti-Quark Systems}},}\ }\href
  {\doibase 10.1016/0370-2693(80)90491-8} {\bibfield  {journal} {\bibinfo
  {journal} {Phys. Lett.}\ }\textbf {\bibinfo {volume} {B95}},\ \bibinfo
  {pages} {299} (\bibinfo {year} {1980}{\natexlab{b}})}\BibitemShut {NoStop}%
\bibitem [{\citenamefont {Mandula}(1984)}]{Mandula:1983wc}%
  \BibitemOpen
  \bibfield  {author} {\bibinfo {author} {\bibfnamefont {Jeffrey~E.}\
  \bibnamefont {Mandula}},\ }\bibfield  {title} {\enquote {\bibinfo {title}
  {{Quark - Anti-quark - Gluon Exotic Fields for Lattice {QCD}}},}\ }\href
  {\doibase 10.1016/0370-2693(84)90473-8} {\bibfield  {journal} {\bibinfo
  {journal} {Phys. Lett.}\ }\textbf {\bibinfo {volume} {B135}},\ \bibinfo
  {pages} {155--158} (\bibinfo {year} {1984})}\BibitemShut {NoStop}%
\bibitem [{\citenamefont {Juge}\ \emph {et~al.}(2004)\citenamefont {Juge},
  \citenamefont {Kuti},\ and\ \citenamefont {Morningstar}}]{Juge:2003qd}%
  \BibitemOpen
  \bibfield  {author} {\bibinfo {author} {\bibfnamefont {K.~Jimmy}\
  \bibnamefont {Juge}}, \bibinfo {author} {\bibfnamefont {Julius}\ \bibnamefont
  {Kuti}}, \ and\ \bibinfo {author} {\bibfnamefont {Colin}\ \bibnamefont
  {Morningstar}},\ }\bibfield  {title} {\enquote {\bibinfo {title} {The
  heavy-quark hybrid meson spectrum in lattice qcd},}\ }\href@noop {}
  {\bibfield  {journal} {\bibinfo  {journal} {AIP Conf. Proc.}\ }\textbf
  {\bibinfo {volume} {688}},\ \bibinfo {pages} {193--207} (\bibinfo {year}
  {2004})},\ \Eprint {http://arxiv.org/abs/nucl-th/0307116} {nucl-th/0307116}
  \BibitemShut {NoStop}%
\bibitem [{\citenamefont {Berwein}\ \emph {et~al.}(2015)\citenamefont
  {Berwein}, \citenamefont {Brambilla}, \citenamefont {Tarrús~Castellà},\
  and\ \citenamefont {Vairo}}]{Berwein:2015vca}%
  \BibitemOpen
  \bibfield  {author} {\bibinfo {author} {\bibfnamefont {Matthias}\
  \bibnamefont {Berwein}}, \bibinfo {author} {\bibfnamefont {Nora}\
  \bibnamefont {Brambilla}}, \bibinfo {author} {\bibfnamefont {Jaume}\
  \bibnamefont {Tarrús~Castellà}}, \ and\ \bibinfo {author} {\bibfnamefont
  {Antonio}\ \bibnamefont {Vairo}},\ }\bibfield  {title} {\enquote {\bibinfo
  {title} {{Quarkonium Hybrids with Nonrelativistic Effective Field
  Theories}},}\ }\href {\doibase 10.1103/PhysRevD.92.114019} {\bibfield
  {journal} {\bibinfo  {journal} {Phys. Rev.}\ }\textbf {\bibinfo {volume}
  {D92}},\ \bibinfo {pages} {114019} (\bibinfo {year} {2015})},\ \Eprint
  {http://arxiv.org/abs/1510.04299} {arXiv:1510.04299 [hep-ph]} \BibitemShut
  {NoStop}%
\bibitem [{\citenamefont {Shifman}\ \emph
  {et~al.}(1979{\natexlab{a}})\citenamefont {Shifman}, \citenamefont
  {Vainshtein},\ and\ \citenamefont {Zakharov}}]{Shifman:1978bx}%
  \BibitemOpen
  \bibfield  {author} {\bibinfo {author} {\bibfnamefont {Mikhail~A.}\
  \bibnamefont {Shifman}}, \bibinfo {author} {\bibfnamefont {A.~I.}\
  \bibnamefont {Vainshtein}}, \ and\ \bibinfo {author} {\bibfnamefont
  {Valentin~I.}\ \bibnamefont {Zakharov}},\ }\bibfield  {title} {\enquote
  {\bibinfo {title} {{QCD and Resonance Physics. Theoretical Foundations}},}\
  }\href {\doibase 10.1016/0550-3213(79)90022-1} {\bibfield  {journal}
  {\bibinfo  {journal} {Nucl. Phys.}\ }\textbf {\bibinfo {volume} {B147}},\
  \bibinfo {pages} {385--447} (\bibinfo {year}
  {1979}{\natexlab{a}})}\BibitemShut {NoStop}%
\bibitem [{\citenamefont {Shifman}\ \emph
  {et~al.}(1979{\natexlab{b}})\citenamefont {Shifman}, \citenamefont
  {Vainshtein},\ and\ \citenamefont {Zakharov}}]{Shifman:1978by}%
  \BibitemOpen
  \bibfield  {author} {\bibinfo {author} {\bibfnamefont {Mikhail~A.}\
  \bibnamefont {Shifman}}, \bibinfo {author} {\bibfnamefont {A.~I.}\
  \bibnamefont {Vainshtein}}, \ and\ \bibinfo {author} {\bibfnamefont
  {Valentin~I.}\ \bibnamefont {Zakharov}},\ }\bibfield  {title} {\enquote
  {\bibinfo {title} {{QCD and Resonance Physics: Applications}},}\ }\href
  {\doibase 10.1016/0550-3213(79)90023-3} {\bibfield  {journal} {\bibinfo
  {journal} {Nucl. Phys.}\ }\textbf {\bibinfo {volume} {B147}},\ \bibinfo
  {pages} {448--518} (\bibinfo {year} {1979}{\natexlab{b}})}\BibitemShut
  {NoStop}%
\bibitem [{\citenamefont {Shifman}\ \emph
  {et~al.}(1979{\natexlab{c}})\citenamefont {Shifman}, \citenamefont
  {Vainshtein},\ and\ \citenamefont {Zakharov}}]{Shifman:1978bw}%
  \BibitemOpen
  \bibfield  {author} {\bibinfo {author} {\bibfnamefont {Mikhail~A.}\
  \bibnamefont {Shifman}}, \bibinfo {author} {\bibfnamefont {A.~I.}\
  \bibnamefont {Vainshtein}}, \ and\ \bibinfo {author} {\bibfnamefont
  {Valentin~I.}\ \bibnamefont {Zakharov}},\ }\bibfield  {title} {\enquote
  {\bibinfo {title} {{QCD and Resonance Physics. The rho-omega Mixing}},}\
  }\href {\doibase 10.1016/0550-3213(79)90024-5} {\bibfield  {journal}
  {\bibinfo  {journal} {Nucl. Phys.}\ }\textbf {\bibinfo {volume} {B147}},\
  \bibinfo {pages} {519--534} (\bibinfo {year}
  {1979}{\natexlab{c}})}\BibitemShut {NoStop}%
\bibitem [{\citenamefont {Matheus}\ \emph {et~al.}(2007)\citenamefont
  {Matheus}, \citenamefont {Narison}, \citenamefont {Nielsen},\ and\
  \citenamefont {Richard}}]{Matheus:2006xi}%
  \BibitemOpen
  \bibfield  {author} {\bibinfo {author} {\bibfnamefont {R.~D.}\ \bibnamefont
  {Matheus}}, \bibinfo {author} {\bibfnamefont {S.}~\bibnamefont {Narison}},
  \bibinfo {author} {\bibfnamefont {M.}~\bibnamefont {Nielsen}}, \ and\
  \bibinfo {author} {\bibfnamefont {J.~M.}\ \bibnamefont {Richard}},\
  }\bibfield  {title} {\enquote {\bibinfo {title} {{Can the $X(3872)$ be a
  $1^{++}$ four-quark state?}}}\ }\href {\doibase 10.1103/PhysRevD.75.014005}
  {\bibfield  {journal} {\bibinfo  {journal} {Phys. Rev.}\ }\textbf {\bibinfo
  {volume} {D75}},\ \bibinfo {pages} {014005} (\bibinfo {year} {2007})},\
  \Eprint {http://arxiv.org/abs/hep-ph/0608297} {arXiv:hep-ph/0608297}
  \BibitemShut {NoStop}%
\bibitem [{\citenamefont {Chen}\ \emph {et~al.}(2008)\citenamefont {Chen},
  \citenamefont {Liu}, \citenamefont {Hosaka},\ and\ \citenamefont
  {Zhu}}]{Chen:2008ej}%
  \BibitemOpen
  \bibfield  {author} {\bibinfo {author} {\bibfnamefont {Hua-Xing}\
  \bibnamefont {Chen}}, \bibinfo {author} {\bibfnamefont {Xiang}\ \bibnamefont
  {Liu}}, \bibinfo {author} {\bibfnamefont {Atsushi}\ \bibnamefont {Hosaka}}, \
  and\ \bibinfo {author} {\bibfnamefont {Shi-Lin}\ \bibnamefont {Zhu}},\
  }\bibfield  {title} {\enquote {\bibinfo {title} {{The Y(2175) State in the
  QCD Sum Rule}},}\ }\href {\doibase 10.1103/PhysRevD.78.034012} {\bibfield
  {journal} {\bibinfo  {journal} {Phys. Rev.}\ }\textbf {\bibinfo {volume}
  {D78}},\ \bibinfo {pages} {034012} (\bibinfo {year} {2008})},\ \Eprint
  {http://arxiv.org/abs/0801.4603} {arXiv:0801.4603 [hep-ph]} \BibitemShut
  {NoStop}%
\bibitem [{\citenamefont {Narison}\ \emph {et~al.}(2010)\citenamefont
  {Narison}, \citenamefont {Navarra},\ and\ \citenamefont
  {Nielsen}}]{Narison:2010ak}%
  \BibitemOpen
  \bibfield  {author} {\bibinfo {author} {\bibfnamefont {Stephan}\ \bibnamefont
  {Narison}}, \bibinfo {author} {\bibfnamefont {Fernando~S.}\ \bibnamefont
  {Navarra}}, \ and\ \bibinfo {author} {\bibfnamefont {Marina}\ \bibnamefont
  {Nielsen}},\ }\bibfield  {title} {\enquote {\bibinfo {title} {{Investigating
  different structures for the $X(3872)$}},}\ }\bibfield  {booktitle} {\emph
  {\bibinfo {booktitle} {{Proceedings, 15th High-Energy Physics International
  Conference on Quantum Chromodynamics (QCD 10)}}},\ }\href {\doibase
  10.1016/j.nuclphysbps.2010.10.064} {\bibfield  {journal} {\bibinfo  {journal}
  {Nucl. Phys. Proc. Suppl.}\ }\textbf {\bibinfo {volume} {207-208}},\ \bibinfo
  {pages} {249--252} (\bibinfo {year} {2010})},\ \Eprint
  {http://arxiv.org/abs/1007.4575} {arXiv:1007.4575 [hep-ph]} \BibitemShut
  {NoStop}%
\bibitem [{\citenamefont {DeTar}(2015)}]{DeTar:2015orc}%
  \BibitemOpen
  \bibfield  {author} {\bibinfo {author} {\bibfnamefont {Carleton}\
  \bibnamefont {DeTar}},\ }\bibfield  {title} {\enquote {\bibinfo {title}
  {{LQCD: Flavor Physics and Spectroscopy}},}\ }\href@noop {} {\  (\bibinfo
  {year} {2015})},\ \bibinfo {note} {{Lepton-Photon 2015, Ljubljana, Slovenia,
  to appear in PoS}},\ \Eprint {http://arxiv.org/abs/1511.06884}
  {arXiv:1511.06884 [hep-lat]} \BibitemShut {NoStop}%
\bibitem [{\citenamefont {Fiebig}\ \emph {et~al.}(2000)\citenamefont {Fiebig},
  \citenamefont {Rabitsch}, \citenamefont {Markum},\ and\ \citenamefont
  {Mihaly}}]{Fiebig:1999hs}%
  \BibitemOpen
  \bibfield  {author} {\bibinfo {author} {\bibfnamefont {H.~R.}\ \bibnamefont
  {Fiebig}}, \bibinfo {author} {\bibfnamefont {K.}~\bibnamefont {Rabitsch}},
  \bibinfo {author} {\bibfnamefont {H.}~\bibnamefont {Markum}}, \ and\ \bibinfo
  {author} {\bibfnamefont {A.}~\bibnamefont {Mihaly}},\ }\bibfield  {title}
  {\enquote {\bibinfo {title} {Exploring the $\pi^+\pi^+$ interaction in
  lattice qcd},}\ }\href@noop {} {\bibfield  {journal} {\bibinfo  {journal}
  {Few Body Syst.}\ }\textbf {\bibinfo {volume} {29}},\ \bibinfo {pages}
  {95--120} (\bibinfo {year} {2000})},\ \Eprint
  {http://arxiv.org/abs/hep-lat/9906002} {hep-lat/9906002} \BibitemShut
  {NoStop}%
\bibitem [{\citenamefont {Inoue}\ \emph {et~al.}(2011)\citenamefont {Inoue}
  \emph {et~al.}}]{Inoue:2010es}%
  \BibitemOpen
  \bibfield  {author} {\bibinfo {author} {\bibfnamefont {Takashi}\ \bibnamefont
  {Inoue}} \emph {et~al.} (\bibinfo {collaboration} {HAL QCD}),\ }\bibfield
  {title} {\enquote {\bibinfo {title} {{Bound H-dibaryon in Flavor SU(3) Limit
  of Lattice QCD}},}\ }\href {\doibase 10.1103/PhysRevLett.106.162002}
  {\bibfield  {journal} {\bibinfo  {journal} {Phys. Rev. Lett.}\ }\textbf
  {\bibinfo {volume} {106}},\ \bibinfo {pages} {162002} (\bibinfo {year}
  {2011})},\ \Eprint {http://arxiv.org/abs/1012.5928} {arXiv:1012.5928
  [hep-lat]} \BibitemShut {NoStop}%
\bibitem [{\citenamefont {Beane}\ \emph {et~al.}(2011)\citenamefont {Beane}
  \emph {et~al.}}]{Beane:2010hg}%
  \BibitemOpen
  \bibfield  {author} {\bibinfo {author} {\bibfnamefont {S.~R.}\ \bibnamefont
  {Beane}} \emph {et~al.} (\bibinfo {collaboration} {NPLQCD}),\ }\bibfield
  {title} {\enquote {\bibinfo {title} {{Evidence for a Bound H-dibaryon from
  Lattice QCD}},}\ }\href {\doibase 10.1103/PhysRevLett.106.162001} {\bibfield
  {journal} {\bibinfo  {journal} {Phys. Rev. Lett.}\ }\textbf {\bibinfo
  {volume} {106}},\ \bibinfo {pages} {162001} (\bibinfo {year} {2011})},\
  \Eprint {http://arxiv.org/abs/1012.3812} {arXiv:1012.3812 [hep-lat]}
  \BibitemShut {NoStop}%
\bibitem [{\citenamefont {Suganuma}\ \emph {et~al.}(2011)\citenamefont
  {Suganuma}, \citenamefont {Iritani}, \citenamefont {Okiharu}, \citenamefont
  {Takahashi},\ and\ \citenamefont {Yamamoto}}]{Suganuma:2011ci}%
  \BibitemOpen
  \bibfield  {author} {\bibinfo {author} {\bibfnamefont {H.}~\bibnamefont
  {Suganuma}}, \bibinfo {author} {\bibfnamefont {T.}~\bibnamefont {Iritani}},
  \bibinfo {author} {\bibfnamefont {F.}~\bibnamefont {Okiharu}}, \bibinfo
  {author} {\bibfnamefont {T.~T.}\ \bibnamefont {Takahashi}}, \ and\ \bibinfo
  {author} {\bibfnamefont {A.}~\bibnamefont {Yamamoto}},\ }\bibfield  {title}
  {\enquote {\bibinfo {title} {{Lattice QCD Study for Confinement in
  Hadrons}},}\ }\href {\doibase 10.1063/1.3647373} {\bibfield  {journal}
  {\bibinfo  {journal} {AIP Conf. Proc.}\ }\textbf {\bibinfo {volume} {1388}},\
  \bibinfo {pages} {195--201} (\bibinfo {year} {2011})},\ \Eprint
  {http://arxiv.org/abs/1103.4015} {arXiv:1103.4015 [hep-lat]} \BibitemShut
  {NoStop}%
\bibitem [{\citenamefont {Aoki}\ \emph {et~al.}(2012)\citenamefont {Aoki},
  \citenamefont {Doi}, \citenamefont {Hatsuda}, \citenamefont {Ikeda},
  \citenamefont {Inoue}, \citenamefont {Ishii}, \citenamefont {Murano},
  \citenamefont {Nemura},\ and\ \citenamefont {Sasaki}}]{Aoki:2012tk}%
  \BibitemOpen
  \bibfield  {author} {\bibinfo {author} {\bibfnamefont {Sinya}\ \bibnamefont
  {Aoki}}, \bibinfo {author} {\bibfnamefont {Takumi}\ \bibnamefont {Doi}},
  \bibinfo {author} {\bibfnamefont {Tetsuo}\ \bibnamefont {Hatsuda}}, \bibinfo
  {author} {\bibfnamefont {Yoichi}\ \bibnamefont {Ikeda}}, \bibinfo {author}
  {\bibfnamefont {Takashi}\ \bibnamefont {Inoue}}, \bibinfo {author}
  {\bibfnamefont {Noriyoshi}\ \bibnamefont {Ishii}}, \bibinfo {author}
  {\bibfnamefont {Keiko}\ \bibnamefont {Murano}}, \bibinfo {author}
  {\bibfnamefont {Hidekatsu}\ \bibnamefont {Nemura}}, \ and\ \bibinfo {author}
  {\bibfnamefont {Kenji}\ \bibnamefont {Sasaki}} (\bibinfo {collaboration} {HAL
  QCD}),\ }\bibfield  {title} {\enquote {\bibinfo {title} {{Lattice QCD
  approach to Nuclear Physics}},}\ }\href {\doibase 10.1093/ptep/pts010}
  {\bibfield  {journal} {\bibinfo  {journal} {PTEP}\ }\textbf {\bibinfo
  {volume} {2012}},\ \bibinfo {pages} {01A105} (\bibinfo {year} {2012})},\
  \Eprint {http://arxiv.org/abs/1206.5088} {arXiv:1206.5088 [hep-lat]}
  \BibitemShut {NoStop}%
\bibitem [{\citenamefont {Kirscher}\ \emph {et~al.}(2015)\citenamefont
  {Kirscher}, \citenamefont {Barnea}, \citenamefont {Gazit}, \citenamefont
  {Pederiva},\ and\ \citenamefont {van Kolck}}]{Kirscher:2015yda}%
  \BibitemOpen
  \bibfield  {author} {\bibinfo {author} {\bibfnamefont {Johannes}\
  \bibnamefont {Kirscher}}, \bibinfo {author} {\bibfnamefont {Nir}\
  \bibnamefont {Barnea}}, \bibinfo {author} {\bibfnamefont {Doron}\
  \bibnamefont {Gazit}}, \bibinfo {author} {\bibfnamefont {Francesco}\
  \bibnamefont {Pederiva}}, \ and\ \bibinfo {author} {\bibfnamefont
  {Ubirajara}\ \bibnamefont {van Kolck}},\ }\bibfield  {title} {\enquote
  {\bibinfo {title} {{Spectra and Scattering of Light Lattice Nuclei from
  Effective Field Theory}},}\ }\href {\doibase 10.1103/PhysRevC.92.054002}
  {\bibfield  {journal} {\bibinfo  {journal} {Phys. Rev.}\ }\textbf {\bibinfo
  {volume} {C92}},\ \bibinfo {pages} {054002} (\bibinfo {year} {2015})},\
  \Eprint {http://arxiv.org/abs/1506.09048} {arXiv:1506.09048 [nucl-th]}
  \BibitemShut {NoStop}%
\bibitem [{\citenamefont {Dudek}(2016)}]{Dudek:2016bxq}%
  \BibitemOpen
  \bibfield  {author} {\bibinfo {author} {\bibfnamefont {Jozef~J.}\
  \bibnamefont {Dudek}},\ }\bibfield  {title} {\enquote {\bibinfo {title}
  {{Hadron scattering and resonances in QCD}},}\ }\bibfield  {booktitle} {\emph
  {\bibinfo {booktitle} {{Proceedings, 16th International Conference on Hadron
  Spectroscopy (Hadron 2015)}}},\ }\href {\doibase 10.1063/1.4949382}
  {\bibfield  {journal} {\bibinfo  {journal} {AIP Conf. Proc.}\ }\textbf
  {\bibinfo {volume} {1735}},\ \bibinfo {pages} {020014} (\bibinfo {year}
  {2016})}\BibitemShut {NoStop}%
\bibitem [{\citenamefont {Francis}\ \emph {et~al.}(2016)\citenamefont
  {Francis}, \citenamefont {Hudspith}, \citenamefont {Lewis},\ and\
  \citenamefont {Maltman}}]{Francis:2016hui}%
  \BibitemOpen
  \bibfield  {author} {\bibinfo {author} {\bibfnamefont {Anthony}\ \bibnamefont
  {Francis}}, \bibinfo {author} {\bibfnamefont {Renwick~J.}\ \bibnamefont
  {Hudspith}}, \bibinfo {author} {\bibfnamefont {Randy}\ \bibnamefont {Lewis}},
  \ and\ \bibinfo {author} {\bibfnamefont {Kim}\ \bibnamefont {Maltman}},\
  }\bibfield  {title} {\enquote {\bibinfo {title} {{Doubly bottom
  strong-interaction stable tetraquarks from lattice QCD}},}\ }\href@noop {} {\
   (\bibinfo {year} {2016})},\ \Eprint {http://arxiv.org/abs/1607.05214}
  {arXiv:1607.05214 [hep-lat]} \BibitemShut {NoStop}%
\bibitem [{\citenamefont {Green}\ and\ \citenamefont
  {Pennanen}(1998)}]{Green:1998nt}%
  \BibitemOpen
  \bibfield  {author} {\bibinfo {author} {\bibfnamefont {Anthony~M.}\
  \bibnamefont {Green}}\ and\ \bibinfo {author} {\bibfnamefont
  {P.}~\bibnamefont {Pennanen}},\ }\bibfield  {title} {\enquote {\bibinfo
  {title} {{An Interquark potential model for multiquark systems}},}\ }\href
  {\doibase 10.1103/PhysRevC.57.3384} {\bibfield  {journal} {\bibinfo
  {journal} {Phys. Rev.}\ }\textbf {\bibinfo {volume} {C57}},\ \bibinfo {pages}
  {3384--3391} (\bibinfo {year} {1998})},\ \Eprint
  {http://arxiv.org/abs/hep-lat/9804003} {arXiv:hep-lat/9804003 [hep-lat]}
  \BibitemShut {NoStop}%
\bibitem [{\citenamefont {Michael}\ and\ \citenamefont
  {Pennanen}(1999)}]{Michael:1999nq}%
  \BibitemOpen
  \bibfield  {author} {\bibinfo {author} {\bibfnamefont {Christopher}\
  \bibnamefont {Michael}}\ and\ \bibinfo {author} {\bibfnamefont
  {P.}~\bibnamefont {Pennanen}} (\bibinfo {collaboration} {UKQCD}),\ }\bibfield
   {title} {\enquote {\bibinfo {title} {{Two heavy - light mesons on a
  lattice}},}\ }\href {\doibase 10.1103/PhysRevD.60.054012} {\bibfield
  {journal} {\bibinfo  {journal} {Phys. Rev.}\ }\textbf {\bibinfo {volume}
  {D60}},\ \bibinfo {pages} {054012} (\bibinfo {year} {1999})},\ \Eprint
  {http://arxiv.org/abs/hep-lat/9901007} {arXiv:hep-lat/9901007 [hep-lat]}
  \BibitemShut {NoStop}%
\bibitem [{\citenamefont {Bali}\ and\ \citenamefont
  {Hetzenegger}(2010)}]{Bali:2010xa}%
  \BibitemOpen
  \bibfield  {author} {\bibinfo {author} {\bibfnamefont {Gunnar}\ \bibnamefont
  {Bali}}\ and\ \bibinfo {author} {\bibfnamefont {Martin}\ \bibnamefont
  {Hetzenegger}},\ }\bibfield  {title} {\enquote {\bibinfo {title}
  {{Static-light meson-meson potentials}},}\ }\href@noop {} {\bibfield
  {journal} {\bibinfo  {journal} {PoS}\ }\textbf {\bibinfo {volume}
  {LATTICE2010}},\ \bibinfo {pages} {142} (\bibinfo {year} {2010})},\ \Eprint
  {http://arxiv.org/abs/1011.0571} {arXiv:1011.0571 [hep-lat]} \BibitemShut
  {NoStop}%
\bibitem [{\citenamefont {Bicudo}\ \emph {et~al.}(2016)\citenamefont {Bicudo},
  \citenamefont {Cichy}, \citenamefont {Peters},\ and\ \citenamefont
  {Wagner}}]{Bicudo:2015kna}%
  \BibitemOpen
  \bibfield  {author} {\bibinfo {author} {\bibfnamefont {Pedro}\ \bibnamefont
  {Bicudo}}, \bibinfo {author} {\bibfnamefont {Krzysztof}\ \bibnamefont
  {Cichy}}, \bibinfo {author} {\bibfnamefont {Antje}\ \bibnamefont {Peters}}, \
  and\ \bibinfo {author} {\bibfnamefont {Marc}\ \bibnamefont {Wagner}},\
  }\bibfield  {title} {\enquote {\bibinfo {title} {{BB interactions with static
  bottom quarks from Lattice QCD}},}\ }\href {\doibase
  10.1103/PhysRevD.93.034501} {\bibfield  {journal} {\bibinfo  {journal} {Phys.
  Rev.}\ }\textbf {\bibinfo {volume} {D93}},\ \bibinfo {pages} {034501}
  (\bibinfo {year} {2016})},\ \Eprint {http://arxiv.org/abs/1510.03441}
  {arXiv:1510.03441 [hep-lat]} \BibitemShut {NoStop}%
\bibitem [{\citenamefont {De~R{\'u}jula}\ \emph {et~al.}(1975)\citenamefont
  {De~R{\'u}jula}, \citenamefont {Georgi},\ and\ \citenamefont
  {Glashow}}]{DeRujula:1975ge}%
  \BibitemOpen
  \bibfield  {author} {\bibinfo {author} {\bibfnamefont {A.}~\bibnamefont
  {De~R{\'u}jula}}, \bibinfo {author} {\bibfnamefont {Howard}\ \bibnamefont
  {Georgi}}, \ and\ \bibinfo {author} {\bibfnamefont {S.~L.}\ \bibnamefont
  {Glashow}},\ }\bibfield  {title} {\enquote {\bibinfo {title}
  {H\lowercase{ADRON MASSES IN A GAUGE THEORY}},}\ }\href@noop {} {\bibfield
  {journal} {\bibinfo  {journal} {Phys. Rev.}\ }\textbf {\bibinfo {volume}
  {D12}},\ \bibinfo {pages} {147--162} (\bibinfo {year} {1975})}\BibitemShut
  {NoStop}%
\bibitem [{\citenamefont {Takahashi}\ \emph {et~al.}(2001)\citenamefont
  {Takahashi} \emph {et~al.}}]{Takahashi:2001nm}%
  \BibitemOpen
  \bibfield  {author} {\bibinfo {author} {\bibfnamefont {H.}~\bibnamefont
  {Takahashi}} \emph {et~al.},\ }\bibfield  {title} {\enquote {\bibinfo {title}
  {{Observation of a $\isotope[6][\Lambda\Lambda]{He}$ double hypernucleus}},}\
  }\href {\doibase 10.1103/PhysRevLett.87.212502} {\bibfield  {journal}
  {\bibinfo  {journal} {Phys. Rev. Lett.}\ }\textbf {\bibinfo {volume} {87}},\
  \bibinfo {pages} {212502} (\bibinfo {year} {2001})}\BibitemShut {NoStop}%
\bibitem [{\citenamefont {Oka}\ \emph {et~al.}(1983)\citenamefont {Oka},
  \citenamefont {Shimizu},\ and\ \citenamefont {Yazaki}}]{Oka:1983ku}%
  \BibitemOpen
  \bibfield  {author} {\bibinfo {author} {\bibfnamefont {M.}~\bibnamefont
  {Oka}}, \bibinfo {author} {\bibfnamefont {K.}~\bibnamefont {Shimizu}}, \ and\
  \bibinfo {author} {\bibfnamefont {K.}~\bibnamefont {Yazaki}},\ }\bibfield
  {title} {\enquote {\bibinfo {title} {{The Dihyperon State in the Quark
  Cluster Model}},}\ }\href {\doibase 10.1016/0370-2693(83)91523-X} {\bibfield
  {journal} {\bibinfo  {journal} {Phys. Lett.}\ }\textbf {\bibinfo {volume}
  {B130}},\ \bibinfo {pages} {365} (\bibinfo {year} {1983})}\BibitemShut
  {NoStop}%
\bibitem [{\citenamefont {Rosner}(1986)}]{Rosner:1985yh}%
  \BibitemOpen
  \bibfield  {author} {\bibinfo {author} {\bibfnamefont {Jonathan~L.}\
  \bibnamefont {Rosner}},\ }\bibfield  {title} {\enquote {\bibinfo {title}
  {{SU(3) Breaking and the $H$ Dibaryon}},}\ }\href {\doibase
  10.1103/PhysRevD.33.2043} {\bibfield  {journal} {\bibinfo  {journal} {Phys.
  Rev.}\ }\textbf {\bibinfo {volume} {D33}},\ \bibinfo {pages} {2043} (\bibinfo
  {year} {1986})}\BibitemShut {NoStop}%
\bibitem [{\citenamefont {Karl}\ and\ \citenamefont
  {Zenczykowski}(1987)}]{Karl:1987cg}%
  \BibitemOpen
  \bibfield  {author} {\bibinfo {author} {\bibfnamefont {G.}~\bibnamefont
  {Karl}}\ and\ \bibinfo {author} {\bibfnamefont {P.}~\bibnamefont
  {Zenczykowski}},\ }\bibfield  {title} {\enquote {\bibinfo {title} {{$H$
  dibaryon spectrocopy}},}\ }\href {\doibase 10.1103/PhysRevD.36.2079}
  {\bibfield  {journal} {\bibinfo  {journal} {Phys. Rev.}\ }\textbf {\bibinfo
  {volume} {D36}},\ \bibinfo {pages} {2079} (\bibinfo {year}
  {1987})}\BibitemShut {NoStop}%
\bibitem [{\citenamefont {Faessler}\ and\ \citenamefont
  {Straub}(1990)}]{Faessler:1989si}%
  \BibitemOpen
  \bibfield  {author} {\bibinfo {author} {\bibfnamefont {A.}~\bibnamefont
  {Faessler}}\ and\ \bibinfo {author} {\bibfnamefont {U.}~\bibnamefont
  {Straub}},\ }\bibfield  {title} {\enquote {\bibinfo {title} {{Baryon baryon
  interaction in the quark model and the $H$ dibaryon}},}\ }\bibfield
  {booktitle} {\emph {\bibinfo {booktitle} {{In Erice 1989, Proceedings, The
  nature of hadrons and nuclei by electron scattering, 323-332.}}},\ }\href
  {\doibase 10.1016/0146-6410(90)90023-W} {\bibfield  {journal} {\bibinfo
  {journal} {Prog. Part. Nucl. Phys.}\ }\textbf {\bibinfo {volume} {24}},\
  \bibinfo {pages} {323--332} (\bibinfo {year} {1990})}\BibitemShut {NoStop}%
\bibitem [{\citenamefont {Leandri}\ and\ \citenamefont
  {Silvestre-Brac}(1989)}]{Leandri:1989su}%
  \BibitemOpen
  \bibfield  {author} {\bibinfo {author} {\bibfnamefont {J.}~\bibnamefont
  {Leandri}}\ and\ \bibinfo {author} {\bibfnamefont {B.}~\bibnamefont
  {Silvestre-Brac}},\ }\bibfield  {title} {\enquote {\bibinfo {title}
  {{Systematics of $\bar{Q} Q^4$ Systems With a Pure Chromomagnetic
  Interaction}},}\ }\href {\doibase 10.1103/PhysRevD.40.2340} {\bibfield
  {journal} {\bibinfo  {journal} {Phys. Rev.}\ }\textbf {\bibinfo {volume}
  {D40}},\ \bibinfo {pages} {2340--2352} (\bibinfo {year} {1989})}\BibitemShut
  {NoStop}%
\bibitem [{\citenamefont {Silvestre-Brac}\ and\ \citenamefont
  {Leandri}(1992)}]{SilvestreBrac:1992yg}%
  \BibitemOpen
  \bibfield  {author} {\bibinfo {author} {\bibfnamefont {B.}~\bibnamefont
  {Silvestre-Brac}}\ and\ \bibinfo {author} {\bibfnamefont {J.}~\bibnamefont
  {Leandri}},\ }\bibfield  {title} {\enquote {\bibinfo {title} {{Systematics of
  $q^6$ systems in a simple chromomagnetic model}},}\ }\href {\doibase
  10.1103/PhysRevD.45.4221} {\bibfield  {journal} {\bibinfo  {journal} {Phys.
  Rev.}\ }\textbf {\bibinfo {volume} {D45}},\ \bibinfo {pages} {4221--4239}
  (\bibinfo {year} {1992})}\BibitemShut {NoStop}%
\bibitem [{\citenamefont {Leandri}\ and\ \citenamefont
  {Silvestre-Brac}(1995)}]{Leandri:1995zm}%
  \BibitemOpen
  \bibfield  {author} {\bibinfo {author} {\bibfnamefont {J.}~\bibnamefont
  {Leandri}}\ and\ \bibinfo {author} {\bibfnamefont {B.}~\bibnamefont
  {Silvestre-Brac}},\ }\bibfield  {title} {\enquote {\bibinfo {title}
  {{Dibaryon states containing two different types of heavy quarks}},}\ }\href
  {\doibase 10.1103/PhysRevD.51.3628} {\bibfield  {journal} {\bibinfo
  {journal} {Phys. Rev.}\ }\textbf {\bibinfo {volume} {D51}},\ \bibinfo {pages}
  {3628--3637} (\bibinfo {year} {1995})}\BibitemShut {NoStop}%
\bibitem [{\citenamefont {Lichtenberg}\ and\ \citenamefont
  {Roncaglia}(1992)}]{Lichtenberg:1992ji}%
  \BibitemOpen
  \bibfield  {author} {\bibinfo {author} {\bibfnamefont {D.~B.}\ \bibnamefont
  {Lichtenberg}}\ and\ \bibinfo {author} {\bibfnamefont {R.}~\bibnamefont
  {Roncaglia}},\ }\bibfield  {title} {\enquote {\bibinfo {title}
  {{Colormagnetic interaction, diquarks, and exotic hadrons}},}\ }in\ \href
  {http://alice.cern.ch/format/showfull?sysnb=0160281} {\emph {\bibinfo
  {booktitle} {{International Workshop on Diquarks, Turin, Italy, November 2-4,
  1992}}}}\ (\bibinfo {year} {1992})\BibitemShut {NoStop}%
\bibitem [{\citenamefont {Buccella}\ \emph {et~al.}(2007)\citenamefont
  {Buccella}, \citenamefont {H\o{}g\aa{}sen}, \citenamefont {Richard},\ and\
  \citenamefont {Sorba}}]{Buccella:2006fn}%
  \BibitemOpen
  \bibfield  {author} {\bibinfo {author} {\bibfnamefont {Franco}\ \bibnamefont
  {Buccella}}, \bibinfo {author} {\bibfnamefont {Hallstein}\ \bibnamefont
  {H\o{}g\aa{}sen}}, \bibinfo {author} {\bibfnamefont {Jean-Marc}\ \bibnamefont
  {Richard}}, \ and\ \bibinfo {author} {\bibfnamefont {Paul}\ \bibnamefont
  {Sorba}},\ }\bibfield  {title} {\enquote {\bibinfo {title} {{Chromomagnetism,
  flavour symmetry breaking and S-wave tetraquarks}},}\ }\href {\doibase
  10.1140/epjc/s10052-006-0142-1} {\bibfield  {journal} {\bibinfo  {journal}
  {Eur. Phys. J.}\ }\textbf {\bibinfo {volume} {C49}},\ \bibinfo {pages}
  {743--754} (\bibinfo {year} {2007})},\ \Eprint
  {http://arxiv.org/abs/hep-ph/0608001} {arXiv:hep-ph/0608001} \BibitemShut
  {NoStop}%
\bibitem [{\citenamefont {Ader}\ \emph {et~al.}(1982)\citenamefont {Ader},
  \citenamefont {Richard},\ and\ \citenamefont {Taxil}}]{Ader:1981db}%
  \BibitemOpen
  \bibfield  {author} {\bibinfo {author} {\bibfnamefont {J.~P.}\ \bibnamefont
  {Ader}}, \bibinfo {author} {\bibfnamefont {J.~M.}\ \bibnamefont {Richard}}, \
  and\ \bibinfo {author} {\bibfnamefont {P.}~\bibnamefont {Taxil}},\ }\bibfield
   {title} {\enquote {\bibinfo {title} {Do narrow heavy multiquark states
  exist?}}\ }\href@noop {} {\bibfield  {journal} {\bibinfo  {journal} {Phys.
  Rev.}\ }\textbf {\bibinfo {volume} {D25}},\ \bibinfo {pages} {2370} (\bibinfo
  {year} {1982})}\BibitemShut {NoStop}%
\bibitem [{\citenamefont {Heller}\ and\ \citenamefont
  {Tjon}(1987)}]{Heller:1986bt}%
  \BibitemOpen
  \bibfield  {author} {\bibinfo {author} {\bibfnamefont {L.}~\bibnamefont
  {Heller}}\ and\ \bibinfo {author} {\bibfnamefont {J.~A.}\ \bibnamefont
  {Tjon}},\ }\bibfield  {title} {\enquote {\bibinfo {title} {On the existence
  of stable dimesons},}\ }\href@noop {} {\bibfield  {journal} {\bibinfo
  {journal} {Phys. Rev.}\ }\textbf {\bibinfo {volume} {D35}},\ \bibinfo {pages}
  {969} (\bibinfo {year} {1987})}\BibitemShut {NoStop}%
\bibitem [{\citenamefont {Zouzou}\ \emph {et~al.}(1986)\citenamefont {Zouzou},
  \citenamefont {Silvestre-Brac}, \citenamefont {Gignoux},\ and\ \citenamefont
  {Richard}}]{Zouzou:1986qh}%
  \BibitemOpen
  \bibfield  {author} {\bibinfo {author} {\bibfnamefont {S.}~\bibnamefont
  {Zouzou}}, \bibinfo {author} {\bibfnamefont {B.}~\bibnamefont
  {Silvestre-Brac}}, \bibinfo {author} {\bibfnamefont {C.}~\bibnamefont
  {Gignoux}}, \ and\ \bibinfo {author} {\bibfnamefont {J.~M.}\ \bibnamefont
  {Richard}},\ }\bibfield  {title} {\enquote {\bibinfo {title} {Four quark
  bound states},}\ }\href@noop {} {\bibfield  {journal} {\bibinfo  {journal}
  {Z. Phys.}\ }\textbf {\bibinfo {volume} {C30}},\ \bibinfo {pages} {457}
  (\bibinfo {year} {1986})}\BibitemShut {NoStop}%
\bibitem [{\citenamefont {Carlson}\ \emph {et~al.}(1988)\citenamefont
  {Carlson}, \citenamefont {Heller},\ and\ \citenamefont
  {Tjon}}]{Carlson:1988hh}%
  \BibitemOpen
  \bibfield  {author} {\bibinfo {author} {\bibfnamefont {J.}~\bibnamefont
  {Carlson}}, \bibinfo {author} {\bibfnamefont {L.}~\bibnamefont {Heller}}, \
  and\ \bibinfo {author} {\bibfnamefont {J.~A.}\ \bibnamefont {Tjon}},\
  }\bibfield  {title} {\enquote {\bibinfo {title} {Stability of dimesons},}\
  }\href@noop {} {\bibfield  {journal} {\bibinfo  {journal} {Phys. Rev.}\
  }\textbf {\bibinfo {volume} {D37}},\ \bibinfo {pages} {744} (\bibinfo {year}
  {1988})}\BibitemShut {NoStop}%
\bibitem [{\citenamefont {Bressanini}\ \emph {et~al.}(1998)\citenamefont
  {Bressanini}, \citenamefont {Mella},\ and\ \citenamefont
  {Morosi}}]{PhysRevA.57.4956}%
  \BibitemOpen
  \bibfield  {author} {\bibinfo {author} {\bibfnamefont {Dario}\ \bibnamefont
  {Bressanini}}, \bibinfo {author} {\bibfnamefont {Massimo}\ \bibnamefont
  {Mella}}, \ and\ \bibinfo {author} {\bibfnamefont {Gabriele}\ \bibnamefont
  {Morosi}},\ }\bibfield  {title} {\enquote {\bibinfo {title} {Stability of
  four-body systems in three and two dimensions: A theoretical and quantum
  monte carlo study of biexciton molecules},}\ }\href {\doibase
  10.1103/PhysRevA.57.4956} {\bibfield  {journal} {\bibinfo  {journal} {Phys.
  Rev. A}\ }\textbf {\bibinfo {volume} {57}},\ \bibinfo {pages} {4956--4959}
  (\bibinfo {year} {1998})}\BibitemShut {NoStop}%
\bibitem [{\citenamefont {{Varga}}\ \emph {et~al.}(1999)\citenamefont
  {{Varga}}, \citenamefont {{Usukura}},\ and\ \citenamefont
  {{Suzuki}}}]{1999fbpp.conf...11V}%
  \BibitemOpen
  \bibfield  {author} {\bibinfo {author} {\bibfnamefont {K.}~\bibnamefont
  {{Varga}}}, \bibinfo {author} {\bibfnamefont {J.}~\bibnamefont {{Usukura}}},
  \ and\ \bibinfo {author} {\bibfnamefont {Y.}~\bibnamefont {{Suzuki}}},\
  }\bibfield  {title} {\enquote {\bibinfo {title} {{Recent applications of the
  stochastic variational method}},}\ }in\ \href@noop {} {\emph {\bibinfo
  {booktitle} {Few-Body Problems in Physics '98}}},\ \bibinfo {editor} {edited
  by\ \bibinfo {editor} {\bibfnamefont {B.}~\bibnamefont {{Desplanques}}},
  \bibinfo {editor} {\bibfnamefont {K.}~\bibnamefont {{Protasov}}}, \bibinfo
  {editor} {\bibfnamefont {B.}~\bibnamefont {{Silvestre-Brac}}}, \ and\
  \bibinfo {editor} {\bibfnamefont {J.}~\bibnamefont {{Carbonell}}}}\ (\bibinfo
  {year} {1999})\ p.~\bibinfo {pages} {11}\BibitemShut {NoStop}%
\bibitem [{\citenamefont {Lloyd}\ and\ \citenamefont
  {Vary}(2004)}]{Lloyd:2003yc}%
  \BibitemOpen
  \bibfield  {author} {\bibinfo {author} {\bibfnamefont {Richard~J.}\
  \bibnamefont {Lloyd}}\ and\ \bibinfo {author} {\bibfnamefont {James~P.}\
  \bibnamefont {Vary}},\ }\bibfield  {title} {\enquote {\bibinfo {title} {{All
  charm tetraquarks}},}\ }\href {\doibase 10.1103/PhysRevD.70.014009}
  {\bibfield  {journal} {\bibinfo  {journal} {Phys. Rev.}\ }\textbf {\bibinfo
  {volume} {D70}},\ \bibinfo {pages} {014009} (\bibinfo {year} {2004})},\
  \Eprint {http://arxiv.org/abs/hep-ph/0311179} {arXiv:hep-ph/0311179 [hep-ph]}
  \BibitemShut {NoStop}%
\bibitem [{\citenamefont {Ay}\ \emph {et~al.}(2009)\citenamefont {Ay},
  \citenamefont {Richard},\ and\ \citenamefont {Rubinstein}}]{Ay:2009zp}%
  \BibitemOpen
  \bibfield  {author} {\bibinfo {author} {\bibfnamefont {Cafer}\ \bibnamefont
  {Ay}}, \bibinfo {author} {\bibfnamefont {Jean-Marc}\ \bibnamefont {Richard}},
  \ and\ \bibinfo {author} {\bibfnamefont {J.~Hyam}\ \bibnamefont
  {Rubinstein}},\ }\bibfield  {title} {\enquote {\bibinfo {title} {{Stability
  of asymmetric tetraquarks in the minimal-path linear potential}},}\ }\href
  {\doibase 10.1016/j.physletb.2009.03.018} {\bibfield  {journal} {\bibinfo
  {journal} {Phys. Lett.}\ }\textbf {\bibinfo {volume} {B674}},\ \bibinfo
  {pages} {227--231} (\bibinfo {year} {2009})},\ \Eprint
  {http://arxiv.org/abs/0901.3022} {arXiv:0901.3022 [math-ph]} \BibitemShut
  {NoStop}%
\bibitem [{\citenamefont {Lenz}\ \emph {et~al.}(1986)\citenamefont {Lenz},
  \citenamefont {Londergan}, \citenamefont {Moniz}, \citenamefont
  {Rosenfelder}, \citenamefont {Stingl},\ and\ \citenamefont
  {Yazaki}}]{Lenz:1985jk}%
  \BibitemOpen
  \bibfield  {author} {\bibinfo {author} {\bibfnamefont {F.}~\bibnamefont
  {Lenz}}, \bibinfo {author} {\bibfnamefont {J.~T.}\ \bibnamefont {Londergan}},
  \bibinfo {author} {\bibfnamefont {E.~J.}\ \bibnamefont {Moniz}}, \bibinfo
  {author} {\bibfnamefont {R.}~\bibnamefont {Rosenfelder}}, \bibinfo {author}
  {\bibfnamefont {M.}~\bibnamefont {Stingl}}, \ and\ \bibinfo {author}
  {\bibfnamefont {K.}~\bibnamefont {Yazaki}},\ }\bibfield  {title} {\enquote
  {\bibinfo {title} {{Quark Confinement and Hadronic Interactions}},}\ }\href
  {\doibase 10.1016/0003-4916(86)90088-6} {\bibfield  {journal} {\bibinfo
  {journal} {Annals Phys.}\ }\textbf {\bibinfo {volume} {170}},\ \bibinfo
  {pages} {65} (\bibinfo {year} {1986})}\BibitemShut {NoStop}%
\bibitem [{\citenamefont {Vijande}\ \emph
  {et~al.}(2007{\natexlab{b}})\citenamefont {Vijande}, \citenamefont
  {Valcarce},\ and\ \citenamefont {Richard}}]{Vijande:2007ix}%
  \BibitemOpen
  \bibfield  {author} {\bibinfo {author} {\bibfnamefont {J.}~\bibnamefont
  {Vijande}}, \bibinfo {author} {\bibfnamefont {A.}~\bibnamefont {Valcarce}}, \
  and\ \bibinfo {author} {\bibfnamefont {J.~M.}\ \bibnamefont {Richard}},\
  }\bibfield  {title} {\enquote {\bibinfo {title} {{Stability of multiquarks in
  a simple string model}},}\ }\href {\doibase 10.1103/PhysRevD.76.114013}
  {\bibfield  {journal} {\bibinfo  {journal} {Phys. Rev.}\ }\textbf {\bibinfo
  {volume} {D76}},\ \bibinfo {pages} {114013} (\bibinfo {year}
  {2007}{\natexlab{b}})},\ \Eprint {http://arxiv.org/abs/0707.3996}
  {arXiv:0707.3996 [hep-ph]} \BibitemShut {NoStop}%
\bibitem [{\citenamefont {Richard}(2010)}]{Richard:2009rp}%
  \BibitemOpen
  \bibfield  {author} {\bibinfo {author} {\bibfnamefont {Jean-Marc}\
  \bibnamefont {Richard}},\ }\bibfield  {title} {\enquote {\bibinfo {title}
  {{Stability of the pentaquark in a naive string model}},}\ }\href {\doibase
  10.1103/PhysRevC.81.015205} {\bibfield  {journal} {\bibinfo  {journal} {Phys.
  Rev.}\ }\textbf {\bibinfo {volume} {C81}},\ \bibinfo {pages} {015205}
  (\bibinfo {year} {2010})},\ \Eprint {http://arxiv.org/abs/0908.2944}
  {arXiv:0908.2944 [hep-ph]} \BibitemShut {NoStop}%
\bibitem [{\citenamefont {Vijande}\ \emph {et~al.}(2012)\citenamefont
  {Vijande}, \citenamefont {Valcarce},\ and\ \citenamefont
  {Richard}}]{Vijande:2011im}%
  \BibitemOpen
  \bibfield  {author} {\bibinfo {author} {\bibfnamefont {J.}~\bibnamefont
  {Vijande}}, \bibinfo {author} {\bibfnamefont {A.}~\bibnamefont {Valcarce}}, \
  and\ \bibinfo {author} {\bibfnamefont {J-M.}\ \bibnamefont {Richard}},\
  }\bibfield  {title} {\enquote {\bibinfo {title} {{Stability of hexaquarks in
  the string limit of confinement}},}\ }\href {\doibase
  10.1103/PhysRevD.85.014019} {\bibfield  {journal} {\bibinfo  {journal}
  {Phys.Rev.}\ }\textbf {\bibinfo {volume} {D85}},\ \bibinfo {pages} {014019}
  (\bibinfo {year} {2012})},\ \Eprint {http://arxiv.org/abs/1111.5921}
  {arXiv:1111.5921 [hep-ph]} \BibitemShut {NoStop}%
\bibitem [{\citenamefont {Vijande}\ \emph {et~al.}(2013)\citenamefont
  {Vijande}, \citenamefont {Valcarce},\ and\ \citenamefont
  {Richard}}]{Vijande:2013qr}%
  \BibitemOpen
  \bibfield  {author} {\bibinfo {author} {\bibfnamefont {Javier}\ \bibnamefont
  {Vijande}}, \bibinfo {author} {\bibfnamefont {Alfredo}\ \bibnamefont
  {Valcarce}}, \ and\ \bibinfo {author} {\bibfnamefont {Jean-Marc}\
  \bibnamefont {Richard}},\ }\bibfield  {title} {\enquote {\bibinfo {title}
  {{Adiabaticity and color mixing in tetraquark spectroscopy}},}\ }\href
  {\doibase 10.1103/PhysRevD.87.034040} {\bibfield  {journal} {\bibinfo
  {journal} {Phys. Rev.}\ }\textbf {\bibinfo {volume} {D87}},\ \bibinfo {pages}
  {034040} (\bibinfo {year} {2013})},\ \Eprint {http://arxiv.org/abs/1301.6212}
  {arXiv:1301.6212 [hep-ph]} \BibitemShut {NoStop}%
\bibitem [{\citenamefont {Silvestre-Brac}\ \emph {et~al.}(2003)\citenamefont
  {Silvestre-Brac}, \citenamefont {Semay}, \citenamefont {Narodetskii},\ and\
  \citenamefont {Veselov}}]{SilvestreBrac:2003sa}%
  \BibitemOpen
  \bibfield  {author} {\bibinfo {author} {\bibfnamefont {Bernard}\ \bibnamefont
  {Silvestre-Brac}}, \bibinfo {author} {\bibfnamefont {Claude}\ \bibnamefont
  {Semay}}, \bibinfo {author} {\bibfnamefont {Ilya~M.}\ \bibnamefont
  {Narodetskii}}, \ and\ \bibinfo {author} {\bibfnamefont {A.~I.}\ \bibnamefont
  {Veselov}},\ }\bibfield  {title} {\enquote {\bibinfo {title} {{The Baryonic
  $Y$-shape confining potential energy and its approximants}},}\ }\href
  {\doibase 10.1140/epjc/s2003-01416-9} {\bibfield  {journal} {\bibinfo
  {journal} {Eur. Phys. J.}\ }\textbf {\bibinfo {volume} {C32}},\ \bibinfo
  {pages} {385--397} (\bibinfo {year} {2003})},\ \Eprint
  {http://arxiv.org/abs/hep-ph/0309247} {arXiv:hep-ph/0309247 [hep-ph]}
  \BibitemShut {NoStop}%
\bibitem [{\citenamefont {Bicudo}\ and\ \citenamefont
  {Cardoso}(2009)}]{Bicudo:2008yr}%
  \BibitemOpen
  \bibfield  {author} {\bibinfo {author} {\bibfnamefont {P.}~\bibnamefont
  {Bicudo}}\ and\ \bibinfo {author} {\bibfnamefont {M.}~\bibnamefont
  {Cardoso}},\ }\bibfield  {title} {\enquote {\bibinfo {title} {{Iterative
  method to compute the Fermat points and Fermat distances of multiquarks}},}\
  }\href {\doibase 10.1016/j.physletb.2009.03.008} {\bibfield  {journal}
  {\bibinfo  {journal} {Phys. Lett.}\ }\textbf {\bibinfo {volume} {B674}},\
  \bibinfo {pages} {98--102} (\bibinfo {year} {2009})},\ \Eprint
  {http://arxiv.org/abs/0812.0777} {arXiv:0812.0777 [physics.comp-ph]}
  \BibitemShut {NoStop}%
\bibitem [{\citenamefont {Dmitra\v{s}inovi\'{c}}\ \emph
  {et~al.}(2009)\citenamefont {Dmitra\v{s}inovi\'{c}}, \citenamefont {Sato},\
  and\ \citenamefont {\v{S}uvakov}}]{Dmitrasinovic:2009dy}%
  \BibitemOpen
  \bibfield  {author} {\bibinfo {author} {\bibfnamefont {V.}~\bibnamefont
  {Dmitra\v{s}inovi\'{c}}}, \bibinfo {author} {\bibfnamefont {Toru}\
  \bibnamefont {Sato}}, \ and\ \bibinfo {author} {\bibfnamefont {Milovan}\
  \bibnamefont {\v{S}uvakov}},\ }\bibfield  {title} {\enquote {\bibinfo {title}
  {{Low-lying spectrum of the $Y$-string three-quark potential using
  hyper-spherical coordinates}},}\ }\href {\doibase
  10.1140/epjc/s10052-009-1050-y} {\bibfield  {journal} {\bibinfo  {journal}
  {Eur. Phys. J.}\ }\textbf {\bibinfo {volume} {C62}},\ \bibinfo {pages}
  {383--397} (\bibinfo {year} {2009})},\ \Eprint
  {http://arxiv.org/abs/0906.2327} {arXiv:0906.2327 [hep-ph]} \BibitemShut
  {NoStop}%
\bibitem [{\citenamefont {Semay}\ and\ \citenamefont
  {Silvestre-Brac}(1994)}]{Semay:1994ht}%
  \BibitemOpen
  \bibfield  {author} {\bibinfo {author} {\bibfnamefont {C.}~\bibnamefont
  {Semay}}\ and\ \bibinfo {author} {\bibfnamefont {B.}~\bibnamefont
  {Silvestre-Brac}},\ }\bibfield  {title} {\enquote {\bibinfo {title}
  {{Diquonia and potential models}},}\ }\href {\doibase 10.1007/BF01413104}
  {\bibfield  {journal} {\bibinfo  {journal} {Z. Phys.}\ }\textbf {\bibinfo
  {volume} {C61}},\ \bibinfo {pages} {271--275} (\bibinfo {year}
  {1994})}\BibitemShut {NoStop}%
\bibitem [{\citenamefont {Janc}\ and\ \citenamefont
  {Rosina}(2004)}]{Janc:2004qn}%
  \BibitemOpen
  \bibfield  {author} {\bibinfo {author} {\bibfnamefont {D.}~\bibnamefont
  {Janc}}\ and\ \bibinfo {author} {\bibfnamefont {M.}~\bibnamefont {Rosina}},\
  }\bibfield  {title} {\enquote {\bibinfo {title} {{The $T_{cc}= D D^*$
  molecular state}},}\ }\href@noop {} {\bibfield  {journal} {\bibinfo
  {journal} {Few Body Syst.}\ }\textbf {\bibinfo {volume} {35}},\ \bibinfo
  {pages} {175--196} (\bibinfo {year} {2004})},\ \Eprint
  {http://arxiv.org/abs/hep-ph/0405208} {hep-ph/0405208} \BibitemShut {NoStop}%
\bibitem [{\citenamefont {Vijande}\ \emph
  {et~al.}(2007{\natexlab{c}})\citenamefont {Vijande}, \citenamefont
  {Weissman}, \citenamefont {Valcarce},\ and\ \citenamefont
  {Barnea}}]{Vijande:2007rf}%
  \BibitemOpen
  \bibfield  {author} {\bibinfo {author} {\bibfnamefont {J.}~\bibnamefont
  {Vijande}}, \bibinfo {author} {\bibfnamefont {E.}~\bibnamefont {Weissman}},
  \bibinfo {author} {\bibfnamefont {A.}~\bibnamefont {Valcarce}}, \ and\
  \bibinfo {author} {\bibfnamefont {N.}~\bibnamefont {Barnea}},\ }\bibfield
  {title} {\enquote {\bibinfo {title} {{Are there compact heavy four-quark
  bound states?}}}\ }\href {\doibase 10.1103/PhysRevD.76.094027} {\bibfield
  {journal} {\bibinfo  {journal} {Phys. Rev.}\ }\textbf {\bibinfo {volume}
  {D76}},\ \bibinfo {pages} {094027} (\bibinfo {year} {2007}{\natexlab{c}})},\
  \Eprint {http://arxiv.org/abs/0710.2516} {arXiv:0710.2516 [hep-ph]}
  \BibitemShut {NoStop}%
\bibitem [{\citenamefont {Vijande}\ \emph {et~al.}(2016)\citenamefont
  {Vijande}, \citenamefont {Valcarce}, \citenamefont {Richard},\ and\
  \citenamefont {Sorba}}]{Vijande:2016nzk}%
  \BibitemOpen
  \bibfield  {author} {\bibinfo {author} {\bibfnamefont {J.}~\bibnamefont
  {Vijande}}, \bibinfo {author} {\bibfnamefont {A.}~\bibnamefont {Valcarce}},
  \bibinfo {author} {\bibfnamefont {J.~M.}\ \bibnamefont {Richard}}, \ and\
  \bibinfo {author} {\bibfnamefont {P.}~\bibnamefont {Sorba}},\ }\bibfield
  {title} {\enquote {\bibinfo {title} {{Search for doubly-heavy dibaryons in a
  quark model}},}\ }\href {\doibase 10.1103/PhysRevD.94.034038} {\bibfield
  {journal} {\bibinfo  {journal} {Phys. Rev.}\ }\textbf {\bibinfo {volume}
  {D94}},\ \bibinfo {pages} {034038} (\bibinfo {year} {2016})},\ \Eprint
  {http://arxiv.org/abs/1608.03982} {arXiv:1608.03982 [hep-ph]} \BibitemShut
  {NoStop}%
\bibitem [{\citenamefont {Hartree}(1957)}]{Hartree:112712}%
  \BibitemOpen
  \bibfield  {author} {\bibinfo {author} {\bibfnamefont {Douglas~Rayner}\
  \bibnamefont {Hartree}},\ }\href {http://cds.cern.ch/record/112712} {\emph
  {\bibinfo {title} {{The calculation of atomic structures}}}}\ (\bibinfo
  {publisher} {Wiley},\ \bibinfo {address} {New York, NY},\ \bibinfo {year}
  {1957})\BibitemShut {NoStop}%
\bibitem [{\citenamefont {Quigg}\ and\ \citenamefont
  {Rosner}(1979)}]{Quigg:1979vr}%
  \BibitemOpen
  \bibfield  {author} {\bibinfo {author} {\bibfnamefont {C.}~\bibnamefont
  {Quigg}}\ and\ \bibinfo {author} {\bibfnamefont {Jonathan~L.}\ \bibnamefont
  {Rosner}},\ }\bibfield  {title} {\enquote {\bibinfo {title} {{Quantum
  Mechanics with Applications to Quarkonium}},}\ }\href {\doibase
  10.1016/0370-1573(79)90095-4} {\bibfield  {journal} {\bibinfo  {journal}
  {Phys.Rept.}\ }\textbf {\bibinfo {volume} {56}},\ \bibinfo {pages} {167--235}
  (\bibinfo {year} {1979})}\BibitemShut {NoStop}%
\bibitem [{\citenamefont {Grosse}\ and\ \citenamefont
  {Martin}(1980)}]{Grosse:1979xm}%
  \BibitemOpen
  \bibfield  {author} {\bibinfo {author} {\bibfnamefont {Harald}\ \bibnamefont
  {Grosse}}\ and\ \bibinfo {author} {\bibfnamefont {Andr{\'e}}\ \bibnamefont
  {Martin}},\ }\bibfield  {title} {\enquote {\bibinfo {title} {{Exact Results
  on Potential Models for Quarkonium Systems}},}\ }\href {\doibase
  10.1016/0370-1573(80)90031-9} {\bibfield  {journal} {\bibinfo  {journal}
  {Phys. Rept.}\ }\textbf {\bibinfo {volume} {60}},\ \bibinfo {pages} {341}
  (\bibinfo {year} {1980})}\BibitemShut {NoStop}%
\bibitem [{\citenamefont {Grosse}\ and\ \citenamefont
  {Martin}(2005)}]{Grosse:847188}%
  \BibitemOpen
  \bibfield  {author} {\bibinfo {author} {\bibfnamefont {Harald}\ \bibnamefont
  {Grosse}}\ and\ \bibinfo {author} {\bibfnamefont {Andr{\'e}}\ \bibnamefont
  {Martin}},\ }\href {http://cds.cern.ch/record/847188} {\emph {\bibinfo
  {title} {{Particle Physics and the Schrödinger Equation; new ed.}}}},\
  Cambridge monographs on particle physics, nuclear physics, and cosmology\
  (\bibinfo  {publisher} {Cambridge Univ.},\ \bibinfo {address} {Cambridge},\
  \bibinfo {year} {2005})\BibitemShut {NoStop}%
\bibitem [{\citenamefont {Richard}\ and\ \citenamefont
  {Taxil}(1990)}]{Richard:1989ra}%
  \BibitemOpen
  \bibfield  {author} {\bibinfo {author} {\bibfnamefont {Jean-Marc}\
  \bibnamefont {Richard}}\ and\ \bibinfo {author} {\bibfnamefont {Pierre}\
  \bibnamefont {Taxil}},\ }\bibfield  {title} {\enquote {\bibinfo {title} {{The
  Ordering of Low Lying Bound States of Three Identical Particles}},}\ }\href
  {\doibase 10.1016/0550-3213(90)90144-3} {\bibfield  {journal} {\bibinfo
  {journal} {Nucl. Phys.}\ }\textbf {\bibinfo {volume} {B329}},\ \bibinfo
  {pages} {310--326} (\bibinfo {year} {1990})}\BibitemShut {NoStop}%
\bibitem [{\citenamefont {Stancu}\ and\ \citenamefont
  {Stassart}(1991)}]{Stancu:1991cz}%
  \BibitemOpen
  \bibfield  {author} {\bibinfo {author} {\bibfnamefont {F.}~\bibnamefont
  {Stancu}}\ and\ \bibinfo {author} {\bibfnamefont {P.}~\bibnamefont
  {Stassart}},\ }\bibfield  {title} {\enquote {\bibinfo {title} {{Negative
  parity nonstrange baryons}},}\ }\href {\doibase 10.1016/0370-2693(91)90163-K}
  {\bibfield  {journal} {\bibinfo  {journal} {Phys. Lett.}\ }\textbf {\bibinfo
  {volume} {B269}},\ \bibinfo {pages} {243--246} (\bibinfo {year}
  {1991})}\BibitemShut {NoStop}%
\bibitem [{\citenamefont {Richard}(1992)}]{Richard:1992uk}%
  \BibitemOpen
  \bibfield  {author} {\bibinfo {author} {\bibfnamefont {J.~M.}\ \bibnamefont
  {Richard}},\ }\bibfield  {title} {\enquote {\bibinfo {title} {{The
  Nonrelativistic three-body problem for baryons}},}\ }\href {\doibase
  10.1016/0370-1573(92)90078-E} {\bibfield  {journal} {\bibinfo  {journal}
  {Phys. Rept.}\ }\textbf {\bibinfo {volume} {212}},\ \bibinfo {pages} {1--76}
  (\bibinfo {year} {1992})}\BibitemShut {NoStop}%
\bibitem [{\citenamefont {Mitroy}\ \emph {et~al.}(2013)\citenamefont {Mitroy},
  \citenamefont {Bubin}, \citenamefont {Horiuchi}, \citenamefont {Suzuki},
  \citenamefont {Adamowicz}, \citenamefont {Cencek}, \citenamefont {Szalewicz},
  \citenamefont {Komasa}, \citenamefont {Blume},\ and\ \citenamefont
  {Varga}}]{RevModPhys.85.693}%
  \BibitemOpen
  \bibfield  {author} {\bibinfo {author} {\bibfnamefont {Jim}\ \bibnamefont
  {Mitroy}}, \bibinfo {author} {\bibfnamefont {Sergiy}\ \bibnamefont {Bubin}},
  \bibinfo {author} {\bibfnamefont {Wataru}\ \bibnamefont {Horiuchi}}, \bibinfo
  {author} {\bibfnamefont {Yasuyuki}\ \bibnamefont {Suzuki}}, \bibinfo {author}
  {\bibfnamefont {Ludwik}\ \bibnamefont {Adamowicz}}, \bibinfo {author}
  {\bibfnamefont {Wojciech}\ \bibnamefont {Cencek}}, \bibinfo {author}
  {\bibfnamefont {Krzysztof}\ \bibnamefont {Szalewicz}}, \bibinfo {author}
  {\bibfnamefont {Jacek}\ \bibnamefont {Komasa}}, \bibinfo {author}
  {\bibfnamefont {D.}~\bibnamefont {Blume}}, \ and\ \bibinfo {author}
  {\bibfnamefont {K\'alm\'an}\ \bibnamefont {Varga}},\ }\bibfield  {title}
  {\enquote {\bibinfo {title} {Theory and application of explicitly correlated
  gaussians},}\ }\href {\doibase 10.1103/RevModPhys.85.693} {\bibfield
  {journal} {\bibinfo  {journal} {Rev. Mod. Phys.}\ }\textbf {\bibinfo {volume}
  {85}},\ \bibinfo {pages} {693--749} (\bibinfo {year} {2013})}\BibitemShut
  {NoStop}%
\bibitem [{\citenamefont {Hiyama}\ \emph {et~al.}(2003)\citenamefont {Hiyama},
  \citenamefont {Kino},\ and\ \citenamefont {Kamimura}}]{Hiyama:2003cu}%
  \BibitemOpen
  \bibfield  {author} {\bibinfo {author} {\bibfnamefont {E.}~\bibnamefont
  {Hiyama}}, \bibinfo {author} {\bibfnamefont {Y.}~\bibnamefont {Kino}}, \ and\
  \bibinfo {author} {\bibfnamefont {M.}~\bibnamefont {Kamimura}},\ }\bibfield
  {title} {\enquote {\bibinfo {title} {{Gaussian expansion method for few-body
  systems}},}\ }\href {\doibase 10.1016/S0146-6410(03)90015-9} {\bibfield
  {journal} {\bibinfo  {journal} {Prog. Part. Nucl. Phys.}\ }\textbf {\bibinfo
  {volume} {51}},\ \bibinfo {pages} {223--307} (\bibinfo {year}
  {2003})}\BibitemShut {NoStop}%
\bibitem [{\citenamefont {Anselmino}\ \emph {et~al.}(1993)\citenamefont
  {Anselmino}, \citenamefont {Predazzi}, \citenamefont {Ekelin}, \citenamefont
  {Fredriksson},\ and\ \citenamefont {Lichtenberg}}]{Anselmino:1992vg}%
  \BibitemOpen
  \bibfield  {author} {\bibinfo {author} {\bibfnamefont {Mauro}\ \bibnamefont
  {Anselmino}}, \bibinfo {author} {\bibfnamefont {Enrico}\ \bibnamefont
  {Predazzi}}, \bibinfo {author} {\bibfnamefont {Svante}\ \bibnamefont
  {Ekelin}}, \bibinfo {author} {\bibfnamefont {Sverker}\ \bibnamefont
  {Fredriksson}}, \ and\ \bibinfo {author} {\bibfnamefont {D.~B.}\ \bibnamefont
  {Lichtenberg}},\ }\bibfield  {title} {\enquote {\bibinfo {title}
  {{Diquarks}},}\ }\href {\doibase 10.1103/RevModPhys.65.1199} {\bibfield
  {journal} {\bibinfo  {journal} {Rev. Mod. Phys.}\ }\textbf {\bibinfo {volume}
  {65}},\ \bibinfo {pages} {1199--1234} (\bibinfo {year} {1993})}\BibitemShut
  {NoStop}%
\bibitem [{\citenamefont {Brodsky}\ \emph {et~al.}(2016)\citenamefont
  {Brodsky}, \citenamefont {de~Téramond}, \citenamefont {Dosch},\ and\
  \citenamefont {Lorcé}}]{Brodsky:2016rvj}%
  \BibitemOpen
  \bibfield  {author} {\bibinfo {author} {\bibfnamefont {Stanley~J.}\
  \bibnamefont {Brodsky}}, \bibinfo {author} {\bibfnamefont {Guy~F.}\
  \bibnamefont {de~Téramond}}, \bibinfo {author} {\bibfnamefont
  {Hans~Günter}\ \bibnamefont {Dosch}}, \ and\ \bibinfo {author}
  {\bibfnamefont {Cédric}\ \bibnamefont {Lorcé}},\ }\bibfield  {title}
  {\enquote {\bibinfo {title} {{Meson/Baryon/Tetraquark Supersymmetry from
  Superconformal Algebra and Light-Front Holography}},}\ }\bibfield
  {booktitle} {\emph {\bibinfo {booktitle} {{Conference on New Physics at the
  Large Hadron Collider Singapore, Singapore, February 29-March 4, 2016}}},\
  }\href {\doibase 10.1142/S0217751X16300295} {\bibfield  {journal} {\bibinfo
  {journal} {Int. J. Mod. Phys.}\ }\textbf {\bibinfo {volume} {A31}},\ \bibinfo
  {pages} {1630029} (\bibinfo {year} {2016})},\ \Eprint
  {http://arxiv.org/abs/1606.04638} {arXiv:1606.04638 [hep-ph]} \BibitemShut
  {NoStop}%
\bibitem [{\citenamefont {Lichtenberg}\ \emph {et~al.}(1997)\citenamefont
  {Lichtenberg}, \citenamefont {Roncaglia},\ and\ \citenamefont
  {Predazzi}}]{Lichtenberg:1997bq}%
  \BibitemOpen
  \bibfield  {author} {\bibinfo {author} {\bibfnamefont {D.~B.}\ \bibnamefont
  {Lichtenberg}}, \bibinfo {author} {\bibfnamefont {R.}~\bibnamefont
  {Roncaglia}}, \ and\ \bibinfo {author} {\bibfnamefont {E.}~\bibnamefont
  {Predazzi}},\ }\bibfield  {title} {\enquote {\bibinfo {title} {{Predicting
  exotic hadron masses from supersymmetry and a quark-diquark model}},}\ }\href
  {\doibase 10.1088/0954-3899/23/8/001} {\bibfield  {journal} {\bibinfo
  {journal} {J. Phys.}\ }\textbf {\bibinfo {volume} {G23}},\ \bibinfo {pages}
  {865--874} (\bibinfo {year} {1997})}\BibitemShut {NoStop}%
\bibitem [{\citenamefont {Grach}\ and\ \citenamefont
  {Narodetsky}(1994)}]{Grach:1993mj}%
  \BibitemOpen
  \bibfield  {author} {\bibinfo {author} {\bibfnamefont {I.~L.}\ \bibnamefont
  {Grach}}\ and\ \bibinfo {author} {\bibfnamefont {I.~M.}\ \bibnamefont
  {Narodetsky}},\ }\bibfield  {title} {\enquote {\bibinfo {title} {{Diquark
  correlations in the proton}},}\ }\href {\doibase 10.1007/BF01344395}
  {\bibfield  {journal} {\bibinfo  {journal} {Few Body Syst.}\ }\textbf
  {\bibinfo {volume} {16}},\ \bibinfo {pages} {151--163} (\bibinfo {year}
  {1994})}\BibitemShut {NoStop}%
\bibitem [{\citenamefont {Fleck}\ \emph {et~al.}(1988)\citenamefont {Fleck},
  \citenamefont {Silvestre-Brac},\ and\ \citenamefont
  {Richard}}]{Fleck:1988vm}%
  \BibitemOpen
  \bibfield  {author} {\bibinfo {author} {\bibfnamefont {S.}~\bibnamefont
  {Fleck}}, \bibinfo {author} {\bibfnamefont {B.}~\bibnamefont
  {Silvestre-Brac}}, \ and\ \bibinfo {author} {\bibfnamefont {J.~M.}\
  \bibnamefont {Richard}},\ }\bibfield  {title} {\enquote {\bibinfo {title}
  {{Search for Diquark Clustering in Baryons}},}\ }\href {\doibase
  10.1103/PhysRevD.38.1519} {\bibfield  {journal} {\bibinfo  {journal} {Phys.
  Rev.}\ }\textbf {\bibinfo {volume} {D38}},\ \bibinfo {pages} {1519--1529}
  (\bibinfo {year} {1988})}\BibitemShut {NoStop}%
\bibitem [{\citenamefont {Martin}(1986)}]{Martin:1985hw}%
  \BibitemOpen
  \bibfield  {author} {\bibinfo {author} {\bibfnamefont {Andr\'{e}}\
  \bibnamefont {Martin}},\ }\bibfield  {title} {\enquote {\bibinfo {title}
  {R\lowercase{EGGE TRAJECTORIES IN THE QUARK MODEL}},}\ }\href@noop {}
  {\bibfield  {journal} {\bibinfo  {journal} {Z. Phys.}\ }\textbf {\bibinfo
  {volume} {C32}},\ \bibinfo {pages} {359} (\bibinfo {year}
  {1986})}\BibitemShut {NoStop}%
\bibitem [{\citenamefont {Braaten}\ \emph {et~al.}(2014)\citenamefont
  {Braaten}, \citenamefont {Langmack},\ and\ \citenamefont
  {Smith}}]{Braaten:2014qka}%
  \BibitemOpen
  \bibfield  {author} {\bibinfo {author} {\bibfnamefont {Eric}\ \bibnamefont
  {Braaten}}, \bibinfo {author} {\bibfnamefont {Christian}\ \bibnamefont
  {Langmack}}, \ and\ \bibinfo {author} {\bibfnamefont {D.~Hudson}\
  \bibnamefont {Smith}},\ }\bibfield  {title} {\enquote {\bibinfo {title}
  {{Born-Oppenheimer Approximation for the $XYZ$ Mesons}},}\ }\href {\doibase
  10.1103/PhysRevD.90.014044} {\bibfield  {journal} {\bibinfo  {journal} {Phys.
  Rev.}\ }\textbf {\bibinfo {volume} {D90}},\ \bibinfo {pages} {014044}
  (\bibinfo {year} {2014})},\ \Eprint {http://arxiv.org/abs/1402.0438}
  {arXiv:1402.0438 [hep-ph]} \BibitemShut {NoStop}%
\bibitem [{\citenamefont {Dubynskiy}\ and\ \citenamefont
  {Voloshin}(2008)}]{Dubynskiy:2008mq}%
  \BibitemOpen
  \bibfield  {author} {\bibinfo {author} {\bibfnamefont {S.}~\bibnamefont
  {Dubynskiy}}\ and\ \bibinfo {author} {\bibfnamefont {M.~B.}\ \bibnamefont
  {Voloshin}},\ }\bibfield  {title} {\enquote {\bibinfo {title}
  {{Hadro-Charmonium}},}\ }\href {\doibase 10.1016/j.physletb.2008.07.086}
  {\bibfield  {journal} {\bibinfo  {journal} {Phys. Lett.}\ }\textbf {\bibinfo
  {volume} {B666}},\ \bibinfo {pages} {344--346} (\bibinfo {year} {2008})},\
  \Eprint {http://arxiv.org/abs/0803.2224} {arXiv:0803.2224 [hep-ph]}
  \BibitemShut {NoStop}%
\bibitem [{\citenamefont {Voloshin}(2013)}]{Voloshin:2013dpa}%
  \BibitemOpen
  \bibfield  {author} {\bibinfo {author} {\bibfnamefont {M.~B.}\ \bibnamefont
  {Voloshin}},\ }\bibfield  {title} {\enquote {\bibinfo {title} {{$Z_c(3900)$ -
  what is inside?}}}\ }\href {\doibase 10.1103/PhysRevD.87.091501} {\bibfield
  {journal} {\bibinfo  {journal} {Phys. Rev.}\ }\textbf {\bibinfo {volume}
  {D87}},\ \bibinfo {pages} {091501} (\bibinfo {year} {2013})},\ \Eprint
  {http://arxiv.org/abs/1304.0380} {arXiv:1304.0380 [hep-ph]} \BibitemShut
  {NoStop}%
\bibitem [{\citenamefont {Wu}\ and\ \citenamefont {Wu}(2015)}]{Wu:2015bfr}%
  \BibitemOpen
  \bibfield  {author} {\bibinfo {author} {\bibfnamefont {Tai~Tsun}\
  \bibnamefont {Wu}}\ and\ \bibinfo {author} {\bibfnamefont {Sau~Lan}\
  \bibnamefont {Wu}},\ }\bibfield  {title} {\enquote {\bibinfo {title}
  {{Augmented standard model and the simplest scenario}},}\ }\href {\doibase
  10.1142/S0217751X15502012} {\bibfield  {journal} {\bibinfo  {journal} {Int.
  J. Mod. Phys.}\ }\textbf {\bibinfo {volume} {A30}},\ \bibinfo {pages}
  {1550201} (\bibinfo {year} {2015})}\BibitemShut {NoStop}%
\bibitem [{\citenamefont {Bigi}\ \emph {et~al.}(1986)\citenamefont {Bigi},
  \citenamefont {Dokshitzer}, \citenamefont {Khoze}, \citenamefont {Kuhn},\
  and\ \citenamefont {Zerwas}}]{Bigi:1986jk}%
  \BibitemOpen
  \bibfield  {author} {\bibinfo {author} {\bibfnamefont {Ikaros I.~Y.}\
  \bibnamefont {Bigi}}, \bibinfo {author} {\bibfnamefont {Yuri~L.}\
  \bibnamefont {Dokshitzer}}, \bibinfo {author} {\bibfnamefont {Valery~A.}\
  \bibnamefont {Khoze}}, \bibinfo {author} {\bibfnamefont {Johann~H.}\
  \bibnamefont {Kuhn}}, \ and\ \bibinfo {author} {\bibfnamefont {Peter~M.}\
  \bibnamefont {Zerwas}},\ }\bibfield  {title} {\enquote {\bibinfo {title}
  {{Production and Decay Properties of Ultraheavy Quarks}},}\ }\href {\doibase
  10.1016/0370-2693(86)91275-X} {\bibfield  {journal} {\bibinfo  {journal}
  {Phys. Lett.}\ }\textbf {\bibinfo {volume} {B181}},\ \bibinfo {pages}
  {157--163} (\bibinfo {year} {1986})}\BibitemShut {NoStop}%
\bibitem [{\citenamefont {Luo}\ \emph {et~al.}(2016)\citenamefont {Luo},
  \citenamefont {Wang}, \citenamefont {Xu}, \citenamefont {Zhang},\ and\
  \citenamefont {Zhu}}]{Luo:2015yio}%
  \BibitemOpen
  \bibfield  {author} {\bibinfo {author} {\bibfnamefont {Ming-xing}\
  \bibnamefont {Luo}}, \bibinfo {author} {\bibfnamefont {Kai}\ \bibnamefont
  {Wang}}, \bibinfo {author} {\bibfnamefont {Tao}\ \bibnamefont {Xu}}, \bibinfo
  {author} {\bibfnamefont {Liangliang}\ \bibnamefont {Zhang}}, \ and\ \bibinfo
  {author} {\bibfnamefont {Guohuai}\ \bibnamefont {Zhu}},\ }\bibfield  {title}
  {\enquote {\bibinfo {title} {{Squarkonium, diquarkonium, and octetonium at
  the LHC and their diphoton decays}},}\ }\href {\doibase
  10.1103/PhysRevD.93.055042} {\bibfield  {journal} {\bibinfo  {journal} {Phys.
  Rev.}\ }\textbf {\bibinfo {volume} {D93}},\ \bibinfo {pages} {055042}
  (\bibinfo {year} {2016})},\ \Eprint {http://arxiv.org/abs/1512.06670}
  {arXiv:1512.06670 [hep-ph]} \BibitemShut {NoStop}%
\bibitem [{\citenamefont {Backovi\'{c}}(2016)}]{Backovic:2016xno}%
  \BibitemOpen
  \bibfield  {author} {\bibinfo {author} {\bibfnamefont {Mihailo}\ \bibnamefont
  {Backovi\'{c}}},\ }\bibfield  {title} {\enquote {\bibinfo {title} {{A Theory
  of Ambulance Chasing}},}\ }\href@noop {} {\  (\bibinfo {year} {2016})},\
  \Eprint {http://arxiv.org/abs/1603.01204} {arXiv:1603.01204 [physics.soc-ph]}
  \BibitemShut {NoStop}%
\bibitem [{\citenamefont {Hey}\ and\ \citenamefont {Kelly}(1983)}]{Hey:1982aj}%
  \BibitemOpen
  \bibfield  {author} {\bibinfo {author} {\bibfnamefont {Anthony J.~G.}\
  \bibnamefont {Hey}}\ and\ \bibinfo {author} {\bibfnamefont {Robert~L.}\
  \bibnamefont {Kelly}},\ }\bibfield  {title} {\enquote {\bibinfo {title}
  {{Baryon spectroscopy}},}\ }\href {\doibase 10.1016/0370-1573(83)90114-X}
  {\bibfield  {journal} {\bibinfo  {journal} {Phys. Rept.}\ }\textbf {\bibinfo
  {volume} {96}},\ \bibinfo {pages} {71} (\bibinfo {year} {1983})}\BibitemShut
  {NoStop}%
\bibitem [{\citenamefont {Dytman}(2004)}]{Dytman:2004kk}%
  \BibitemOpen
  \bibfield  {author} {\bibinfo {author} {\bibfnamefont {S.~A.}\ \bibnamefont
  {Dytman}},\ }\bibfield  {title} {\enquote {\bibinfo {title} {{Impact of
  recent data on N* structure}},}\ }\bibfield  {booktitle} {\emph {\bibinfo
  {booktitle} {{ICTP 4th International Conference on Perspectives in Hadronic
  Physics Trieste, Italy, May 12-16, 2003}}},\ }\href {\doibase
  10.1140/epjad/s2004-03-010-4} {\bibfield  {journal} {\bibinfo  {journal}
  {Eur. Phys. J.}\ }\textbf {\bibinfo {volume} {A19}},\ \bibinfo {pages}
  {SUPPL161--65} (\bibinfo {year} {2004})},\ \bibinfo {note}
  {[,61(2004)]}\BibitemShut {NoStop}%
\bibitem [{\citenamefont {Briceno}\ \emph {et~al.}(2016)\citenamefont {Briceno}
  \emph {et~al.}}]{Briceno:2015rlt}%
  \BibitemOpen
  \bibfield  {author} {\bibinfo {author} {\bibfnamefont {R.~A.}\ \bibnamefont
  {Briceno}} \emph {et~al.},\ }\bibfield  {title} {\enquote {\bibinfo {title}
  {{Issues and Opportunities in Exotic Hadrons}},}\ }\href {\doibase
  10.1088/1674-1137/40/4/042001} {\bibfield  {journal} {\bibinfo  {journal}
  {Chin. Phys.}\ }\textbf {\bibinfo {volume} {C40}},\ \bibinfo {pages} {042001}
  (\bibinfo {year} {2016})},\ \Eprint {http://arxiv.org/abs/1511.06779}
  {arXiv:1511.06779 [hep-ph]} \BibitemShut {NoStop}%
\bibitem [{\citenamefont {Jaffe}\ and\ \citenamefont
  {Low}(1979)}]{Jaffe:1978bu}%
  \BibitemOpen
  \bibfield  {author} {\bibinfo {author} {\bibfnamefont {R.~L.}\ \bibnamefont
  {Jaffe}}\ and\ \bibinfo {author} {\bibfnamefont {F.~E.}\ \bibnamefont
  {Low}},\ }\bibfield  {title} {\enquote {\bibinfo {title} {{The Connection
  Between Quark Model Eigenstates and Low- Energy Scattering}},}\ }\href
  {\doibase 10.1103/PhysRevD.19.2105} {\bibfield  {journal} {\bibinfo
  {journal} {Phys. Rev.}\ }\textbf {\bibinfo {volume} {D19}},\ \bibinfo {pages}
  {2105--2118} (\bibinfo {year} {1979})}\BibitemShut {NoStop}%
\bibitem [{\citenamefont {Fredriksson}\ and\ \citenamefont
  {J{\"a}ndel}(1982)}]{Fredriksson:1981mh}%
  \BibitemOpen
  \bibfield  {author} {\bibinfo {author} {\bibfnamefont {Sverker}\ \bibnamefont
  {Fredriksson}}\ and\ \bibinfo {author} {\bibfnamefont {Magnus}\ \bibnamefont
  {J{\"a}ndel}},\ }\bibfield  {title} {\enquote {\bibinfo {title} {{The diquark
  deuteron}},}\ }\href {\doibase 10.1103/PhysRevLett.48.14} {\bibfield
  {journal} {\bibinfo  {journal} {Phys. Rev. Lett.}\ }\textbf {\bibinfo
  {volume} {48}},\ \bibinfo {pages} {14} (\bibinfo {year} {1982})}\BibitemShut
  {NoStop}%
\bibitem [{\citenamefont {Maiani}\ \emph {et~al.}(2015)\citenamefont {Maiani},
  \citenamefont {Polosa},\ and\ \citenamefont {Riquer}}]{Maiani:2015iaa}%
  \BibitemOpen
  \bibfield  {author} {\bibinfo {author} {\bibfnamefont {L.}~\bibnamefont
  {Maiani}}, \bibinfo {author} {\bibfnamefont {A.~D.}\ \bibnamefont {Polosa}},
  \ and\ \bibinfo {author} {\bibfnamefont {V.}~\bibnamefont {Riquer}},\
  }\bibfield  {title} {\enquote {\bibinfo {title} {{From pentaquarks to
  dibaryons in $\Lambda_b$(5620) decays}},}\ }\href {\doibase
  10.1016/j.physletb.2015.08.049} {\bibfield  {journal} {\bibinfo  {journal}
  {Phys. Lett.}\ }\textbf {\bibinfo {volume} {B750}},\ \bibinfo {pages}
  {37--38} (\bibinfo {year} {2015})},\ \Eprint
  {http://arxiv.org/abs/1508.04459} {arXiv:1508.04459 [hep-ph]} \BibitemShut
  {NoStop}%
\bibitem [{\citenamefont {Amsler}\ \emph
  {et~al.}(2008{\natexlab{b}})\citenamefont {Amsler} \emph
  {et~al.}}]{Amsler:2008zz}%
  \BibitemOpen
  \bibfield  {author} {\bibinfo {author} {\bibfnamefont {C.}~\bibnamefont
  {Amsler}} \emph {et~al.},\ }\bibfield  {title} {\enquote {\bibinfo {title}
  {{Review of particle physics}},}\ }\href {\doibase
  10.1016/j.physletb.2008.07.018} {\bibfield  {journal} {\bibinfo  {journal}
  {Phys. Lett.}\ }\textbf {\bibinfo {volume} {B667}},\ \bibinfo {pages} {1}
  (\bibinfo {year} {2008}{\natexlab{b}})}\BibitemShut {NoStop}%
\end{thebibliography}
%

\end{document}